\documentclass[a4paper,12pt]{report}

\usepackage[utf8]{inputenc}

\usepackage{amsfonts,amsmath,amssymb,amsthm,bbm,color,hyperref,setspace,young,multirow}

\usepackage{graphicx}

\usepackage{tikz}
\usetikzlibrary{decorations} 
\usetikzlibrary{decorations.pathmorphing} 
\usetikzlibrary{calc}

\usepackage{makeidx}
\makeindex

\numberwithin{equation}{chapter}

\addtocounter{secnumdepth}{1}

\newcounter{appendices}

\PassOptionsToPackage{hyphens}{url}

\makeatletter
\newif\if@display

\everydisplay{\@displaytrue}
\everymath{\@displayfalse}

\@displayfalse

\makeatother

\usepackage[vcentermath,enableskew]{youngtab}

\newcommand{\bea}{\begin{eqnarray}}
\newcommand{\eea}{\end{eqnarray}}
\newcommand{\beq}{\begin{equation}}
\newcommand{\eeq}{\end{equation}}

\newcommand{\be}{\begin{equation}}
\newcommand{\ee}{\end{equation}}

\def\no{\nonumber}
\newcommand{\half}{\frac{1}{2}}

\newcommand{\p}{\partial}
\newcommand{\N}{\mathcal{N}}

\def\qth{\tau}
\newcommand{\tr}{\textrm{Tr}}
\newcommand{\ch}{\textrm{ch}}
\newcommand{\sh}{\textrm{sh}}

\newcommand{\lp}{\left(}
\newcommand{\rp}{\right)}

\newcommand{\rf}[1]{(\ref{#1})}

\newcommand{\bC}{\ensuremath{\mathbb{C}}}

\newcommand{\bP}{\ensuremath{\mathbb{P}}}
\newcommand{\bQ}{\ensuremath{\mathbb{Q}}}
\newcommand{\bR}{\ensuremath{\mathbb{R}}}

\newcommand{\bZ}{\ensuremath{\mathbb{Z}}}

\newcommand{\fN}{\ensuremath{\mathbf{N}}}

\newcommand{\scA}{\ensuremath{\mathcal{A}}}

\newcommand{\scD}{\ensuremath{\mathcal{D}}}

\newcommand{\scL}{\ensuremath{\mathcal{L}}}

\newcommand{\scN}{\ensuremath{\mathcal{N}}}
\newcommand{\scO}{\ensuremath{\mathcal{O}}}

\newcommand{\scT}{\ensuremath{\mathcal{T}}}

\newcommand{\scW}{\ensuremath{\mathcal{W}}}

\newcommand{\scZ}{\ensuremath{\mathcal{Z}}}

\title{{%
     Dualité Holographique pour Théories des Champs Super-Conformes en 3 Dimensions}}
\author{%
B. Assel}

\begin{document}

\begin{titlepage}

 \center{ \includegraphics[width=.15\textwidth]{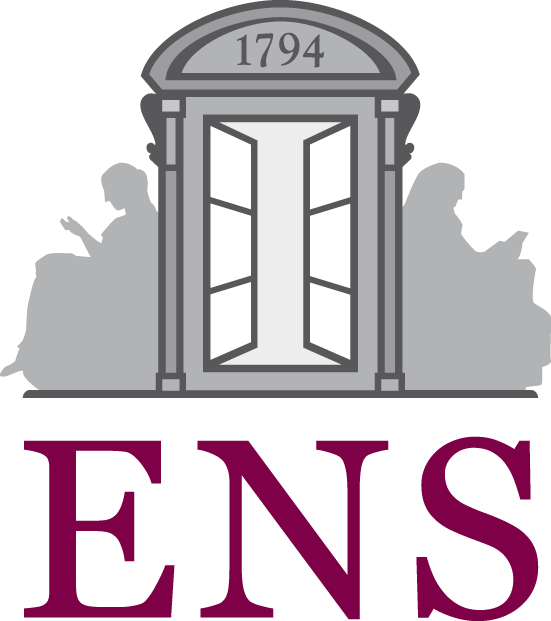} }
  \begin{center}

    {\Large \textsc{Thèse de doctorat de \\ l'Ecole Normale Supérieure}}\\
\vspace{1cm}
    {\large Spécialité : Physique théorique}\\
\vspace{2cm}
    {\large Présentée par}\\
{\Large Benjamin ASSEL}\\
\vspace{2cm}
    {\large pour obtenir le grade de}\\
    {\Large \textsc{Docteur de l'Ecole Normale Supérieure}}\\
  \end{center}
\vspace{1cm}
{\large Sujet :} {\center{ \Large \textbf{Dualité Holographique pour Théories Superconformes en trois dimensions} } }
\vspace{2cm}\\
{\large Soutenance le 5 juillet 2013, devant le jury composé de}\\
\begin{center}
  \begin{tabular}{ll}
    Henning \textsc{Samtleben}&Rapporteur\\
    Amihay \textsc{Hanany}&Rapporteur\\
    Michela \textsc{Petrini}&Presidente\\
    Dario \textsc{Martelli}&Examinateur\\
    Jan \textsc{Troost}&Examinateur\\
    Costas \textsc{Bachas}&Directeur de thèse\\
  \end{tabular}
\end{center}
 ~\\
Thèse délivrée par l'\textit{Ecole Normale Supérieure}, 45 rue d'Ulm, 75005 Paris \textit{FRANCE},\\
suite à des recherches effectuées au \textit{LPT-ENS}, 24 rue Lhomond, 75005 Paris \textit{FRANCE}, \\
dans le cadre de l'\textit{Ecole Doctorale de Physique de la Région Parisienne} ED107.
\end{titlepage}
 \addtocounter{page}{1}

{\bf Résumé}

\vspace{1cm}

L'objet principal de la thèse consiste à exhiber et étudier une large classe de nouveaux exemples de correspondances  holographiques de type AdS/CFT entre des théories de jauge super-conformes en trois dimensions avec $\N=4$ supersymétrie et la théorie des cordes de type IIB sur un espace $AdS_4 \times K_6$ (produit de AdS à quatre dimension et d'un espace compact à six dimensions). Les théories de jauge superconformes en question sont obtenues comme points fixes infrarouges de théories de type Yang-Mills ``quivers'' (càd avec des produits de groupe unitaires comme groupe de jauge et un certain contenu en matière). Dans cette limite infrarouge le couplage de Yang-Mills est renormalisé et diverge ce qui rend ces théories inaccessibles à tout calcul perturbatif, d'où l'intérêt d'en avoir une description holographique. \\
Une large partie du manuscrit de thèse est consacrée à la présentation des solutions de supergravité et à l'établissement du dictionnaire avec les théories super-conformes. Les cas des quivers linéaires et circulaires sont traités, ainsi que les solutions de ``domain wall'', qui correspondent à des théories  Super-Yang-Mills à quatre dimensions couplées à un ``défaut'' à trois dimensions. \\
Plusieurs vérifications des correspondances sont données, notamment par le calcul de l'énergie libre qui est calculée du côté théorie de jauge, en utilisant certains résultats issus des techniques de localisation d'intégrales de chemin, et qui est comparée avec succès à l'action de supergravité. 

\vspace{4cm}

{\bf Mots clés } : Correspondance AdS/CFT,  supergravité, supersymétrie, théorie des cordes, D-branes, dualités, énergie libre, modèles de matrice.

\newpage

\section*{\centering Holographic Duality for three-dimensional Super-conformal Field Theories}

\vspace{2cm}

{\bf Abstract}

\vspace{1cm}

We present a large class of new holographic dualities relating three-dimensional $\N=4$ super-conformal field theories and type IIB string theory on supergravity backgrounds, which have  $AdS_4 \times K_6$ metric. The superconformal theories arise as infrared fixed points of Yang-Mills quiver gauge theories (the gauge group is a product of unitary groups and the matter content is made of fundamental and bifundamental matter). In the infrared limit the Yang-Mills coupling diverges, so that the theories are  strongly coupled and hence inaccessible to perturbative computations. This is a motivation for finding their dual holographic description. \\
The main part of the manuscript is devoted to the exposition of the supergravity solutions and to establishing the holographic dictionary. The cases of linear and circular quivers is covered entirely, as well as the domain wall solutions that are dual to  4d Super-Yang-Mills theory coupled to a half-BPS 3-dimensional defect field theory. \\
Several checks of the correspondences are given. Particularly we compute the free energy of the gauge theories, using the techniques of localization of path integral, and compare it successfully with the supergravity action.

\vspace{4cm}

{\bf Key words } : AdS/CFT correspondence,  supergravity, supersymmetry, string theory, D-branes, free energy, matrix models.

\tableofcontents

\newpage

\section*{Acknowledgments}
\addcontentsline{toc}{section}{Acknowledgments}
\label{sec:acknow}

I would like to thank Costas Bachas. His guidance along these three years has been enlightning and rigorous. I have learned a lot from him. I particularly appreciated his pedagogical capacities in the highly interesting discussions we had about physics. Besides doing a great job as advisor, he was always very friendly and responsive to my demands. He made me feel as a real collaborator rather than a student. Working with him during these three years was an honor and a pleasure. I hope we can stay in contact in the future.

I am very gratefull to Jaume Gomis, whose level of involvment in the success of my PhD makes him a natural co-advisor. I am indebted to him for his hospitality at Perimeter Institute. Thanks to him I have been able to access another domain of the theory, which is more gauge theory oriented and provides a very rich and interesting complement to my PhD. I have been very happy to meet him and to work with him.

I also want to thank John Estes and Masahito Yamazaki. We had a very pleasant and fruitfull collaboration.

\smallskip

I thank the LPT-ENS and its members for their kindness and help all along these three years. I thank the secretaries Beatrice Fixois and Viviane Sébille for their sympathy and efficiency.
I thank Perimeter Insitute for hospitality during my PhD.

I thank the members of the jury for their interest and the time spent on reading this manuscript.

I thank my office mates Fabien Nugier, Konstantina Kontoudi, Sophie Rosay and Evgeny Sobko for the friendly atmosphere in which we have spent these short three years.

I thank all my family and my friends for their support and encouragments, especially the swimmers-kebab team.

Last but not least, I thank my parents. They gave me the education that stimulates curiosity and openness to the world. All the credit, if granted, should go to them.

\chapter*{Presentation}
\addcontentsline{toc}{chapter}{ Presentation }
\label{chap:intro}

At the core of this work lies the holographic AdS/CFT correspondence and it would be appropriate to start with a few words reasserting its importance in view of the challenges of theoretical physics today. Two crucial features of the correspondence are the establishment of dualities between strongly coupled and weakly coupled field theories and the proposal of a relation between quantum field theories without gravity interactions and a quantum theory with gravity, which is string theory.  This opens a window to the two major problems encountered in theoretical particle physics, which are the accessibility to the strong coupling regime of gauge theories, especially QCD, and the understanding of quantum gravity. Besides the intrinsic mathematical beauty of the correspondence and the interest it might raise on its own, the AdS/CFT-type dualities proposed in the last fifteen years have come closer to phenomenological issues. It started with the original setup of Maldacena relating $\
N=4$ Super-Yang-Mills theory with $SU(N)$ gauge group to type IIB string theory on $AdS_5 \times S^5$, which has no connection to known physical theories and it has come now for instance to AdS/Condensed Matter Theories propositions, higher spin/vector model correspondence,  AdS/Chern-Simons-Matter dualities, ... , which offer connections with  condensed matter physics. There also exist attempts to describe QCD from a dual AdS side. Some of these dualities are speculative while others are established on firmer grounds. The main difficulties on the road to phenomenology are the presence of supersymmetry and the necessary large N limit.

The AdS/CFT dictionnary has been growing in various directions, but essentially with the purpose of understanding quantum field theories in the different language of the gravity side. The converse study, which consists in formulating supergravity problems on the gauge theory side, is less developed, however the question of the reconstruction of AdS supergravity solutions from the field theory data received some attention recently. AdS/CFT also offers an interesting connection to the mysterious M-theory, which is the gravity side of many known AdS/CFT dualities. 

The research on $AdS_4/CFT_3$ dualities has seen a major progress with the discovery in \cite{Aharony:2008ug} of the famous duality between the ABJM Chern-Simons gauge theory and M-theory on $AdS_4 \times S^7/\bZ_k$ background, which are two descriptions of the low-energy physics of M2-branes placed at the origin of a $\bC^4/\bZ_k$ orbifold. It corresponds to the maximally supersymmetric cases in three dimensions with $\N=8$ for $k=1$ and $\N=6$ otherwise. Following this breakthrough many $AdS_4/CFT_3$ dualities were found, relating $\N=2$ Chern-Simons theories with fundamental and bifundamental matter fields to M-theory on $AdS_4 \times Y_7$, where $Y_7$ is a Sasaki-Einstein manifold (\cite{Martelli:2008si,Martelli:2009ga,Jafferis:2008qz,Imamura:2008nn,Benini:2009qs,Hanany:2008fj}). These are supposed to describe the low-energy physics of M2-branes placed at the origin of the cone Calabi-Yau four-fold $X_8 = C(Y_7)$, which is a cone with base $Y_7$. Generically $Y_7$ has orbifold singularities. In most cases the gauge theories considered have similar features : Chern-Simons kinetic terms, $U(N)^p$ gauge group with bifundamental matter forming a circular ``chain'' (the $U(N)$ are the nodes and the bifundamental multiplets are the links), and the gravity duals are in M-theory.

\vspace{8mm}

The research presented here contains the new proposals of $AdS_4/CFT_3$ correspondences for a very large class (if not all) of $\N=4$ CFTs, that we derived in \cite{Assel:2011xz,Assel:2012cj}, the tests of the correspondences through free energy computations that we shown in \cite{Assel:2012cp}, plus an extension to the holographic dictionnary of defect SCFTs and some new remarks about applications to the F-theorem.\\
On the gauge theory side the conformal field theories are strongly interacting infrared fixed points that arise from the RG-flow of three-dimensional $\N=4$ quiver gauge theories. 
In \cite{Gaiotto:2008ak} Gaiotto and Witten discussed a large class of 3-dimensional super-conformal field theories with $\N=4$ supersymmetry that arise as IR fixed points of Yang-Mills $\N=4$ linear quiver gauge theories. In 3 dimensions the Yang-Mills coupling $g_{YM}$  is dimensionful, $g_{YM}^2$ has the dimension of a mass, so that in the infrared limit $g_{YM}$ is expected to be renormalized and to diverge (\cite{Hanany:1996ie,deBoer:1996mp,Gaiotto:2008sa,Gaiotto:2008ak}). This means that the IR fixed points are infinitely strongly coupled and there is no possibility to perform perturbative calculations to get some insights into their properties.\\
One possibility to gain informations about this rich group of SCFTs  is to use the techniques of localization of path integrals developed recently (\cite{Kapustin:2009kz}) for $\N\geq 2$ supersymmetric field theories on the 3-sphere. This technique is based on the property that one can deform path integrals computing supersymmetric observables by ``Q-exact'' terms without changing their values. In the limit of very large deformation the path integrals reduce to one-loop contributions which capture the full non-perturbative results. For the theories on $S^3$ the path integrals are reduced to simple enough matrix models. We will use these exact results in the presentation to provide quantitative checks of the AdS/CFT proposals.\\
Another possibility to understand the properties of these 3-dimensional SCFTs is via the AdS/CFT correspondence. For instance, although it is technically involved, one can compute correlation functions from the gravity side in a regime of parameters corresponding to the supergravity regime. The main object of our work consisted in exhibing the gravity duals of all $d=3$ $\N=4$ SCFTs arising from the RG-flow of linear and circular quivers, as well as all $\half$-BPS defect SCFTs in $d=4$ $\N=4$ Super-Yang-Mills theories.

Besides the interest that they have as dual descriptions of strongly interacting SCFTs, the type IIB supergravity solutions  that we study are interesting in their own right. They are $AdS_4 \times S^2 \times S^2 \ltimes \Sigma$ warped geometries, where $\Sigma$ is a Riemann surface. When $\Sigma$ is compact (disk or annulus), string theory on these backgrounds provides an effective realization of 4-dimensional quantum gravity. It is not directly relevant for phenomenology as the backgrounds are supersymmetric and the effective 4-dimensional cosmological constants are negative. It becomes more interesting when $\Sigma$ is non-compact, namely when it has one infinite direction. These geometries are domain walls interpolating between to two different $AdS_5 \times S^5$ asymptotic regions and correspond to the defect-SCFTs. They provide an explicit realization in string theory of the Karch-Randall scenario (\cite{Karch:2000ct,Karch:2001cw}). This model contains a 4d ``thin brane'' embedded in a $AdS_4$ slice of 
$AdS_5$ spacetime.  The effective 4-dimensionnal graviton modes are organized in a tower of massive excitations. The lowest mode is, in a good limit, nearly massless and its wavefunction is localized near the``thin brane''. Moreover the graviton spectrum has a large mass gap between the first mode and the other modes. This model provides an effective realization of 4d (AdS) gravity with a non-compact internal space. The domain wall solutions presented here are the string theory arena to test the possibility of this model. These interesting aspects will not be addressed in the main chapters but we provide in appendix \ref{app:KR} a short analysis (mainly qualitative) of the graviton mass spectrum in a simple domain wall background. We find that the good features of the Karch-Randall model are not reproduced in this simple case.

\vspace{8mm}
Let's describe the super-conformal field theories we study in more details. 
In three dimensions the field content of $\N=4$ gauge theories is organized in multiplets with 4 real bosonic fields. The vector multiplet contains a $\N=2$ vector multiplets and a chiral multiplet in the adjoint representation of the gauge group. The hyper-multiplet contains two chiral multiplets in conjugate representation of the gauge group. The quiver theories considered here have a gauge group which is a product of unitary gauge {\it nodes} $U(N_1)\times U(N_2) \times \cdots \times U(N_P)$. Each $U(N_i)$ node is associated to a $\N=4$ vector multiplet. The matter content is made of bifundamental hyper-multiplets for adjacent nodes $U(N_i)\times U(N_{i+1})$ and fundamental hypermultiplets in each $U(N_i)$ node. This describes {\it linear quivers}. The {\it circular quivers}
are obtained by adding a $U(N_P) \times U(N_1)$ bifundamental hypermultiplet connecting the first and last nodes. The kinetic terms for the vector fields are Yang-Mills terms with (dimensionful) gauge couplings $g_{YM}^{(i)}$ for each node.
We propose a AdS/CFT dictionnary for all such IR fixed points of linear and circular quivers.\\
Moreover we extend the correspondence to all the $\half$-BPS defect SCFTs with the same $OSp(4|4)$ supergroup of symmetries (\cite{DeWolfe:2001pq,Erdmenger:2002ex,Gaiotto:2008sa,Gaiotto:2008sd,Gaiotto:2008ak}).
The defect SCFTs are four-dimensional Super-Yang-Mills gauge theories coupled to a three-dimensional defect, supporting (the IR fixed point of) a 3d linear quiver gauge theory. The defect splits the four dimensional space in two half-spaces where live different $\N=4$ SYM theories. The couplings to the defect fields through bifundamental hypermultiplets preserve half of the supersymmetries. The general $\half$-BPS boundary conditions for the 4d fields are described in \cite{Gaiotto:2008sa}. 

All these gauge theories can be understood as the low-energy description of the worldvolume gauge theories of D3-branes in type IIB string theory, as in the case of pure $d=4$ $\N=4$ Super-Yang-Mills (which is the ``minimal case'' of defect SCFTs). The D3-branes can intersect D5-branes and end on NS5-branes. 

\begin{table}[h]
\label{tab:orientations}
\begin{center}
\begin{tabular}{|c||c|c|c|c|c|c|c|c|c|c||}
  \hline
      & 0 & 1 & 2 & 3 & 4 & 5 & 6 & 7 & 8 & 9 \\ \hline
  D3  & X & X & X & X &   &   &   &   &   &   \\
  D5  & X & X & X &   & X & X & X &   &   &   \\
  NS5 & X & X & X &   &   &   &   & X & X & X \\ \hline
\end{tabular}
\caption{\footnotesize Brane array for three-dimensional quiver gauge theories and defect gauge theories.}
\end{center}
\end{table}

The branes orientation preserves one quarter of the 32 supersymmetries of ten-dimensional spacetime (see table). They all share $2+1$ dimensions. For the 3d quiver theories the D3-branes have a finite extent in the $x^3$ ``transverse'' direction : they end on NS5-branes. In the low energy limit the excitations in the $x^3$ directions are suppressed and the theory is effectively three-dimensional. For the circular quivers the $x^3$ direction is a circle, allowing for D3-branes wrapping it without ending on any NS5-branes. For the defect gauge theories, the brane configurations have semi-infinite D3-branes (or even complete D3-branes) and the infrared worldvolume theory remains four-dimensional. 

The essential picture is that $N_i$ D3-branes suspended between two NS5-branes support a vector multiplets for a $U(N_i)$ gauge node, strings stretched between $N_i$ and $N_{i+1}$ D3-branes across a NS5-brane excite a hypermultiplet in the bifundamental representation of $U(N_i)\times U(N_{i+1})$ and $M_i$ D5-branes intersecting $N_i$ D3-branes add $M_i$ hypermultiplets in the fundamental representation of $U(N_i)$. With these basic ingredients it is easy to derive the brane configuration corresponding to any linear or circular quivers. The situation of the defect theories consists in adding semi-infinite D3-branes ending on NS5-branes or D5-branes on the left and on the right of a linear quiver brane configuration.

The relation to the brane picture is crucial for establishing the duality with the supergravity solutions. These solutions were derived in \cite{D'Hoker:2007xy} as the most general type IIB backgrounds preserving 16 real supersymmetries with a $SO(2,3)\times SO(3)\times SO(3) \subset OSp(4|4)$ ansatz. The metric is a warp product $AdS_4 \times S^2 \times S^2 \ltimes \Sigma$, where $\Sigma$ is a two-dimensional manifold. The whole solutions are determined in terms of two real harmonic functions $h_1,h_2$ on $\Sigma$. In the companion paper \cite{D'Hoker:2007xz} the conditions on $h_1,h_2$ for the regularity of the solutions were derived, with the allowed D5-brane and NS5-brane singularities. In \cite{Assel:2011xz} we found the limit of compactification of the internal space, which amounts to closing asymptotic $AdS_5 \times S^5$ regions, and established the precise dictionnary between these supergravity solutions and the fixed points of linear quivers. In \cite{Assel:2012cj} we found new solutions by periodic 
identifications along one direction in $\Sigma$. We found that these solutions ``on the annulus'' correspond to the fixed points of circular quivers and gave again the explicit dictionnary.
In this presentation we complete the picture by giving the holographic map for the defect SCFTs. The common features of all supergravity solutions are the presence of D3-brane charges (non-zero 5-form flux), D5-brane singularities on one boundary of $\Sigma$ (supporting $F_3$ flux) and NS5-brane singularities on the other boundary of $\Sigma$ (supporting $H_3$ flux), $\Sigma$ being either an infinite strip or an annulus. The quantized fluxes contain the data describing the solutions and corresponding quiver theories. 

All the solutions provide an elegant holographic realization of the mirror symmetry of three dimensional $\N=4$ super-conformal gauge theories in terms of Type IIB S-duality, which exchanges D5-branes and NS5-branes. The holographic dictonnary also confirms the prediction of \cite{Gaiotto:2008ak} for the existence of {\it irreducible} infrared fixed points for quiver theories with matter contents respecting specific inequalities.

\vspace{6mm}

Apart from the detailed exposition of the holographic dualities, we provide a number of consistency checks of the correspondences. As an important piece of work, we compute the free energy of linear quiver gauge theories in the large $N$ limit, using the exact results of \cite{Nishioka:2011dq} for the partition function, obtained from localization techniques on the 3-sphere (\cite{Kapustin:2009kz}). We compare it with the evaluation of the supergravity action on the solutions and found agreement (this was done in \cite{Assel:2012cp}). Along the road we derived a nice formula for the regularized IIB action in terms of the harmonic functions $h_1,h_2$. As a bonus we found inequalities between the free energy of different theories that have an interpretation in terms of F-theorem.

Finally we realized that new solutions can be found by acting with the $SL(2,\bR)$ symmetries of type IIB supergravity. The previous solutions correspond to background with vanishing axion field and appropriately quantized brane-charges. Acting with $SL(2,\bZ)$ generates solutions with non-zero axion that are different descriptions of the same quantum theory. However it does not cover the whole set of solutions. Acting with general $SL(2,\bR)$ transformations and quantizing the brane-charges of the new solutions leads to the full set of string theory backgrounds. Only part of those are related to the vanishing-axion solutions by $SL(2,\bZ)$ duality. The others are new solutions that contain generically two (and only two) types of $(p,q)$-5branes. The general inequivalent solutions are classified by a collection of NS5-branes on one part of the boundary of $\Sigma$ and a collection of $(p,q)$-5branes, with $0 \leq p \leq |q|$, on the other part of the boundary of $\Sigma$. The case of vanishing axion field 
corresponds to $(p,q)=(0,1)$.\\
The gauge theory duals of these more general holographic backgrounds are not easily described (see however \cite{Gaiotto:2008ak}). In the simpler case of NS5-branes and $(1,k)$-5branes the gauge theories are understood as Chern-Simons-Matter gauge super-conformal theories with enhanced $\N=4$ supersymmetry, such as ABJM gauge theory (which has even $\N=6$ supersymmetry). The $SL(2,\bQ)$ classical symmetry of type IIB supergravity translates into an ``orbifold'' symmetry for gauge theories, in which ``untwisted'' observables can be mapped in the large $N$ limit.

\vspace{10mm}

The presentation is organized as follows. In chapter \ref{chap:AdSCFT} we review the basics of AdS/CFT and its original derivation in terms of dual descriptions of D3-branes. We also review the principles of the holographic regularization of the gravity action. We remind a few properties of $d=3$ $\N=4$ (super-conformal) gauge theories in chapter \ref{chap:quivers}, we describe the quiver and defect gauge theories and relate them to the (important) brane configurations. In chapter \ref{chap:sugra} we expose the supergravity solutions on the strip and on the annulus, compute the brane-charges and establish 
the holographic dictionnary. The computation of the free energy of linear quivers in the large $N$ limit and the match with the supergravity action are given chapter \ref{chap:GKPW} and the results are shown to support the F-theorem. Finally in chapter \ref{chap:SL2R} we generalize the solutions to non-vanishing axion backgrounds, propose the ``orbifold'' equivalence and check it on the gauge theory side with matrix model computations of the free energy in the large $N$ limit.\\
A few computations have been placed in the appendices \ref{appendices}. Appendix \ref{app:KR} is devoted to the study of the Karch-Randall scenario in domain wall supergravity backgrounds.

\vspace{8mm}

This presentation is essentially based on the three papers \cite{Assel:2011xz,Assel:2012cp,Assel:2012cj}. The new (unpublished) parts are the precise holographic dictionnary for defect SCFTs (\S\ref{sec:defectsugra}), the discussion on the supergravity regimes of parameters (\S\ref{subsec:parameters}) and the explanation of the free energy inequalities in term of the (speculative) F-theorem (\S\ref{sec:F-theorem}).

\newpage

\section*{Résumé (français)}
\addcontentsline{toc}{section}{ Résumé (français)}
\label{sec:french}

\vspace{8mm}

L'objet principal de cette thèse est l'établissement de nouvelles dualités holographiques reliant des théories super-conformes à trois dimensions et supersymétrie $\N=4$ à des théories des cordes sur des solutions de supergravité de type IIB.
 Ces propositions constituent une large extension des correspondances $AdS_4/CFT_3$ actuellement connues.

Rappelons que la découverte de la correspondence AdS/CFT par Maldacena à la fin du XXème sciècle a eu un impact important sur la recherche en physique théorique (des hautes énergies) et suscité un intéret qui n'a fait que s'accroître depuis. La correspondence fait le lien entre des théories quantiques conformes des champs sans interactions gravitationnelles en dimension $D$ et des théories avec interaction gravitationnelles, qui sont théories des cordes sur des espaces $AdS_{D+1} \times K_{9-D}$, où $AdS_{D+1}$ est l'espace de courbure négative Anti-de Sitter à $D+1$ dimensions et $K_{9-D}$ est un espace compact à $9-D$ dimensions. La correspondence originelle relie la théorie conforme Super-Yang-Mills à quatre dimensions et supersymétrie $\N=4$ et groupe de jauge $SU(N)$, à la théorie des cordes de type IIB sur l'espace $AdS_5 \times S^5$. L'intérêt de la correspondance, et ce qui rend difficile sa vérification, est qu'elle relie une théorie dans un régime de couplage fort à l'autre théorie dans un régime 
de 
couplage faible. Elle offre donc une description accessible perturbativement de théories des champs dans la limite de grand couplage, ce qui est un des problèmes majeurs de la QCD aujourd'hui. D'un autre côté elle met à jour la nature holographique de la gravité quantique dans les espaces Anti-de Sitter, ce qui est aussi une avancée importante dans la compréhension de la gravité quantique.

Les efforts fournis au cours des années qui suivirent ont mis à jour de nombreux autres exemples de correspondances, avec un rapprochement vers des théories physiques phénoménologiques, notament vers la physique de la matière condensée qui peut être décrite en terme de théorie des champs. Les difficultés essentielles consistent à étendre la correspondance AdS/CFT à des théories non-supersymmétriques et à ``$N$ fini'' (habituellement la correspondence n'est utilisable que dans une certaine limite où le ``paramètre $N$'' est très grand). Déjà l'extension à des théories de jauge non-conformes est comprise avec un dual gravitationnel dont la métrique est seulement asymptotiquement AdS. Les théories des champs à température finie par exemple correspondent à des solutions de trou noir AdS. Récemment des calculs de supergravité ont été capable de reproduire certaines propriétés des supraconducteurs. Les efforts pour trouver une description holographique pour la QCD existent mais se heurtent encore à un certain 
nombre de difficultés. 

\bigskip

L'intérêt essentiel des nouvelles dualités AdS/CFT décrites dans cette thèse, au delà de l'enrichement des connaissances sur la correspondence en elle-même, est de fournir pour la première fois une description (holographique) de théories de jauges infiniment fortement couplés. En effet les théories superconformes que nous étudions sont obtenues comme point fixe infrarouge de théories de Yang-Mills $\N=4$ à trois dimensions. La constante de couplage $g_{YM}$ diverge dans l'infrarouge rendant impossible tout calcul perturbatif, d'où l'intérêt d'en avoir une description holographique.\\
Le contenu en champs des théories $\N=4$ $d=3$ se compose de multiplets à huit degree de liberté réels bosonics. Le multiplet vectoriel rassemble un multiplet vectoriel $\N=2$ et un multiplet chiral adjoint, tandis que l'hyper-multiplet rassemble deux multiplets chiraux transformant dans des représentations conjuguées du groupe de jauge.
Les théories de jauge en question sont de type ``quiver'', c'est-à-dire que leur groupe de jauge est un produit de groupes unitaires $U(N_1)\times U(N_2) \times \cdots \times U(N_P)$, avec un multiplet vectoriel pour chaque {\it noeud} $U(N_i)$. Le contenu en matière est donné par des hyper-multiplets bifundamentaux pour chaque paire de noeuds adjacents $U(N_i)\times U(N_{i+1}$, plus $M_i$ hyper-multiplets fondamentaux pour chaque noeud $U(N_i)$. Ceci décrit les quivers linéaires. Les quivers circulaires sont obtenus an ajoutant un hypermultiplet bifondamental pour le couple $U(N_P)\times U(N_1)$.

Comme extension nous proposons aussi les duaux holographiques de théories de type ``defect-SCFT'' qui sont des théories de jauge à quatre dimensions $\N=4$ Super-Yang-Mills couplées à un défaut à trois dimensions sur lequel vivent les champs trois-dimensionnels d'un quiver linéaire. Les solutions de supergravité correspondentes sont de simples extensions des solutions duales aux quiver linéaires.

L'établissement du dictionnaire AdS/CFT repose de manière cruciale sur la compréhension des théories de quiver en termes de limite de basse énergie de champs vivants sur des D3-branes en théorie des cordes IIB. Les configurations de branes en question rassemblent des D3-branes, des D5-branes et des NS5-branes orientées de manière à preserver 8 supercharges sur 32. L'orientation des branes est donnée dans le tableau.

\begin{table}[h]
\label{tab:orient2}
\begin{center}
\begin{tabular}{|c||c|c|c|c|c|c|c|c|c|c||}
  \hline
      & 0 & 1 & 2 & 3 & 4 & 5 & 6 & 7 & 8 & 9 \\ \hline
  D3  & X & X & X & X &   &   &   &   &   &   \\
  D5  & X & X & X &   & X & X & X &   &   &   \\
  NS5 & X & X & X &   &   &   &   & X & X & X \\ \hline
\end{tabular}
\caption{\footnotesize Orientations des branes correspondant aux quivers $\N=4$ à trois dimensions.}
\end{center}
\end{table}

Le contenu en champs des théories de quivers correspondent aux excitations non-massive de cordes fondamentales ouvertes dont les deux bout sont fixés sur les branes. $N_i$ D3-branes étendues entre deux NS5-branes correspondent à un multiplet vectoriel pour un groupe de jauge $U(N_i)$, $M_i$ D5-branes croisant ces $N_i$ D3-branes correspondent à $M_i$ hypermultiplets fondamentaux four ce groupe de jauge $U(N_i)$. Quand $N_i$ D3-branes terminent sur la gauche d'une NS5-brane et $N_{i+1}$ D3-branes terminent sur sa droite, les cordes étendues entre les $N_i$ et $N_{i+1}$ D3-branes excitent un hypermultiplet bifondamental $U(N_i) \times U(N_{i+1})$ (\cite{Hanany:1996ie}). 

Ainsi des assemblages de branes avec une succesion de NS5-branes et D5-branes traversées par des D3-branes le long de la direction $x^3$ reproduisent les théories de quiver à basse énergie. Les quivers linéaire ont des configurations de branes où les D3-branes sont toutes étendues entre deux NS5-branes dans la direction $x^3$. Dans la limite de basse énergie les fluctuations selon $x^3$ sont suprimées et la théorie vivant sur les D3-branes est de manière effective trois-dimensionnelle. C'est assi le cas des quivers circulaires qui sont obtenus en compactifiant la direction $x^3$ sur un cercle. En revanche les théories de type ``defect'' correspondent à des configurations de branes avec des D3-branes semi-infinies dans la direction $x^3$ à droite et à gauche des 5-branes et la théorie des champs vivant sur les D3-branes est bien quatre-dimensionnelle.

\vspace{6mm}

Les relations entre quiver théories et configurations branaires en théories des cordes de type IIB sont cruciales pour établir le dictionnaire avec les solutions de supergravité.
Ces solutions ont été trouvées dans \cite{D'Hoker:2007xy} en temps que solutions de la supergravité de type IIB préservant 16 supersymétries et possédant les isométries $SO(2,3) \times SO(3) \times SO(3) \subset OSp(4|4)$. La métrique est une fibration $AdS_4 \times S^2 \times S^2 \ltimes \Sigma$, où $\Sigma$ est une surface. Les différents champs d'une solution sont donnés de manière générale par deux fonctions réelle harmoniques $h_1, h_2$ sur $\Sigma$. Dans \cite{D'Hoker:2007xz} les conditions
sur $h_1, h_2$ de régularité de la solutions sont présentées, ainsi que les singularités admissibles de type D5-brane et NS5-branes sur le bord de $\Sigma$. Dans \cite{Assel:2011xz} nous avons obtenu les solutions correspondant aux quiver linéaires en prenant une limite de fermeture des régions asymptotiques $AdS_5 \times S^5$, qui rend l'espace interne compact, et nous avonc établi le dictionnaire AdS/CFT. Dans \cite{Assel:2012cj} nous avons obtenu les solutions correspondant aux quivers circulaires en identifiant périodiquement des solutions le long d'une direction infinie sur $\Sigma$, qui devient alors un anneau. Les solutions correspondant aux defect-quiver théories sont les solutions initiales avec deux régions asymptotiques $AdS_5 \times S^5$. Toutes ces solutions sont caratérisées par les flux quantifiés de D3, D5 et NS-branes, qui à travers l'image des configurations de branes, sont reliés aux données définissant les théories de quiver.

Ces solutions fournissent une réalisation naturelle de la symétrie miroir des théories $\N=4$ à trois dimension à travers la S-dualité de la théorie des cordes de type IIB, qui échange les D5-branes avec les NS5-branes. Elles donnent aussi une preuve holographique de la conjecture de \cite{Gaiotto:2008ak}, qui prédit l'existence de points fixes {\it irréductibles} infrarouges pour les théories de quiver vérifiant certaines inégalités.

\vspace{8mm}

Une large partie du travail de thèse est consacré à l'exposition des solutions de supergravité et à l'établissement du dictionnaire holographique. Ce travail est complété par un certain nombre de vérifications, notamment nous procédons au calcul de l'énergie libre des théories de quiver linéaires dans la limite de grand $N$ en utilisant des résultats issus de calcul de technique de localisation sur la 3-sphere (\cite{Kapustin:2009kz,Nishioka:2011dq}), et comparons avec le calcul de l'action de type IIB évaluée sur les solutions correspondantes. Nous montrons l'accord entre les deux calculs (ceci a été fait dans \cite{Assel:2012cj}). En passant nous établissons un formule générale élégante pour l'action de supergravité régularisée directement en fonction des fonction $h_1$ et $h_2$ définissant les solutions et expliquons les inégalités obtenues entre les énergies libres des différentes théories conformes en terme du supposé théorème F.

Nous présentons aussi une extension des solutions de supergravité à des solutions avec axion non-nul en utilisant la symmétrie $SL(2,\bR)$ de la supergravité IIB. Les solutions reliées par les transformations $SL(2,\bZ)$ sont équivalentes car 
le groupe $SL(2,\bZ)$ est un groupe de symmétrie de la théorie des cordes IIB. Cependant, par des transformations $SL(2,\bR)$ dont on quantifie les flux on obtient de nouvelles solutions contenant des $(p,q)$ 5-branes. Les solutions inéquivalentes sont classifiées par la donnée de singularités de NS5-branes sur un bord de $\Sigma$ et de singularités de $(p,q)$ 5-branes, avec $0 \le p \le |q|$, sur l'autre bord de $\Sigma$. Les solutions avec D5-branes correspondent à $(p,q)=(0,1)$. Les théories de jauges superconformes duales ne sont pas aisément descriptibles (voir cependant \cite{Gaiotto:2008ak}). Dans le cas simple où les singularités sont de type NS5-branes et $(1,k)$ 5-branes, il est possible de décrire les théories superconformes en termes de théories de Chern-Simons à trois dimensions avec supersymétrie étandue $\N=4$, où $\pm k$ correspond au niveau de Chern-Simons de certain noeuds unitaires du groupe de jauge, comme c'est le cas de la célèbre théorie ABJM. Les symétries $SL(2,\bR)$ de la 
supergravité classique se traduisent du côté théories de jauge par des équivalences ``orbifold'' entre différentes théories, qui prédit l'égalité entre observables ``untwisted'' dans la limite de grand $N$.

\vspace{8mm}

L'essentiel du matériel présenté ici est issu des articles \cite{Assel:2011xz,Assel:2012cp,Assel:2012cj}. Nous résumons maintenant les différents chapitres du manuscrit.

\vspace{10mm}

{\large {\bf I. Elements sur la correspondance AdS/CFT}}

\vspace{8mm}

Dans ce chapitre nous rappelons les fondements de la correspondance AdS/CFT de Maldacena (\cite{Maldacena:1997re}), ainsi quelques relations  de base qui définissent la dualité (voir \cite{Witten:1998qj,Aharony:1999ti}). Nous présentons aussi la méthode de renormalisation holographique (\cite{deHaro:2000xn}) qui permet de régulariser l'action de gravité.

\vspace{6mm}

L'idée de la correspondance a son origine en théorie des cordes, où l'on peut décrire de deux manières la limite de basse énergie d'un paquet de D3-branes. Les D3-branes sont des objets solitoniques à $3+1$ dimensions définis par la propriété que les bouts des cordes ouvertes y sont attachés (voir figure \ref{fig1_1}). Les D3-branes peuvent être décrites par la théorie des champs vivant sur leur ``worldvolume'' quatre-dimensionnel, ou bien en temps qu'objet solitonique dans les 10 dimensions de la théorie des cordes de type IIB. 

Le contexte originel de Maldacena consiste à considérer un paquet de $N$ D3-branes coincidentes dans l'espace de Minkovski à 10 dimensions. La théorie de basses énergies (c-à-d contenant seulement les champs de masse nulle) vivant sur le worldvolume des D3-branes est la théorie Super-Yang-Mills $\N=4$ à 4 dimensions avec groupe de jauge $U(N)$. Cette théorie est superconforme et est déterminée par le paramètre de jauge $N$ et la constante de couplage adimensionnée $g_{YM}$.


D'un autre côté la limite de basse énergie de la théorie des cordes de type IIB en présence de D3-branes/solitons consiste à ne garder que les fluctuations infiniment proches de l'horizon (ou position) des branes. On peut en avoir une description en ``zoomant'' sur les branes. La solution de supergravité obtenue dans cette limite est appelée limite de ``near-horizon'' et correspond à la métrique de $AdS_5 \times S^5$ avec rayon $L$ identique pour les deux facteurs. Les parametres qui définisent la solution sont le rayon $L$ et le dilaton $g_s = e^{\phi}$ qui est constant.

\vspace{6mm}

L'expression générale de la correspondance est alors la suivante :

\begin{center}

\fbox{
\begin{minipage}{12cm}
\begin{center}
\vspace{4mm}

{\it  $\N=4$ Super-Yang-Mills on $\bR^{1,3}$ with gauge group $U(N)$}

\vspace{4mm}

 $\Updownarrow$ 

\vspace{4mm}

{\it Type IIB string theory on $AdS_5 \times S^5$ with radius $L$ }. 
\vspace{4mm}

\end{center}
\end{minipage}
}
\no
\end{center}

Et les paramètres sont identifiés selon 
\begin{align}
 g_{YM}^2 = g_s  \quad , \quad \frac{L^4}{l_s^4} &= 4\pi g_s N  \ . \no
\end{align}

Le régime dans lequel la théorie Super-Yang-Mills est faiblement couplées est $\lambda \equiv g_{YM}^2 N << 1$, alors que le régime de supergravité classique est donné par $\lambda = g_s N >> 1$ et $g_s << 1$. Ces deux régimes sont incompatibles se qui rend la correspondence à la fois très utile et très difficile à prouver. 

Une version plus faible de la correspondance consiste limiter le postulat de dualité à la limite de grand $N$, dans laquelle l'expansion perturbative des amplitudes de théorie des champs prend la forme d'une expansion topologique identique à celle de la théorie des cordes. C'est ce qu'on appelle la limite de 't Hooft : $N >> 1$ et $\lambda$ constant. 

\vspace{6mm}

La correspondence exprime que les symétries des deux théories sont les mêmes. Il s'agit dans ce cas du groupe superconforme de symétries $SU(2,2|4)$. Il existe aussi un isomorphisme entre les operateurs invariants de jauge et les champs vivants dans l'espace $AdS$ : à un opérateur $\scO_{\Delta}$ de dimension conforme $\Delta$ correspond un champs $\phi_m$ de masse $m$ d'AdS avec une certaine relation entre $\Delta$ et $m$ qui dépend du spin du champs en question.

Un élément central de la correspondence est la relation GKPW (\cite{Gubser:1998bc,Witten:1998qj}), qui montre que la théorie des champs peut être imaginée comme vivant sur le bord (à l'infini) de l'espace $AdS$. La relation GKPW est donnée par 
\begin{align}
\label{GKPWintro}
 \left< e^{\int d^4 x \, \phi_0(x^{\mu}) \scO(x^{\mu}) } \right>_{CFT} = Z_{string} \Big[ \phi(x^{\mu},u=0) = \phi_0(x^{\mu}) \Big] \ , \no
\end{align}
où le terme de gauche correspond à la génératrice des fonctions de corrélation de l'opérateur $\scO$, avec $\phi_0$ la source, et le terme de droite est la fonction de partition de la théorie des cordes sur l'espace $AdS$ avec les conditions aux bords (à l'infini) pour le champs $\phi$ associé à $\scO$, $\phi = \phi_0$.

Les dérivées fonctionnelles par rapport à $\phi_0$ du terme de gauche génèrent les fonctions à $n$ points de $\scO$. En utilisant cette relation, on peut traduire le calcul de ces fonctions de corrélation du côté théorie des cordes. Dans la limite de supergravité, ces calculs se traduisent par une expansion perturbative en {\it diagrammes de Witten} qui sont analogues aux diagrammes de Feynman en théorie des champs.

\vspace{6mm}

De nombreuses autres correspondences AdS/CFT ont été mises à jour. Un exemple important (\cite{Aharony:2008ug}) est la dualité entre la théorie ABJM, qui est une théorie de Chern-Simons à trois dimensions $\N=6$ superconforme avec groupe de jauge $U(N)\times U(N)$, niveau de Chern-Simons $+k$ pour un $U(N)$,$-k$ pour l'autre, et deux hypermultiplets bifondamentaux, et du côté gravité la théorie M sur $AdS_4 \times S^7/\bZ_k$. Les deux théories sont deux descriptions de basse énergie d'un paquet de $N$ M2-branes coincidentes placées au sommet d'un certain orbifold $\bZ_k$ en théorie M. 

\vspace{6mm}

Pour finir nous présentons le calcul de regularisation holographique de l'action de gravité (\cite{deHaro:2000xn}).
L'idée générale est que le volume d'AdS étant infini, l'action de (super)gravité est généralement divergente et qu'il est possible de régulariser cette action en imposant d'abord un cut-off infrarouge, càd en considérant l'espace tronqué à un certain rayon $r$, et en ajoutant un contre-terme qui est un terme de bord universel (le même pour toute les solutions asymptotiquement AdS), de manière que la limite $r \rightarrow \infty$ donne une action finie.

\vspace{10mm}

{\large {\bf II. 3d $\N=4$ théries de quiver et réalisation branaires}}

\vspace{8mm}

Dans ce chapitre nous détaillons the contenu en champs des théories de jauge en trois dimensions avec supersymétrie $\N=4$, nous donnons les Lagrangiens de chaque multiplet et nous rappelons quelques propriétés des points fixes superconformes infrarouges, telle que la symétrie miroir. Puis nous présentons les quiver linéaires, circulaires et defect quivers.
Finalement nous donnons l'expression exacte de la fonction de partition avec paramètres de déformation postulée dans \cite{Nishioka:2011dq} à partir des techniques de localisation d'intégrales de chemin (\cite{Kapustin:2009kz}).

\vspace{6mm}

Les théories des champs $d=3$ $\N=4$ csont composés de multiplets à quatre champs bosoniques réels. Le multiplet vectoriel rassemble un multiplet vectoriel $\N=2$ et un multiplet chiral adjoint, tandis que l'hyper-multiplet contient deux multiplets chiraux transformant dans des représentations conjuguées du groupe de jauge. Le Lagrangien associé est fixé par la supersymmétrie. Il contient les termes cinétiques standard $\N=2$ (Yang-Mills pour le champs vectoriel) et couplages aux champs de jauge minimaux pour les multiplets chiraux, plus le superpotentiel de la supersymmétrie $\N=4$. Le Langrangien peut être déformé (en préservant la supersymmétrie $\N=4$, par des paramètres de masse pour les hypermultiplets et des paramètres de Fayet-Iliopoulos pour chaque $U(1) \subset U(N_i)$ diagonal. \\
Le groupe de R-symétrie de ces théories est $SU(2)_L \times SU(2)_R$.

\smallskip

Ces théories admettent un large espace de modules, ou espace des vides, qui comprend deux ensembles distincts : la branche de Coulomb où les scalaires des multiplets vectoriels ont des vevs non-nulles et la branche de Higgs où ce sont les scalaires des hypermultiplets qui ont des vevs non-nulles.
Les points fixes infrarouge de ces théories sont à couplage (infiniment) fort.
A l'intersection de la branche de Higgs et de la branche de Coulomb vivent (dans l'infrarouge) des théories superconformes non-triviales. La symétrie miroir en trois dimension est une dualité entre ces points fixes infrarouges qui échange la branche de Higgs et la branche de Coulomb. De manière générale la symétrie miroir échange les rôles de $SU(2)_L$ et $SU(2)_R$. Les paramètres de masse et de Fayet-Iliopoulos sont aussi échangés.

\vspace{8mm}

Les théories de jauge de quiver ont un groupe de jauge qui est un produit de groupes unitaires $U(N_1)\times U(N_2) \times \cdots \times U(N_P)$, avec un multiplet vectoriel pour chaque {\it noeud} $U(N_i)$. Le contenu en matière est donné par des hyper-multiplets bifundamentaux pour chaque paire de noeuds adjacents $U(N_i)\times U(N_{i+1})$, plus $M_i$ hyper-multiplets fondamentaux pour chaque noeud $U(N_i)$. Ceci décrit les quivers linéaires. Les quivers circulaires sont obtenus an ajoutant un hypermultiplet bifondamental pour le couple $U(N_P)\times U(N_1)$. La description d'un quiver est résumé dans un petit diagramme où les noeuds sont symbolisés par des ronds indiquant le rang $N_i$, les hypermultiplets fondamentaux par des carrés indiquants leur nombre $M_i$ et les hypermultiplets bifondamentaux par des lignes reliants les ronds, comme sur les figures \ref{fig:linquiv},\ref{fig:circquiv}.

%

D'après la prédiction de \cite{Gaiotto:2008ak}, ces théories de quivers possèdent un point fixe (théorie limite) infrarouge {\it irreductible}, au sens où il n'existe pas champs qui découplent, à la condition que pour chaque noeud on ait $2 N_i \le M_i + N_{i+1} + N_{i-1}$. Les duaux gravitationnels que nous proposons seront en bijection avec les quivers qui vérifient ces conditions. Les théories de quiver qui ne vérifient pas ces conditions ont une limite infrarouge qui doit contenir une partie en interaction équivalente à celle d'un quiver qui vérifie les conditions, plus un certain nombre d'hypermultiplets libres (non-couplés).

\vspace{8mm}

Comme expliqué en introduction, les théories de quiver peuvent être réalisés comme théorie vivant sur le worldvolume de D3-branes étendues entre des NS5-branes et croisant des D5-branes. Les configurations branaires pour les quivers linéaires et circulaires sont présentées dans les figures \ref{fig:linquivbrane} \ref{fig:circquivbrane}.

Les paramètres caractérisant les quivers linéaires peuvent être rearrangés en {\it linking numbers} associés aux 5-branes. 
Les {\it linking numbers} pour la $i$-ème D5-brane et la $j$-ème NS5-brane sont définis par
\begin{align}\
l_i &= - n_i + R_i^{\rm NS5} \qquad (i=1,...,k) \no\\
\hat l_j &= \hat n_j + L_j^{\rm D5} \qquad (j=1,...,\hat k) \ , \no
\end{align}
où  $n_i$ ($\hat n_j$) est le nombre de D3-branes terminant sur la droite de $i$ème D5-brane ($j$ème NS5-brane) moins le nombre terminant sur sa gauche, $R_i^{\rm NS5}$ est le nombre de NS5-branes placées à droite de la $i$ème D5-brane et $L_j^{\rm D5}$est le nmbre de D5-branesplacées à gauche de la $j$ème NS5-brane.  Ces nombres sont invariants par rapport aux mouvement de Hanany-Witten (\cite{Hanany:1996ie}), où une D5-brane croise une NS5-brane, créant une D3-brane étendue entre elles.

Dans une configuration de quiver linéaire les linking numbers des D5-branes $l_i$ sont automatiquement positifs et ordonnés, constituant une partition $\rho$ d'un certain entier $N$. Les conditions d'irréducibilité du point fixe infrarouge impliquent que les linking numbers des NS5-branes $\hat l_j$ sont aussi positifs et ordonnés. Ils constituent en fait une deuxième partition $\hat \rho$ du même entier $N$. Les deux partitions $\rho,\hat\rho$ caractérisent entièrement le quiver linéaire et la théorie infrarouge est notée $T^{\rho}_{\hat\rho}(SU(N))$.
\smallskip

Le cas des quivers circulaires est similaire, bien que plus technique. Les paramètres du quiver sont réarrangés en deux partitions de $N$ contenant les linking numbers des 5-branes, plus un nouveau paramètre $L$ qui caractérise le nombre de D3-branes enroulées autour du cercle. Les points fixes infrarouges correspondants sont notés $C_{\hat\rho}^{\rho}(SU(N), L)$.

\smallskip
Les théories de type defect quivers sont des théories de jauge à quatre dimensions $\N=4$ Super-Yang-Mills couplées à un défaut à trois dimensions, sur lequel vivent les champs trois-dimensionnels d'un quiver linéaire. Le défaut est couplé aux champs à 4d par des hypermultiplets bifondamentaux qui transforment selon un des deux noeuds extérieurs du quiver linéaire et selon le groupe de jauge induit sur le défaut d'une théorie (droite ou gauche) SYM. Les conditions aux bords sur le défaut des champs à 4d préservent la moitié des supersymmétries de SYM $\N=4$ et sont classifiées dans \cite{Gaiotto:2008ak}. \\
Les configuration branaires des defect quivers sont identiques à celles des quivers linéaires, à ceci près que l'on a des D3-branes semi-infinies à droite $N_R$ et à gauche $N_L$ de la configuration de branes. Ces D3-branes semi-infinies peuvent se terminer sur des D5-branes ou des NS5-branes, décrivant alors des conditions aux bords sur le défaut spécifiques. ces configurations de branes sont données en figure \ref{fig:defectquivbrane} et les quivers associés en figure \ref{fig:defectquiv}. Les théories de defects sont classifiées par la donnée une partition $\rho$ de $N-N_R$, une partition $\hat\rho$ de $N- N_L$, les rangs $N_L$ et $N_R$ des groupes de jauge SYM et les couplages de Yang-Mills $g_{YM}^{(L)},g_{YM}^{(R)}$. Le point fixe infrarouge correspondant est noté $\scD(\rho,\hat \rho,N_L,N_R,g_{YM}^{(L)},g_{YM}^{(R)})$. Les linking numbers des partitions peuvent cette fois être négatifs.

\vspace{10mm}

{\large {\bf III. Solutions de supergravité et correspondance holographique}}

\vspace{8mm}

Dans ce chapitre nous présentons les solutions de supergravité duales aux points fixes infrarouges des quivers du chapitre précédent, nous établissons le dictionnaire AdS/CFT et nous étudions plusieurs limites intéressantes des paramètres.

\vspace{6mm}

Les solutions que nous présentons ont été trouvées dans \cite{D'Hoker:2007xy} en temps que solutions de la supergravité de type IIB préservant 16 supersymétries et possédant les isométries $SO(2,3) \times SO(3) \times SO(3) \subset OSp(4|4)$. La métrique est une fibration $AdS_4 \times S^2 \times S^2 \ltimes \Sigma$, où $\Sigma$ est une surface.
\begin{align}
ds^2 = f_4^2 ds^2_{AdS_4} + f_1^2 ds^2_{S^2_1} + f_2^2 ds^2_{S^2_2} + 4 \rho^2 dz d\bar z\ , \no
\end{align}
où le complexe $z$ paramétrise la surface $\Sigma$.

 Les différents champs d'une solution sont donnés par deux fonctions réelles harmoniques $h_1, h_2$ sur $\Sigma$ par les formules \ref{W}, \ref{metric}, \ref{dil}, \ref{3forms0},\ref{3forms1}, \ref{5form0}, \ref{calF}. Dans \cite{D'Hoker:2007xz} les conditions sur $h_1, h_2$ de régularité de la solutions sont présentées.
 Les singularités admissibles car ayant une interprétation en théorie des cordes correspondent à des D5-branes et des NS5-branes localisées sur le bord de $\Sigma$. 

\vspace{6mm}

Les solutions correspondant aux quivers linéraires et aux defect quivers sont données par les fonctions harmoniques
\begin{align}
h_1  &= -i \alpha \sinh(z-\beta) \  - \sum_{a=1}^{p} \gamma_{a} \ln\, \tanh\left(
{\small  {\frac{i\pi}{4}-\frac{z-\delta_a}{2}
}}
\right) \ + \ c.c. \ ,  \no\\
h_2 &= \hat \alpha \cosh(z-\hat \beta) \   - \sum_{b=1}^{\hat{p}} \hat{\gamma}_{b} \ln \tanh\left(\frac{ z-\hat{\delta}_{b} }{2}\right)   \ + \ c.c. \ . \no
\end{align}
où $z$ paramétrise un bandeau $\Sigma = \bR +i [0,\pi/2]$. Tous les paramètres de la solution sont réels et de plus $\alpha$ et les $\gamma_a$ ont tous le même signe, de même que $\hat \alpha$ et les $\hat\gamma_b$ ont tous le même signe.

Si $\alpha \neq 0$ ou $\hat\alpha \neq 0$ la solution possède deux régions asymptotiques $Re(z) \rightarrow \pm \infty$ dont la géométrie est celle d'$AdS_5 \times S^5$. Il s'agit donc de solutions de domain-wall interpolant entre les géométries duales de deux $\N=4$ Super-Yang-Mills. ce sont toutes les solutions de defect quivers. 

Pour $\alpha = \hat\alpha = 0$ les deux régions asymptotiques ``se ferment'' et l'espace interne devient compact. Ces solutions correspondent aux quivers linéaires.

Ces solutions de supergravité sur le bandeau sont caractérisées par des singularités de type NS5 aux positions $\hat\delta_b$ sur le bord inférieur de $\Sigma$ et des singularités de type D5 aux positions $\delta_a + i\frac{\pi}{2}$ sur le bord supérieur de $\Sigma$, comme représenté sur la figure \ref{Disk}. \\
La singularité de type NS5 en $\hat\delta_b$ est la source d'un flux $\hat N_5^{(b)}$ de 3-forme $H_3$ proportionnel à $\hat\gamma_b$, mesurant un  nombre de NS5-branes, et d'un flux $\hat N_3^{(b)}$ de 5-forme (dont la définition est subtile) lié à la position $\hat\delta_b$, mesurant le nombre de D3-branes terminant sur le paquet de NS5-branes. De manière similaire la singularité de type D5 en $\delta_4 + i \pi/2$ est la source d'un flux $N_5^{(a)}$ de 3-forme $F_3$ proportionnel à $\gamma_a$, mesurant un nombre de D5-branes, et un flux $N_3^{(a)}$ de 5-forme mesurant un nombre de D3-branes, lié à $\delta_a$.\\
Ainsi les paramètres $\gamma_a, \delta_a, \hat\gamma_b, \hat\delta_b$ d'une solution peuvent être utilisés pour définir deux partitions $\rho, \hat\rho$ selon 
\begin{align}
\rho &= \Big( \overbrace{l^{(1)},l^{(1)},..,l^{(1)}}^{N_{5}^{(1)}},\ \overbrace{l^{(2)},l^{(2)},..,l^{(2)}}^{N_{5}^{(2)}},\ ...\ ,
\ \overbrace{l^{(p)},l^{(p)},..,l^{(p)}}^{N_{5}^{(p)}} \Big) \ ,   \no \\
\hat \rho &=
 \Big( \overbrace{\hat l^{(1)},\hat l^{(1)},..,\hat l^{(1)}}^{\hat N_{5}^{(1)}},\ \overbrace{\hat l^{(2)},\hat l^{(2)},..,\hat l^{(2)}}^{\hat N_{5}^{(2)}},\ ...\ ,\
 \overbrace{\hat l^{(\hat p)},\hat l^{(\hat p)},..,\hat l^{(\hat p)}}^{\hat N_{5}^{(\hat p)}} \Big) \ , \no
\end{align}
où l'on a défini
\begin{align}
 l^{(a)} = \frac{N_3^{(a)} }{N_5^{(a)} } \quad , \quad  \hat l^{(b)} = - \frac{\hat N_3^{(b)}}{\hat N_5^{(b)} } \ . \no
\end{align}
(Le signe négatif pour $\hat l^{(b)} $ vient du fait que dans nos conventions $\hat N_3^{(b)}$ est négatif) Les expressions exactes des flux en termes des paramètres de la solution sont donnés par \ref{ginvN5} \ref{ginvN3} pour les solutions de quivers linéaires.

Les nombres (flux quantifiés) $l^{(a)}, \hat l^{(b)}$ correspondent exactement aux linking numbers des 5-branes pour une configuration branaire associée à un quiver linéaire. Le point fixe de quiver linéaire qui est le dual holographique de la solution de supergravité décrite par $\rho$ et $\hat\rho$ est simplement $T^{\rho}_{\hat\rho}(SU(N))$.

\bigskip

Pour les solutions de domain-wall on a quatre paramètres additionnels, qui sont les deux rayons $L^{\pm}$ des régions asymptotiques  $AdS_5 \times S^5$ et les valeurs asymptotiques du dilaton $g_{\pm}= e^{2\phi^{\pm}}$, donnés par les formules \ref{asymp}. Ces quatres paramètres supplémentaires sont à mettre en lien avec les quatre paramètres $\alpha,\beta,\hat\alpha,\hat\beta$ des fonctions harmoniques. Ils correspondent, à travers la dualité AdS/CFT aux paramètres $N_L, g_{YM}^L, N_R, g_{YM}^R$ décrivant les deux théories Super-Yang-Mills occupant les demi-espaces de part et d'autre du défaut à trois dimensions.

Les expressions explicites des flux de D3-branes décrivant les partitions $\rho$ et $\hat\rho$, ainsi que les paramètres asymptotiques, sont données par \ref{defectN3}, \ref{asympN}. La correspondance avec les ``defect'' quivers découle là aussi de l'image branaire : la solution de domain wall décrite par les paramètres quantifiés $\rho,\hat\rho, N_{\pm},g_{\pm}$ correspont au point fixe infrarouge $\scD(\rho,\hat \rho,-N_-,N_+,g_{-}^{1/2},g_{+}^{1/2})$ \footnote{Ici encore le signe négatif devant $N_-$ est du à un choix de convention qui fixe $N_- <0$.}.

\vspace{10mm}

Les solutions de supergravités correspondant aux quivers circulaires ont été obtenus dans \cite{Assel:2012cj}. l'idée étant de considérer une solution de quiver linéaire sur le bandeau $\Sigma$, contenant une infinité de singularités de type D5-branes sur le bord supérieur et une infinité de singularités de type NS5-branes sur le bord inférieur, réparties de manière périodique le long de la direction ``infinie'' $x$ de $\Sigma$. Les fonctions harmoniques sont alors des séries infinies qui convergent, leur limites étant données par des expressions simples faisant intervenir les fonctions elliptiques $\theta_i$ (\cite{bateman2007higher}). La solution peut alors être tronquée pour ne garder qu'une partie du bandeau correspondant à une période $2t$ dans la direction $x$, les deux bords en $x=0$ et $x= 2t$ étant identifiés. On obtient une solution de supergravité sur l'anneau $\Sigma$. Avec $\tau = it/\pi$, les fonctions harmoniques sont données par 
 \begin{align}
h_1 &= -  \sum_{a =1}^p \gamma_a \ln \bigg[ \frac{\vartheta_{1}\left(\nu_a\vert \qth  \right)}{\vartheta_{2}\left(\nu_a\vert \qth  \right)} \bigg]
 + c.c.    \  , \  \qquad {\rm with}\ \ \
  i\, \nu_a = - \frac{z-\delta_a}{2 \pi } + \frac{i}{4} \ , \no\\
h_2 &= - \sum_{b=1}^{\hat p} \hat \gamma_b \ln \bigg[ \frac{\vartheta_{1}\left(  \hat \nu_b\vert \qth \right)}{\vartheta_{2}\left(  \hat \nu_b\vert \qth \right)} \bigg]
+ c.c.  \ ,
\ \  \qquad {\rm with}\ \ \    i\,  \hat \nu_b =   \frac{z - \hat \delta_b}{2 \pi } \ . \no
\end{align}

Ces solutions ont globalement les mêmes caratéristiques que les quivers linéaires : elles possèdent des singularités ponctuelles de type D5-branes sur le bord supérieur de $\Sigma$ et des singularités ponctuelles de type NS5-branes sur le bord inférieur (voir figure \ref{Annulus}), avec des flux de 3-formes $N_5^{(a)},\hat N_5^{(b)}$ donnés par \ref{ginvN5} et des flux de 5-forme $N_3^{(a)},\hat N_3^{(b)}$ donnés par \ref{emanateD5}, \ref{emanateNS5}. Ces solutions possèdent un flux de 5-forme $L$ indépendant supplémentaire qui correspont au flux circulant autour de l'anneau, lié au paramètre additionnel $t$ et donné par \ref{Lcharge}. Ces flux quantifiés réorganisent les paramètres $\gamma_a, \delta_a, \hat\gamma_b, \hat\delta_b, t$ d'une solution et la caractérisent entièrement. Ils permettent de définir deux partitions $\rho,\hat\rho$ avec la même définition que pour les solutions sur la bandeau (ci-dessus). 
\smallskip

La correspondance avec les quivers circulaires est alors naturelle au vu de la réalisation branaire des quiver circulaires : la solution de supergravité donnée par les partitions $\rho \hat\rho$ et le flux de D3-brane $L$ enroulant l'anneau correspond au point fixe infrarouge $C_{\hat\rho}^{\rho}(SU(N), L)$. Dans la réalisation branaire du quiver, $L$ est logiquement le nombre de D3-branes enroulant la direction compacte $x^3$.

\vspace{5mm}

Une bonne partie de la présentation des solutions de quivers circulaires est consacrée aux subtilités liées aux choix de jauge possibles pour les 2-formes $B_2$ et $C_2$, qui introduisent une ambigu\"ité dans les flux de 5-formes s'échappant des singularités et enroulant l'anneau. Nous montrons comment cette ambigu\"ité est liée au mouvements (de 5-branes) de Hanany-Witten (\cite{Hanany:1996ie}) autour de la direction compacte $x^3$ dans la configuration branaire du quiver circulaire. Ces mouvements de 5-branes créent des D3-branes supplémentaires et changent donc les charges de D3-branes, sans que le point fixe infrarouge en soit modifié. Les différents choix de jauge dans la solutions de supergravité reproduisent exactement les modifications de charges de D3-branes associées au mouvements de Hanany-Witten.  

\vspace{6mm}

Un des premiers tests des correspondence AdS/CFT proposées est la vérification de certaines inégalités sur les partitions $\rho$ et $\hat\rho$. Ces inégalités assurent du côté théorie de quiver que les rangs $N_i$ des noeuds $U(N_i)$ sont positifs et non-nuls. Du côté supergravité, on montre que ces inégalités sont satisfaites dans l'appendice \ref{app:ineq2}.

\vspace{12mm}

La dernière partie du chapitre traite de géométries obtenues dans certaines limites des paramètres et détaille les régimes de paramètres dans lesquels la supergravité de type IIB peut être utilisée de manière perturbative.

\vspace{6mm}

Une des limites décrite, appelée limite de ``wormbrane'', consiste à séparer les paquets de 5-branes en deux groupes très éloignés dans la direction $x$ du bandeau, pour des solutions de quiver linéaires. On obtient alors une géométrie avec deux régions séparées par une région centrale qui s'approche de $AdS_5 \times S^5$ avec un rayon très petit (la géométrie ressemble à la région centrale de $AdS_5$ tronquée à un certain rayon), d'où le nom de ``wormbrane'', qui évoque un trou de ver (``wormhole'') en dimension supérieure. Un schéma qualitatif est présenté en figure \ref{factorize}.  Dans la limite d'une séparation infinie, la région centrale disparait et l'on obtient deux solutions séparées de supergravité sur le bandeau.

Du côté théorie de jauge, cette limite correspond à avoir un rang $N_i$ pour un noeud $U(N_i)$ qui tend vers zero $N_i \rightarrow 0$. Les rang étant des entiers, cette limite peut être vue comme une limite de grand rangs $N_j >> N_i$, pour $j \neq i$, appelée limite de ``noeud faible''. Il est plus aisé pour la discussion d'oublier la quantification des paramètres du quiver pour un moment et de simplement considérer la limite $N_i \rightarrow 0$. Dans cette limite le quiver linéaire se sépare en deux quivers linéaires distincts (sauf cas spéciaux où ce sont les noeuds des extrémités du quiver qui disparaîssent). Nous vérifions explicitement qu'alors les points fixes infrarouges de ces deux quivers linéaires ont pour duaux gravitationnels les deux solutions de supergravité obtenues dans la limite de wormbrane correspondante.

\vspace{6mm}

Dans le cas de l'anneau la limite de worbrane existe et correspond une très grande demi-période de l'anneau $t >> 1$ avec tout les paquets de 5-branes situés dans une région de l'anneau de taille petite devant $t$. La grande région ``vide'' de l'anneau tend vers la géométrie de wormbrane ($AdS_5 \times S^5$ de petit rayon) et dans la limite $t = \infty$, l'anneau devient un bandeau. On peut voir cette limite comme une limite de ``pincement'' où les deux bords de l'anneau se rapprochent en un point et finissent par se toucher, transformant l'anneau en disque, qui est topologiquement identique au bandeau des solutions de quiver linéaires.

Cette limite pour la théorie de quiver circulaire associée correspond là aussi à un ``noeud faible'' $N_i \rightarrow 0$. Le quiver circulaire devient alors un quiver linéaire. La solution de supergravité associée au point fixe infrarouge de ce quiver linéaire est donnée par la limite de wormbrane (ou de pincement) correspondante où l'anneau dégénère en un bandeau.

Une image de cette limite de wormbrane sur l'anneau et la limite correspondante pour le quiver circulaire est donnée figure \ref{degenerate}.

\vspace{8mm}

L'autre limite discutée dans cette partie est la limite $t << 1$ des solutions sur l'anneau, où limite de ``gros anneau''. Cette limite correspond à avoir un grand flux de D3-branes enroulant l'anneau  $L >>1$. Dans cette limite les paquets de 5-branes sont lissés de manière effective dans la direction $x$, qui devient une isométrie de la solutions. La dépendance dans la majaure partie des paramètres disparaît. Ne restent que les paramètres donnant le nombre total de D5-branes $k$, le nombre total de NS5-branes $\hat k$ et la période $t$. Après un changement de coordonnées $2\pi z= 2t  x +i \pi^2  y$, les fonctions harmoniques prennent la forme remarquablement simple 
\begin{align}
h_1  &=   k  \,  \frac{\pi^2 y}{2t}   \no \\
h_2  &=  \hat k   \,  \frac{\pi^2 (1-y)}{2t}  \ \ . \no
\end{align}

Cette limite de grand $L$ est très instructive car elle permet de faire le lien avec les solutions de supergravité de type IIA et de M-théorie. L'isométrie dans la direction compacte $x$ permet de T-dualiser la solution et d'obtenir la solution de type IIA correspondante, puis de calculer la solution de M-théorie (supergravité à 11 dimensions) associée (voir appendice \ref{Tduality}). La solutiona de M-théorie obtenue est purement géométrique (pas de présence de M5-branes) et est donnée par une géométrie $AdS_4 \times S^7/(\bZ_k \times \bZ_{\hat k})$ , où les orbifolds $\bZ_k$ et $\bZ_{\hat k}$ agissent de manière indépendante sur les deux 3-spheres de la fibration $S^7=  S^3 \times S^3 \ltimes I$ ($I$ est un intervalle). Cette géométrie rappelle celle du dual d'ABJM et déjà est connue comme géométrie de M-théorie duale au quivers circulaires dans le cas où les rangs des noeuds sont égaux (à $L$) et très grands (\cite{Imamura:2008nn}). Le cas des quivers circulaires avec rangs différents pour les noeuds a 
aussi été abordé dans \cite{Imamura:2008ji} où les données décrivant le quiver circulaire sont mises en lien avec les holonomies possibles du potentiel $C_3$ sur les différents 3-cycles existants dans la géométrie d'orbifold.

L'étude de la limite de ``lissage`` de grand $L$ met le doigt sur la question plus difficile des dualités avec la supergravité de type IIA et la M-théorie pour les solutions non-lissées. La réalisation présice de ces dualités, notament l'interpretation de la localisation des singularités de 5-branes sur la surface $\Sigma$, semble compliquée et mériterait un travail beaucoup plus approfondi (voir \cite{Tong:2002rq} pour des pistes intéressantes faisant intervenir les instantons de worldsheet dans la T-dualité).

\vspace{8mm}

Enfin nous présentons les régimes de paramètres dans lesquels la supergravité est valide, c'est-à-dire que le rayon de courbure est grand devant la longueur de Planck et la constant de couplage de la corde est faible. Cela revient à avoir 
$R_{r.c.} >> 1$ et $e^{2\phi} << 1$, où $R_{r.c.}$ est le rayon de courbure en unité de longueur de la corde $l_s$ et $\phi$ est le dilaton. Le rayon de courbure et le dillaton varient sur la surface $\Sigma$ et notament divergent au niveau des singularités de 5-branes, rendant la solution de supergravité a priori inadéquate quels que soient les valeurs des paramètres. Cependant on sait que ces divergences doivent être résolues par des corrections de théorie des cordes. il est alors raisonable de penser que la supergravité est utilisable dans un régime de paramètres où les zones de petit rayon de courbure et de grand dilaton sont confinées aux voisinages immédiat des singularités de 5-branes. 

Pour les solutions de quiver linéaires notre analyse montre que le régime de paramètres de supergravité IIB est donné par
\begin{align}
 N >>1 \ , \ k >> \hat k \ , \no
\end{align}
où $k$ est le nombre total de D5-branes et $\hat k$ le nombre total de NS5-branes. 
$N >>1$ assure un grand rayon de courbure $R_{r.c.} >> 1$ et $k >> \hat k$ assure $e^{2\phi} << 1$ dans la majeure partie de la géométrie.

Les solutions de quiver circulaires on un régime de supergravité plus compliqué, du au fait que les direction $x$ et $y$ de l'anneau on des courbure différentes. Le régime est donné par
\begin{align}
 \frac{\hat k}{k} << 1 \quad , \quad 1 << \frac{L \hat k}{k} << \frac{\hat k^5}{k}  \quad  . \no
\end{align}

Lorsque $k << \hat k$ il est possible d'utiliser la solution de supergravité IIB qui est S-duale et qui échange $k$ et $\hat k$.

\vspace{10mm}

{\large {\bf IV. Energie libre dans la limite de grand $N$}}

\vspace{8mm}

Dans ce chapitre nous testons la correspondance AdS/CFT pour les solutions duales des points fixes de quivers linéaires, en vérifiant la relation GKPW
\begin{align}
\left | Z_{\rm CFT} \right |& =e^{-S_{\rm gravity}} \ , \quad
\textrm{i.e.}
\quad
F_{\rm CFT}=S_{\rm gravity} \ , \no
\end{align}
 qui relie l'énergie libre $F_{\rm CFT}=-\ln |Z_{\rm CFT}|$ des théories superconformes à l'action de la supergravité évaluée sur les solutions correspondantes.

Les résultats présentés sont issus de \cite{Assel:2012cp}, sauf pour les commentaires sur le théorème F qui sont nouveaux.

Nous nous concentrons sur une classe de point fixes de quiver linéaires 
$T^{\rho}_{\hat{\rho}}[SU(N)]$ dans la limite de grand $N$, telle que les nombres de 5-branes sont proportionnels à des puissances fractionnaires (positives) de $N$, c'est-à-dire qu'ils sont très grands eux aussi.

\vspace{6mm}

Du côté théories de jauge, nous considérons la limite de grand $N$ des fonctions de partition sur la 3-sphère $S^3$ calculée dans \cite{Benvenuti:2011ga,Nishioka:2011dq} et évaluée dans la limite superconforme (parametres de déformation à zero). L'expression utilisée pour la fonction partition est exacte et issue des techniques de localisation d'intégrales de chemin pour les théories des champs supersymmétriques sur $S^3$ (\cite{Kapustin:2009kz}). Elle dépend des paramètres de déformation de masses et de Fayet-Iliopoulos qui doivent être nuls au point conforme. La limite dans laquelle ces paramètres tendent vers zero dans l'expression de la fonction de partition $Z$ n'est pas simple et nous la calculons uniquement pour les théories conformes de type $T^{[11...1]}_{\hat\rho}(SU(N))$. Le résultats est donc obtenu d'abord pour $N$ fini, puis en calculant le premier terme de l'expansion de grand $N$. 

\vspace{6mm}

Du côté supergravité nous évaluons l'action pour les solutions correspondantes. Une grande simplification des calculs vient du fait que la solution à 10 dimensions peut être tronquée (''consistent truncation``) à une solution de pure gravité à 4-dimensions sur $AdS_4$ avec un certain rayon qui dépend des fonctions harmoniques $h_1$ et $h_2$ de la solution de départ. Evaluer l'action de supergravité revient alors à évaluer l'action de Einstein-Hilbert à 4-dimensions avec constante cosmologique négative. Cette action est divergente car l'espace $AdS_4$ possède un volume infini. Elle est régularisée par les techniques connues de renormalisation holographiques (\cite{deHaro:2000xn}), qui consitent à ajouter un contreterme sur le bord de l'espace. Les détails de cette régularisation sont détaillés dans le premier chapitre introductif. le volume régularisé de l'espace euclidien $AdS_4$ de rayon $L$ est ${\rm vol}_{AdS_4} = (4/3) \pi^2 L^4$

L'expression explicite (et remarquablement simple) que nous trouvons pour l'action d'une solution de supergravité est 
\begin{align}
S_{\rm eff} = -\frac{1}{(2\pi)^7 (\alpha')^4} {\rm vol}_6 \left( \frac{4}{3} \pi^2 \right) (-6) \ , \no
\end{align}
avec 
\begin{align}
{\rm vol}_6 = 32 (4 \pi)^2 \int_{\Sigma} d^2x (-W) h_1 h_2 \ , \no
\end{align}
où $W = \p \bar \p(h_1 h_2)$.

Dans la limite de grand $N$ les fonction harmoniques $h_1, h_2$ prennent une forme relativement simple et cette formule permet d'évaluer le terme dominant de l'action.

\vspace{6mm}

Nous trouvons dans les deux cas une contribution principale à l'énergie libre dans la limite de  grand $N$ qui se comporte en
\begin{equation*}
F\sim N^2 \ln N+{\cal O}(N^2) \ .
\end{equation*}

Du côté théorie de conforme $N^2 \ln N$ vient du comportement asymptotique de la fonction $\ln G(N)$, où $G$ est la fonction de Barnes (voir appendice \ref{app:Barnes}). Du côté gravité le facteur $N^2$ vient du comportement des champs à grand $N$, et le facteur $\ln N$ vient de la taille de l'espace compact.

\vspace{8mm}

Les résultats sont les suivants :

\begin{itemize}
\item l'exemple le plus simple est la théorie super-conforme $T[SU(N)]$, qui est la théorie $T^{\rho}_{\hat{\rho}}[SU(N))]$ avec
\beq
\rho = \hat\rho = \big[\overbrace{1,1,...,1}^{N}\big] \ . \no
\eeq

Le calcul de la fonction de partition au point conforme donne
\beq
Z_{\rm CFT}=\frac{1}{(N-1)!(N-2)! \ldots 2!
1!}\left(\frac{1}{2\pi}\right)^{\frac{N(N-1)}{2}} =\frac{1}{G(N+1)} \left(\frac{1}{2\pi}\right)^{\frac{N(N-1)}{2}} \ .
\no
\eeq

la solution de supergravité possède un paquet de $N$ D5-branes et un paquet de $N$ NS5-branes, séparés par une distance $2\delta \simeq \ln N$ dans la limite de grand $N$ (voir figure \ref{strip_tsun}). La géométrie possède alors trois régions distinctes : une région centrale $-\delta < x < \delta$ entre les paquets de 5-branes qui contient la contribution dominante à l'action et qui est responsable de l'apparition du facteur $\ln N$, et deux régions externes $|x| > \delta$ dont les contributions sont sous-dominantes. Après un changement de variable $z = \delta x + i y$ les fonction harmoniques dans la région centrale $-1 < x <1$ sont données par
\begin{align}
h_1 &\simeq 4 \sin(y) N\, e^{\delta(x-1)} \ ,\no\\
h_2 &\simeq 4 \cos(y) N\, e^{-\delta(1+x)} \ . \no
\end{align}

Dans ce cas, nous trouvons
\beq
F_{\rm CFT}=S_{\rm gravity}=\frac{1}{2} N^2 \ln N +\scO(N^2) \ . \no
\eeq

\item Plus généralement nous considérons les cas où l'on a un seul paquet de NS5-branes (ou un seul paquet de D5-branes), i.e.,
\begin{align}
\begin{split}
\rho &= \Big[ \overbrace{l^{(1)},l^{(1)},..,l^{(1)}}^{N_{5}^{(1)}},\ \overbrace{l^{(2)},l^{(2)},..,l^{(2)}}^{N_{5}^{(2)}},\ ...\ ,
\ \overbrace{l^{(p)},l^{(p)},..,l^{(p)}}^{N_{5}^{(p)}} \Big] \ ,   \no\\
\hat \rho &=
\Big[ \overbrace{\hat l,\hat l,..,\hat l}^{\hat N_5} \Big] \ . \no
\end{split}
\end{align}
On choisit aussi les dépendances en $N$ suivantes
\beq
N_5^{(a)} =N^{1-\kappa_a} \gamma_a,\quad l^{(a)} = N^{\kappa_a}
\lambda^{(a)},\quad
\hat N_5 = N \hat \gamma \ . \no
\eeq
On étudie la limite de grand $N$ à $\kappa_a, \lambda^{(a)}, \gamma_a, \hat{\gamma}$ fixés et
on impose aussi
\beq
\kappa_{a-1}\ge \kappa_a,\quad  0\le \kappa_a<1,   \quad \textrm{pour tout  } a
\ .
\no
\eeq
La première condition est nécessaire pour que $\rho$ soit une partition de $N$ avec des linking numbers décroissants, et la seconde assure que les $N_5^{(a)}$ deviennent larges, ce qui rend le calcul réalisable.

Le calcul de la fonction de partition n'est donné que pour le cas où $\hat{l}=1$.

La solution de supergravité possède un paquet de NS5-branes et plusieurs paquets de D5-branes, tous étant séparés par les distances d'ordre $\ln N$. Le bandeau $\Sigma$ est alors divisé en plusieurs régions de taille d'ordre $\ln N$ où les fonctions harmoniques prennent des formes simples comme dans le cas de $T(SU(N))$. Les régions centrales contribuent toutes à l'ordre dominant à l'action, tandis que les deux régions externes sont sous-dominantes.

Dans ce cas plus général on trouve :
\begin{align}
F_{\rm CFT}=S_{\rm gravity}= \frac{1}{2} N^2\ln N \left[  (1-\kappa_1)
+ \sum_{i=2}^p \left( \sum_{a=i}^p \gamma_a \lambda^{(a)} \right)^2\left( \kappa_{i-1}-\kappa_i\right)
\right] + {\cal O}(N^2).
\no
\end{align}

\end{itemize}

Nos résultats confirment les prédictions de la correspondence AdS/CFT.

\vspace{8mm}

Pour finir on fait le lien avec le théorème F, qui est encore une conjecture et qui stipule que deux théories conformes $\scT_{UV}$ et $\scT_{IR}$ reliées par un flow de renormalisation (de l'ultraviolet UV à l'infrarouge IR) ont des énergies libres qui vérifient $F_{UV} > F_{IR}$ (\cite{Jafferis:2010un,Jafferis:2011zi}). 

Nos résultats indiquent que l'énergie libre de la théorie $T(SU(N))$ est la plus grande parmis les théories que nous considérons. Nous sommes ammenés à postuler
\beq
F_{T^{\rho}_{\hat{\rho}}[SU(N)]}\le F_{T[SU(N)]} \ ,
\no
\eeq
pour toute théorie $T^{\rho}_{\hat{\rho}}[SU(N)]$.

Il est possible d'expliquer ces résultats à l'aide du théorème F. Nous montrons en nous appuyant sur la representation branaire des théories $T^{\rho}_{\hat{\rho}}[SU(N)]$, comment il est possible d'initier un flow de renormalisation entre $T[SU(N)]$ et une théorie infrarouge $T^{\rho}_{\hat{\rho}}[SU(N)]$ quelconque, en se déplaçant sur la branche de Coulomb et la branche de Higgs de l'espace des modules de $T[SU(N)]$. Nos considérations nous amènent aussi à la conjecture 
\begin{align}
\left\{
\begin{array}{c}
\rho_1 \geq \rho_2 \\
\hat{\rho}_1 \geq \hat{\rho}_2 
\end{array}
\right.
  \quad  \Longrightarrow  \quad   
  F_{T^{\rho_1}_{\hat{\rho}_1}[SU(N)]}  \leq F_{T^{\rho_2}_{\hat{\rho}_2}[SU(N)]}   \no
\end{align}
qui est en accord avec nos résultats. Nous fournissons donc par nos calculs un élément supplémentaire qui accrédite le théorème F.

\vspace{10mm}

{\large {\bf V. Solutions avec $(p,q)$ 5-branes et théories de Chern-Simons}}

\vspace{8mm}

Dans ce chapitre nous présentons une extension des solutions de supergravité à des solutions avec axion non-nul en utilisant la symmétrie $SL(2,\bR)$ de la supergravité IIB. Les solutions reliées par les transformations $SL(2,\bZ)$ sont équivalentes au niveau quantique car le groupe $SL(2,\bZ)$ est un groupe de symmétrie de la théorie des cordes IIB. 
Les transformations $SL(2,\bR)$ génèrent des solutions équivalente de la supergravité IIB classique mais ne sont valides pour la théorie quantique sous-jacente. Par des transformations $SL(2,\bR)$ des solutions avec axion nul (qui sont les solutions étudiées jusqu'ici) ont peut générer des solutions de supergravité correspondant à d'autres théories superconformes. 

Les transformation de $SL(2,\bR)$ sont données par 
\begin{align}
 S^\prime =  {a S +  b\over  c S + d}\ , \qquad
  \left(   \begin{array}{c}
\, \, H_{(3)}^\prime  \\
F_{(3)}^\prime
 \end{array}
  \right) =
   \left(   \begin{array}{cc}
 d & -c \\ - b & a
  \end{array}
  \right)
  \left(   \begin{array}{c}
\, H_{(3)}  \\
F_{(3)}
 \end{array}
  \right) \ , \no
 \end{align}
où $ad-bc=1$ et $S = \chi + i e^{-2 \phi}$ est l'axion-dilaton.

La stratégie pour trouver toutes les solutions possibles consiste à appliquer une transformation générale de $SL(2,\bR)$ à la solution d'axion nul avec des paramètres non-quantifiés, puis à quantifier les flux dans un second temps. Pour obtenir l'ensemble des solutions inéquivalentes, ont se ramène à des solutions ''canoniques`` par des transformations de $SL(2,\bZ)$.
De cette manière on obtient de nouvelles solutions contenant des $(p,q)$ 5-branes. Les solutions inéquivalentes sont classifiées par la donnée de singularités de NS5-branes sur un bord de $\Sigma$ et de singularités de $(p,q)$ 5-branes, avec $0 \le p \le |q|$, sur l'autre bord de $\Sigma$ (voir figure \ref{annulusSL2R}). Les solutions avec D5-branes correspondent à $(p,q)=(0,1)$.  

\vspace{6mm}

Les théories de jauges superconformes duales ne sont pas aisément descriptibles (voir \cite{Gaiotto:2008ak}). Dans le cas simple où les singularités sont de type NS5-branes et $(1,k)$ 5-branes, il est possible de décrire les théories superconformes en termes de théories de Chern-Simons à trois dimensions avec supersymétrie étandue $\N=4$, où $\pm k$ correspond au niveau de Chern-Simons de certain noeuds unitaires du groupe de jauge.

A titre d'exemple on donne le dual de supergravité de la théorie ABJM, qui est une solution sur l'anneau avec une NS5-brane et une $(1,k)$ 5-brane. Cette théorie n'est pas $SL(2,\bZ)$-équivalente à une théorie d'axion nul, sauf dans le cas $k=1$.

\vspace{8mm}

Les symétries $SL(2,\bR)$ de la supergravité classique se traduisent du côté théories de jauge par des équivalences ``orbifold'' entre différentes théories.
Cette pseudo-équivalence prédit l'égalité entre observables de théories de jauge différentes (non-équivalentes) dont les quantités associées du côté gravité sont invariantes par les transformations $SL(2,\bR)$. L'égalité entre ces observables ``untwisted'' n'existe a priori que  dans un régime des paramètres où les calculs de supergravité sont corrects (corrections de théorie des cordes négligeables), ce qui implique une limite de grand $N$. la terminologie d'équivalences ``orbifold'' vient de résultats analogues de pseudo-équivalence entre des théories conformes dont les duaux de M-théorie sont reliés par l'action de certains orbifolds. Dans notre contexte il n'y a pas d'orbifolds.

\vspace{6mm}

Pour finir ce chapitre nous testons notre proposition de correspondence orbifold dans la limite de grand $N$ sur les théories conformes données par les quivers circulaires suivant (voir figure \ref{slQ}):

\begin{itemize}
 \item Quiver composé d'une chaîne (circulaire) de $\hat k$ noeuds $U(N)$ et $M$ hyper-multiplets fondamentaux pour chaque noeud. La configuration branaire associée a $N$ D3-branes enroulées sur la direction compacte $x^3$, croisant $\hat k$ NS5-branes et $M$ D5-branes entre chaque paire de NS5-branes.

 \item Quiver composé d'une chaine cicrulaire de $2 \hat k$ noeuds $U(N)$ avec termes de Chern-Simons pour chaque noeud alternant entre les niveau $+ M$ et $-M$ le long de la chaine. La configuration de branes associée contient $N$ D3-branes enroulées sur la direction compacte $x^3$, croisant $\hat k$ NS5-branes et une $(1,M)$ 5-brane entre chaque paire de NS5-branes, càd que les $M$ D5-branes ont été remplacée par une $(1,M)$ 5-brane.
\end{itemize}
 
La transformation reliant les duaux de supergravité est donnée par la matrice de $SL(2,\bR)$ :
\bea
   \left(   \begin{array}{cc}
  1  &    M^{-1}  \\  0   &  1
  \end{array}
  \right)\qquad {\rm with}\quad M\in \mathbb{N} \, . \no
\eea

Nous étudions le model de matrice associé à chaque quiver dans la limite de grand $N$ qui correspond à un grand nombre de valeur propres (variables d'intégrations). Dans cette limite on peut remplcer l'intégrale matricielle par une intégrale sur une densité continue de valeurs propres et résoudre plus simplement les équations du point scelle qui donnent le comportement dominant de la fonction de partition (ou directement de l'énergie libre). Le calcul pour la théorie de Chern-Simons a déjà été présenté dans \cite{Herzog:2010hf}. Nous complétons per le calcul de l'énergie libre de l'autre théorie impliquée dans la dualité.

Nous trouvons un accord entre les deux résultats 
 \begin{align}
F_{CFT}  = \frac{\pi \sqrt{2}}{3}\hat k \sqrt{M} N^{\frac{3}{2}}\ . \no
\end{align}
Ce résultats est aussi reproduit par le calcul de l'action de supergravité, confirmant encore la correspondence holographique.

\vspace{10mm}

{\large {\bf Perspectives futures}}

\vspace{8mm}

Le travail de thèse présenté apporte une extension significative et précise des correspondences $AdS_4/CFT_3$ mettant en jeu les théories superconformes $\N=4$ à trois dimensions. Il semble cependant que certaines théories conformes nous échappent encore. Ces théories décrites dans \cite{Benini:2010uu} prennent la forme de ``quivers étoilés'' et possèdent des noeuds $SU(N)$ attachés à trois hypermultiplets bifondamentaux. Il est possible que des solutions de supergravité analogues à celles que nous avons présentées soient duales à ce type de théories superconformes. Il s'agirait alors de trouver des fonctions harmoniques $h_1,h_2$ sur un disque $\Sigma$ dont le bord est divisé en plus que deux segments, càd que le bord de $\Sigma$ présenterait une séquence de segments avec des paquets de D5-branes et de NS5-branes. Il pourrait aussi s'agir de solutions où $\Sigma$ est une surface de plus grand genus. Jusqu'à présent la recherche de telles solutions s'est heurtée à la présence de singularités ponctuelles 
coniques à l'intérieure de $\Sigma$, pour lesquelles nous n'avons pas d'interprétation (en théorie des cordes).

Une autre voie que nous avons explorée, mais qui n'a pas encore fournit ses conclusions, concerne l'étude du scénario de Karch-Randall dans les géométries de domain-wall (\cite{Karch:2000ct,Karch:2001cw}). L'idée est qu'une géométrie obtenue à partir de la configuration de brane faite de D3-branes intersectant un paquet de D5-branes pourrait conduire au phénomène de localisation de la gravité. Plus précisément le spectre du graviton à 4-dimensions (dans $AdS_4$) aurait un mode zero de très petite masse comparée au reste du spectre du graviton et dont la fonction d'onde dans l'espace interne non-compact serait localisée au voisinage du paquet de D5-branes. Ce modèle est le seul (à notre connaissance) qui reproduit une gravité à quatre dimensions avec un espace interne non-compact. Les solutions de supergravité étudiées dans cette thèse correspondent exactement aux géométries candidates pour le scénario de Karch-Randall, avec la possibilité d'enrichir l'image par la présence de plusieurs paquets de D5-branes 
et NS5-branes. Le spectre de gravitons pour les géométries de domain-wall de type Janus (sans 5-branes) a été étudié dans \cite{Bachas:2011xa}, où les auteurs ont montrés que les éléments du modèle de Karch-Randall n'étaient pas réunis. L'analyse des solutions de domain-wall avec 5-brane n'a pas encore donné de conclusions définitives, même si les indications obtenues jusqu'ici tendent à montrer la localization de la gravité n'est pas reproduite dans les situations les plus simples.

\chapter{Elements of AdS/CFT correspondence}
\addcontentsline{lot}{chapter}{ Elements of AdS/CFT correspondence }
\label{chap:AdSCFT}


The purpose of this introductory section is to remind some elements of the celebrated AdS/CFT correspondence. Especially we emphasize the derivation of the correspondence between 4-dimensional $\N=4$ Super-Yang-Mills gauge theory and type IIB string theory on $AdS_5\times S^5$ in the original setup of Maldacena \cite{Maldacena:1997re}, using the low-energy descriptions of stacks of D3-branes.

Many details are eluded. We focus on the general ideas that are important for this presentation. We refer to the reviews \cite{Aharony:1999ti,D'Hoker:2002aw} for a pedagogical introduction to the AdS/CFT correspondence. \\
We also assume that the reader has a background knowledge in string theory and supersymmetric gauge theories in various dimensions. The standard textbooks are \cite{polchinski1998string,kiritsis2011string} for string theory and \cite{wess1992supersymmetry} for supersymmetry. For D-branes we recommend \cite{johnson2006d}.

\section{Low-energy descriptions of D3-branes}
\label{sec:Msetup}

\begin{figure}[!h]
\centering
\includegraphics[height=8cm,width=16cm]{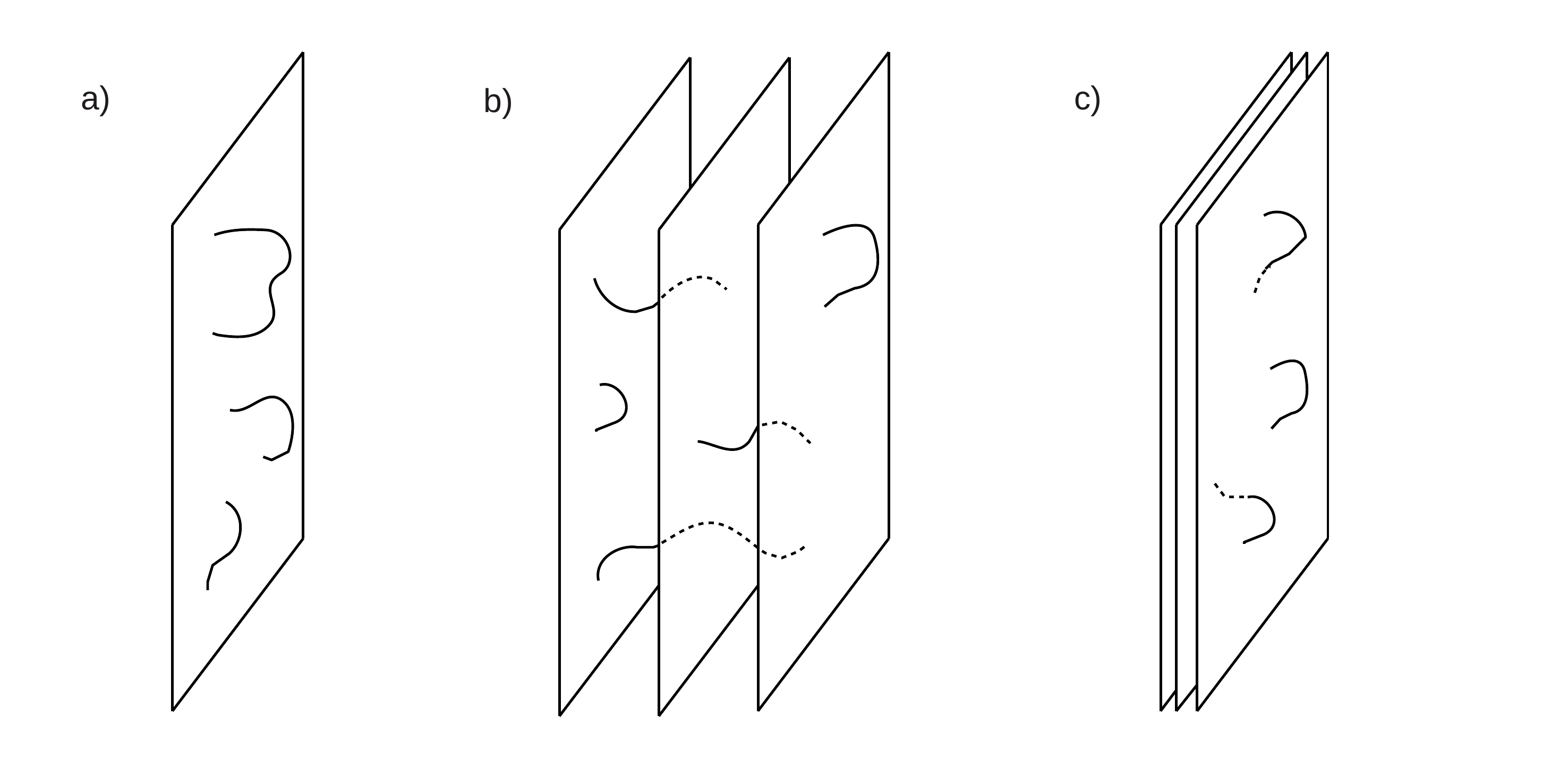}
\caption{\footnotesize{a) A single D$p$-brane with open strings attached. b) Several parallel D$p$-branes. The open strings can end on different branes. c) A stack of coincident D$p$-branes with enhanced worldvolume gauge symmetry.}}
\label{fig1_1}
\end{figure}

The story begins by considering D3-branes in string theory. D$p$-branes are solitonic objects defined as boundary conditions for open strings.\\
 If $X^{M}(\sigma,\tau)$, $M = 0,1, \cdots , 9$, denote the target space coordinates of the open string and $\sigma \in [0,\pi]$, $\tau \in \bR$ are the worldsheet coordinates, the boundary conditions
\begin{align}
 \p_{\sigma}X^{\mu}(0,\tau) &=\p_{\sigma}X^{\mu}(\pi,\tau) = 0 \quad  \textrm{for} \quad \mu = 0,1, \cdots, p  \no\\  \p_{\tau}X^{i}(0,\tau) &= \p_{\tau}X^{i}(\pi,\tau) = 0  \quad  \textrm{for} \quad  i= p+1, \cdots, 9 
\end{align}
define a D$p$-brane.\\
Saying it more simply, the D$p$-brane is a flat $p+1$ dimensional objects where the endpoints of open strings are attached, as pictured in figure \ref{fig1_1}. These endpoints are sources for a $U(1)$ gauge field on the $p+1$ dimensional worldvolume of the brane.\\
The D$p$-branes with $p$ odd are $\half$-BPS solitons in type IIB string theory, which means that they preserve 16 out of the 32 real supercharges of the 10-dimensional $\N=1$ Poincaré superalgebra. \footnote{The D$p$-branes with even $p$ breaks all supersymmetry in type IIB string theory. The situation is inversed in type IIA string theory where even $p$ means $\half$-BPS while odd $p$ means non-supersymmetric.}

\vspace{5mm}

\underline{\bf{D-branes described with open strings :}}

\vspace{5mm}

With $N$ parallel D$p$-branes as in figure \ref{fig1_1} the open string spectrum contains generically $N$ copies of $U(1)$ gauge fields coupled through massive excitations corresponding to strings stretched between different D$p$-branes. In the low energy limit the $N$ worldvolume theories decouple. However if the $N$ D$p$-branes are on top of each other, the lowest modes of open strings stretched between different D$p$-branes become massless and the worldvolume gauge symmetry is enhanced to $U(N)$.

\vspace{2mm}

Let's consider a stack of $N$ coincident D3-branes. The worldvolume $U(N)$ gauge theory from open strings is 4-dimensional and preserve 16 supercharges, corresponding to $\N=4$ supersymmetry. The low-energy limit is obtained by keeping only the massless fields living on the brane and corresponds to the well-known $\N=4$ Super-Yang-Mills gauge theory with gauge group $U(N)$.
\smallskip

Let's give a rapid description of the theory. It contains only an $\N=4$ vector multiplet $(A^{\mu}, \lambda^a, X^i)$, so all fields are in the adjoint representation of $U(N)$. The fields charged under the overall $U(1)$ of $U(N)=U(1)\times SU(N)$ (the trace of the matrices) decouple from the theory and we usually consider only the $SU(N)$ gauge theory.\\
The bosonic fields are a vector field $A^{\mu}$ and 6 real scalars $X^i$ corresponding to the position of the stack of D3-branes in the 6 transverse dimensions. The fermionic fields are 4 Weyl fermions $\lambda^a$.
\smallskip

The $\N=4$ SYM theory is superconformal,its full supergroup of symmetries is $SU(2,2|4)$. 
In particular the bosonic symmetries are the spacetime $SO(2,4) \sim SU(2,2)$ combining the 4-dimensional Poincaré symmetries (translations, rotations, Lorentz boosts) and the conformal symmetries (dilatation, special conformal transformations), and the $SO(6)_R \sim SU(4)_R$ R-symmetry under which the fields transform as $(A^{\mu}, \lambda^a, X^i) = (\mathbf{1},\mathbf{4},\mathbf{6})$. The fermionic symmetries are 16 Poincaré supersymmetries and 16 conformal supersymmetries.
\smallskip

There is a single complex coupling $\tau = \frac{\theta_{YM}}{2\pi} + \frac{4 \pi i}{g_{YM}^2}$. The Lagrangian is given by

\begin{align}
\scL & = \tr \biggl \{
-{1 \over 2 g_{YM}^2} F_{\mu \nu} F^{\mu \nu}
+ {\theta_{YM} \over 8 \pi ^2} F_{\mu \nu} \tilde F^{\mu \nu}
- \sum _a i \bar \lambda ^a \bar \sigma ^\mu D_\mu \lambda _a
- \sum _i D_\mu X^i D^\mu X^i
\no \\
 & \quad
+  \sum _{a,b,i} g_{YM} C^{ab} _i \lambda _a [X^i, \lambda _b] +
\sum _{a,b,i} g_{YM} \bar C_{iab}  \bar \lambda ^a [X^i, \bar \lambda ^b]
+ {g_{YM}^2 \over 2} \sum _{i,j} [X^i , X^j]^2 \biggr \}
\end{align}
where the constants $C^{ab}_i$ and $C_{iab}$ are related to the Clifford
Dirac matrices for $SO(6) _R \sim SU(4)_R$.
\smallskip

The quantum theory enjoys an $SL(2,\bZ)$ group of dualities under which the $\tau$ parameter transforms as
\begin{align}
 \tau \rightarrow \frac{a\tau + b}{c\tau + d} \quad , \quad ad - bc = 1 \quad , \quad a,b,c,d \in \bZ \ .
\end{align}

The low energy description of $N$ coincident D3-branes in type IIB string theory contains on one side the low excitations of open strings attached to the D3-branes, which reduce to the 4-dimensional $\N=4$ Super-Yang-Mills conformal gauge theory \footnote{This implies sending the string length $l_s$ to zero, suppressing higher derivative terms.}, and on the other side the low excitations of closed strings which is the 10-dimensional flat space IIB supergravity. In the low energy limit (and string length $l_s \rightarrow 0$) these two pieces decouple because the interaction terms are proportional to positive powers of the supergraviy Newton constant (which tends to zero).

\vspace{5mm}

\underline{\bf{D-branes as solitons in 10-dimensions :}}

\vspace{5mm}

The D3-branes have a dual description in string theory as extended objects in 10 dimensions which are sources for supergravity fields. The $\half$-BPS soliton corresponding to a D$p$-brane in IIB supergravity is the extremal $p$-brane whose metric and dilaton are given by
\begin{align}
 ds^2 &= H(y)^{-\half} \ dx^{\mu}dx_{\mu} + H(y)^{\half} \ d\vec y^2 \quad , \quad e^{\Phi} = g_s  \, H(y)^{\frac{3-p}{4}} \no\\
  H(y) &= 1 + \frac{L^{7-p}}{y^{7-p}} \quad , \quad y \equiv \sqrt{\vec y^2}
\end{align}
where $x^{\mu}$, $\mu = 0,1, \cdots , p$ parametrize the coordinates parallel to the $p$-brane and $\vec y =(y^i)$, $i=p+1, \cdots , 9$ are the transverse coordinates. This metric corresponds to a $p$-brane located at $\vec y = \vec 0$. It has $SO(1,p)\times SO(9-p)\times \bR^{p+1}$ isometries. The (extremal) $p$-brane solution also has non-vanishing $8-p$-form flux $F_{p+2}$ sourced by the $p$-brane, depending on the harmonic function $H(y)$.
\smallskip

The radius $L$ of the $p$-brane is related to the string coupling $g_s$ and the string length $l_s = \alpha^{' \half}$ through the relation
\begin{align}
 L^{7-p} = (4 \pi)^{\frac{5-p}{2}} \Gamma \left(\frac{7-p}{2}\right) \ g_s N \, l_s^{7-p} \ ,
\end{align}
where $N$ corresponds to the number of coincident D$p$-branes in the string picture
\begin{align}
 N = \frac{1}{2 \kappa_{10}^2 \, T_p} \int_{S^{8-p}} \star F_{p+2}  \ ,
\end{align}
with $2\kappa_{10}^2 = (2\pi)^7 l_s^8 g_s^2$ and $T_p = [ (2\pi)^p l_s^{p+1} g_s ]^{-1}$ is the D$p$-brane tension setting the unit in which the flux is quantized.

\vspace{2mm}

Specializing to a stack of $N$ D3-branes, the 10-dimensional backreacted geometry in IIB supergravity is
\begin{align}
\label{3brane}
 ds^2 &= \left( 1+ \frac{L^4}{y^4} \right)^{-\half} dx^{\mu}dx_{\mu} 
+ \left( 1+ \frac{L^4}{y^4} \right)^{\half} (d y^2 + y^2 d\Omega_5^2) \no\\
e^{\Phi} &= g_s \quad , \quad C \ \textrm{constant} \ ,\\
F_5 &= (1+\star)dx^0 dx^1 dx^2 dx^3 d(H(y)^{-1})  \no\\
\frac{L^4}{l_s^4} &= 4\pi g_s N \ , \no
\end{align}
where $d\Omega_5^2$ is the metric of the unit radius 5-sphere and the axion field $C$ is also non-zero (it is constant).

\vspace{2mm}

The coefficient $g_{00}$ of the metric varies along the radial direction $y$ in such a way that the energy of an object at radial position $y$ of the geometry measured by an observer at infinity goes to zero as the object approaches the center $lim_{y \rightarrow 0} E(y) = 0$. So the low-energy limit of IIB string theory on this background contains excitations localized near $y=0$, plus the very large wavelength excitations that are those of type IIB flat spacetime supergravity. The two sectors decouple essentially because the large wavelength modes cannot probe the near horizon region.

In the limit $y\rightarrow 0$ the geometry \ref{3brane} asymptotes to
\begin{align}
 ds^2 = L^2 \left( \frac{1}{u^2} \, dx^{\mu}dx_{\mu} + \frac{du^2}{u^2} + d\Omega_5^2 \right)
\end{align}
with $u = L^2/y$. The limit geometry is regular everywhere. This is actually the famous $AdS_5 \times S^5$ spacetime with equal radius $L$ for the $AdS_5$ and $S^5$ part. 
Thus the sector of the theory describing modes localized near $y=0$ or $u = \infty$ is type IIB string theory on $AdS_5 \times S^5$ background.

\vspace{6mm}

The last step to reach the Maldacena's proposal of AdS/CFT correspondence is to identify the two descriptions that we have summarized and to drop the decoupling flat 10-dimensional type IIB supergravity that appears in both descriptions.

\vspace{4mm}

The identification of the two reamining pieces leads to the AdS/CFT conjecture :

\vspace{5mm}

\begin{center}

\fbox{
\begin{minipage}{12cm}
\begin{center}
\vspace{4mm}

{\it  $\N=4$ Super-Yang-Mills on $\bR^{1,3}$ with gauge group $SU(N)$}

\vspace{4mm}

 $\Updownarrow$ 

\vspace{4mm}

{\it Type IIB string theory on $AdS_5 \times S^5$ with radius $L$ }. 
\vspace{4mm}

\end{center}
\end{minipage}
}

\end{center}

\vspace{8mm}

The parameters of the two theories are identified as follows
\begin{align}
\label{match}
 g_{YM}^2 = g_s \quad , \quad \theta_{YM} = C \quad , \quad \frac{L^4}{l_s^4} &= 4\pi g_s N  \ .
\end{align}
The meaning of this correspondence will be explained in the next subsection.

This form of the conjecture is the strongest as it is meant for any values of $N$ and $g_{YM}$, however it can be tested in practice only in some regimes of parameters where both sides of the correspondence are tractable.\\
On the SYM side we can use perturbation theory in the weak coupling limit. Allowing for large values of $N$, the effective coupling is $\lambda = g_{YM}^2 N$, known as the 't Hooft coupling. In the 't Hooft limit, where $\lambda$ is fixed and $N$ is large, the perturbation expansion of Feynman diagrams in powers of $\frac{1}{N}$ becomes topological. This means that the diagrams are weighted by $N^{\chi}$, where $\chi$ is the Euler characteristic of the surface on which the diagram can be drawn. The dominant contribution comes for planar diagrams which are the diagrams one can put on a 2-sphere, the next contribution comes from the diagram one can put on a torus, ...etc. This topological expansion is similar to the perturbative expansion of closed string amplitudes. In this planar limit, the loop expansion on the sphere is an expansion in powers of $\lambda$ (sigma-model loop expansion), so perturbative computations can be done only for small $\lambda$.

On the string theory side the tractable supergravity description is obtained in the limit of large $L/l_s$ and the weak coupling regime correspond to small $g_s$. Looking back at \ref{match} it means $g_s << 1$ and $g_s N >> 1$. This is possible only if $N >> 1$.

We conclude that the supergravity limit is obtained for $N >> 1$ and large $\lambda$, while the weak coupling limit of SYM in the planar limit corresponds to small $\lambda$. These two regimes are incompatible, expaining why the conjectured correspondence is difficult to check. The regime of parameters that is mostly studied is this 't Hooft limit or planar limit, 
\begin{align}
 N \rightarrow \infty \quad , \quad \lambda= g_s N \quad \textrm{fixed} \ ,
\end{align}
for which integrability techniques can be used to study $\N=4$ SYM theory for arbitrary $\lambda$ .

\section{Elements of correspondence}
\label{sec:corresp}

The first prediction of the AdS/CFT duality is that the global symmetries of both sides should match. This is the case.
Let's compare the bosonic symmetries. The isometry group of $AdS_5$ is $SO(2,4) \sim SU(2,2)$ corresponding to the 4-dimensional conformal group of SYM and the isometry group of $S^5$ is $SO(6) \sim SU(4)$ corresponding to the R-symmetry of SYM.\\
The discrete $SL(2,\bZ)$ symmetry of SYM is mapped to the $SL(2,\bZ)$ symmetry of type IIB string theory which is preserved by the D3-branes.

\vspace{5mm}

 The next prediction is that the spectrum of (gauge invariant) operators in SYM theory should be in one-to-one correspondence with the $AdS_5$ fields, obtained by expanding the 10d fields in harmonics of $S^5$. More precisely the representations of $SU(2,2|4)$ should be mapped and masses $m$ of $AdS_5$ fields are related to scalling dimensions $\Delta$ of SYM operators. \\
It is a very difficult problem to find the match in general, especially because the full IIB string theory spectrum on $AdS_5\times S^5$ is not known.\\
Among the gauge multiplets a special role is played by the {\it chiral} multiplets or BPS multiplets whose primary operators are anihilated by at least one supercharge. The chiral multiplets thus belong to shorten representations and have the property that their scaling dimension is not renormalized by quantum corrections. The relation to $AdS_5$ fields is then easier to find. Generically single trace operators in SYM correspond to single particle (canonical) fields in $AdS_5$ (\cite{Maldacena:1997re,Andrianopoli:1998jh}). 

\vspace{5mm}

The correspondence between $AdS$ fields and gauge theory operator is the key ingredient to relate the two dual theories. In \cite{Gubser:1998bc,Witten:1998qj} it was argued that the gauge theory could be thought of as living on the boundary at infinity of $AdS_5$ with the asymptotic values of the $AdS$ fields $\phi(x^{\mu},u=0)$ playing the role of sources for their dual operator $\scO(x^{\mu})$ in the gauge theory. This lead to the crucial GKPW relation 
\begin{align}
\label{GKPW}
 \left< e^{\int d^4 x \, \phi_0(x^{\mu}) \scO(x^{\mu}) } \right>_{CFT} = Z_{string} \Big[ \phi(x^{\mu},u=0) = \phi_0(x^{\mu}) \Big] \ ,
\end{align}
where the left-hand side is the generating functional of correlation functions for the operator $\scO$ in the gauge theory and
the right-hand side is the string theory partition function with asymptotic values $\phi_0$ for the $AdS$ field $\phi$.

In the supergravity regime the right-hand side can be approximated by the saddle point
\begin{align}
\label{GKPW2}
 \left< e^{\int d^4 x \, \phi_0(x^{\mu}) \scO(x^{\mu}) } \right>_{CFT} \simeq e^{-S_{SUGRA}[\phi_0]} \ .
\end{align}
It is then possible to compute gauge theory correlation functions as functional derivatives of the right-hand side with respect to the boundary values $\phi_0(x^{\mu})$. The 5-dimensional action used in such a computation comes from dimensional reduction of the $S^5$ part and the effective 5d gravitational constant is $G_5 = \pi/(4 N^2)$.\\
In the large $N$ limit the computations can be organized in a perturbative expansion in $1/N$, using the so-called Witten diagrams. The endpoints of Witten diagrams lie on the boundary on $AdS$ and the building blocks are {\it boundary-to-bulk} propagators and {\it bulk-to-bulk} propagators for each AdS field. The perturbative expansion is again a loop expansion.
\bigskip

The simplest prediction from the GKPW relation \ref{GKPW} is the equality between the free energy $F_{CFT} \equiv - \log|Z_{CFT}|$ and the action $S_{sugra}$ evaluated on the dual supergravity background $g_{\mu\nu}^{(0)}, \phi^{(0)}, ...$ 
\begin{align}
\label{GKPW3}
 F_{CFT} = S_{sugra}[g_{\mu\nu}^{(0)}, \phi^{(0)}, ...] \ .
\end{align}

The UV divergences one encounters on the gauge theory side have a counterpart on the gravity side as IR divergences related to the infinite size of the AdS spacetime. Imposing a UV cutoff in the gauge theory translates into imposing a radial cut-off in AdS. The regularization techniques of the gravity computations go under the name of {\it  holographic renormalization} \cite{deHaro:2000xn, Emparan:1999pm, Skenderis:2002wp} and amounts to adding universal covariant boundary counterterms to the action.
We will review the holographic renormalization of the pure gravity action in the next subsection and we will use the results for the regularized (euclidean) AdS volume in the core of the presentation.

\vspace{5mm}

\underline{\bf{Generalizations of the AdS/CFT correspondence}}

\vspace{5mm}

Up to now we have only presented the original AdS/CFT correspondence between $\N=4$ SYM gauge theory and Type IIB string theory on $AdS_5 \times S^5$. One of its most surprizing feature is that it relates a theory without gravity and a theory containing  gravity. Moreover in the supergravity regime one can match classical gravity computations with quantum computations on the gauge theory side. These general features are expected to hold for more general dualities involving a $d$-dimensional CFT and quantum theory of gravity on $AdS_{d+1} \times K$ background, where $K$ is a compact space whose dimension is $9-d$ for string theories and $10-d$ for the (mysterious) M-theory. 
\smallskip

A first generalization consists in orbifolding the 5-sphere to obtain a duality between Type IIB string theory on $AdS_5 \times S^5/\Gamma$ and 4d SYM with $\N=4,2,1,0$ supersymmetry depending on the orbifold \cite{Kachru:1998ys}. A simple example is the duality between $\N=4$ SYM with gauge group $SO(N)$ or $Sp(N/2)$ and IIB string theory on $AdS_5 \times \bR\bP^5$ \cite{Witten:1998xy}, with $\bR\bP^5 = S^5/\bZ_2$ the 5d real projective space.

\vspace{5mm}

For $AdS_4/CFT_3$ dualities, which are the main topic here, the most famous and well-understood example relates M-theory on $AdS_4 \times S^7/\bZ_k$ and the so-called ABJM gauge theory \cite{Aharony:2008ug}. Let's describe it in detail as it is of particular interest for us.
\bigskip

The gauge theory side is a 3-dimensional SCFT with $\N=6$ supersymmetry with $U(N) \times U(N)$ gauge group with level $k$ and $-k$ Chern-Simons terms for the two unitary nodes respectively. The matter content is made of two ($\N=4$) bifundamental hypermultiplets, that is one $(\fN,\bar \fN)$ chiral multiplet and one $(\bar \fN,\fN)$ chiral multiplet for each. The Lagrangian has also a $\N=4$ superpotential. The theory has a priori only $\N=3$ supersymmetry due to the Chern-Simons terms, however one can show the presence of an $SO(6)_R$ R-symmetry ensuring $\N=6$ supersymmetry.\\
When $k=1$ or $k=2$ the supersymmetry is enhanced to $\N=8$ and the theory is known as the BLG theory \cite{Bagger:2007jr}.
 The 't Hooft coupling is $\lambda = N/k$ so that the theory is weakly coupled when $N/k << 1$.  

\bigskip

The ABJM SCFT is supposed to be the low-energy worldvolume description of a stack of $N$ M2-branes placed at the tip of a $\bC^4/\bZ_k$ orbifold in M-theory.\\
In the regime $N >> k^5$ the theory is correctly described by 11-dimensional supergravity on $AdS_4 \times S^7/\bZ_k$ with metric
\begin{align}
\label{ABJMMth}
 ds^2 &= \frac{L^2}{4} \, ds^2_{AdS_4} + L^2 \, ds^2_{S^7/\bZ_k}  \\
 L^6 &= 2^5 \pi^2 k N \, l_p^6  \no
\end{align}
where $l_p$ is the eleven-dimensional Planck length. The unit seven-sphere can be embedded in $\bC^4$ as $|z_1|^2 + |z_2|^2 +|z_3|^2 +|z_4|^2 =1$ and the orbifold action is $z_i \sim e^{\frac{2 i \pi}{k}} \, z_i$ (which has no fixed point). The 11d supergravity solution also has $N$ units of four-form flux along the $AdS_4$ factor.\\
$S^7/\bZ_k$ can be described as an $S^1$ fibered of $\bC\bP^3$ with the orbifold acting only on the $S^1$ angle. When $N^{1/5} << k << N$ the $S^1$ circle shrinks and the theory is well-described as Type IIA string theory on $AdS_4 \times \bC\bP^3$
\begin{align}
\label{ABJMIIA}
 ds^2_{IIA} &= R^2_{str} \left( \frac{1}{4} \, ds^2_{AdS_4} + ds^2_{\bC\bP^3} \right) \\
 R^2_{str} &= 2^{5/2} \pi (N/k)^{\half} \, l_s^2. \no
\end{align}
The ABJM correspondence admits a generalization to $U(N)_{k} \times U(M)_{-k}$ gauge group with $|M-N|\le |k|$, corresponding to have discrete torsion flux of the 3-form potential $C^3$ along $S^3/\bZ_k \subset S^7/\bZ_k$ in M-theory or discrete $B_2$ holonomy along $\bC\bP^1 \subset \bC\bP^3$ in type IIA \cite{Aharony:2008gk}.
\bigskip

\begin{figure}[!h]
\centering
\includegraphics[height=7cm,width=7cm]{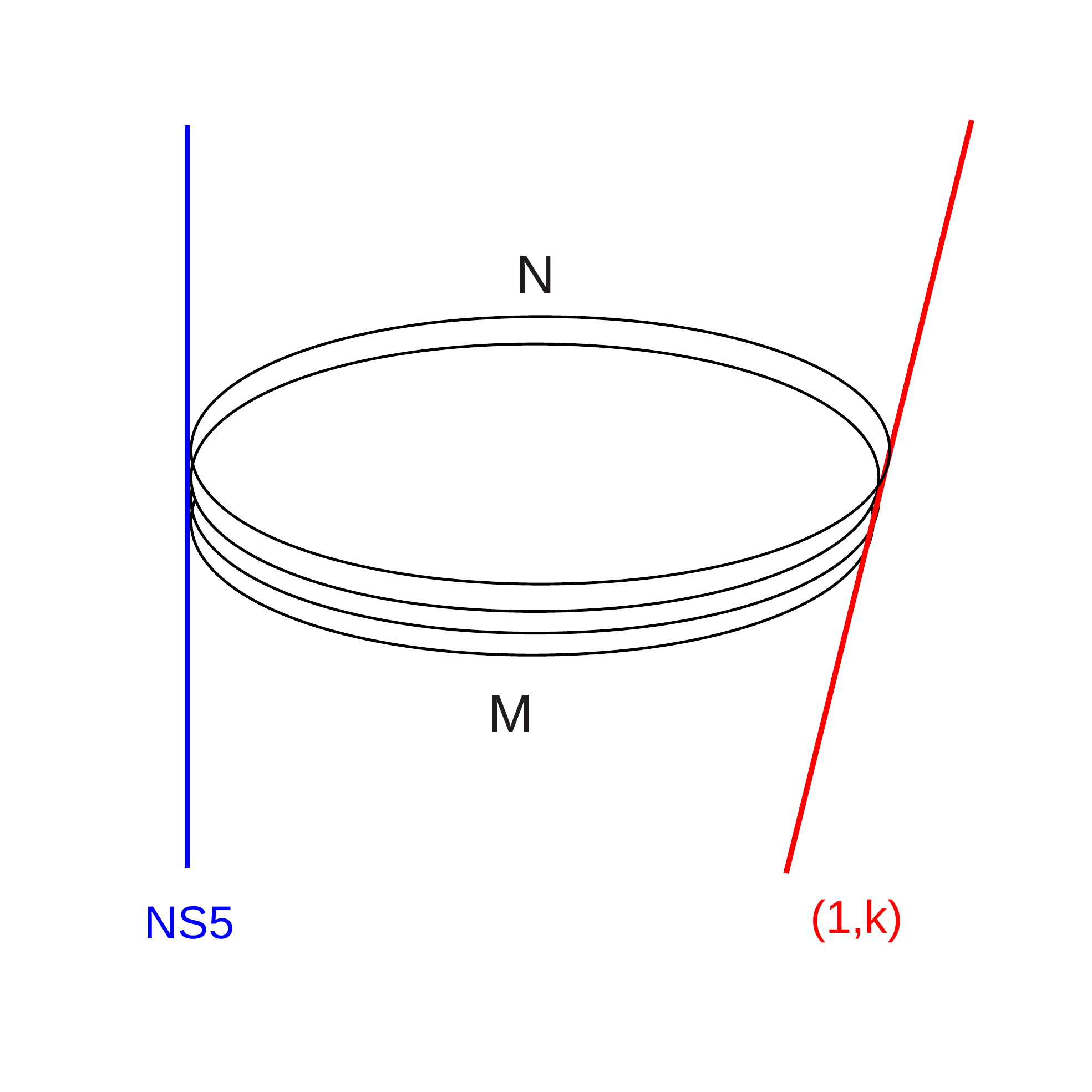}
\caption{\footnotesize{IIB Brane configuration for ABJ(M) gauge theory with $M > N$. $N$ D3-branes wrap a circle and cross a NS5-brane and a $(1,k)$ 5-brane. $M-N$ D3-branes are stretched between the 5-branes.}}
\label{ABJM}
\end{figure}

T-dualizing to Type IIB string theory the ABJM gauge theory can be thought of as the low-energy theory living on D3-branes in a brane configuration where the $N$ D3-branes wrap a circle (T-duality circle) and cross a NS5-brane and a $(1,k)$ 5-brane. The generalization includes $M-N$ additional D3-branes stretched between the 5-branes, as in figure \ref{ABJM}. This is an example of circular quiver. We will encounter this kind of brane configurations all along the presentation and will give much more details.

\vspace{5mm}

Today many AdS/CFT duals have been proposed in various dimensions 
. One remaining  challenge is to understand the mysterious $\N = (2,0)$ 6-dimensional SCFT that is supposed to live on the worldvolume of a stack of $N$ M5-branes in M-theory.

\smallskip

The AdS/CFT correspondence is actually expected to hold for spacetime whose metric is only asymptotically AdS (AAdS), the generic case being a blackhole with asymptotic AdS metric (AdS blackhole), which is related to a quantum field theory at finite temperature. The deviation from AdS spacetime along the radial direction when going from the boundary to the bulk is now understood as the RG flow from a UV theory ($\N=4$ SYM in general) perturbed by a relevant operator to the IR. We refer to the reviews \cite{Hubeny:2010ry,Hartnoll:2009sz} for presentations of some extensions of AdS/CFT correspondence.

\vspace{5mm}

\section{Holographic renormalization of the gravity action}
\label{sec:HR}

In this subsection we summarize the derivation of the renormalized pure gravity action in Eucildean AdS spacetime. A good  review presenting the holographic renormalization techniques is \cite{Skenderis:2002wp}. Our very brief presentation is based on the review \cite{Marino:2011nm} (section 5), which adresses the question of the regularization of the euclidean $AdS$ gravity action in great details. Some of the original papers are \cite{deHaro:2000xn, Emparan:1999pm,Balasubramanian:1999re}.

\vspace{5mm}

The gravitational Einstein-Hilbert action in Euclidean $n+1$ dimensional space with cosmological constant $\Lambda$ is the sum of a bulk term and a boundary term named Gibbons-Hawking term
\begin{align}
\label{action}
S \quad &= S_{bulk} + S_{GH}  \no\\
S_{bulk} &=  - \frac{1}{16 \pi G_N} \int_M d^{n+1}x \, |G|^{\half}(R-2\Lambda)\\
S_{GH} &=  -\frac{1}{8 \pi G_N} \int_{\p M} d^n x \, K\, |\gamma|^{\half}  \no
\end{align}
where $G_N$ is the Newton's constant, $G_{\mu\nu}$ is the metric on the $n+1$-dimensional manifold $M$, $\p M$ is the boundary of $M$, $\gamma_{ij}$ is the induced metric on $\p M$ and $K$ is the extrinsic curvature of $\p M$ satisfying $|\gamma|^{\half} K = \scL_{\vec s}|\gamma|^{\half}$ with $\scL_{\vec s}$ the Lie derivative along the unit vector $\vec s$ normal to $\p M$.
\smallskip

The cosmological constant for $AdS_{n+1}$ space with radius $L$ is
\begin{align}
 \Lambda = - \frac{n(n-1)}{2 \, L^2}
\end{align}
and the Ricci tensor and scalar are given by
\begin{align}
 R_{\mu\nu} &= -\frac{n}{L^2} \ G_{\mu\nu}  \quad , \quad R = -\frac{n(n+1)}{L^2} \ .
\end{align}

Both terms in the action \ref{action} diverge when evaluated on (Euclidean) AdS space because of its infinite size.

\bigskip

A metric asymptotically AdS can be written
\begin{align}
 ds^2 &= L^2 \left( \frac{du^2}{u^2} + \frac{1}{u^2} \, g_{ij}(u^2,x) \, dx^i dx^j \right)
\end{align}
and the $g_{ij}$ can be expanded in a power series of $u^2$ near the boundary $u=0$ of $AdS_{n+1}$
\begin{align}
\label{gexpand}
 g_{ij}(u^2,x) &= g^{(0)}_{ij}(x) + u^2 g^{(2)}_{ij}(x) + \cdots + u^{n} \lp g^{(n)}_{ij}(x) + h^{(n)}_{ij}(x) \log(\sqrt{u}) \rp + \cdots
\end{align}
where the term $h^{(n)}$ is present only for even $n$. The coefficients $g^{(2)},g^{(4)}, \cdots, g^{(n-2)}$ (or $g^{(n-1)}$) and $h^{(n)}$ can be expressed in terms of $g^{(0)}$ recursively by plugging \ref{gexpand} in Einstein's equations. $g^{(n)}$ is independent of $g^{(0)}$ and is related to the one-point function of the boundary stress-energy tensor.
\smallskip

As mentioned above the regularization consists in truncating the manifold $M$ to the manifold $M_{\epsilon}$ with $u \geq \epsilon$, so that the boundary $\p M_{\epsilon}$ is at finite distance.

\noindent Evaluating the action \ref{action} on $M_{\epsilon}$ yields a regulated action with the structure
\begin{align}
S_{\epsilon} \  &= \frac{L^{n-1}}{16 \pi G_N} \int d^n x \, \sqrt{g^{(0)}} \left( \epsilon^{-n} a_{(0)} + \epsilon^{-n+2} a_{(2)} + \cdots + \epsilon^{-2} a_{(n-2)} - 2 \log(\epsilon) a_{(n)}    \right)  \quad +O(\epsilon^0)  \no\\
 & a_{(0)} = 2(1-n)  \no\\
 & a_{(2)} = - \frac{(n-4)(n-1)}{n-2} \ \tr \lp g^{(0)-1}g^{(2)} \rp  \quad (n>2)   \\
 & a_{(4)} = \cdots  \quad (n > 4) \no
\end{align}
where the logarithm appears only for even $n$.

\noindent The procedure consists simply in adding a counterterm equal to minus the divergent part of the action as $\epsilon \rightarrow 0$ and rewrite it in terms of covariant quantities on the boundary with metric $\gamma$
\begin{align}
 S_{ct} &= - \frac{L^{n-1}}{16 \pi G_N} \int d^n x \, \sqrt{g^{(0)}} \left( \epsilon^{-n} a_{(0)} + \epsilon^{-n+2} a_{(2)} + \cdots + \epsilon^{-2} a_{(n-2)} - 2 \log(\epsilon) a_{(n)}    \right)  \no\\
 &= \frac{1}{8 \pi G_N} \int d^n \sqrt{\gamma} \lp \frac{n-1}{L} + \frac{L}{2(n-2)} \, R[\gamma] + \cdots + 2 a_{(n)}[\gamma] \log(\epsilon)  \rp \ .
\end{align}

\noindent The total action on $M_{\epsilon}$ is then 
\begin{align}
\label{Sreg}
 S_{reg} = S_{bulk} + S_{GH} + S_{ct} \ .
\end{align}
It is finite in the limit $\epsilon \rightarrow 0$ by construction, so the regularized action is computed using \ref{Sreg} at finite $\epsilon$ and then putting $\epsilon = 0$.

\vspace{5mm}

This procedure can be applied to Euclidean AdS space in several coordinate systems with a variety of topological boundaries leading to different results (see \cite{Emparan:1999pm}). For our purpose we can extract from the action a regularized $AdS$ volume, which is directly proportional to the pure gravity AdS action. Let's just mention the result that we will use, namely the regularized volume of pure $AdS_4$ Euclidean space with the 3-sphere $S^3$ as conformal boundary
\begin{align}
 \textrm{Volume}(AdS_4)_{reg} &= \frac{16 \pi G_N \, L^2}{6} \  S_{reg} \ =  \ \frac{4}{3} \pi^2 L^4 \ .
\end{align}

\chapter{ 3d $\N=4$ quivers and brane realizations}
\addcontentsline{lot}{chapter}{  3d $\N=4$ quivers and brane realizations }
\label{chap:quivers}

In this chapter we describe the class of Super-Conformal field theories for which we will propose Type IIB holographic duals. These gauge theories arise as strongly interacting infrared fixed points of 3d quiver gauge theories with $\N=4$ supersymmetries. The relevant supergroup of symmetries is $OSp(4|4)$. The  bosonic symmetries are the 3-dimensional conformal group $SO(2,3) \sim USp(4)$ and the $SU(2)_L \times SU(2)_R \sim SO(4)$ R-symmetry.

 As in the original setup of Maldacena (see section \ref{sec:Msetup}), they  can be understood as the low energy limit of the worlvolume theories of D3-branes, but this time the brane configurations involve also D5-branes and NS5-branes. The relation to the brane configurations will prove crucial when we come to the supergravity duals in the next chapter.

\vspace{5mm}

\section{$\N=4$ supersymmetric gauge theories in 3 dimensions}
\label{sec:3dN4}

The following brief presentation of $\N=4$ $d=3$ (super-conformal) gauge  theories and their known properties is freely inspired by \cite{Aharony:1997bx, deBoer:1996mp,deBoer:1996ck,deBoer:1997kr,Seiberg:1996nz,Intriligator:1996ex,Kapustin:2009kz,Kapustin:2010xq,Jafferis:2010un,Hama:2010av,Hama:2011ea,Marino:2011nm,Yaakov:2012usa}, which also describe many interesting features about the dynamics of the $\N=2$ theories.  

\subsection{$\N=4$ supersymmetry in 3 dimensions and Lagrangian}

The $\N=4$ supersymmetry algebra in 2+1 dimensions has 8 real supercharges. This is four times the minimal supersymmetry (2 supercharges : $\N=1$) and half the maximal amount (16 supercharges : $\N=8$). The algebra can be obtained by reducing the 4-dimensional $\N=2$ supersymmetry algebra to 3 dimensions. It can be written in terms of 4 real spinor generators $Q^A$, $A=1,2,3,4$.
\begin{align}
\{Q_{\alpha}^A , Q_{\beta}^B \} = 2 \ \sigma^{\mu}_{\alpha\beta} \, \delta^{AB} \, P_{\mu} + 2 \, \epsilon_{\alpha\beta} Z^{AB} \ ,
\end{align}
with $A,B = 1,2,3,4$ , $\mu = 0,1,2$, $\{\sigma^{\mu}\}$ are a set of generators of the 3-dimensional Clifford algebra, $\epsilon_{\alpha\beta}$ is a conventional anti-hermitian matrix used for lowering spinor indices and $Z^{AB}$ is a real antisymmetric matrix of central charges that commutes with all the generators of the algebra. $Z^{AB}$ has two independent components, that are derived from the real central charge in $d=4$ $\N=2$ and the momentum generator $P^3$ in the reduced dimension.

The super-algebra has a $SU(2)_{L} \times SU(2)_R \simeq SO(4)$ group of R-symmetry (automorphisms of the algebra) that rotates the supercharges $Q^A$ as the $\mathbf{4}$ of $SO(4)$.

Generally the super-algebra for $\N$ supersymmetry is the same with $A,B = 1, 2 , \cdots, \N$ and the R-symmetry group that rotates the supercharges is $SO(\N)$.

\bigskip

The superconformal extension is given by the super-group $OSp(4|4)$. It contains the conformal extension of the Poincaré group in 3 dimensions $SO(2,3) \simeq USp(4)$ and it has 8 additional real conformal supercharges, so 16 (real) supercharges in total. In Euclidean signature the conformal group is $SO(1,4) \simeq USp(2,2)$ and the super-conformal group is named $OSp(4|2,2)$.

\bigskip

$\N=4$ $d=3$ of quiver gauge theories contain vector multiplets and hypermultiplets, defined in turn in terms of $\N=2$ multiplets, for which there is a superspace formulation.

The field contents of the $\N=4$ vector multiplet and hypermultiplet are summarized in table \ref{tab:multiplets}, with their transformation under the R-symmetry $SU(2)_{L}\times SU(2)_{R}$ and gauge group $G$ indicated. The notations for the various fields and auxiliary fields are pretty standard and should not bring confusion. All fermions are two-component complex spinors. The scalar $\sigma$ and auxiliary scalar $D$ in the vector multiplet are real, while the scalars ``$\phi$'' and auxiliary scalars ``$F$'' in each $\N=2$ chiral multiplet are complex.

\begin{table}
\begin{tabular}{|l|l|c|c|l|}
  \hline
$\mathcal{N}=4$ & $\mathcal{N}=2$ (superfield) & Components & $SU(2)_{L}\times SU(2)_{R}$ & $G$ \\
\hline\hline
vector & vector & $A_{\mu}$ &  & \multirow{7}{*}{adjoint} \\ 
 multiplet & multiplet ($V$) & $\lambda_{\alpha}$ & & \\
& & $\sigma$ &  $\big \{ \sigma, Re\varphi, Im\varphi \big \}  \quad \textrm{in}\quad \left(1,0\right)$ &  \\
& &  $D$  & $\big \{\lambda_{\alpha},\xi_{\alpha} \big \}  \quad \textrm{in}\quad \left(\frac{1}{2},\frac{1}{2}\right)$ &  \\
 \cline{2-3}
 & chiral & $\varphi$ &  $\big \{ D, ReF_{\Phi}, ImF_{\Phi} \big \}  \quad \textrm{in}\quad \left(0,1\right)$ & \\
 & multiplet ($\Phi$) & $\xi_{\alpha}$ & & \\
&  & $F_{\Phi}$ & & \\
\hline
hyper & chiral & $\phi$ &  & \multirow{3}{*}{$R$} \\
 multiplet & multiplet ($\phi$) & $\psi_{\alpha}$ &  & \\
&  & $F$ & $\big \{ \phi^{\dag},\tilde{\phi}  \big \}  \quad \textrm{in}\quad \left(0,\frac{1}{2}\right)$ & \\
 \cline{2-3} \cline{5-5}
 &  chiral &  $\tilde{\phi}$  &  $\big\{ \psi_{\alpha},\tilde{\psi}_{\alpha}  \big \}  \quad \textrm{in}\quad \left(\frac{1}{2},0 \right)$ & \multirow{3}{*}{$R^{*}$} \\
 & multiplet ($\tilde{\phi}$) & $\tilde{\psi}_{\alpha}$  & \quad $F,\tilde{F} \quad$ integrated out & \\
&  & $\tilde{F}$ & & \\
\hline
\end{tabular}
\caption{Field content and R-charges of the $\N=4$ supermultiplets.}
\label{tab:multiplets}
\end{table}

\vspace{8mm}

A $\N=2$ chiral multiplet can be recast in terms of a chiral superfield $\Phi$ with $\bar D_{\alpha} \Phi =0$
\begin{align}
 \Phi = \phi + \sqrt{2} \theta \, \psi + \theta ^2 F
\end{align}
with $\theta$ a two components grassmann variable.\\
The $\N=2$ vector multiplet can be recast in terms of a real superfield $V$ with $V^{\dagger} = V$, which reads in Wess-Zumino gauge
\begin{align}
 V = - \theta^{\alpha} \sigma^{\mu}_{\alpha\beta} \bar\theta^{\beta} A^{\mu} - \theta\bar\theta \sigma + i \theta \theta \bar\theta \bar\lambda - i \bar\theta\bar\theta \theta \lambda + \half \theta \theta \bar\theta \bar\theta D \ ,
\end{align}
where the components are adjoint valued matrices. The chiral field strength is defined via $W_{\alpha} = - \frac{1}{4} \bar D \bar D e^{-V} D_{\alpha} e^{V}$.

\vspace{8mm}

The flat space euclidean action for the $\mathcal{N}=4$ quiver theories is composed of the following $\N=2$ superspace pieces.

\begin{itemize}

\item A Yang-Mills action for each node in the gauge group. The gauge couplings for the different nodes need not be the same but all flow to strong coupling in the IR.
\begin{align}
 S_{vector}^{\N=4} 
&= \frac{1}{g^2} \int d^3 x d^2\theta d^2\bar \theta  \ \tr \lp W_{\alpha}^2  - \Phi ^\dag e^{2V}\Phi \rp  + h.c. \no
\end{align}
where $W$ is the chiral field strength of the $\N=2$ vector superfield $V$ and $\Phi$ is the adjoint chiral superfield. In components we have\\
 $S_{vector}^{\N=4} = S_{vector}^{\N=2} + \frac{1}{g^2} \, S_{adj\ chiral}^{\N=2}$ with
\begin{align}
 S_{vector}^{\N=2} &= \frac{1}{2 g^2} \int d^3 x \ \tr \lp \half F_{\mu\nu}F^{\mu\nu} + D_{\mu}\sigma D^{\mu}\sigma + D^2 + i \bar\lambda \gamma^{\mu} D_{\mu} \lambda + i \bar\lambda [\sigma,\lambda] \rp \no\\
 S_{chiral}^{\N=2} &= -\int d^3 x \  \lp D_{\mu}\bar\phi D^{\mu}\phi + \bar\phi \sigma^2 \phi + i \bar\phi D \phi + \bar F F - i \bar\psi \gamma^{\mu}D_{\mu} \psi + i \bar\psi \sigma \psi + i \bar\psi \lambda \phi- i \bar\phi \bar\lambda\psi \rp \no 
\end{align}

\item A kinetic term and gauge coupling for each hypermultiplet.
\begin{align}
S_{hyper}^{\N=4}
 = - \int {d^3 x {d^2}\theta {d^2}\bar \theta } \sum\limits_{matter} {({\phi ^\dag }{e^{2V}}\phi }  + {{\tilde \phi }^\dag }{e^{ - 2V}}\tilde \phi ) \no
\end{align}
where $\phi, \tilde \phi$ are two chiral superfields in conjugate representations.

\item A $\N=4$ superpotential. 
\begin{align}
S_{spot}^{\N=4} =  - i\sqrt 2 \int d^3 x d^2\theta \ \sum\limits_{matter} (\tilde \phi \Phi \phi ) + h.c \no
\end{align}
where the sum runs over all matter charged under the gauge symmetry associated with $\Phi$.

\end{itemize} 

The gauge theories enjoy two possible deformations :

\begin{itemize}

\item Real and complex mass terms for the hypermultiplets. The 3 real parameters transform as a triplet of $SU(2)_R$ and can be viewed as the lowest components of a background $\mathcal{N}=4$ abelian vector multiplet coupled to the flavor symmetry currents.
\begin{align}
S_{mass}^{\N=4} =  - \int d^3 x d^2\theta d^2\bar \theta  \sum\limits_{matter} (\phi ^{\dag} e^{2{V_m}}\phi   + \tilde \phi^{\dag} e^{ - 2{V_m}} \tilde \phi ) - \lp i \sqrt{2} \int d^3 x d^2\theta \sum\limits_{matter} (\tilde \phi {\Phi _m}\phi ) + h.c \rp   \no
\end{align}
The Lagrangian (with the complex mass to rotated to zero) is then obtained by setting $V_{m}\propto m\bar{\theta}\theta$ and $\Phi_{m}=0$, where $m$ is the real mass parameter. This ensures the vanishing of the fermion variations of the background multiplet. In components it reads
\begin{align}
S_{mass}^{\N=2}(\phi,m)  = \int d^3 x \  \lp D_{\mu}\bar\phi D^{\mu}\phi + m^2 \, \bar\phi \phi + \bar F F - i \bar\psi \gamma^{\mu}D_{\mu} \psi + i \, m \,  \bar\psi \psi \rp \no 
\end{align}
and $S_{mass}^{\N=4}= S_{mass}^{\N=2}(\phi,m)  + S_{mass}^{\N=2}(\tilde\phi,-m) $.

\item Fayet-Iliopoulos (FI) terms for the $U(1)$ factors of the gauge group. The three real parameters transform as a triplet of $SU(2)_{L}$. They can be viewed as the lowest components of a background twisted $\mathcal{N}=4$ abelian vector multiplet coupled to the topological currents associated with the $U(1)$ gauge factors by a $BF$ type coupling.
\begin{align}
S_{FI}^{\N=4}  =  \int d^3 x d^2\theta d^2\bar \theta \ \tr \lp \Sigma \, \hat V_{FI} \rp \ + \ \int d^3 x d^2\theta  \ \tr \lp \Phi \hat\Phi _{FI} + h.c \rp \no
\end{align}
where the $\tr$ picks out the central $U(1)$ factor of the gauge group (we have a FI deformation for each gauge node of a quiver theory). Again the deformed Lagrangian (with two deformation parameters rotated to zero) is obtained by setting  ${\hat V}_{FI}\propto \eta\bar{\theta}\theta,\hat\Phi_{FI}=0$, leading to
\begin{align}
 S_{FI} = i \, \eta \ \int d^3 x \ D  \no
\end{align}

\end{itemize}

It is also possible to add a Chern-Simons term of level $k \in \bZ$ to the action, which breaks $\N=4$ to $\N=3$ :
\begin{align}
 S_{CS}^{\N=3} &= S_{CS}^{\N=2} - \frac{k}{8\pi}  \int d^3 x d^2\theta \tr \lp \Phi^2 + h. c. \rp   \\
 S_{CS}^{\N=2} &= \frac{k}{4\pi} \int d^3x \ \tr \lp \epsilon^{\mu\nu\rho} \lp A_{\mu} \p_{\nu} A_{\rho} + \frac{2i}{3} A_{\mu}A_{\nu}A^{\rho} \rp  - \lambda \bar\lambda + 2 D \sigma \rp \no
\end{align}
where $\Phi$ is an adjoint chiral superfield.

\bigskip

To close this introductory part, we mention the duality between abelian vector field and scalar field in 3 dimensions, represented by the relation
\begin{align}
 F_{\mu\nu} = \epsilon_{\mu\nu\sigma} \p^{\sigma} \gamma \ ,
\end{align}
where $\gamma$ is a periodic ($\gamma \simeq \gamma + g_{YM}$) real scalar called {\it dual photon}. The whole $\N=2$ vector multiplet can be dualized to a chiral multiplet with lowest component $\sigma + i\gamma$ (see \cite{deBoer:1997kr}).

\subsection{Mirror symmetry}

The 3d $\N=4$
 quiver gauge theories moduli space of vacua of the quiver gauge theories is obtained by minimizing the scalar potential in the Lagrangian. This implies solving the D-term and F-term constraints, which are scalar potentials arising after integrating out the auxiliary scalars $D$ and $F$.
Generically the moduli space is decomposed into two branches \cite{Seiberg:1996nz,Intriligator:1996ex,deBoer:1996mp} :
\begin{itemize}
 \item The Coulomb branch corresponding to giving vevs to the scalars $\phi_i$, $i=1,2,3$ in the vector multiplet preserving the vanishing potential condition $\sum_{i<j} Tr ([\phi_i , \phi_j]^2) = 0$. The scalar vevs $<\phi_i>$ are given by diagonal matrices breaking the gauge group to its maximal torus $U(1)^{r}$. The full Coulomb branch is obtained by giving vevs to the $r$ dual photons associated to the surviving $r$ abelian vector fields. The Coulomb branch (of $\N=4$ theories) is a hyper-K\"ahler manifold of (real) dimension $4r$. The metric on the Coulomb branch receives quantum corrections due to 3-dimensional monopoles (also named instantons in the litterature as they are codimension 3 solutions of BPS equations), but its dimension is unaffected.

 \item The Higgs branch corresponding to giving vevs to the scalars $\phi_i$, $i=1,2,3,4$ in the hyper-multiplets. The equations defining the Higgs branch may allow for a complete gauge symmetry breaking. For instance for a gauge group $U(N_c)$ with $N_f$ hypermultiplets the conditions for complete higgsing is $N_f \geq 2 N_c$. When a complete higgsing is not possible, one expect that the infrared theory contains free vector multiplets (\cite{Gaiotto:2008ak}). The Higgs branch (of $\N=4$ theories) is a hyper-K\"aler manifold. Promoting the gauge coupling constant to a superfield, one can show that the scalars of the superfield do not transform under the $SU(2)_R$ R-symmetry, so that they can appear only in the Coulomb branch (see \cite{Argyres:1996eh,Intriligator:1996ex}). It implies that the Higgs branch does not receive quantum corrections.
\end{itemize}

Mirror symmetry (\cite{Intriligator:1996ex}) is a duality between $d=3$ $N=4$ supersymmetric gauge theories, with different gauge groups and matter contents, which exchanges the Higgs branch and Coulomb branch of vacua, exchanges the mass and Fayet-Iliopoulos parameters and the $SU(2)_L$ and $SU(2)_R$ R-symmetries. This implies that the quantum corrections of the Coulomb branch are contained in the purely classical Higgs branch of the mirror dual theory (\cite{deBoer:1996mp,Hanany:2011db}).
\smallskip

The 3d Yang-Mills gauge theories are super-renormalizable : they flow to free theories in the ultra-violet. Conversely they are infinitely strongly coupled in the infrared where the Lagrangian description breaks down. The prediction from mirror symmetry is that the IR strongly interacting fixed points of mirror theories arising from the RG flow at the intersection/origin of the Coulomb and Higgs branches are the same SCFTs. By deforming the SCFTs with mass and Fayet-Iliopoulos parameters (that are not renormalized due to supersymmetry) one gets a duality between non-conformal supersymmetric theories.

One prediction of mirror symmetry is the emergence of global symmetries at the super-conformal fixed point (intersection of Higgs and Coulomb branch). The SCFTs have global non-abelian flavor symmetries, which should appear as different global symmetries in the IR limit of the dual theory. More precisely the gauge theories have abelian topological global symmetries associated with each $U(1)$ factor in the gauge group (one for each $U(N_i)$ node), with conserved current $\star F$. At the super-conformal fixed point the topological symmetries can be enhanced to non-abelian symmetries, that are exchanged with the flavor symmetries under mirror symmetry.

\vspace{6mm}

The fact that the SCFTs are infinitely strongly coupled makes it difficult to test mirror symmetry and until recently the only checks concerned moduli spaces. The techniques of localization of path integrals on the 4-sphere developped in \cite{Pestun:2007rz} and pursued in \cite{Kapustin:2009kz} for supersymmetric theories on the 3-sphere have rendered possible  to compute supersymmetric observables by reducing the whole path integral to finite dimensional matrix integrals (matrix models). The resulting matrix models do not depend on the Yang-Mills coupling and are consequently directly related to the observables of the IR fixed points.

 Using the matrix models \cite{Kapustin:2010xq} provided further evidences of mirror symmetry by matching partition functions of mirror pairs, under the exchange of mass and FI parameters. 

While we were completing this manuscript, \cite{Gaiotto:2013bwa} appeared which uses mirror symmetry for 3d $\N=4$ linear quivers and the map between their space of vacua and the eigenvalues of quantum integrable spin chain Hamiltonians to derive new dualities between these integrable models, called bispectral dualities.

On the holographic side that we study, mirror symmetry will be naturally implemented by the S-duality of type IIB string theory.

\subsection{Matrix models}

As we will use the matrix models obtained from the localization of path integrals of $\N=4$ gauge theories on the 3-sphere, we give here a short summary of these matrix models, obtained in \cite{Kapustin:2009kz} for $\N=2$ theories. Further details about $d=3$ supersymmetric gauge theories on $S^3$ can be found in \cite{Hama:2010av}. A good review of the localization on $S^3$ is \cite{Marino:2011nm}.

The localisation on $S^3$ of the partition function $Z_{S^3}$ reduces the whole path integral to an integration over the Cartan subalgebra of the gauge group, divided by the order of the Weyl group $|\scW|$. We give here explicit formulas for a $U(N)$ gauge group.
\begin{align}
Z_{S^3} &= \frac{1}{|\scW|} \int_{\textrm{Cartan}} d\sigma \ \Big( ... \Big) \ = \  \frac{1}{N!} \int \prod_{i}^N d\sigma_i  \ \Big( ... \Big) \ .
\label{3dZ}
\end{align}
The integrand (the dots in \eqref{3dZ}) is a product of several contributions .

The $\N=4$ vector multiplet gives a factor 
\footnote{the matrix factor corresponds actually to a $\N=2$ vector multiplet, but the $\N=2$ adjoint chiral multiplet of the $\N=4$ vector multiplet does not contribute (this is related to the fact that it has conformal dimension equal to one).}
\begin{align}
  \textrm{det}_{\rm Adj} \Big( \sh (\sigma) \Big)  \ = \  \prod_{i<j}^N \sh(\sigma_i - \sigma_j)^2  \ ,
\end{align}
a hyper-multipet in a representation $R$ of the gauge group with mass $m$ gives a factor
\begin{align}
\textrm{det}_{R}\Big( \frac{1}{\ch(\sigma - m)} \Big) &= \prod_j^N \ch(\sigma_j - m) \ \textrm{for the fundamental rep. of } \, U(N)  \\
 &=   \prod_{i,j}^{N,M} \ch(\sigma_i- \tilde \sigma_j - m)  \ \textrm{for the bifundamental rep. of} \, U(N)\times U(M)  \  ,  \no
\end{align}
a Chern-Simons term with level $k$ contributes a factor
\begin{align}
 \textrm{det}_F \Big( e^{ i\pi k \sigma^2} \Big) \ = \ e^{i\pi k \sum_j^N \sigma_j^2  } \  ,
\end{align}
and a Fayet-Iliopoulos deformation $\eta$ produces a term
\begin{align}
\textrm{det}_F \Big( e^{2 i \pi \eta \sigma} \Big) \ = \  e^{2i\pi \eta \sum_j^N \sigma_j } \ .
\end{align}
Here $\det_R$ (and below $\tr_R$) is the
the determinant (the trace) in the representation $R$. The indices $F$ and
${\rm Adj}$ will refer to fundamental and adjoint representations respectively.

\bigskip

To close this introduction we would like to mention that the localization techniques have been applied to the richer cases of $d=3$ $\N=2$ gauge theories with non-canonical R-charge assignments \cite{Jafferis:2010un,Hama:2010av} and on the squashed 3-sphere \cite{Hama:2011ea}, and that the superconformal index (partition function on $S^1 \times  S^2$) was also studied in \cite{Imamura:2011su,Krattenthaler:2011da,Kapustin:2011jm,Dimofte:2011py}.

We also mention that recently $\N=4$ Seiberg-like dualities relating the quiver with $U(N_c)$ gauge group and $N_f$ hypermultiplets, with $N_f \geq 2 N_c$, and the quiver with $U(N_f-N_c)$ gauge group and $N_f$ hypermultiplets plus $N_f - 2 N_c$ free hypermultiplets (corresponding to monopoles operators with special R-charges asignments) have been proposed in \cite{Yaakov:2013fza}. Further details and comments on this duality immediatly followed in \cite{Gaiotto:2013bwa,Bashkirov:2013dda}.\\
This duality is perfectly consistent with the holographic descriptions that we propose in this presentation, in the sense that we have only one supergravity solution for dual theories.

\section{Linear and circular quivers}
\label{sec:quiv}

\begin{figure}[h]
\centering
\includegraphics[height=6cm,width=8.5cm]{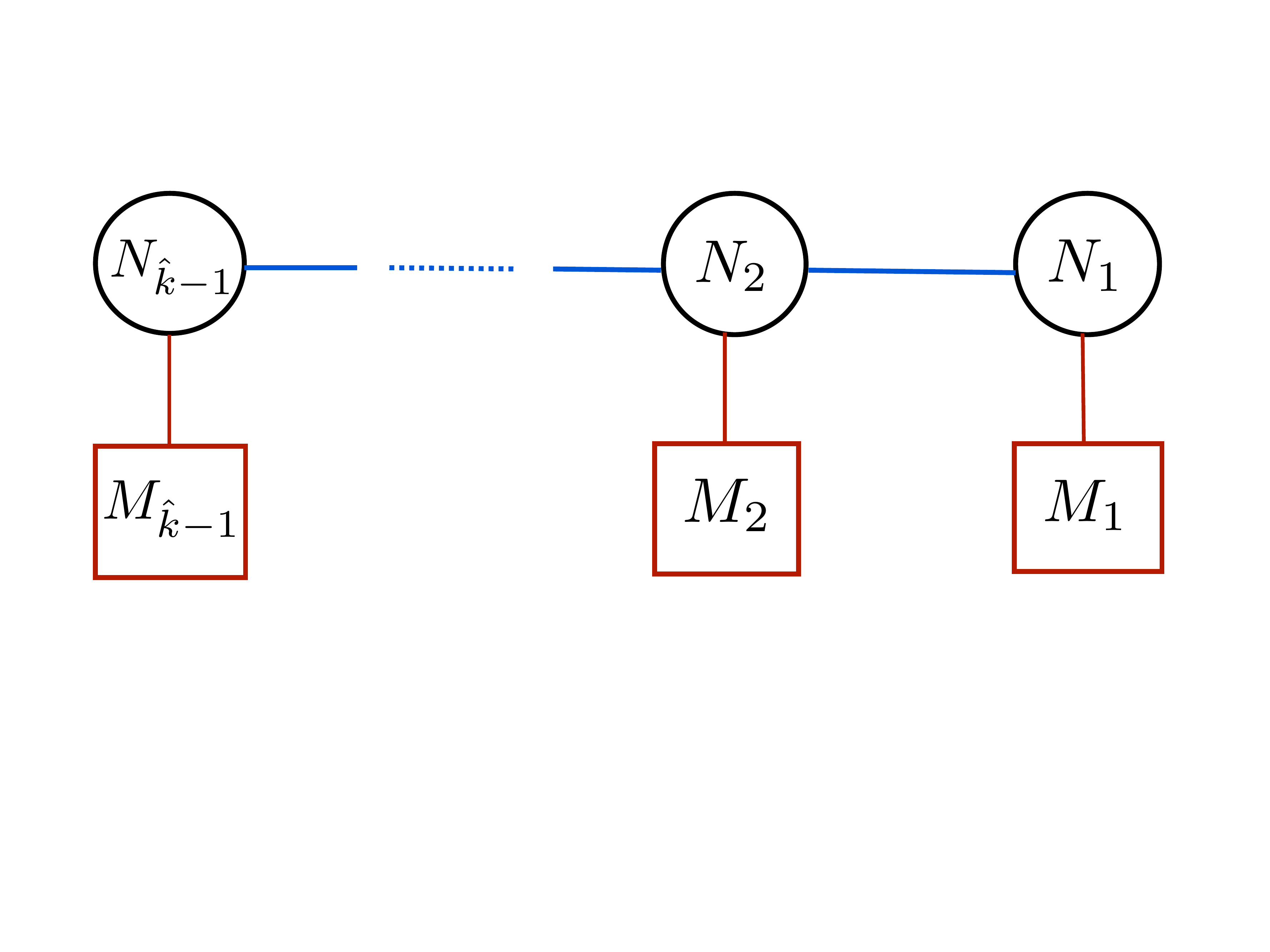}
\vskip -1cm
\caption{\footnotesize A  linear quiver with $\hat k-1$ gauge-group factors $U(N_1)\times U(N_2)\times\cdots$. The red boxes indicate the
numbers of  hypermultiplets in the fundamental representation of each gauge-group  factor.}
\label{fig:linquiv}
\end{figure}

\begin{figure}[h]
\centering
\includegraphics[height=6cm,width=8.5cm]{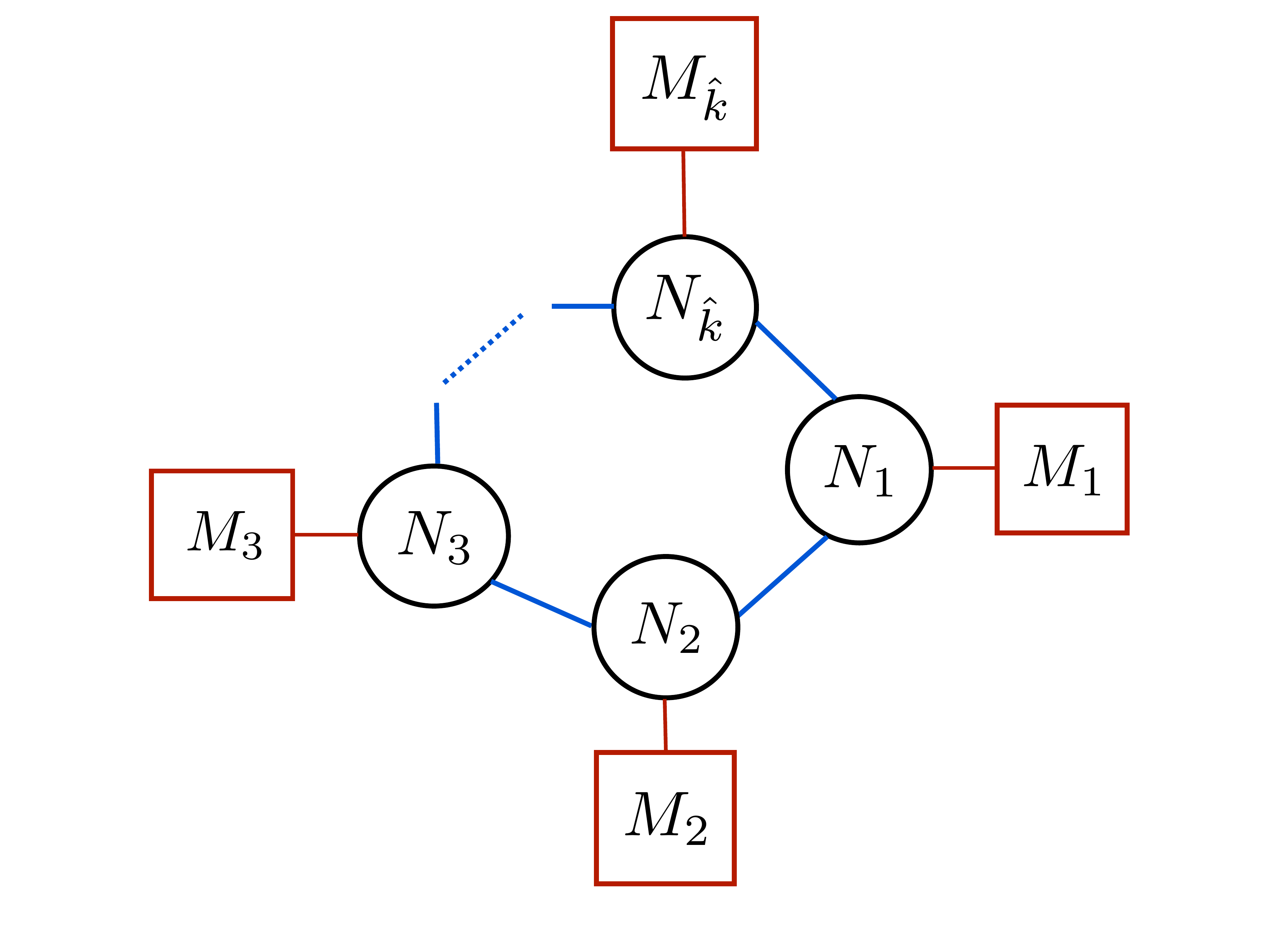}
\caption{\footnotesize A circular quiver with $\hat k$ gauge-group factors. The $U(N_i)$  theories
  interact via  bifundamental hypermultiplets (the blue lines)
which form a circular chain, as opposed to the linear chain of figure 1.}
\label{fig:circquiv}
\end{figure}

The three-dimensional $\N=4$ superconformal field theories considered in this paper arise as
non-trivial infrared  fixed points of three-dimensional  quiver gauge theories with
$\N=4$ supersymmetries. Their field content and  their microscopic
Lagrangians are  succinctly summarized by a quiver diagram \cite{Douglas:1996sw}.
In our case  the diagrams will have  either linear or circular topology (see figures \ref{fig:linquiv} and \ref{fig:circquiv}). We  refer to
 the corresponding quivers as linear and circular respectively.


The quiver gauge theories contain a vector multiplet for each unitary node of the gauge group
\beq
 U(N_1) \times U(N_2) \times ... \  U(N_i) \times ... \, .
\eeq
Moreover, these theories contain a hypermultiplet transforming in the bifundamental representation of each consecutive pair of gauge groups $U(N_i)\times U(N_{i+1})$. For linear quivers $1\leq i \leq \hat k-2$, while for the circular quivers $1\leq i\leq \hat k$  with the convention that  $U(N_{\hat k+1})\equiv U(N_1)$.
 Finally, there are $M_i$ hypermultiplets in the fundamental representation of the group $U(N_i)$.\footnote{In the special case $\hat k=1$ the circular quiver has a single
   gauge-group factor,  $U(N_1)$,  and the  bifundamental hypermultiplet  is a hypermultiplet in  the adjoint representation of   $U(N_1)$. }

\vspace{8mm}

{\bf Irreducible infrared SCFTs}
\vspace{5mm}

A central question about the dynamics of these gauge theories is  to understand the nature of the fixed point of the renormalization group  in the  infrared.
 Since massive fields decouple in the infrared, we will assume that  hypermultiplet masses  and  Fayet-Iliopoulos terms are set to zero.
 The  quiver data and the  extended $\N=4$  supersymmetry   specify  then completely  the  microscopic,
  renormalizable Lagrangian.

   We are interested in  ``irreducible superconformal field theories", i.e.  theories  containing no decoupled sector with free vector multiplets and/or free 
   hypermultiplets.\footnote{On general grounds, we do not expect the bulk dual of a strictly free field theory to be describable by supergravity. Such a theory would require, due to the existence of  higher spin conserved currents,   higher spin fields propagating in the bulk.}
  It has been conjectured by Gaiotto and Witten \cite{Gaiotto:2008ak}  that a
  necessary and sufficient condition for a  gauge theory to flow to an irreducible superconformal field theory is
  \beq
N_{F,i}\geq 2N_i\,.
\label{Higgsing}
\eeq
In words, each  gauge-group  factor $U(N_i)$ should have at  least   $2N_i$ hypermultiplets transforming  in the fundamental representation.
Recall  that a hypermultiplet in the fundamental and anti-fundamental representation are equivalent.
 Therefore, for a quiver gauge theory,  the above requirement of irreducibility in the infrared  imposes the following inequalities on the quiver data
 \beq
M_i+N_{i-1}+N_{i+1}\geq 2N_i\,.
\label{condiSCFT}
\eeq

One way to argue for  the above
 conditions is that when they are obeyed the  gauge group can be completely Higgsed  \cite{deBoer:1996ck},
 and there exists a singularity at the origin of  the Higgs branch, from which the Coulomb branch emanates.
 A non-trivial superconformal field theory  appears in the infrared limit of the gauge theory around that vacuum.
 Conversely, when complete Higgsing is not possible, decoupled multiplets remain in the infrared, thus yielding non-irreducible theories.\\
Confirming this picture, Yaakov recently argued (\cite{Yaakov:2013fza}) that the 3d $\N=4$ gauge theories with a single $U(N_c)$ node and $N_f$ hypermultiplets in the range $N_c \leq N_f < 2 N_c$ flows in the infrared to a non-trivial fixed point with two decoupled pieces : one piece is an SCFT dual to the fixed point of a $U(N_f - N_c)$ gauge theory with $N_f$ hypermultiplets, which satisfies \ref{Higgsing}, and the other piece contains $2 N_c -N_f$ free hypermultiplets. This generalizes Seiberg duality to 3d $\N=4$ Yang-Mills gauge theories.

\vspace{8mm}

The quiver data that characterizes the irreducible superconformal field theories can be repackaged in a convenient way
in terms of two partitions,  $\rho$ and $\hat\rho$,  of the same number N  (this is explained below).  As usual,  one can associate
a Young tableau to each partition. The quiver theory can be described by the following data

\medskip\smallskip
$\bullet\ $   for linear quivers :   $(\rho,\hat\rho)$
 subject to the  constraints
\beq
    \rho^T > \hat \rho \quad  ,
  \label{fixedpoint}
\eeq

\medskip\smallskip
 $\bullet\ $ for   circular quivers:   $(\rho,\hat\rho,  L )$   subject to the constraints
 \medskip\smallskip
  \beq
  \rho^T \geq     \hat\rho \ , \qquad  L>0
   \,  .
  \label{fixedpointcirc}
\eeq
Here $\rho$ and $\hat\rho$ denote the two  partitions of  $N$,  and $L$  is a  positive  integer.  Transposition, noted $^T$, interchanges the columns and rows of a Young tableau.\\
The partitions inequality $a>b$ with $a=(a_1,a_2,...,a_r)$, $a_1 \geq a_2 \geq ... \geq a_r$, and $b = (b_1, b_2, ..., b_s)$,  $b_1 \geq b_2 \geq ... \geq b_s$, is defined as
\begin{align}
a > b  \quad \leftrightarrow \quad  \sum_{j=1}^{k} a_j > \sum_{j=1}^{k} b_j \quad , \textrm{for} \quad 1 \leq j < r \ .
\end{align}
$a \geq b$ is defined similarly with $\geq$ instead of $>$.
 The inequality \eqref{fixedpoint} has
 appeared previously  in different contexts related to solutions of Nahm's equations, see e.g. \cite{Kronheimer:1990ay,Bachas:2000dx}.

\smallskip

We denote the linear-quiver   theory  associated to  $(\rho,\hat\rho)$  by
$T_{\hat\rho}^\rho(SU(N))$,   and the circular-quiver theory with data  $(\rho,\hat\rho,  L )$
 by  $C_{\hat\rho}^\rho(SU(N),  L )$.
\smallskip

  When relating the partitions to the quiver data (see below), it turns out  that the above  Young-tableaux constraints  are automatically satisfied
 if  the ranks of all the gauge groups of the ultraviolet  theories are positive, that is if  all $N_i>0$.
 If  some Young-tableaux inequalities were saturated for a linear quiver, the quiver would break down to decoupled
 quivers plus free hypermultiplets. Circular quivers,  on the other hand,
  degenerate to linear quivers when $L=0$.

\smallskip
 As we shall see, this   data also completely encodes the field content of the ultraviolet mirror pair \cite{Intriligator:1996ex} of quiver gauge theories which flow to the
same fixed point  in the infrared.
Mirror symmetry  for this class of quiver gauge theories is realized very simply by the exchange of the two partitions
 \beq
\text{mirror symmetry}:\qquad \rho\longleftrightarrow \hat\rho\,   .
\eeq
 Therefore, $T_{\hat\rho}^\rho(SU(N))$ and $T_{\rho}^{\hat\rho}(SU(N))$ are mirror linear-quiver gauge theories,
  while $C_{\hat\rho}^\rho(SU(N), L)$
 and $C_{\rho}^{\hat\rho}(SU(N),  L)$  are mirror circular quivers. The
 Young tableaux constraints  are symmetric under the exchange of $\rho$ and $\hat\rho$,
 see appendix \ref{app:ineq},  and are therefore  consistent with  mirror symmetry.

\vspace{8mm}

{\bf Global symmetries}
\vspace{5mm}

These infrared  superconformal field theories are believed to have  a rich pattern of global symmetries,
  inherited from the symmetries acting on the Higgs and Coulomb branch of the
    quiver gauge theory from which the fixed point is reached in the infrared.  Since mirror symmetry exchanges
     the Higgs and Coulomb branches of mirror pairs, we conclude that the
global symmetry at the fixed point is
\beq
H \times \hat H \,,
\eeq
where
   \beq
    H =\prod_i U(M_i)  \qquad {\rm and}\qquad  \hat H = \prod_i U(\hat M_i) \,.
    \eeq
$H$  is the symmetry  that  rotates  the fundamental hypermultiplets    of
$T^{\rho}_{\hat \rho}(SU(N))$ or $ C_{\hat\rho}^\rho(SU(N), L)$, while $\hat H$ rotates the fundamental hypermultiplets of  their
 mirror duals. The two symmetries coexist at the superconformal fixed point.

\smallskip

The main object of this review is to discuss the AdS duals of the irreducible three dimensional $\N=4$ superconformal theories to which linear and circular quiver gauge theories of the above type  flow in the infrared. 


 \section{Brane Realization}
\label{sec:branes}

The above three-dimensional $\N=4$ supersymmetric linear and circular quiver gauge theories admit an elegant realization as the low-energy limit of brane configurations in type-IIB string theory  \cite{Hanany:1996ie}.  The brane configuration consists of an array of D3, D5 and NS5 branes oriented as
shown in the table.\footnote{For more details of these brane constructions
 see  \cite{Hanany:1996ie}\cite{Gaiotto:2008ak}.}
\smallskip\smallskip

\begin{table}[h]
\label{tab:probeconfig}
\begin{center}
\begin{tabular}{|c||c|c|c|c|c|c|c|c|c|c||}
  \hline
      & 0 & 1 & 2 & 3 & 4 & 5 & 6 & 7 & 8 & 9 \\ \hline
  D3  & X & X & X & X &   &   &   &   &   &   \\
  D5  & X & X & X &   & X & X & X &   &   &   \\
  NS5 & X & X & X &   &   &   &   & X & X & X \\ \hline
\end{tabular}
\caption{\footnotesize Brane array for three-dimensional quiver gauge theories}
\end{center}
\end{table}

\noindent The D3 branes span a finite interval along the $x^3$ direction and terminate on the NS5-branes.
 For circular quivers $x^3$ parametrizes a circle.

\vspace{5mm}

\noindent \underline{{\bf Linear Quivers}}

\vspace{5mm}

\begin{figure}[th]
\centering
\includegraphics[height=6cm,width=11cm]{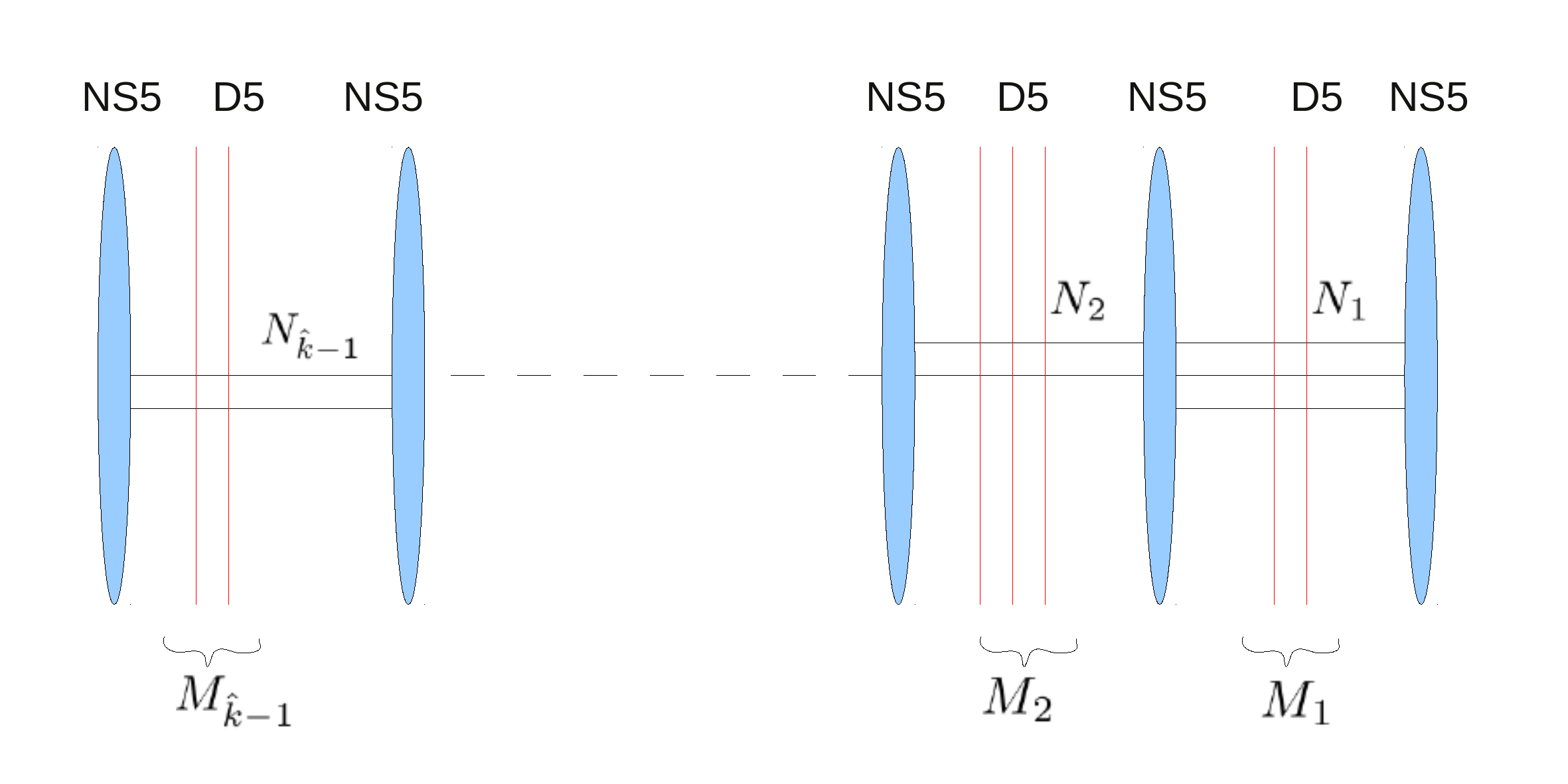}
\caption{\footnotesize Brane realization of linear quivers}
\label{fig:linquivbrane}
\end{figure}

The brane configuration corresponding to the linear quiver gauge theory of figure \ref{fig:linquiv} is depicted in figure \ref{fig:linquivbrane}.
An  invariant way of encoding a brane configuration -- and the corresponding quiver gauge theory -- is by specifying the {\it linking numbers} of the five-branes.\\
We adopt the convention that {\bf D5-branes are labelled from left to right} (the first is on the left) while {\bf NS5-branes are labelled from right to left} (the first is on the right).
Then the {\it linking numbers} for $i$-th D5-brane and the $j$-th NS5-brane can be defined as follows
\begin{align}\
l_i &= - n_i + R_i^{\rm NS5} \qquad (i=1,...,k) \cr
\hat l_j &= \hat n_j + L_j^{\rm D5} \qquad (j=1,...,\hat k) \ ,
\label{defnlinking}
\end{align}
where $n_i$ is the number of D3 branes ending on the $i$th D5 brane from the right minus the number ending from the left, $\hat n_j$ is the same quantity for the $j$th NS5 brane, $R_i^{\rm NS5}$ is the number of NS5 branes lying to the right of the $i$th D5 brane and $L_j^{\rm D5}$ is the number of D5 branes lying to the left of the $j$th NS5 brane.  These numbers  are {\it invariant} under Hanany-Witten moves \cite{Hanany:1996ie},   when a D5 brane crosses a NS5 brane. In such a process the D3-branes that were stretched between the D5 and the NS5 turn into anti-D3-branes (annihilating other D3-branes of the configuration) and one extra D3-brane is created between the two 5-branes. In total the linking numbers of the two 5-branes are conserved in the process.

 Since the extreme infrared limit is expected to be  insensitive to these moves,  it is   convenient to label the infrared dynamics in terms of the linking numbers.

\begin{figure}[th]
\centering
\includegraphics[height=6cm,width=11cm]{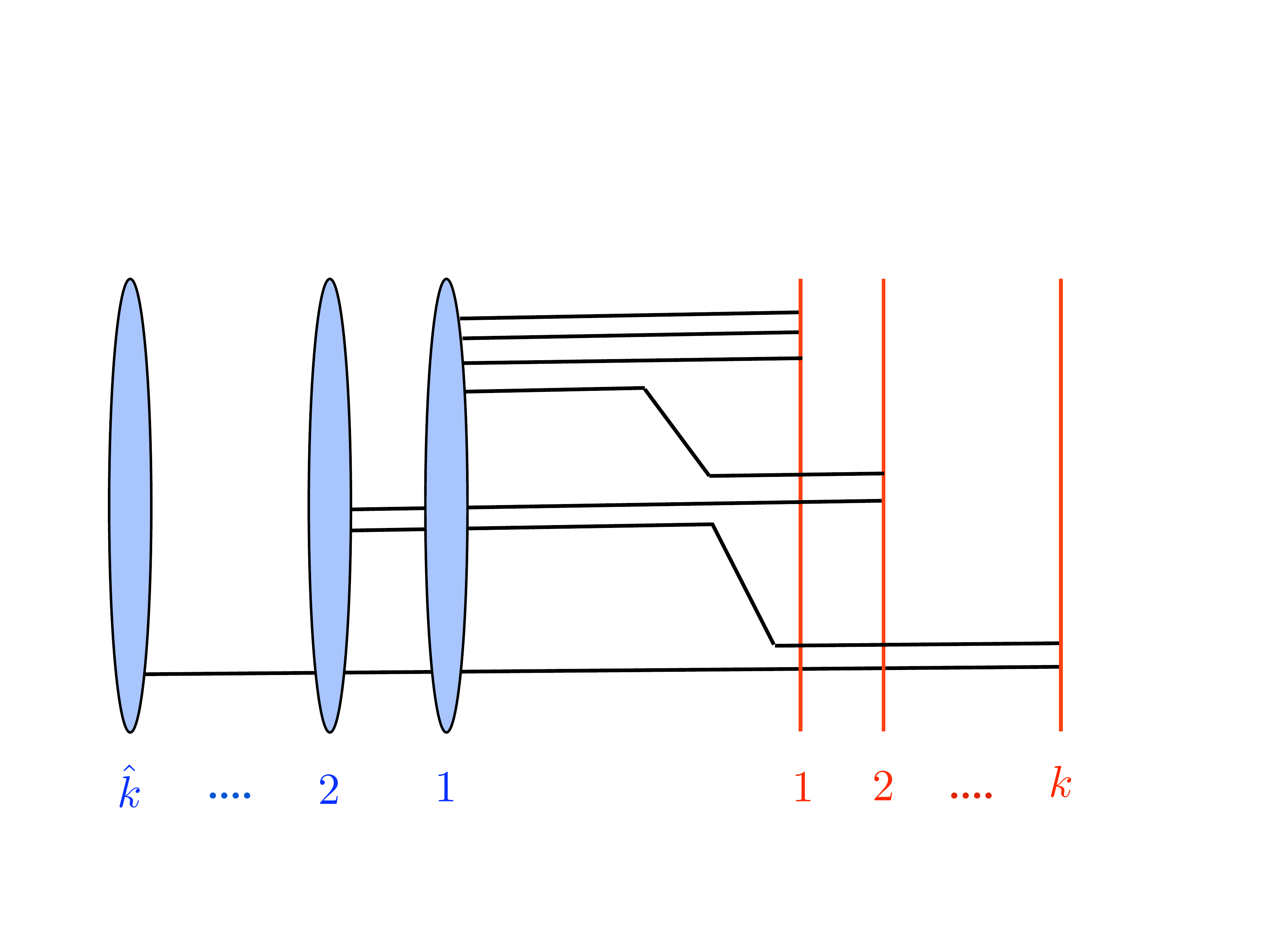}
\caption{\footnotesize Pushing all D5-branes to the right of all NS5-branes makes it easy to read the linking numbers, as
the net number of D3-branes ending on each five-brane.
In this example  $\rho = (3,2, \cdots 2)$ and $\hat\rho = (4, 2, \cdots 1)$. }
\label{separate}
\end{figure}

The brane construction of the linear quivers shown in Figure \ref{fig:linquivbrane} is characterized by the following linking numbers
\bea
l_i &=& j \quad \textrm{for a D5-brane between the $j$-th and $(j+1)$-th NS5 brane} \\\hat l_j &=& N_{j-1} - N_{j} + \sum_{s=j}^{\hat k-1} M_s \quad \textrm{for} \quad j=1,..,\hat k. \quad (N_{0}=N_{\hat k}=0)\,.\label{linkNS}
\eea
We may move all the NS5-branes to the left and all the D5-branes to the right, noting that a new D3-brane is created every time that a
D5  crosses a NS5. In the end, all the D3-branes will be suspended between  a  NS5-brane on the left and
 a D5-brane on the right, so  that the linking numbers satisfy the sum rule
\beq\label{sumrule}
\sum_{i=1}^{k} l_i = \sum_{j=1}^{\hat k} \hat l_j    \equiv N \,,
\eeq
where $N$ is the total number of  suspended D3 branes. This implies that the
  two sets of five-brane linking numbers define two partitions of $N$
   \bea\label{partitions}
  \rho: \qquad  ~N &=&  l_1+\ldots+l_{k}  \no  \\
 &=&   \underbrace{1+\ldots+1}_{M_1}\, +\, \underbrace{2+\ldots+2}_{M_2}\, +\, \ldots+ \ldots \\
  \hat \rho: \qquad  ~N &=& \hat l_1+\ldots +\hat l_{\hat k}\no  \\
  &=&   \underbrace{1+\ldots+1}_{\hat M_1}\, +\, \underbrace{2+\ldots+2}_{\hat M_2}\, +\, \ldots+ \ldots \,.
 \eea
 This is the repackaging of the quiver data in terms of partitions of $N$, mentioned above.
 It is illustrated by Figure  \ref{separate}.

 \vskip 1mm

In the original configuration of Figure  \ref{fig:linquiv}
 the
 D5 brane linking numbers are, by construction, positive and
     non-increasing, i.e.  $l_1\geq \cdots \geq  l_i\geq l_{i+1}\cdots \geq l_k>0$,
but this is not automatic for the linking numbers of the NS5 branes.   Requiring that the
     NS5 brane linking numbers be non-increasing, that is $\hat l_1 \geq \cdots \geq \hat l_i\geq \hat l_{i+1}\cdots \geq \hat l_{\hat k} = N_{\hat k - 1}$,
 is equivalent,  as follows from \rf{linkNS},   to
\beq\label{goodtheories}
M_i+N_{i-1}+N_{i+1}\geq 2N_i\,.
\eeq
This is the same as  \rf{condiSCFT}, the
 necessary and sufficient conditions for the corresponding (`good') quiver gauge theories to flow to an irreducible superconformal
  field theory in the infrared. Notice that if  these
   conditions  are not obeyed the linking numbers of the NS5 branes  need not even be positive integers.
   Furthermore, for the good theories that obey \eqref{goodtheories},
     it  follows from the expressions \rf{linkNS}  that the
   Young tableaux conditions  $  \rho^T >  \hat \rho$
are  automatically satisfied as long as the rank  of each  gauge-group factor  in the quiver diagram is positive.
\smallskip

In the configuration of  Figure \ref{separate} on the other hand, the
 meaning of the above conditions  changes. The ordering and positivity of all  linking numbers is  now automatic (more precisely, it can
 be trivially arranged by  moving 5-branes of the same type past each other).
   The constraints $  \rho^T >  \hat \rho$ on the other hand  are non-trivial;  they are the ones that   guarantee
  that a supersymmetric configuration like the one of Figure \ref{fig:linquiv} can be reached by a sequence of Hanany-Witten moves.  The two types of  configuration shown in the figures are in one-to-one correspondence when {\it all}
  these inequalities are satisfied by the five-brane linking numbers.

  \smallskip

Summarizing, the linear-quiver ${\cal N} = 4$
gauge theories  conjectured in  \cite{Gaiotto:2008ak} to  flow to irreducible fixed points in the infrared (without extra free decoupled multiplets)
 are labeled in an invariant way by two ordered partitions of $N$, with associated Young tableaux $\rho$ and $\hat\rho$
 subject to the conditions  $  \rho^T >  \hat \rho$.

\vspace{8mm}

\noindent \underline{{\bf Circular Quivers}}

\vspace{5mm}

\begin{figure}[th1]
\centering
\includegraphics[height=7.3cm,width=11cm]{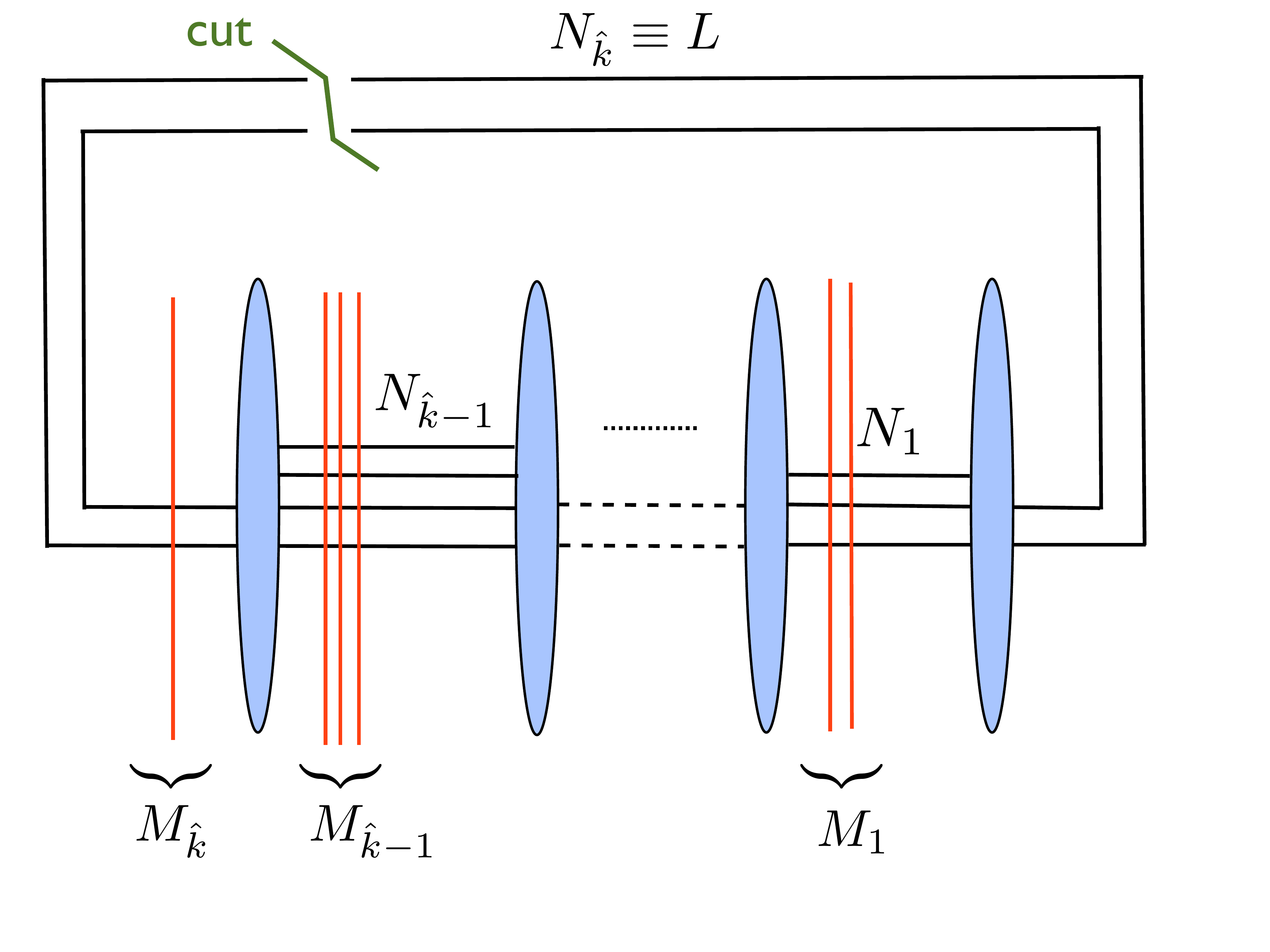}
\caption{\footnotesize Brane realization of circular quivers. To attribute linking numbers to the five-branes we cut open
the $k$-th stack of D3 branes, and place the $k$-th D5 stack at the left-most end.}
\label{fig:circquivbrane}
\end{figure}

The brane configuration corresponding to the circular-quiver gauge theory of Figure \ref{fig:circquiv} is given  in Figure \ref{fig:circquivbrane}.
In this case  the $x^3$ coordinate along the D3 branes is periodic.  Compared to the linear case,   there are
$N_{\hat k} >0$   additional D3 branes extended between the first and the $\hat k$th NS5 branes that  close the circle. There can be, as well,
  $M_{\hat k}\geq 0 $  extra D5 branes giving rise to  fundamental hypermultiplets.

We can associate linking numbers to the five-branes by cutting open the circular
quiver  along one of the suspended D3-brane stacks, say the $k$-th stack.
We also choose to place the
 $\hat k$-th stack of D5 branes   at the left-most end of the open chain, as shown  in  Figure \ref{fig:circquivbrane}.
The linking numbers are gauge-variant quantities, and the above choices amount to fixing partially a gauge.
 In this  gauge the   linking numbers read:
\bea
l_i &=& j \quad \textrm{for the $j$-th stack of D5 branes}  \, ,   \\
\hat l_j &=& N_{j-1} - N_{j} + \sum_{s=j}^{\hat k } M_s\ ,  \quad \textrm{with} \quad j=1,..,\hat k\ . \quad  (N_0 = N_{\hat k})
\label{circularlinking}
\eea
As in the case of linear quivers, we    label the  NS5 branes in order of appearance
 from right to left,  and   the D5 branes  from left to right.

Defined as above, the linking numbers obey the sum rule \eqref{sumrule} with $N\equiv \sum_{s=1}^{\hat k} sM_s$.
Furthermore  the   linking numbers of the D5 branes  are by construction  non-increasing,   positive and bounded by
the number of NS5 branes, i.e.
\beq\label{orderedD5}
\hat k \geq l_1\geq  \cdots l_i \geq l_{i+1} \cdots \geq l_k >0\ .
\eeq
 What  about the linking numbers of NS5 branes? For  linear quivers,
    imposing that  the $\hat l_j$ be non-increasing  was equivalent to the  Higgsing conditions \eqref{Higgsing}
     that singled out  the  `good theories',  i.e. those believed to flow  to  an irreducible
 superconformal fixed point in the infrared.  Now, the Higgsing conditions can be written as
 \bea\label{upperconds}
0 &\leq& N_{j+1} + N_{j-1} - 2 N_j + M_j =
 \hat l_j - \hat l_{j+1}  \quad   \textrm{for} \, \   j = 1,..,\hat k-1 \\
0 &< &  N_{1} + N_{\hat k -1} - 2 N_{\hat k}  + M_{\hat k} =
 \hat l_{\hat k} - \hat l_1 + \sum_{s = 1}^{\hat k} M_s  \,.
 \label{strict}
\eea
The second line, which gives the condition for Higgsing of the $\hat k$-th gauge-group factor,
needs explaining. We have assumed that,  for this factor,  the inequality \eqref{Higgsing} is strict.
A good circular quiver  always has  at least one   such  gauge-group  factor   because,    if all the  inequalities \eqref{Higgsing}
were saturated,  it can be shown  that  all the $N_j$  are equal,  and all $M_j=0$.  So, in this case,  there would be
 only bi-fundamental hypermultiplets, but these  cannot
break completely  the gauge group  since they are neutral  under the diagonal $U(1)$. This
possibility must thus be excluded, i.e. one or more of the inequalities \eqref{Higgsing} must be strict.
 We choose to cut open the circular quiver at a D3-brane stack  for which $N_{F,j} > 2N_j$. Without loss of generality
  this is  the $k$-th stack.
\smallskip

The conditions \eqref{upperconds}  tell  us   that the NS5-brane  linking numbers are  non-increasing.
If we want them to be positive, we must   impose that
\beq\label{extracond1}
\hat l_{\hat k} =  N_{\hat k -1}  - N_{\hat k}  + M_{\hat k} > 0\ .
\eeq
If we furthermore want  our gauge condition  to respect  mirror symmetry  we must impose  the analog of the first inequality
\eqref{orderedD5}, namely
\beq\label{extracond2}
\hat l_1 =    N_{\hat k} - N_1 + k  \leq k \ .
\eeq
Together  \eqref{extracond1} and \eqref{extracond2} imply   \eqref{strict},  but not the other way around.
Fortunately,  these conditions can be always satisfied in  good quivers, for example by choosing  a gauge factor whose  rank
 is locally  minimum along the chain  (i.e.   $N_{\hat k} < N_1, N_{\hat k - 1}$).  With this choice   we
finally   have
  \beq\label{NS5ordered}
 k \geq \hat l_1\geq  \cdots \hat l_j \geq \hat l_{j+1} \cdots \geq \hat l_{\hat k} >0\ ,
\eeq
 so that the NS5-brane   and the D5-brane linking numbers are on equal footing. They define two
 partitions, $\hat \rho$ and $\rho$ of the same number $N$.
  \smallskip

  Contrary to the case of linear quivers, here  the partitions do not  fully determine the brane configuration.
  The reason is that the number,  $N_{\hat k} \equiv L >0 $, of D3 branes in the $k$-th stack is  still free to vary. We can change it,
  without changing the linking numbers of the five-branes,  by adding or removing D3 branes that wrap
   the  circle (thus increasing or decreasing uniformly  all gauge-group ranks). It follows from
   \eqref{circularlinking} that  the condition for  all gauge-group factors
 to  have positive rank  now reads
     \beq
  L+ \rho^T  >    \hat\rho\     \,  .
  \label{fixedpointcircA}
\eeq
To understand this constraint   intuitively, note that  removing   $L$ winding D3 branes  may convert some stacks of D3 branes to stacks of  anti-D3 branes.
In the case of linear quivers  the inequality $\rho^T  >    \hat\rho$  guarantees  the absence of anti-D3 branes.
  Here anti-D3 branes are tolerated, as long as their number is less than $L$.
  \smallskip

   To any data $(\rho, \hat \rho, L)$ subject to the constraints \eqref{fixedpointcircA},  together with the additional conditions
   $l_1\leq \hat k$ and $\hat l_1 \leq k$, there corresponds a `good' circular-quiver gauge theory, i.e. one conjectured to  flow  to
   an irreducible superconformal theory in the infrared. This description is, however, highly redundant because of the arbitrariness
   in choosing at which  D3-brane stack  to cut open the quiver.  A generic circular quiver will have many gauge-group factors
   for which  \eqref{extracond1} and \eqref{extracond2} are satisfied, so
     many different triplets $(\rho, \hat \rho, L)$ would describe the same  SCFT.

   \smallskip

   To remove  this redundancy, one can impose the extra condition that
  the cut-open segment be  of   minimal rank \emph{globally}, i.e. that
    $L \leq N_j$ for all $j$.\footnote{If there are several gauge factors of globally-minimal rank,
  there will  remain some redundancy in our description of the circular quiver.
  This   is however a non-generic case.}
   This condition is compatible with the earlier ones;  it amounts to further fixing the gauge.  Now removing  $L$
  winding D3-branes does not  create any anti-D3 branes, since $L$ was the absolutely minimal   rank.
  The two partitions thus  obey the stronger inequality
      \beq
   \rho^T \geq \hat\rho\     \,  .
 \eeq
As a bonus, the conditions $l_1\leq \hat k$ and $\hat l_1 \leq k$ are now also automatically satisfied. Note  that
  linear-quiver theories saturating  some of the inequalities $\rho^T \geq \hat\rho$  broke down into smaller decoupled linear
  quivers plus free hypermultiplets. For circular quivers, on the other hand, these disjoint pieces are reconnected by the
  $L>0$ winding D3 branes, giving   irreducible  theories in the infrared.

 \smallskip

Summarizing, the circular-quiver gauge theories conjectured to  flow to irreducible superconformal field theories in the infrared
can be  labeled by a positive integer $L$, and by
two ordered partitions $\rho$ and $\hat\rho$  subject to the condition $\rho^T \geq \hat\rho$.
  An alternative but redundant description is in terms of a triplet $(\rho, \hat\rho, L)$  subject to the looser
  conditions \eqref{fixedpointcircA},  together with the additional constraints $l_1\leq \hat k$ and $\hat l_1 \leq k$.
  Both  descriptions are manifestly  mirror-symmetric.
  As we will later discuss,  in  the dual supergravity theory
  these two descriptions correspond to a complete,  or to a partial gauge fixing of  the 2-form potentials.

\vspace{1cm}

\section{Defect SCFTs}
\label{sec:defects}

The IR fixed points of the 3d $\N=4$ linear quiver theories described above have a natural extension as defect SCFTs . The theories in question arise as the IR fixed points of $\N=4$ $d=4$ Super-Yang-Mills theory interacting with a 3d linear quiver gauge theory living on a 3d defect. The bulk-boundary couplings are such that half of the bulk supersymmetries are conserved, hence the name of $\half$-BPS defect SCFTs. Such defect SCFTs with a single 3d hypermultiplet living on the defect and their holography have been considered in \cite{DeWolfe:2001pq,Erdmenger:2002ex}, where the superconformal action was given \footnote{ In the case of a single defect hypermultiplet the theory is directly (super)conformal. Defects supporting 3d quiver gauge theory on the other hand are not conformal theories, the SCFTs arise as their infrared fixed points. }. The more general $\half$-BPS defects have been classified in \cite{Gaiotto:2008sa,Gaiotto:2008sd,Gaiotto:2008ak}
 where they were understood, as in the last subsection, from brane configurations with D3, D5 and NS5 branes.\\
These general defect SCFTs consist in having a 3d $\N=4$ linear quiver gauge theory living on the defect with supersymmetric couplings to the bulk fields through "mixed"  bifundamental hypermultiplets. To describe the gauge theories we consider it it simpler to start with the brane picture.

\begin{figure}[th1]
\centering
\includegraphics[scale=0.5]{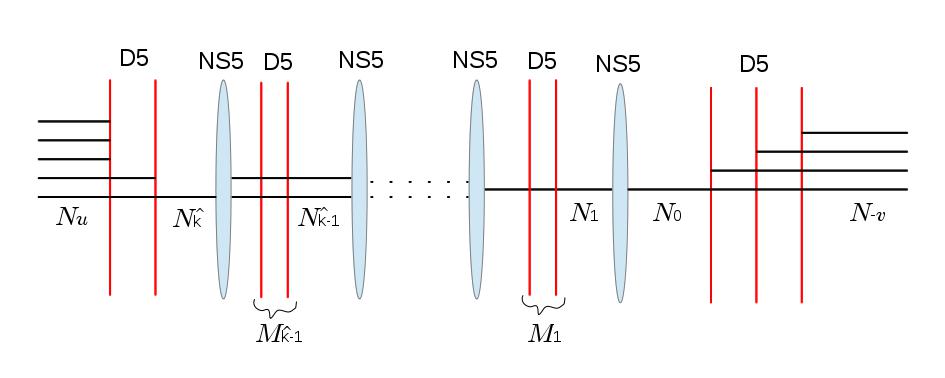}
\caption{\footnotesize Brane realization of a defect quiver (flowing to a $\half$-BPS defect SCFT). The $N_L=N_u$ ($N_R = N_{-v}$) D3-branes on the left (right) are semi infinite in the $x^3$ direction.}
\label{fig:defectquivbrane}
\end{figure}
\vspace{5mm}

The brane setup is the same as the one we considered for linear quivers in section \ref{sec:branes} but now we allow for semi-infinite D3-branes ending on the left NS5-brane and semi-infinite D3-branes ending on the right NS5-brane. This is not actually the most general case. We also allow for the semi-infinite D3-branes, say those coming from the left, to end on D5-branes that are placed on the left of the brane configuration, and similarly for the semi-infinite D3-branes coming from the right.  A generic brane configuration is given in figure \ref{fig:defectquivbrane}.

\begin{figure}[th1]
\centering
\includegraphics[scale=0.5]{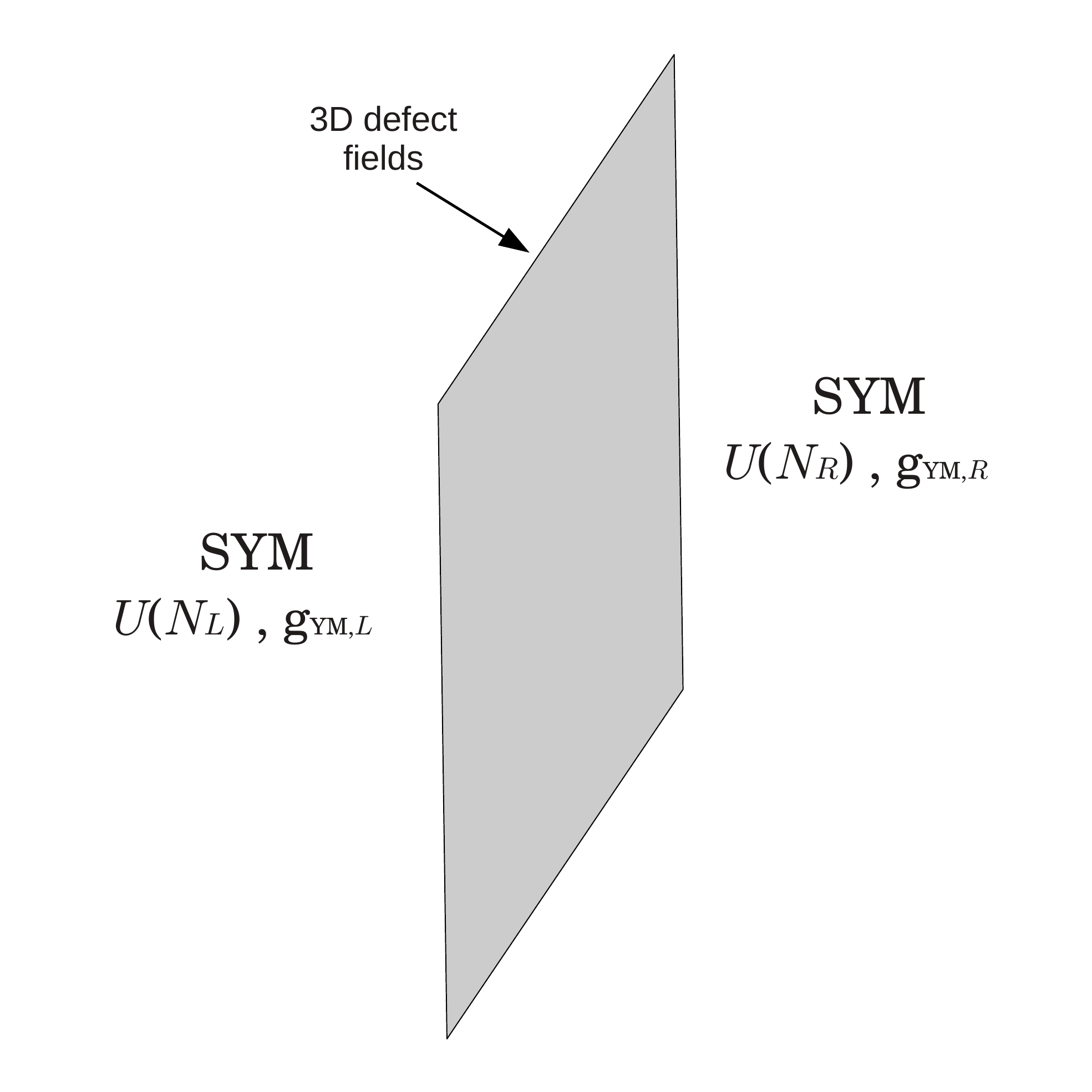}
\caption{\footnotesize 4d $\N=4$ SYM with 3-dimensional gauge theory living on a defect. The left and right bulk regions have independent gauge groups and gauge couplings.}
\label{fig:defect}
\end{figure}

\vspace{5mm}

Let's consider the simple cases when there is no extra D5-branes on the left and right of the brane configuration. In this case there are $N_L$ semi-infinite D3-branes ending on the left NS5-brane and  $N_R$ semi-infinite D3-branes ending on the right NS5-brane. The defect gauge theory is easily understood.

\begin{figure}[th1]
\centering
\includegraphics[scale=0.35]{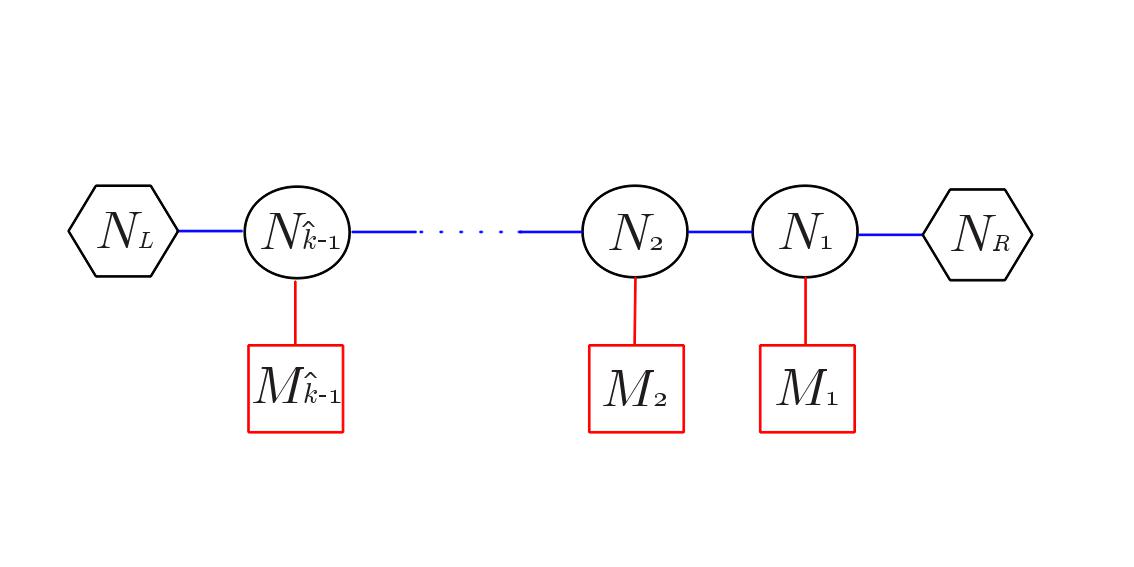}
\caption{\footnotesize Quiver picture of a 3d defect field theory with {\it NS5-like} boundary conditions for the bulk fields (without extra D5-branes) . The meaning of each element is the same as for a 3d linear quiver except for the two external nodes (hexagons) corresponding to $U(N_L)$ and $U(N_R)$ 4d $\N=4$ vector multiplets (SYM) on the two half spaces and the blue line connecting them to the 3d quiver corresponding to the mixed ''bulk-defect`` bifundamental 3d hypermultiplets.}
\label{fig:dCFTs}
\end{figure}

The linear quiver theory obtained form the brane configuration when ignoring the semi-infinite D3-branes is the field theory living on the 3-dimensional defect. This defect cuts the 4d space in two regions (see figure \ref{fig:defect}), say the {\it left} and {\it right} regions, and can be seen as a boundary for each of these two half-spaces. On the {\it left} region (resp. {\it right } region) lives a $\N=4$ $U(N_L)$  (resp. $U(N_R)$) SYM theory, with independent gauge couplings $g_{YM}^L$ and $g_{YM}^R$.
To describe the boundary conditions for the bulk fields, focusing on bosonic fields only, one has to decompose the 4d-fields of SYM into $(A_{\mu}, X^{1,2,3})$, $\mu = 0,1,2$, and $(A^3, Y^{1,2,3})$, where the $X^i$ and $Y^i$ are the six scalars. This decomposition corresponds to vector- and hyper-multiplets from a 3d point of view.\\
 Then the bulk fields obey {\it  NS5-like } boundary conditions on the defect, which means Neumann boundary conditions for $(A_{\mu}, X^{1,2,3})$ and Dirichlet boundary conditions for $(A^3, Y^{1,2,3})$ (see \cite{Gaiotto:2008sa}) .
 The defect fields are coupled to the bulk fields of the {\it left} region through an additional defect hypermultiplet transforming in the fundamental of the left node $U(N_{\hat k -1})$ of the quiver and in the fundamental of the boundary $U(N_L)$, which means a 3d $\N=4$ coupling to the 3d multiplet $(A^{\mu}_L, X^{1,2,3}_L)$  induced on the boundary. The coupling to the 4d fields of the {\it right} region is similar :  {\it  NS5-like } boundary conditions for the { \it right} bulk fields  and another "mixed" defect hypermultiplet transforming in the fundamental of the right node $U(N_1)$ of the quiver and in the fundamental of the boundary $U(N_R)$. 
The theory content is summarized by the quiver diagram of \ref{fig:dCFTs} where the end-nodes correspond to the {\it left} and {\it right} 4d $\N=4$ SYM with their respective gauge groups $U(N_L)$ and $U(N_R)$, and the links from these nodes to the linear quiver nodes correspond to the additional "mixed bulk-defect`` bifundamental hypermultiplets.

\bigskip 

The more general brane configuration when semi-infinite D3-branes end on D5-branes placed at the left or on the right of all the NS5-branes correspond to having more general $\half$-BPS  boundary conditions for the bulk fields. These general boundary conditions are described in detail in \cite{Gaiotto:2008sa} and we will only recall here their main features. They break the gauge symmetry $U(N_L)$ (resp. $U(N_R)$) to a smaller $U(N_{\hat k})$ (resp. $U(N_0)$) on the defect and involve mixed {\it NS5-like} and {\it D5-like} boundary conditions 
\footnote{ The  {\it D5-like} boundary conditions are Dirichlet for $(A_{\mu}, Y^{1,2,3})$ and ''modified`` Neumann for $(A^3, X^{1,2,3})$ on the defect (see \cite{Gaiotto:2008sa}). }
 for the bulk fields. The general boundary conditions  are given by Nahm's equation, which allow pole singularities on the boundary for the scalar fields (Nahm poles). The solutions to Nahm's equations depend on the linking numbers of the external D5-branes, that define an embedding of $su(2)$ into $su(N_L)$ ($su(N_R)$), and additional moduli that should be taken to zero in our context, to preserve conformal symmetry (this is the same as being at the origin of the Higgs branch in a sense).

The presence of these additional D5-branes change the boundary conditions for the bulk fields but not the matter content of the field theory living on the defect, except in the case of an external D5-branes without D3-brane ending on it. Such a D5-brane corresponds to a defect hypermultiplet transforming in the fundamental representation of the boundary $U(N_0)$ (or $U(N_{\hat k})$).

\begin{figure}[th1]
\centering
\includegraphics[scale=0.3]{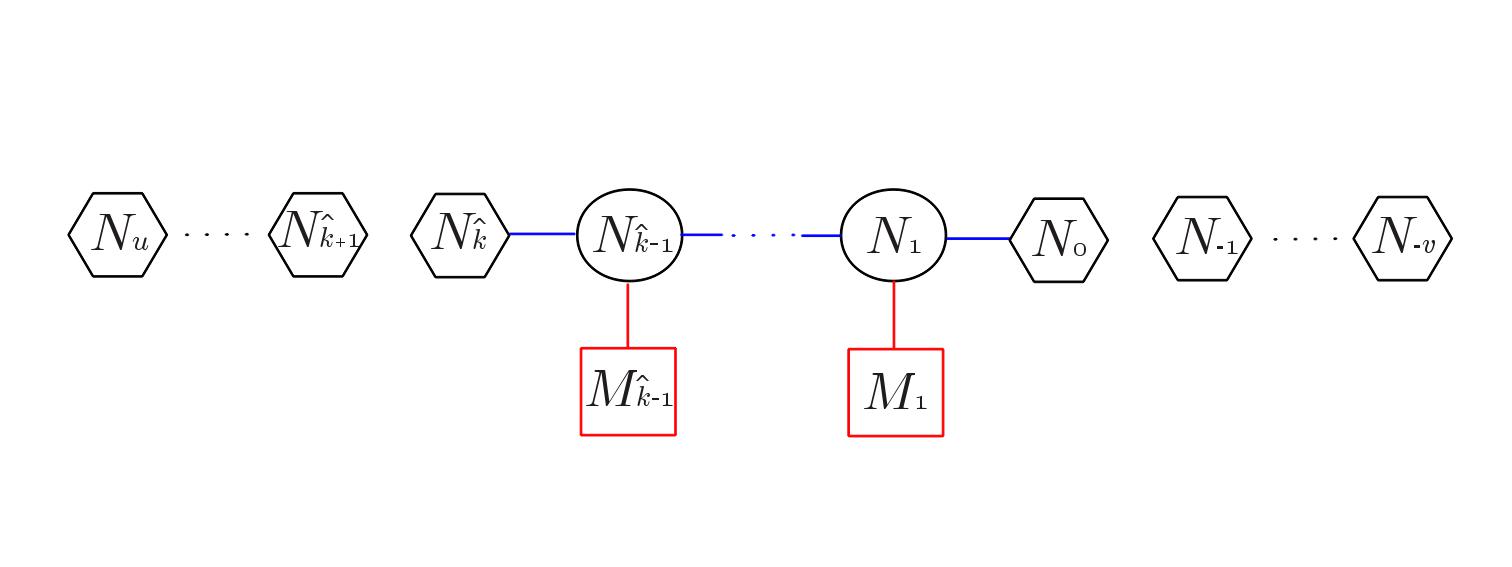}
\caption{\footnotesize Quiver picture of a general $\half$-BPS 3d-defect field theory. The hexagonal nodes denote the number of D3-branes in the segments inbetween the external D5-branes,encoding to the boundary conditions for the bulk fields (\cite{Gaiotto:2008sa}) with $N_u = N_L$ and $N_{-v}=N_R$.}
\label{fig:defectquiv}
\end{figure}

The general $\half$-BPS defect SCFTs can be summarized in the quiver picture of figure \ref{fig:defectquiv}. Compared to \ref{fig:dCFTs} there are extra nodes on the left and on the right, corresponding to the numbers of D3-branes in the segments between the external adjacent D5-branes, accounting for the left and right  boundary conditions.

\bigskip

From figure \ref{fig:defectquivbrane} we can associate a linking number to each 5-brane with the same definition \ref{defnlinking} as for linear quivers, defining two non-standard partitions $\rho, \hat\rho$. We call them ''non-standard`` because there may be negative linking numbers both for the D5 and NS5-branes.

As in the case of linear quivers, the conditions of irreducibility of the IR SCFT are given by the ordering of the linking numbers in non-decreasing order from left to right for NS5-brane and from right to left for the D5-branes. The ordering of the NS5-brane linking numbers is again equivalent to the Gaiotto-Witten conditions \ref{Higgsing}. The ordering of D5-brane linking number was automatic for linear quivers but it is not the case here because of the additional left and right D5-branes. Ordering the linking numbers of the D5-branes amounts to say that the number of (semi-infinite) D3-branes ending on the left-most D5-brane is bigger than the number of D3-branes ending on the second left-most D5-brane, ... etc, and similarly for the right-most D5-branes. According to \cite{Gaiotto:2008ak} it ensures again that there is no decoupling hypermultiplets in the infrared limit.
\smallskip

Moving the NS5-branes on the left and the D5-branes on the right, it is easy to see that $\rho$ defines now an ordered partition of $N-N_R$, while $\hat\rho$ defines now an ordered partition of $N-N_L$, with $N$ the total number of D3-branes inbetween the D5 and NS5-branes. The partitions may contain negative numbers.

\vspace{6mm}

To summarize, the defect quivers are given by the data $(\rho,\hat \rho,N_L,N_R,g_{YM}^{(L)},g_{YM}^{(R)})$,  where $g_{YM}^{(L)}$ and $g_{YM}^{(R)}$ are the Yang-Mills couplings of SYM on the left and right half-spaces. As for linear quivers, this data obeys the sumrule 
\begin{align}
\label{defectsumrule}
\sum_{i=1}^k l_i + N_R = \sum_{j=1}^{\hat k} \hat l_j + N_L  \quad .
\end{align}
We can denote $\scD(\rho,\hat \rho,N_L,N_R,g_{YM}^{(L)},g_{YM}^{(R)})$ the corresponding infrared fixed point SCFT.
\vspace{5mm}

As for linear and circular quivers the partitions $\rho$ and $\hat\rho$ have to satisfy some inequalities, so that they define a defect quiver with positive ranks.
The positivity of the ranks translate in this case into the inequalities
\begin{align}
\label{ineqdefect}
 N_{0} +  \rho^{T}_{>0}  > \hat\rho \quad , \quad  N_0 > 0 \ ,
\end{align}
where $\rho^{T}_{>0}$ is the transposed Young tableau of the partition $\rho$ truncated to its positive components  \footnote{For instance : $\rho =(4,2,1,0,-2)$, $\rho_{>0} = (4,2,1)$. } .\\
Note that the condition $N_0 > 0$ (resp. $N_{\hat k}>0$) ensures $N_{i} > 0$ for $i<0$ (resp. $N_{i} > 0$ for $i> \hat k$) because of the ordering of the linking numbers. 

When $N_R = N_0$ (no D5-branes on the right of the quiver) these inequalities reduce to
\begin{align}
\label{ineqdefect2}
 N_R+  \rho^{T}  > \hat\rho \quad , \quad  N_R > 0 \ .
\end{align}

The saturation of one inequality corresponds again to having a node with zero rank. The defect SCFT then breaks into two independant {\it boundary } SCFTs, which are defect SCFTs with $N_L=0$ or $N_R=0$ : SYM theory on a half-space coupled to a 3d boundary fields.

\vspace{10mm}

\section{Partition function of deformed linear quivers}
\label{sec:Zexpr}

In this section we review the proposal of \cite{Nishioka:2011dq} for the explicit analytic expression for the linear quivers $T^{\rho}_{\hat\rho}(SU(N))$ deformed by real masses $m_j$ and Fayet-Iliopoulos parameters $\eta_j \equiv \xi_j - \xi_{j+1}$ on the 3-sphere.
This explicit expression was derived from the matrix models obtained by the techniques of localization (see section \ref{sec:3dN4} above) and make mirror symmetry manifest. The matrix models do not depend on the Yang-Mills coupling $g_{YM}$,  which means that it computes directly the partition function in the infrared limit $g_{YM} \rightarrow +\infty$.
We will use it in chapter \ref{chap:GKPW} when we compute free energies at the superconformal point (zero mass and FI terms). 

\vspace{8mm}

To express the analytic expression for the partition function we need to introduce what we call the deformed partitions and the deformation $N$-vectors.
\smallskip

To each 5-brane in the brane picture we have associated a linking number $l_i$ or $\hat l_j$. We can associate also to each 5-brane a deformation parameter which corresponds to its position in transverse space.\\
The D5-branes can be displaced along $(x_7,x_8,x_9)$ (without breaking additional supersymmetries) giving three mass parameters (real and complex masses) to the corresponding fundamental hypermultiplet. The masses transform as a triplet of the $SU(2)_{R}\simeq SO(3)$ rotation group of $(x_7,x_8,x_9)$ and two of them can be set to zero (see \S\ref{sec:3dN4}). So we associate only one deformation parameter $m_i$ to each D5-branes and it corresponds to giving a real mass $m_i$ to the fundamental hypermultiplets.\\
Similarly the NS5-branes can be displaced along $(x_4,x_5,x_6)$ (without breaking additional supersymmetries), giving three real parameters that transform as a triplet of the $SU(2)_{L}\simeq SO(3)$ rotation group of $(x_4,x_5,x_6)$. Again two of them are set to zero by using this rotation symmetry. The non-vanishing parameters $\xi_j$ are associated to the NS5-branes (one for each) and they are related to the Fayet-Iliopoulos parameters $\eta_j$ of the gauge nodes as $\eta_j=\xi_{j+1}-\xi_j$. To summarize we have one linking number and one real deformation parameter for each 5-brane.
\smallskip

We define the deformed partitions, that we call again $\rho$ and $\hat \rho$, as 
\begin{align}
 \rho & := \Big( (l_1,m_1),(l_2,m_2), \ ... \ ,(l_k,m_k) \Big) \no\\
 \hat \rho & := \Big( (\hat l_1,\xi_1),(\hat l_2,\xi_2), \ ... \ ,(\hat l_{\hat k},\xi_{\hat k}) \Big)
\end{align}
and the deformation $N$-vectors 
\begin{align}
\label{Nvec}
 \vec m &:= \big( \ \textrm{coord}( \vec m_1) \  , \ \textrm{coord}( \vec m_2) \ , \ ... \ , \ \textrm{coord}( \vec m_k) \ \big) \no \\
   & \textrm{with} \quad \vec m_j = \left\{ \ m_j + i \Big(\frac{l_j+1}{2}-1 \Big), \ m_j + i \Big(\frac{l_j+1}{2}-2 \Big), \ ... \ , \  m_j + i \Big(\frac{l_j+1}{2}-l_j \Big) \ \right\} \no \\
 \vec \xi &:= \big( \ \textrm{coord}( \vec \xi_1) \  , \ \textrm{coord}( \vec \xi_2) \ , \ ... \ , \ \textrm{coord}( \vec \xi_{\hat k}) \  \big) \\
   & \textrm{with} \quad \vec \xi_j = \bigg\{ \ \xi_j + i \Big(\frac{\hat l_j+1}{2}-1\Big), \ \xi_j + i \Big(\frac{\hat l_j+1}{2}-2\Big), \ ... \ , \ \xi_j + i\Big(\frac{\hat l_j+1}{2}- \hat l_j\Big) \ \bigg\}  \ , \no
\end{align}
where coord$(\vec v) = v_1 , v_2 , v_3, ... , v_p$  for a vector $\vec v$ with $p$ coordinates.\\
Note that $\vec m$ and $\vec \xi$ are vectors with $N$ coordinates, while $\vec m_j$ and $\vec \xi_j$ are vectors with $l_j$, resp. $\hat l_j$, coordinates.
\smallskip

For instance for the linear quiver in figure \ref{fig:T(2111,221)brane}, we have
\begin{align}
\label{Nvec_ex}
 \rho & = \Big( (2,m_1),(1,m_2), (1,m_3) ,(1,m_4) \Big) \no\\
 \hat \rho & = \Big( (2,\xi_1),(2,\xi_2),(1,\xi_3) \Big) \no\\
 \vec m & = \left( m_1 + \frac{i}{2} \ , \ m_1 - \frac{i}{2} \ , \ m_2 \ , \ m_3 \ , \ m_4  \right) \\
 \vec \xi & = \left( \xi_1 + \frac{i}{2} \ , \ \xi_1 - \frac{i}{2} \ , \ \xi_2 + \frac{i}{2} \ , \ \xi_2 - \frac{i}{2}  \ , \ \xi_3 \right) \ . \no
\end{align}

\begin{figure}[h]
\centering
\includegraphics[scale=0.6]{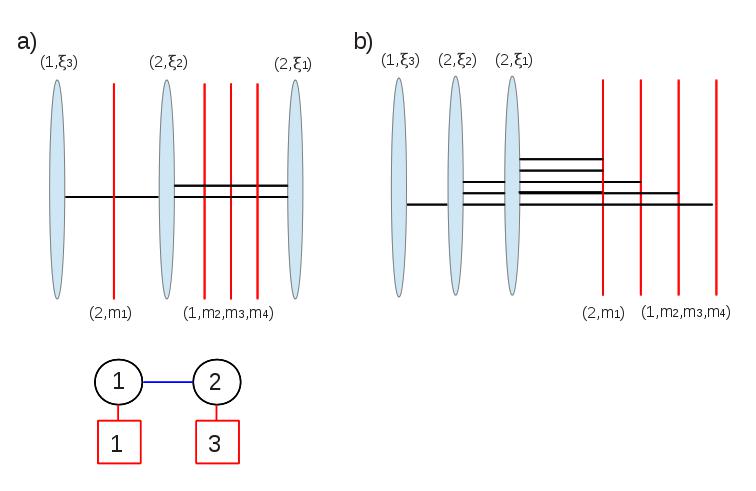}
\caption{\footnotesize{ a) Brane configuration and quiver for the theory $T^{(2111)}_{(221)}(SU(5))$. b) Brane configuration with separated NS5-branes and D5-branes. Below each 5-brane is indicated the linking number and deformation parameter associated to it.  }}
\label{fig:T(2111,221)brane}
\end{figure}

The conjecture of \cite{Nishioka:2011dq} is that the partition function for the deformed $T^{\rho}_{\hat \rho}(SU(N))$ is given up to a phase by
\begin{align}
\label{Zformula}
 Z = \frac{ \scZ }{ \Delta_{\hat\rho}(\vec \xi) \Delta_{\rho} (\vec m) }  \quad ,  \quad
 \scZ = \sum_{w \in \mathfrak{S}^N} (-1)^w \ e^{2 i \pi \, \vec \xi . w(\vec{m})}
\end{align}
with $\mathfrak{S}^N$ the group of permutations of $N$ elements, $(-1)^w$ the signature of a permutation $w$ and $\vec \xi . w(\vec{m}) = \sum_{j=1}^N \xi_j m_{w(j)}$. The definition of the determinants $\Delta_{\hat\rho}(\vec \xi)$ and $\Delta _{\rho}(\vec m)$ requires to arrange the deformation $N$-vectors $\vec \xi$ and $\vec m$ as tableaux $[\xi]=([\xi]_{ab})$ and $[m]=([m]_{ab})$ whose lines are the vectors $\vec \xi_1, \vec\xi_2,...,\vec\xi_{\hat k}$, resp. $\vec m_1, \vec m_2,..., \vec m_k$ and we have 
\begin{align}
\label{Delta}
\Delta_{\hat\rho}(\vec \xi) &=  \prod_{b=1}^{\hat l_{\hat k}} \prod_{a<a'} \sh([\xi]_{ab} -[\xi]_{a'b})  \no\\
\Delta_{\rho}(\vec m) &=  \prod_{b=1}^{l_k} \prod_{a<a'} \sh([m]_{ab} -[m]_{a'b}) \ .
\end{align}
For instance for \ref{Nvec_ex} we have
\begin{align}
 [\xi] &= 
\begin{bmatrix}
\xi_1 + \frac{i}{2} & \xi_1 - \frac{i}{2} \\
\xi_2 + \frac{i}{2} & \xi_2 - \frac{i}{2} \\
\xi_3 & .
\end{bmatrix}
\quad , \quad  
[m] = 
\begin{bmatrix}
m_1 + \frac{i}{2} & m_1 - \frac{i}{2} \\
m_2 & . \\
m_3 & . \\
m_4 & . 
\end{bmatrix}
\end{align}
where a square with ''.`` is not a box of the tableau (these tableaux have five boxes each), in particular it does not produce a $\sh$-factors in \ref{Delta}.\\
And the determinant are
\begin{align}
\Delta_{\hat\rho}(\vec \xi) &= - \ \ch(\xi_1- \xi_2) \ \ch(\xi_1- \xi_3) \ \sh(\xi_2- \xi_3)^2 \\
\Delta_{\rho}(\vec m) &= -i \ \sh(m_1-m_2) \ \sh(m_2-m_3) \ \sh(m_1-m_3) \ \ch(m_1-m_4) \ \ch(m_2-m_4) \ \ch(m_3-m_4) \ . \no
\end{align}

Let's make a few comments about this formula.

First, the partition function \eqref{Zformula} is manifestly invariant
under the simultaneous exchange of the deformed partitions $\rho$ and $\hat{\rho}$ (exchange of the partitions and of the parameters $m$ and $\xi$).This is a manifestation of the 3d mirror symmetry.

Second, \eqref{Zformula} vanishes unless $\rho^T \ge \hat{\rho}$ 
 \cite{Nishioka:2011dq}. This is
consistent with the condition \eqref{fixedpoint}
for the existence of a non-trivial IR SCFT.
\footnote{The case of saturation of one inequality is here accepted as it corresponds to having a node with zero rank, so that the quiver breaks into two pieces. In this case the IR SCFT is not irreducible : the partition function should be the product of the two decoupled SCFT. The vanishing of $Z$ is only expected when supersymmetry is broken (negative rank node).}

Third, the expression \eqref{Zformula} is mysteriously complexe and moreover has a (parameter-independent) phase ambiguity. We will concentrate (in chapter \ref{chap:GKPW}) on the absolute value of the $S^3$ partition function and define the free energy as $F = - \log |Z|$.

\vspace{0.8cm}

\underline{\textbf{$T(SU(N))$ gauge theory}} :
\bigskip

The simplest linear quiver theory one can study turns out to be the so-called $T(SU(N))$ SCFT and its deformed version. The linear quiver is given by the two partitions
\begin{align}
\rho = \hat\rho = \big[\overbrace{1,1,...,1}^{N}\big] \ .
\end{align}

\begin{figure}[h]
\centering
\includegraphics[scale=0.6]{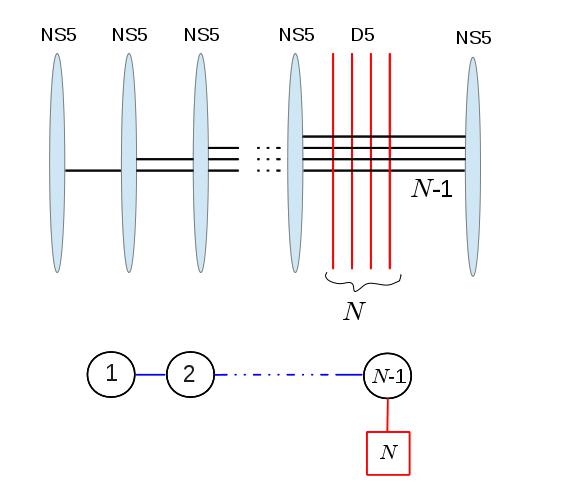}
\caption{\footnotesize{ Brane configuration and quiver of the $T(SU(N))$ theory.}}
\label{fig:T(SU(N))brane}
\end{figure}

The gauge group is $U(1)\times U(2) \times ... \times U(N-1)$ and the matter content, on top of the bifundamental hypermultiplets consists of $N$ hypermultiplets transforming in the fundamental representation of the $U(N-1)$ node. The quiver and the brane configuration are presented in figure \ref{fig:T(SU(N))brane}. Strictly speaking $T(SU(N))$ is the IR fixed point SCFT of the linear quiver.\\
 The theory has a group of global symmetry $SU(N)_F\times SU(N)_J$ with $SU(N)_F$ acting on the $N$ fundamental hypermultiplets (which transform again in the fundamental representation) and the $SU(N)_J$ arising as an enhancement of the topological $U(1)^{N-1}$ symmetry at the IR fixed point. $T(SU(N))$ is known to be invariant under mirror symmetry, which exchanges the two $SU(N)$ global symmetries. This is the larger group of global symmetries accessible for a $T^{\rho}_{\hat\rho}(SU(N))$ linear quiver.

The theory has deformation parameters which can be thought as vev for scalars in background $U(1)$ vector multiplets (meaning vector coupling to $U(1)$ global symmetry current). These parameters are $N$ masses $m_j$ for $N$ fundamental hypermultiplets and $N-1$ FI parameters $\eta_j$ for the $N-1$ nodes. These FI parameters are conveniently recast in $N$ parameters $\xi_j$ defined by $ \eta_j = \xi_j - \xi_{j+1}$ as explained above.\\

The deformed partitions for $T(SU(N))$ are
\begin{align}
 \rho & := \Big( (1,m_1), (1,m_2), \ ... \ , (1,m_N) \Big) \no\\
 \hat \rho & := \Big( (1,\xi_1), (1,\xi_2), \ ... \ , (1,\xi_N) \Big) \ ,
\end{align}
 and the deformation $N$-vectors and the determinants are simply
\begin{align}
\vec m &= (m_1 \ , \  m_2, \ ... \ , \  m_N)  \no\\
\vec \xi &= (\xi_1 \ , \  \xi_2, \ ... \ , \  \xi_N)  \no\\
\Delta_{\rho}(\vec m) &=  \prod_{i<j}^N  \sh(m_i - m_j)   \\
\Delta_{\hat\rho}(\vec \xi) &=  \prod_{i<j}^N  \sh(\xi_i - \xi_j)  \no
\end{align}

The matrix model giving the partition function was computed in \cite{Benvenuti:2011ga, Nishioka:2011dq} and turned out to be one of the simplest non-abelian partition function. 
\begin{align}
\label{ZTSUN}
 Z^{T(SU(N))} &= (-i)^{\frac{N(N-1)}{2}} e^{-2i\pi \xi_N \sum_j^N m_j} \frac{ \sum_{w \in  \mathfrak{S}_N} (-1)^w \, e^{2i\pi \sum_j^N \xi_j m_{w(j)}}}{\prod_{j<k}^N \sh(\xi_j - \xi_k)\sh(m_j - m_k)} 
\end{align}
with $\mathfrak{S}^N$ the group of permutations of $N$ elements. This reproduces, up to a phase, the prediction of \ref{Zformula}.

As will be shown in chapter \ref{chap:GKPW}, the $T(SU(N))$ SCFT plays a distinguished role among the $T^{\rho}_{\hat\rho}(SU(N))$ theories as the SCFT with maximal free energy.

\chapter{Supergravity solutions and the correspondence}
\addcontentsline{lot}{chapter}{  Supergravity solutions and the correspondence}
\label{chap:sugra}

%
%
%
%
%
%
%
%
%
%
%
%
%
%
%
%

In this chapter we present the large class of type IIB supergravity solutions with $OSp(4|4)$ symmetry that we constructed. We expose our AdS/CFT dictionnary with all ``irreducible'' infrared super-conformal fixed points of $d=3$, $\N=4$ linear quivers, circular quivers and $d=4$ SYM $\half$-BPS defect theories. We emphasize the relation with the rich brane picture. Interesting limiting geometries are also discussed.

\section{History of the type IIB supergravity solution with $OSp(4|4)$ symmetry}
\label{history}

The IIB supergravity solutions that we study in this presentation have a local structure that was derived by D'Hoker, Estes and Gutperle in
\cite{D'Hoker:2007xy,D'Hoker:2007xz}. These solutions have been searched as gravity duals of $\half$-BPS 3d-defect SCFTs or infinitely thin BPS domain walls in 4d $\N=4$ super-Yang-Mills theory (described in section \ref{sec:defects}). Before the explicit solutions were derived, \cite{Gomis:2006cu} already classified these $\half$-BPS defect SCFTs and the corresponding supergravity solutions, that are domain walls interpolating between two $AdS_5 \times S^5$ asymptotic regions.
In \cite{Gomis:2006cu} the domain wall solutions were already described as $AdS_4 \times S^2 \times S^2 \ltimes \Sigma_2$ warped geometries where $\Sigma_2$ is a strip with D5-brane singularities on one boundary and NS5-brane singularities on the other boundary. The data describing the solutions was predicted to be encoded in the fluxes escaping the singularities. This is exactly the features of the solutions we describe in this chapter,  realizing the geometric transitions between branes and fluxes. Other type IIB gemetries dual to $\half$-BPS operators in $\N=4$ SYM were found previously (\cite{Lin:2004nb,Liu:2006pd,Lunin:2006xr,Yamaguchi:2006te}). Preliminary work studying the $\half$-BPS $AdS_4$ embedding of probe D5-branes into $AdS_5 \times S^5$ can be found in \cite{Karch:2000gx,Skenderis:2002vf}.

The strategy of \cite{D'Hoker:2007xy} to derive the relevant solutions is to encode the $SO(2,3) \times SO(3) \times SO(3)$ bosonic symmetries of $OSp(4|4)$ in the ansatz
\begin{align}
ds^2 = f_4^2(z,\bar z) \, ds^2_{AdS_4} +   f_1^2(z,\bar z) \, ds^2_{S^2_{(1)}} +   f_2^2(z,\bar z) \, ds^2_{S^2_{(2)}} + 4 \rho(z,\bar z)^2 \, dzd\bar z \ .
\end{align}
with $f_4, f_1, f_2, \rho$ real functions of $z,\bar z$,
 and to look for solutions with 16 supersymmetries (8 Poincaré + 8 super-conformal), the 16 independent supersymmetry generators being Killing spinors of $AdS_4 \times S^2_{(1)} \times S^2_{(2)}$. The BPS equations provide first order diferential equations. The solutions are found to be expressed in terms of two real harmonic functions $h_1, h_2$ on $\Sigma$, describing the cases of vanishing axion field. The solutions with non-vanishing axion are obtained by $SL(2,\bR)$ transformations that are symmetries of type IIB supergravity.

In \cite{D'Hoker:2007xz} the authors exposed the conditions for the regularity of the solutions,  putting severe constraints on the harmonic functions $h_1,h_2$. An important constraint is that on the boundary of $\Sigma$ one harmonic function and the normal derivative of the other should vanish, so that the boundary is partitioned in segments where $h_1 = \p_{\bot}h_2 = 0$ or $h_2 = \p_{\bot}h_1 = 0$. A consequence of this condition is that one 2-sphere or the other vanishes at each point of the boundary and, combining with the normal direction, has the local topology of a 3-sphere. Thus the boundary points of $\Sigma$ are actually interior points of the geometry. The regularity conditions allow for very specific kinds of point singularities on the boundary of $\Sigma$ that we detail below.

In \cite{Assel:2011xz}, we showed that the asymptotic $AdS_5 \times S^5$ regions can be removed from the geometry by taking the limit of zero asymptotic radius, without destroying the solutions. The resulting geometries are dual to the IR fixed point of 3d $\N=4$ linear quiver that have the same  $OSp(4|4)$ symmetries.

\section{Solutions of IIB supergravity : the case of the strip}
\label{sec:gravitysolns}

We will now exhibit the solutions of type-IIB supergravity that are holographic duals of  superconformal field theories
discussed above (chapter \ref{chap:quivers}). 

 First we review the  general local solutions of type IIB supergravity that have the appropriate $OSp(4|4)$ superconformal symmetries found in \cite{D'Hoker:2007xy,D'Hoker:2007xz}. Then we present the solutions on the strip that are related to 
the (IR fixed point of) linear quivers of section \ref{sec:quiv} and to the defect SCFT of section \ref{sec:defects}. We show how the "closure" of the asymptotic regions leads to the supergravity solutions that will be dual to the linear quivers.
The solutions we present also contain, in some limit, the supergravity backgrounds dual to 4-dimensional $\N=4$ SYM with a 3-dimensional boundary CFT \cite{Aharony:2011yc}.

\subsection{Local solutions}
\label{s:localsolutions}

References \cite{D'Hoker:2007xy,D'Hoker:2007xz} give the general local solutions of type-IIB supergravity preserving the superconformal symmetry $OSp(4|4)$.  This
group is the supergroup of the 3d $\N=4$ SCFTs. 
 The solutions are parameterized by a choice of a 2-dimensional Riemann surface $\Sigma$ with boundary
 and by  two real  harmonic functions on $\Sigma$,  $h_1$ and $h_2$.  We then define auxiliary functions on $\Sigma$
\begin{align}\label{W}
W = \p \bar \p(h_1 h_2) \ , \qquad N_{j} = 2 h_1 h_2 |\p h_{j}|^2 - h_{j}^2 W,
\end{align}
with $\p \equiv \p_z$ , $\bar\p \equiv \p_{\bar z}$, the complex $z$ parametrizing $\Sigma$.\\
In terms of these auxiliary functions the metric in Einstein frame can be written as
\begin{align}
ds^2 = f_4^2 ds^2_{AdS_4} + f_1^2 ds^2_{S^2_1} + f_2^2 ds^2_{S^2_2} + 4 \rho^2 dz d\bar z\ ,
\end{align}
where the warp factors are given by
\begin{align}\label{metric}
f_4^8 = 16 \frac{N_1 N_2}{W^2} \ ,
\ \ \
f_{1}^8 = 16 h_{1}^8 \frac{N_{2} W^2}{N_{1}^3} \ ,
\ \ \
f_{2}^8 = 16 h_{2}^8 \frac{N_{1} W^2}{N_{2}^3} \ ,
\ \ \
\rho^8 = \frac{N_1 N_2 W^2}{h_1^4 h_2^4} \ .
\end{align}

This geometry is supported by non-vanishing ``matter"  fields, which include the  (in general complex)  dilaton-axion field
\begin{align}\label{dil}
S = \chi + i e^{2 \phi} = i \sqrt{N_2 \over  N_1 }\ ,
\end{align}
in addition to  3-form and 5-form backgrounds. To specify the corresponding gauge potentials
one needs the dual harmonic functions,  defined by
\begin{align}
\label{dualharm}
h_1 = -i ({\cal A}_1 - \bar {\cal A}_1) \qquad &\rightarrow \qquad h_1^D = {\cal A}_1 + \bar {\cal A}_1 \ , \cr
h_2 = {\cal A}_2 + \bar {\cal A}_2 \qquad &\rightarrow \qquad h_2^D = i ({\cal A}_2 - \bar {\cal A}_2) \ .
\end{align}
The constant ambiguity in the definition of the  dual functions is related to changes of the background
fields under large gauge transformations.
The NS-NS and R-R three forms can be written  as
\begin{align}
\label{3forms0}
H_{(3)}  =   \omega^{\, 45}\wedge db_1  \qquad {\rm and} \qquad  F_{(3)} =   \omega^{\, 67}\wedge db_2   \ ,
\end{align}
where $ \omega^{\, 45}$ and $ \omega^{\, 67}$ are the volume forms of the unit-radius  spheres  ${\rm S}_1^{2}$ and ${\rm S}_2^{2}$, while
\begin{align}
\label{3forms1}
b_1 &= 2 i h_1 {h_1 h_2 (\p  h_1\bar  \p  h_2 -\bar \p  h_1 \p  h_2) \over N_1} + 2  h_2^D \ ,  \cr
b_2 &= 2 i h_2 {h_1 h_2 (\p  h_1 \bar\p  h_2 - \bar\p  h_1 \p  h_2) \over N_2} - 2  h_1^D \ .
\end{align}
The expression for the gauge-invariant self-dual  5-form is a little  more involved:
\begin{align}
\label{5form0}
F_{(5)}  = - 4\,  f_4^{\, 4}\,  \omega^{\, 0123} \wedge {\cal F} + 4\, f_1^{\, 2}f_2^{\, 2} \,  \omega^{\, 45}\wedge \omega^{\, 67}
 \wedge (*_2  {\cal F})   \ ,
\end{align}
where $ \omega^{\, 0123}$ is the volume form of the unit-radius ${\rm AdS}_4$,
${\cal F}$ is a 1-form on $\Sigma$ with the property that $f_4^{\, 4} {\cal F}$ is closed,
and $*_2 $ denotes Poincar\' e duality with respect to the $\Sigma$ metric.
The explicit expression for ${\cal F}$  is given by
\begin{align}\label{calF}
f_4^{\, 4} {\cal F} = d j_1\   \qquad {\rm with} \qquad j_1 =
3 {\cal C} + 3 \bar  {\cal C}  - 3 {\cal D}+ i \frac{h_1 h_2}{W}\,   (\p  h_1 \bar\p  h_2 -\bar \p h_1 \p h_2) \ ,
\end{align}
where ${\cal C}$ and ${\cal D}$ are defined by  $\p {\cal C} = {\cal A}_1 \p  {\cal A}_2 - {\cal A}_2 \p  {\cal A}_1$
and  ${\cal D} = \bar {\cal A}_1 {\cal A}_2 + {\cal A}_1 \bar {\cal A}_2$.
\vskip 1mm

 For any choice of $h_1$ and $h_2$,
 equations \eqref{W} to \eqref{calF} give  local  solutions of the supergravity equations which are invariant
 under $OSp(4|4)$. Global consistency puts  severe constraints on these harmonic functions and on the surface $\Sigma$.
  There is  no complete classification of all consistent choices for this data. 

What has been shown  \cite{D'Hoker:2007xy,D'Hoker:2007xz}
  is that the most general type-IIB solution  with the $OSp(4|4)$  symmetry can be brought to the above form
   by  an  $SL(2,\mathbb{R})$ transformation. This acts as follows on the dilaton-axion and 3-form fields:
 \begin{align}
 S \to {aS+b\over cS+d}\ , \qquad
  \left(   \begin{array}{c}
H_{(3)}  \\
F_{(3)}
 \end{array}
  \right) \to
   \left(   \begin{array}{cc}
 d & -c \\ -b & a
  \end{array}
  \right)
  \left(   \begin{array}{c}
H_{(3)}  \\
F_{(3)}
 \end{array}
  \right) \ .
 \end{align}
 The Einstein-frame metric and the 5-form $F_{(5)}$ are left unchanged.


\subsection{Admissible singularities}
\label{s:admissible}

  The holomorphic functions ${\cal A}_1$ and ${\cal A}_2$ are analytic in the interior of $\Sigma$,
  but can have singularities on its boundary.  Refs. \cite{D'Hoker:2007xy,D'Hoker:2007xz} identified  three  kinds of   ``admissible"  singularities, i.e. singularities that
  can be interpreted  as brane sources in   string theory. Two of these  are logarithmic-cut  singularities and correspond to the two
  elementary  kinds of five-brane.
     In local coordinates,  in which the boundary of  $\Sigma$ is the real axis,  these singularities read
     \begin{align}
\underline{\rm D5}:   \qquad {\cal A}_1 =  - i \gamma\,  {\rm log  } w + \cdots  \ , \qquad {\cal A}_2 =  - i   c + \cdots \ ,
 \end{align} \vskip -7mm
 \begin{align} \label{sings}
\underline{\rm NS5}:  \qquad {\cal A}_1 =  -  \hat c + \cdots  \ , \qquad   {\cal A}_2 =   - \hat\gamma\,  {\rm log  } w + \cdots \ .
 \end{align}
 Here $\gamma, \hat\gamma, c, \hat c$ are real parameters  related to the brane charges, and the dots denote
  subleading terms, which are analytic at $w=0$ and have the same reality properties on the boundary as the leading terms.
 These reality properties imply that,   in the case of the D5-brane,    $h_1$ and $h_2$  obey respectively Neumann and Dirichlet boundary conditions,
   i.e.  $(\partial - \bar\partial) h_1 = h_2 = 0$ on the boundary of  $\Sigma$.   For the NS5 brane
 the  roles of  the two harmonic functions are exchanged.
  \vskip 1mm

\begin{figure}
\centering
\includegraphics[height=7.5cm,width=12cm]{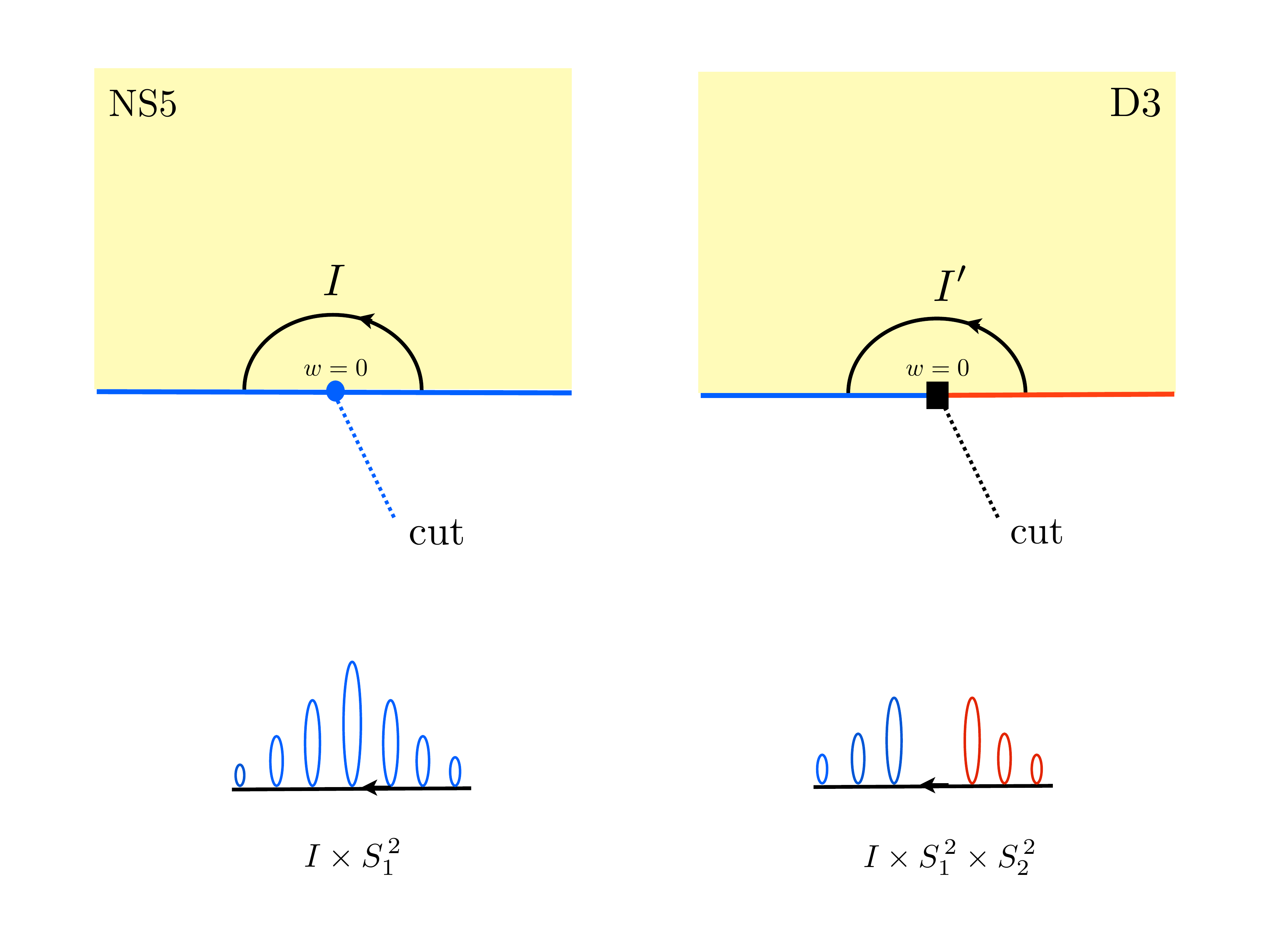}
\vskip -5mm
\caption{\footnotesize  Local singularities corresponding to  a NS5-brane (left) and a D3-brane (right),  as explained in the text.
The boundary of $\Sigma$ is colored  red or blue  according to  which of the two 2-spheres, $S_1^{\, 2}$ or $S_2^{\, 2}$,
shrinks at this part of the boundary to zero.  The  non-contractible cycles
supporting the brane charges are  $I\times S_1^{\, 2}$,  $I  \times S_1^{\, 2}\times S_2^{\, 2}$ and
$I^\prime \times S_1^{\, 2}\times S_2^{\, 2}$,  with  $I$ and $I^\prime$ the (oriented) solid semicircles of the figure. These cycles are topologically
equivalent to a 3-sphere, a 3-sphere times a 2-sphere, and a 5-sphere. The broken lines indicate the logarithmic (on the left) and square root (on the right) branch cuts. }
\label{noncontractI}
\end{figure}

The vanishing of the harmonic function $h_j$ implies that the corresponding 2-sphere $S_j^{\, 2}$ shrinks to a point.
This ensures that the  points on  the boundary of $\Sigma$,  away from the singularities,  correspond  to regular  interior points
of the ten-dimensional geometry.
Non-contractible cycles,   which
 support non-zero brane charges,  are obtained by the fibration of one or both 2-spheres
over any curve  that (semi)circles the singularity on $\partial \Sigma$.
  For instance in the case of the NS5-brane  $I \times  S_1^{\, 2}$,
with  $I$   the interval shown in figure \ref{noncontractI},  is topologically a non-contractible  3-sphere.  The appropriately
normalized
flux of $H_{(3)}$ through this cycle is the number of NS5-branes:\footnote{The
five-brane charge  is quantized in units of $2 \kappa_0^2 T_5$,  where $2 \kappa_0^2 = (2 \pi)^7 (\alpha^\prime)^4$
  is the gravitational coupling constant, and $T_5 = 1/[(2 \pi)^5 (\alpha^\prime)^{3}]$ is the five-brane tension. Note that since we have
 kept the dilaton arbitrary,  we are free to  set the string coupling $g_s=1$; the tension of the NS5-branes and the D5-branes is thus the same,
 while the D3-brane tension and charge is $T_3 = 1/[(2 \pi)^3 (\alpha^\prime)^{2}]$.}
 \begin{align}\label{N5hat}
 \hat N_{5} \, =\,   {1\over 4\pi^2\alpha^\prime } \int_{I\times S_1^{\, 2}} H_{(3)} = {2\over \pi\alpha^\prime} h_2^D\Bigl\vert_{\partial I}
 \qquad \Longrightarrow \ \   \hat N_{5} \,
 =\  {4\over \alpha^\prime}\,  \hat\gamma\ .
\end{align}
In evaluating the flux we have taken $I$ to be   infinitesimally small, and we used the fact that in
 the expression \eqref{3forms1}   only $h_2^D$ is discontinuous
across the singularity on the real axis.
We also  assumed
that the logarithmic cut lies outside the surface $\Sigma$, so that fields in the interior of $\Sigma$ are all continuous (see figure \ref{noncontractI}).
\vskip 1mm

In addition to 5-brane charge, the singularities \eqref{sings} also carry D3-brane charge. The corresponding flux
threads  the 5-cycle
$I \times  S_1^{\, 2}\times S_2^{\, 2}$, which is topologically the product of a 3-sphere with a 2-sphere.
There is a well-known subtlety in the definition of this charge, because of
the  Chern-Simons term in the IIB supergravity action \cite{Assel:2011xz, Page:1984qv, Marolf:2000cb}. In the case at hand
  the conserved flux  is the integral of the gauge-variant 5-form  $F_{(5)}' = F_{(5)} + C_{(2)}\wedge H_{(3)}$, which obeys
 a non-anomalous Bianchi identity $dF_{(5)}'=0$ in a region without brane sources. Further details about the conservation of 5-form flux and the choice of relevant 5-form to integrate will be detailed below when we consider the annulus case (to avoid redundancy).
   The  number of D3-branes   inside the NS5-brane stack
is thus given by
 \begin{align}\label{insideNS5}
  \hat N_{3} \, =\,
{1\over (4\pi^2\alpha^\prime)^2 } \int_{I\times S_1^{\, 2}\times S_2^{\, 2}} [ F_{(5)} + C_{(2)}\wedge H_{(3)} ]
\, = \,
 -  {2\over \pi \alpha^\prime}\, \hat N_{5} \,  h_1^D\Bigl\vert_{w=0}  \ .
\end{align}
It can be checked,  by taking again $I$ arbitrarily small,
 that $F_{(5)}$,  as well as all terms    in the expression for $C_{(2)}$ other than $h_1^D$,
do not contribute to the above flux. This explains the second equality,  leading finally  to
  \begin{align}\label{N3hat}
  \hat N_3 \,
 =\      \left( {4\over \alpha^\prime} \right)^2 \,  \left(  {\hat\gamma\, \hat c \over \pi} \right) \ .
\end{align}
 Note  that $\hat N_3$  depends on the   potential $C_{(2)}$ at the  position of the
5-brane singularity, and may change  under large gauge transformations. This is related to the Hanany-Witten effect \cite{Hanany:1996ie},
an issue  to which we will return in the  next subsection.

 \vskip 1mm

 In principle, using $SL(2, \mathbb{R})$ transformations one can convert the NS5-brane solution to a
 more general $(p,q)$ fivebrane solution.  Such transformations generate, however,  a
 non-trivial Ramond-Ramond axion background, so $(p,q)$ fivebranes cannot coexist with the NS5-brane
 in the IIB solutions described above for which the axion vanishes. The $(p,q)$-5-branes will reappear later when we consider the action of $SL(2, \mathbb{R})$ transformations on the solutions we are presently studying.  There is one exception to the rule: the
  S-duality transformation converts the NS5-brane to a D5-brane without generating an axion background.
  Combined with an exchange of the two 2-spheres, S-duality acts as follows on
   the harmonic functions:
         \begin{align}
 \left(
 \begin{array}{c}
i{\cal A}_2  \\
-{\cal A}_1
 \end{array}
  \right)
 \ \xrightarrow{ S} \
 \left(
 \begin{array}{c}
 {\cal A}_1  \\
i{\cal A}_2
 \end{array}
  \right)\ .
 \end{align}
  This gives
  the D5-brane singularity  anticipated already in equation \eqref{sings}.
The  integer D5-brane  and D3-brane charges read
\begin{align}\label{N5}
 N_5 =   {4\over \alpha^\prime}  \,   \gamma\, , \qquad  N_3 \,
 =\      \left( {  4\over   \alpha^\prime}  \right)^{  2}  \,   \left( {\gamma\,   c\over \pi}\right) \ .
\end{align}
Note that  the D3-brane charge is here  the flux of the  5-form
$F_{(5)} - B_{(2)}\wedge F_{(3)}$, which is the S-duality transform of
 $F_{(5)} + C_{(2)}\wedge H_{(3)}$. This gauge-variant form  is
 well-defined in any  patch around the D5-brane singularity  as long as this patch does not
 contain NS5-brane sources.

  \vskip 1mm

    The last kind of  singularity,  which can coexist with D5- and NS5-brane singularities, is the one   describing free D3-branes,   with no associated
     fivebrane charge.  In this case the holomorphic functions  have  square-root rather than logarithmic cuts \cite{D'Hoker:2007xz}
   \begin{align}\label{D3sing}
\underline{\rm D3}:   \qquad {\cal A}_1 =  {1 \over \sqrt{w}} (a_1   + b_1 w + \cdots  ) \ , \qquad
{\cal A}_2 =   {1 \over \sqrt{w}} (a_2   + b_2 w + \cdots  )
\ .
\end{align}
 Such singularities change the boundary condition of  $h_1$ from Neumann to Dirichlet, and the boundary condition  of
 $h_2$ from Dirichlet to Neumann. This is illustrated in the right part of figure \ref{noncontractI}.
  The integer  D3-brane charge is given by
  \begin{align}
   n_{3} \, =\,
{1\over \left(4\pi^2\alpha^\prime\right)^2 } \int_{I^\prime \times S_1^{\, 2}\times S_2^{\, 2}}  F_{(5)}
\ = \  \left( {4\over \alpha^\prime} \right)^2 \,  \, { (a_1b_2 - a_2 b_1) \over 2\pi}\ .
\end{align}
 The ten-dimensional geometry near the D3-brane singularity is an $AdS_5\times S^5$
 throat  with radius $L$ given by $L^4 = 4\pi \alpha^{\prime\, 2} \vert n_3 \vert$.


\subsection{Linear-quiver geometries and AdS/CFT dictionary}
\label{sec:linsugra}

\begin{figure}
\centering
\includegraphics[height=8cm,width=12cm]{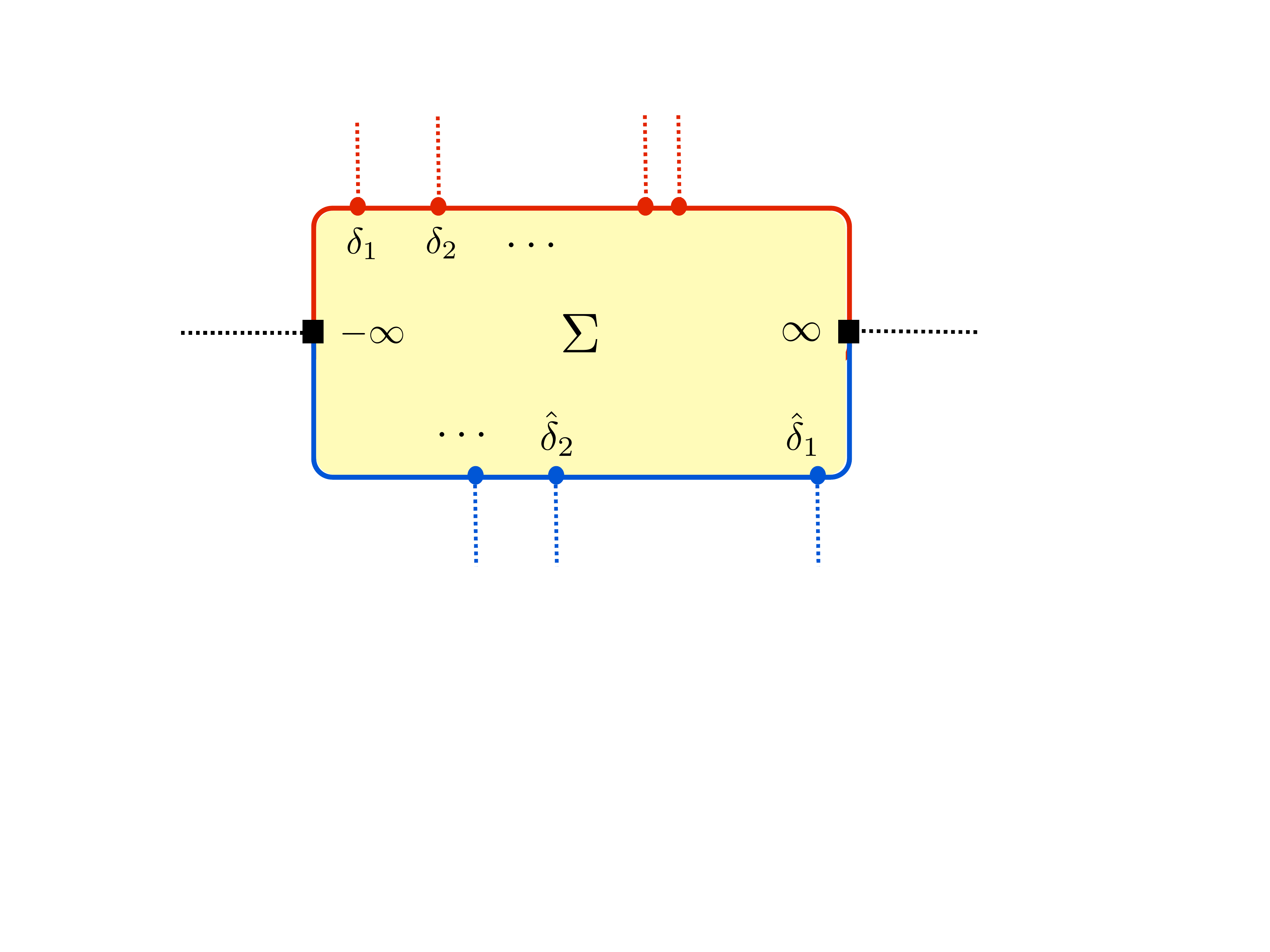}
\vskip -1.9cm
\caption{\footnotesize   Structure of singularities of the harmonic functions for the linear-quiver geometries.
The two boundary-changing singularities at $\pm\infty$, corresponding to $AdS_5\times S^5$ throats, can be capped off by
choosing $a_1=a_2=0$,  as described in the text. They  become regular interior points of the ten-dimensional geometry.}
\label{Disk}
\end{figure}

Consider two  harmonic functions with the singularity structure shown in figure \ref{Disk}.
 The corresponding geometries have the field-theory interpretation of
 superconformal domain walls in ${\cal N} = 4, D=4$ Super Yang Mills \cite{Gaiotto:2008ak}, breaking $SU(2,2|4)$ to $OSp(4|4)$.
 If $n_3^\pm$ are the D3-brane charges of the two boundary-changing (black-box) singularities, then
 the domain wall separates two  gauge theories with gauge groups $U(n_3^-)$ and $U(n_3^+)$.
 One may decouple  the three-dimensional
 SCFT that lives on the domain wall from the  bulk  four-dimensional Yang-Mills theories  by setting
 $a_j^\pm = 0$. Equation \eqref{D3sing} shows that  in this case  $n_3^+ = n_3^- =0$.  The square-root
 singularities of the harmonic functions are then simply  coordinate singularities, while the infinite $AdS_5\times S^5$ throats
 are replaced by regular interior points in  ten-dimensions.

\vspace{4mm}

In \cite{Aharony:2011yc} another limit of these harmonic functions was taken, namely the limit in which only one $AdS_5 \times S^5$ region is capped off $n^+_3=0$. The geometries then correspond to coupling $\N=4$ D=4 Super-Yang-Mills to a 3-dimensional boundary CFT.

\smallskip

Following references \cite{Bachas:2011xa, Assel:2011xz}, we choose $\Sigma$ to be the infinite strip parametrized by a complex $z=x+iy$ , $x \in (-\infty , +\infty)$ , $y \in [0, \frac{\pi}{2}]$, and the
harmonic functions to be given by
\begin{align}
\label{harm0}
{\cal A}_1  &= \alpha \sinh(z-\beta) \  - i \sum_{a=1}^{p} \gamma_{a} \ln\, \tanh\left(
{\small  {\frac{i\pi}{4}-\frac{z-\delta_a}{2}
}}
\right)  \ ,  \no\\
{\cal A}_2 &= \hat \alpha \cosh(z-\hat \beta) \   - \sum_{b=1}^{\hat{p}} \hat{\gamma}_{b} \ln \tanh\left(\frac{ z-\hat{\delta}_{b} }{2}\right)    \ .
\end{align}
The parameters ($\alpha,\beta,\gamma_a,\delta_a$)  and ($\hat\alpha,\hat\beta,\hat\gamma_b,\hat\delta_b$)
are all real. The only other condition on this set of parameters,
explained in  \cite{D'Hoker:2007xz},  is that $\alpha,\gamma_1,\cdots , \gamma_{p}$ must be all positive or all negative and similarly for $\hat\alpha,\hat\gamma_1, \cdots, \hat\gamma_{\hat p}$.
If not,  the solution has curvature singularities supported  on a one-dimensional curve in the interior of $\Sigma$,
which have no interpretation in string theory.\\
 Here   $\delta_1 < \delta_2 < ... < \delta_p$ are the positions of the D5-brane singularities on the upper boundary of the
 strip, whereas
   $\hat \delta_1 > \hat \delta_2 > ... > \hat \delta_{\hat p}$ are the positions of the NS5-brane singularities on the lower
   boundary. It can be checked that on these two boundaries $h_1$ obeys, respectively, Neumann and Dirichlet conditions,
   while  $h_2$ has Dirichlet and Neumann conditions.

\vspace{5mm}

Here we focus on the solutions with $n_3^+ = n_3^-=0$ when the two asymptotic $AdS_5 \times S^5$ regions at $\pm \infty$ are capped off. This correspond to taking $\alpha=\hat\alpha=0$ and the solutions are simply given by 
\begin{align}
\label{harm1}
{\cal A}_1  &=  - i \sum_{a=1}^{p} \gamma_{a} \ln\, \tanh\left(
{\small  {\frac{i\pi}{4}-\frac{z-\delta_a}{2}
}}
\right)  \quad , \quad
{\cal A}_2 &=  - \sum_{b=1}^{\hat{p}} \hat{\gamma}_{b} \ln \tanh\left(\frac{ z-\hat{\delta}_{b} }{2}\right)    \ .
\end{align}

 The boundary-changing square-root singularities are at $z= \pm\infty$. In the local
   coordinate $w = e^{\mp z}$ one can verify easily  that $a_j^\pm = 0$,  so these points at infinity correspond to regular interior points
  of the ten-dimensional geometry.
 \smallskip

 To simplify the formulae we will adopt from now on the (non-standard)  convention $\alpha^\prime = 4$.
Equations \eqref{N5} and \eqref{N5hat}   give the numbers of NS5-branes and D5-branes for each   fivebrane
singularity:
\begin{align}
\label{ginvN5}
 N_5^{(a)} = \gamma_a\ , \qquad  \hat N_5^{(b)} = \hat\gamma_b \ .
\end{align}
Unbroken supersymmetry requires that there are only branes (or only anti-branes) of each kind. Thus
all the $ \gamma_a$ must have the same sign, and  likewise for all the
 $ \hat\gamma_b$. This agrees with the regularity condition mentioned above . 
We choose to take all $\gamma_a>0$ and all $\hat \gamma_b >0$. The other possibilities are obtained by charge conjugations and do not lead to different CFT duals.
Dirac quantization forces furthermore  these parameters to be integer.
\vskip 1mm

Next let us consider the D3-brane charge.  Inserting the harmonic functions \eqref{harm1} inside the
expressions \eqref{N5} and \eqref{N3hat} gives
\begin{align}
&N_3^{(a)} =  N_5^{(a)} \sum_{b=1}^{\hat p}  \hat N_5^{(b)} \, {2\over\pi} {\rm arctan}(e^{\hat\delta_b -\delta_a}) \ ,
\nonumber  \\
&\hat N_3^{(b)} = - \hat N_5^{(b)} \sum_{a=1}^{p}    N_5^{(a)} \, {2\over\pi} {\rm arctan}(e^{\hat\delta_b -\delta_a})\ ,
\label{ginvN3}
 \end{align}
 where we used the identity $i\,$log\,tanh$({i\pi\over 4} - {x\over 2} )=  -$2\,arctan($e^x$).
As already noted in the previous subsection, this calculation of the  D3-brane charge
 depends on the  2-form potentials $B_{(2)}$ and $C_{(2)}$ and is, a priori, ambiguous.
One may indeed add a real constant
 to ${\cal A}_1$, or an imaginary constant to   ${\cal A}_2$,  thereby changing $h_j^D$ without affecting $h_j$.
This gauge ambiguity is also reflected in the arbitrary choice of Riemann sheet  for the logarithmic functions that enter in
equations  \eqref{harm1}.

\smallskip
We fix this ambiguity by placing all logarithmic cuts outside $\Sigma$, as in figure \ref{Disk},
and by choosing the sheet so that   the imaginary part of  (${\rm ln\, tanh} \, z$)   vanishes
when $z$ goes to $+\infty$ on the real axis. This implies that the arctangent functions take values in the interval $[0, \pi/2]$.
Our  choice  of gauge  is continuous in the interior of $\Sigma$ (which is covered by a single patch),  and sets
$B_{(2)}=0$ at $+\infty$  and  $C_{(2)}=0$
at $-\infty$.  With this  choice, D5-branes at $\delta=+\infty$ and NS5-branes at $\hat\delta = -\infty$ do not contribute to
the D3-brane charge. Placing, on the other hand,  one NS5-brane at $\hat\delta = +\infty$ adds one unit of D3-brane charge to
each D5-brane, while placing one D5-brane at $\delta = -\infty$ adds one unit of charge to each NS5-brane.
This can be understood as a holographic manifestation of the Hanany-Witten effect.

\smallskip

Since this story will be important to us later, let us explain it a little more.  The 2-form potential $B_{(2)}$ is proportional
to the volume form ($\omega^{45}$)  of the sphere $S_1^{\, 2}$, which shrinks to a point in the lower boundary of the strip
(the blue line in figure \ref{Disk}).
The solution \ref{harm1} is such that $B_{(2)}$ is constant on the lower boundary intervals between the NS5-singularities and jumps accross each singularity proportionally to the NS5-flux it supports. 
 When  $B_{(2)}\not= 0$ on a boundary interval, this interval corresponds to a Dirac singularity
of codimension 3 in (the 9-dimensional)  space.  This   is unobservable if
\begin{align}
{1\over 2\pi\alpha^\prime} \int_{S_1^{\, 2}} B_{(2)} \in 2\pi \mathbb{Z}\ \Longrightarrow\
B_{(2)}\Bigl \vert_{{\rm Im} z = 0}\  =\  \pi\alpha^\prime\omega^{45} \times ({\rm integer})\ .
\end{align}
With our choice of gauge,
\begin{align}
B_{(2)}\Bigl \vert_{{\rm Im} z = 0}\  =\  \pi\alpha^\prime\omega^{45} \times  \sum_{b=1}^\beta  \hat N_5^{(b)}
\qquad  {\rm for} \ \ \   \hat\delta_{\beta+1} < {\rm Re}z  <  \hat\delta_{\beta }  \ .
\end{align}
Large gauge transformations  change  $B_{(2)}$ everywhere in the strip by a multiple of    $\pi\alpha^\prime\omega^{45}$,
and can remove the Dirac sheet in one of the   intervals of the boundary. For us this was the interval $( \hat\delta_1, \infty)$.
A similar story holds also for the upper (red) boundary and the 2-form $C_{(2)}$. The  D3-brane charges with our  choice of gauge  agree  with
the  invariant linking numbers  defined  in  chapter \ref{chap:quivers}  section \ref{sec:branes}.

\smallskip

     The brane  engineering of the dual gauge field theories \cite{Hanany:1996ie,Gaiotto:2008ak} involves
     $N$ D3-branes
      suspended between $\hat k$ NS5-branes on the left and $k$ D5-branes on the right.  In the IIB supergravity
      the corresponding numbers are:
   \begin{align}\label{conserv}
   N =  \sum_{a=1}^p N_3^{(a)} = - \sum_{b=1}^{\hat p} \hat N_3^{(b)}\ , \qquad
   k = \sum_{a=1}^p N_5^{(a)}\ , \qquad \hat k = \sum_{b=1}^{\hat p} \hat N_5^{(b)}\ .
   \end{align}
  The way in which the D3-branes are suspended to the five-branes is given by two partitions $\rho$ and $\hat\rho$,
   which define  the linear-quiver gauge  theory and its IR fixed point  $T^{\rho}_{\hat\rho}(SU(N))$. These partitions are given in terms of the linking numbers:
\begin{align}
\label{linpartitions}
\rho &= \Big( \overbrace{l^{(1)},l^{(1)},..,l^{(1)}}^{N_{5}^{(1)}},\ \overbrace{l^{(2)},l^{(2)},..,l^{(2)}}^{N_{5}^{(2)}},\ ...\ ,
\ \overbrace{l^{(p)},l^{(p)},..,l^{(p)}}^{N_{5}^{(p)}} \Big) \ ,   \nonumber \\
\hat \rho &=
 \Big( \overbrace{\hat l^{(1)},\hat l^{(1)},..,\hat l^{(1)}}^{\hat N_{5}^{(1)}},\ \overbrace{\hat l^{(2)},\hat l^{(2)},..,\hat l^{(2)}}^{\hat N_{5}^{(2)}},\ ...\ ,\
 \overbrace{\hat l^{(\hat p)},\hat l^{(\hat p)},..,\hat l^{(\hat p)}}^{\hat N_{5}^{(\hat p)}} \Big) \ ,
\end{align}
where
\begin{align}
 l^{(a)} = \frac{N_3^{(a)} }{N_5^{(a)} } \quad , \quad  \hat l^{(b)} = - \frac{\hat N_3^{(b)}}{\hat N_5^{(b)} } \ .
\end{align}

Here $l^{(a)}$ is the number of D3-branes ending on  each D5-brane in the $a$th stack, while   $\hat l^{(b)}$ is the
number of D3-branes emanating from each NS5-brane in the $b$th stack.
Because these numbers must be integers, the parameters $\delta_a$ and $\hat\delta_b$ are quantized.\footnote{The relations
between the integer brane charges and the supergravity parameters are not easily inverted. To express the latter in terms
of the brane charges one must solve a system of  transcendental equations.
}
In all one has $2p + 2\hat p -1$  parameters, since a global translation of all the $\delta_a$ and $\hat\delta_b$ does not
change the solution. The parameters of the quiver are $N_{5}^{(a)}, l^{(a)}, \hat N_{5}^{(b)}, \hat l^{(b)}$ subject
to one constraint  \eqref{conserv},  which expresses the
 conservation of D3-brane charge. The two parameter counts therefore match.

\smallskip

The linking numbers of the supergravity solutions obey the inequalities $\rho^T > \hat\rho$, which
were the conditions for the existence of an  infrared fixed point of the quiver gauge theory \cite{Gaiotto:2008ak}, see
\ref{sec:branes}. On the supergravity side,  the  inequalities
follow  from the fact that $0< {\rm arctan}(x) < \pi/2$ for positive $x$. The details of the computations are presented in appendix \ref{app:ineq2}.  This is a non-trivial
check of the AdS/CFT correspondence.

\subsection{Defect SCFT solutions}
\label{sec:defectsugra}

To complete the discussion on supergravity solutions on the strip, let's mention more general solutions \ref{harm0} with arbitrary $\alpha, \beta, \hat\alpha, \hat\beta$. These geometries have already been studied in \cite{Bachas:2011xa} in the contect of the search for graviton zero mode localized on the strip.\\
In this case the internal geometry is non-compact and we have two asymptotic $AdS_5 \times S^5$ regions at $x = \pm \infty$. These solutions are believed to be dual to $D=4$ $\N=4$ Super-Yang-Mills coupled to a $\half$-BPS 3-dimensional defect CFT \cite{DeWolfe:2001pq} describe in section \ref{sec:defects}. 
The defect splits the 4d space in two regions with independent gauge groups $U(N_{-})$ and $U(N_{+})$ and Yang-Mills gauge couplings $g_{-}$ and $g_{+}$ that correspond to the two asymptotic $AdS_5 \times S^5$ regions, as in the original Maldacena setup (see section \ref{sec:Msetup}).

\vspace{5mm}

The asymptotics of the solutions \ref{harm0} when $x \rightarrow \pm \infty$ are given by the $AdS_5 \times S^5$ supergravity solution with different radii $L_{\pm}$ and dilaton values $e^{2\phi_{\pm}}$.

\begin{align}
 ds^2 &= L_{\pm}^2 \lp dx^2 + \cosh^2(x) \, ds^2_{AdS_4} + dy^2 + \sin^2(y) \, ds^2_{S_1^2} + \cos^2(y) \, ds^2_{S_2^2}   \rp \no\\
F^5 &= -4 L_{\pm}^4 (1+ \star) \omega^{4567y}  \ ,
\end{align}
where $\omega^{4567y}$ is the volume form of the unit 5-sphere. In terms of the supergravity parameters, the asymptotic data reads
\begin{align}\label{asymp}
 L_{\pm}^4 &=  16 \,  \lp \alpha \hat\alpha \, \cosh(\beta - \hat \beta) + \sum_{a=1}^p 2 \hat\alpha \gamma_a \, e^{\pm(\delta_a - \hat \beta)}  + \sum_{b=1}^{\hat p} 2 \alpha \hat \gamma_b \, e^{\pm (\hat\delta_b - \beta)} \rp \no\\
 e^{2\phi_{\pm}} &= \frac{\hat\alpha \, e^{\pm \hat\beta} + \sum_{b=1}^{\hat p} 4 \hat\gamma_b \, e^{\pm \hat\delta_b}}{\alpha \, e^{\pm \beta} + \sum_{a=1}^{p} 4 \gamma_a \, e^{\pm \delta_a}} \ .
\end{align}
the relations to the gauge theory data are the usual \footnote{There is no $g_{\pm}$ in the formula for $L_{\pm}^4$ because the asymptotic metrics are given in Einstein frame.}
\begin{align}
L_{\pm}^4 &= 4\pi |N_{\pm}| \alpha'^{2} \no\\
e^{2\phi_{\pm}} &= g_{\pm}  \ .\no
\end{align}

The 5-brane charges are the same as for the linear quiver geometries ($\alpha' = 4$)
\begin{align}
 N_5^{(a)} = \gamma_a\ , \qquad  \hat N_5^{(b)} = \hat\gamma_b \ .
\end{align}

The (quantized) D3-brane charges follow from the general formulas \ref{insideNS5}, \ref{N3hat}, \ref{N5}, and are given in the usual gauge $C_2 =0$ on the upper boundary segment $(-\infty, \delta_1]$, $B_2 =0$ on the lower boundary segment $[\hat \delta_1, + \infty)$ by
\begin{align}
&N_3^{(a)} =  \frac{2}{\pi} \, N_5^{(a)} \lp - \frac{\hat\alpha}{2} \sinh(\delta_a - \hat\beta) + \sum_{b=1}^{\hat p}  \hat N_5^{(b)} \,  \textrm{arctan}(e^{\hat\delta_b -\delta_a}) \rp \ ,
\nonumber  \\
&\hat N_3^{(b)} = - \frac{2}{\pi} \,\hat N_5^{(b)} \lp  \frac{\alpha}{2} \sinh(\hat\delta_b - \beta) + \sum_{a=1}^{p}    N_5^{(a)} \, \textrm{arctan}(e^{\hat\delta_b -\delta_a})  \rp \ .
\label{defectN3}
 \end{align}
The asymptotic (quantized) D3-charges $N_{\pm}$ measured with approriate orientation are
\begin{align}
\label{asympN}
 N_{\pm} &=  \pm \frac{1}{4\pi} \,  \lp \alpha \hat\alpha \, \cosh(\beta - \hat \beta) + \sum_{a=1}^p 2 \hat\alpha \, N_5^{(a)} \, e^{\pm(\delta_a - \hat \beta)}  + \sum_{b=1}^{\hat p} 2 \alpha \, \hat N_5^{(b)} \, e^{\pm (\hat\delta_b - \beta)} \rp \ . 
\end{align}

One can check that the D3-charge is conserved in the geometry
\begin{align}
\label{sumrule3}
 N_{+} + N_{-} + \sum_{a=1}^p N_5^{(a)} + \sum_{b=1}^{\hat p} \hat N_5^{(b)} = 0 \ .
\end{align}

The $2(p + \hat p) +3$ parameters of the supergravity solution \ref{harm0} are now recast in terms of the data $(N_{+}, N_{-},
N_5^{(a)}, \hat N_5^{(b)}, g_{+}, g_{-})$ which obey \ref{sumrule3}. 

\vspace{8mm}

The relation to the $\half$-BPS defect SCFTs described in section \ref{sec:defects} is very close to what we had in the case of linear quivers.
The two partitions $\rho$ and $\hat\rho$ are again defined by
\begin{align}
\rho &= \Big( \overbrace{l^{(1)},l^{(1)},..,l^{(1)}}^{N_{5}^{(1)}},\ \overbrace{l^{(2)},l^{(2)},..,l^{(2)}}^{N_{5}^{(2)}},\ ...\ ,
\ \overbrace{l^{(p)},l^{(p)},..,l^{(p)}}^{N_{5}^{(p)}} \Big) \ ,   \nonumber \\
\hat \rho &=
 \Big( \overbrace{\hat l^{(1)},\hat l^{(1)},..,\hat l^{(1)}}^{\hat N_{5}^{(1)}},\ \overbrace{\hat l^{(2)},\hat l^{(2)},..,\hat l^{(2)}}^{\hat N_{5}^{(2)}},\ ...\ ,\
 \overbrace{\hat l^{(\hat p)},\hat l^{(\hat p)},..,\hat l^{(\hat p)}}^{\hat N_{5}^{(\hat p)}} \Big) \ ,
\end{align}
where
\begin{align}
 l^{(a)} = \frac{N_3^{(a)} }{N_5^{(a)} } \quad , \quad  \hat l^{(b)} = - \frac{\hat N_3^{(b)}}{\hat N_5^{(b)} } \ .
\end{align}
And the bulk SYM data are simply related to the asymptotic data of the supergravity solution
\begin{align}
 N_L &= -N_{-} \quad , \quad g_{YM}^{(L) \ 2} = g_{-}  \ , \no\\
 N_R &= N_{+} \qquad , \quad g_{YM}^{(R) \ 2} = g_{+} \ .
\end{align}
The corresponding $\half$-BPS defect SCFT is then $\scD(\rho,\hat \rho,-N_{-},N_{+},g_{-}^{1/2},g_{+}^{1/2})$.

Note the $l^{(a)}$ and $\hat l^{(b)}$ may be negative as the linking numbers of the gauge theory description and that $\sum_{a=1}^{p} l^{(a)} N_5^{(a)} = N - N_R$ and $\sum_{b=1}^{\hat p} \hat l^{(b)} \hat N_5^{(b)} = N - N_L$ with 
\begin{align}
 N &= \frac{1}{4\pi} \,  \Big( \alpha \hat\alpha \, \cosh(\beta - \hat \beta) + \sum_{a=1}^p 2 \hat\alpha \, N_5^{(a)} \, e^{-\delta_a + \hat \beta}  + \sum_{b=1}^{\hat p} 2 \alpha \, \hat N_5^{(b)} \, e^{\hat\delta_b - \beta} \no\\
& \hspace{20mm} + \ 8 \sum_{a=1}^p \sum_{b=1}^{\hat p} \hat N_5^{(a)} N_5^{(b)} \, \textrm{arctan}(e^{\hat\delta_b -\delta_a})   \Big) \ ,
\end{align}
consistently with the fact that $\rho$ is a partition of $N-N_R$, while $\hat\rho$ is a partition of $N-N_L$.

\smallskip

The two partitions have to satisfy the inequalities \ref{ineqdefect} which ensures that no anti-D3-branes will appear in the brane picture corresponding to the defect SCFT. The cases of negative linking number for D5-branes make the inequalities difficult to express on the supergravity side, so we have only checked them for solutions with only positive D5-linking numbers. In this case the inequalities takes the simpler form \ref{ineqdefect2} and the explicit computations of appendix \ref{app:ineq2} can be easily adapted to the case of the domain wall solutions, using the explicit formula for the charges \ref{defectN3}, \ref{asympN}, to show that the inequalities \ref{ineqdefect2} are satisfied (it comes down to the trivial inequalities $ 2\sinh(x) < e^x$ and $\frac{2}{\pi}\arctan(e^x) < 1$).

\vspace{8mm}

All the formulas for the linear quiver geometries are obtained by setting $\alpha = \hat\alpha = 0$, corresponding to $N_{\pm} = 0$.

\vspace{10mm}


\section{Circular quiver geometries : from  strip to  annulus}
\label{s:striptoannulus}

We present now how the geometries on the strip lead under periodic identification to the supergravity solutions  where $\Sigma$ is an annulus and we show how to relate them to the SCFTs arising from the flow of the circular quivers presented in chapter \ref{chap:quivers}.

A class of three-dimensional ${\cal N}=4$ superconformal field theories that arise from circular quivers are known to admit an M-theory description in terms of  orbifolds of the seven-sphere  \cite{Hosomichi:2008jd,Benna:2008zy,Imamura:2008ji}. 
By taking a certain (large $L$) smearing limit of our solutions, T-dualizing the periodic coordinate of the annulus 
 and   lifting the resulting type-IIA background to eleven dimensions,  we reproduce the relevant M-theory geometries  ${\rm AdS}_4\times {\rm S}^7/(\mathbb{Z}_k\times \mathbb{Z}_{\hat k})$. In  the process one looses however  the dependence on 
  the full quiver data $(\rho,\hat\rho,L)$.  This data can be in principle encoded  in the non-contractible 3-cycles of the compact space and the associated 3-form fluxes  \cite{Imamura:2008ji, Witten:2009xu,Aharony:2009fc,Dey:2011pt}.  
 The  3-cycles  degenerate however  in the orbifold limit and  we are not aware of any solutions of eleven-dimensional supergravity that resolve the singularity on the M-theory side. By contrast in our IIB solutions, the full  data  is encoded in the positions of  five-brane throats along the annulus circle.

\subsection{Solutions on the annulus}

  The strategy for constructing holographic IIB duals  for the circular quivers is the following:  one starts from
  the  linear-quiver solutions  that we have just described,
 and arranges the  five-branes   in infinite  pedrodic arrays along the $x$ axis. The holomorphic  functions ${\cal A}_j$   become logarithms of quasi-periodic
 elliptic  functions.
Modding out  by
 discrete translations corresponding to the period of the array then converts the  strip domain, $\Sigma$,   to an annulus,
 and the dual  linear-quiver theories to  theories based on circular quivers.
\vskip 1mm

\begin{figure}
\centering
\includegraphics[height=8cm,width=12cm]{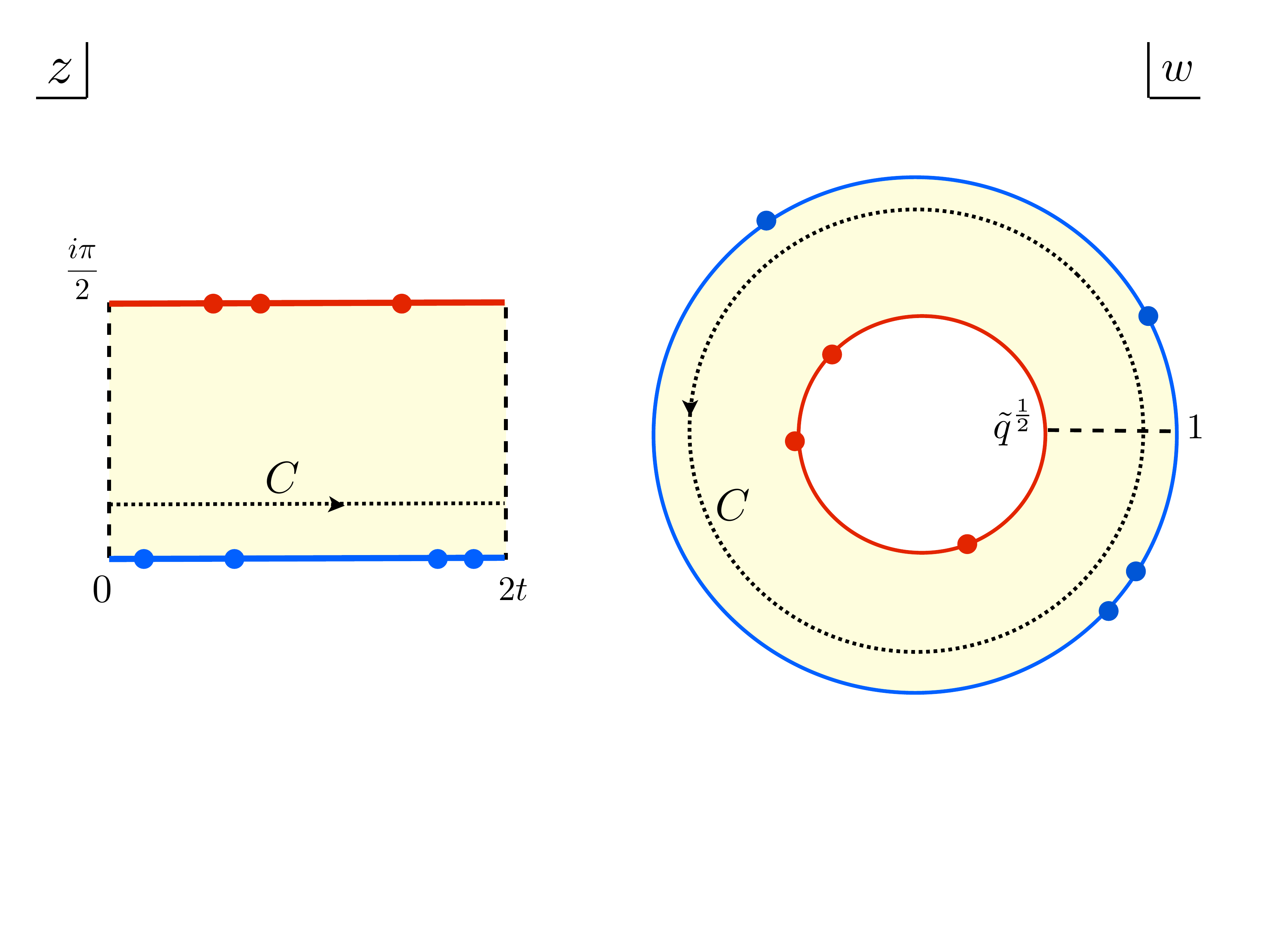}
\vskip -1.6cm
\caption{\footnotesize  The annulus $\Sigma$   for the type-IIB solutions that are
 dual to  $D=3, {\cal N} = 4$  circular-quiver  theories.   $\Sigma$ is the infinite strip in the $z$ plane modulo
 the translations $z\to z+2t$ (left), or the annular domain in the $w = {\rm exp}(i\pi z/t)$ plane (right). The radius
 of the inner boundary of the annulus  is   $\tilde q^{1/2}$ where $\tilde q  =  {\rm exp}(-\pi^2/t)$   is the exponentiated dual modulus
 of the elliptic $\vartheta$-functions. The monodromies of $h_j^D$ around the curve $C$ give the total number
 of NS5 and D5-branes, as explained in the main text. }
\label{Annulus}
\end{figure}

 More explicitly, given a set of fivebrane singularities at $\delta_a$ and $\hat \delta_b$, we may always pick a
 positive parameter $t$ such that, after a rigid translation,    $0  \leq \delta_a \leq 2 t$ and $0 \leq \hat \delta_b \leq 2t$.
Replicating the fivebrane sources with  periodicity $2t$ then leads to the following harmonic functions
\begin{align}
\label{ellipticharm}
h_1  &= -  \sum_{a =1}^p  \gamma_a \ln \bigg[ \prod_{n = -\infty}^{\infty} \tanh \bigg( \frac{i \pi}{4} - \frac{z - (\delta_a + 2 n t)}{2} \bigg) \bigg] + c.c. \ , \cr
h_2 &= - \sum_{b=1}^{\hat p} \hat \gamma_b \ln \bigg[ \prod_{n = -\infty}^{\infty} \tanh \bigg( \frac{z - (\hat \delta_b + 2 n t)}{2} \bigg) \bigg] + c.c. \ .
\end{align}
These functions are  manifestly periodic  under translations by $2t$, so we are free to identify  $z\equiv z+2t$  thereby converting the strip
  $\Sigma$ to  an annulus.
 Figure \ref{Annulus} depicts this annular domain in the $w$-plane, where $w= {\rm exp}({i\pi z/t})$.
 \smallskip

\smallskip
To see that the infinite products in the above expressions converge, we will rewrite them in terms of elliptic
$\vartheta$-functions  (we use the conventions of  reference \cite{Green:1987mn}).  This can be done with the help of the identity
\begin{align}
\label{thident}
\Bigl\vert \frac{\vartheta_1(\nu\vert \qth)}{\vartheta_2(\nu\vert \qth)} \Bigr\vert \,
=\,   \left\vert   \prod_{n=-\infty}^{\infty}  \tanh(i\pi \nu  +  nt )\,  \right\vert \ ,
 \qquad {\rm where} \quad e^{i \pi \qth} = e^{-t}  \ .
\end{align}
The proof of this identity  follows from  the product formulae for the $\vartheta$-functions
\begin{align}\label{productf}
\vartheta_{1}(\nu\vert \qth) = 2 e^{i \pi \qth /4} \sin (\pi \nu) \prod_{n=1}^\infty (1-e^{2n i \pi \qth})(1 - e^{2n i \pi \qth} e^{2 \pi i \nu})(1 - e^{2n i \pi \qth} e^{-2 \pi i \nu}) \ , \cr
\vartheta_{2}(\nu\vert \qth) = 2 e^{i \pi \qth /4} \cos (\pi \nu) \prod_{n=1}^\infty (1-e^{2n i \pi \qth})(1 + e^{2n i \pi \qth} e^{2 \pi i \nu})(1 + e^{2n i \pi \qth} e^{-2 \pi i \nu}) \ .
\end{align}
Note that the modular parameter is $\tau = it/\pi$,  because the hyperbolic tangents are periodic under  $z\to z+2\pi i$.
Inserting the   identity  \eqref{thident}  in  \eqref{ellipticharm}
leads to the following expressions for $h_1$ and $h_2$:
 \begin{align}
\label{hmany}
h_1 &= -  \sum_{a =1}^p \gamma_a \ln \bigg[ \frac{\vartheta_{1}\left(\nu_a\vert \qth  \right)}{\vartheta_{2}\left(\nu_a\vert \qth  \right)} \bigg]
 + c.c.    \  , \  \qquad {\rm with}\ \ \
  i\, \nu_a = - \frac{z-\delta_a}{2 \pi } + \frac{i}{4} \ , \cr
h_2 &= - \sum_{b=1}^{\hat p} \hat \gamma_b \ln \bigg[ \frac{\vartheta_{1}\left(  \hat \nu_b\vert \qth \right)}{\vartheta_{2}\left(  \hat \nu_b\vert \qth \right)} \bigg]
+ c.c.  \ ,
\ \  \qquad {\rm with}\ \ \    i\,  \hat \nu_b =   \frac{z - \hat \delta_b}{2 \pi } \ .
\end{align}
These harmonic functions are well-defined everywhere inside the annulus. They have logarithmic singularities on the boundaries,
wherever $\nu_a$ or $\hat\nu_b$ vanish.

\smallskip

   Decomposing $h_j$ into holomorphic and anti-holomorphic parts requires, as in the previous subsection, a choice of gauge.
A convenient choice is to make the  ${\cal A}_j$ analytic in the
interior of the covering strip,  {before}  the periodic identification of $z$. This amounts to placing again all logarithmic branch cuts outside the strip.
With this understanding,  and recalling that the Jacobi $\vartheta$-functions are holomorphic,   we have
   \begin{align}
\label{Amany}
{\cal A}_1  = - i \sum_{a =1}^p \gamma_a \ln \bigg( \frac{\vartheta_{1}\left(\nu_a \vert \qth  \right)}{\vartheta_{2}\left( \nu_a\vert \qth \right)} \bigg)
 + \varphi_1   \  , \  \qquad
{\cal A}_2  = - \sum_{b=1}^{\hat p} \hat \gamma_b \ln \bigg( \frac{\vartheta_{1}\left(\hat \nu_b\vert \qth  \right)}{\vartheta_{2}\left(\hat \nu_b\vert \qth\right)} \bigg)
+ i \varphi_2  \
  ,
\end{align}
where the constant phases $\varphi_1$ and $\varphi_2$ are residual quantized gauge degrees of  freedom,
corresponding to large gauge transformations of the 2-form potentials. As in the case of the linear quiver, we may use this residual  freedom to
enforce the absence of Dirac singularities in one interval on each annulus  boundary.
\smallskip

Unlike $h_j$, the above holomorphic functions and  the dual harmonic functions $h_j^D$  are {\it not }
 periodic under $z\to z+2t$.
Their  gauge-invariant holonomies (or Wilson lines)  give  the total fivebrane charges.
  To see why, note that translating  $z\to z+2t$ changes all the arguments $\nu_a$ by $it/\pi$ (and all the $\hat\nu_b$ by $-it/ \pi$).
 From the product formulae \eqref{productf} one finds that
 under these  translations the $\vartheta$-functions are quasi-periodic:
 \begin{align}
\vartheta_1(\nu+ {it\over \pi}\vert \qth ) = -e^{-2\pi i \nu + t} \vartheta_1(\nu\vert \qth)  \ , \qquad
\vartheta_2(\nu+ {it\over \pi}\vert \qth) =  e^{-2\pi i \nu + t} \vartheta_2(\nu\vert \qth) \ .
\end{align}
 The
  ratio  $\vartheta_1/\vartheta_2$ changes  only by a minus sign.
 Thus ln($\vartheta_1/\vartheta_2)\to$ln($\vartheta_1/\vartheta_2) \mp i\pi$ when $\nu \to \nu\pm it/\pi$, from which we conclude
   \begin{align}\label{holonomy}
 {\cal A}_1(z+ 2t) = {\cal A}_1(z) -  \pi \sum_{a =1}^p \gamma_a  \  \, , \qquad
  {\cal A}_2(z+ 2t) = {\cal A}_2(z) - i\pi    \sum_{b=1}^{\hat p} \hat \gamma_b  \ .
 \end{align}
The meaning of these holonomies becomes clear if one integrates the 3-form field strengths over the
3-cycles $C\times S_j^{\, 2}$, where $C$ is the   dotted curve in figure \ref{Annulus}.
Consider for example the $H_{(3)}$ flux through  $C\times S_1^{\, 2}$. From  equations
  \eqref{3forms0} and  \eqref{3forms1} we deduce  that this   is proportional to
  \begin{align}\label{b2hol}
  \oint_C db_1 = 2 \oint_C d h_2^D\,  =\,  4i \, [ {\cal A}_2(z+2t) - {\cal A}_2(z) ]\ ,
  \end{align}
where in the first step we used the fact that  $(db_1 - 2 d h_2^D)$  is an exact differential which, therefore,   integrates to  zero.
Since the  total flux is conserved, the right-hand-side of  \eqref{b2hol}  must be  $z$-independent.
One finds that the integrated flux  is proportional to the total number of NS5-branes.
 The holonomy of ${\cal A}_1$ is likewise determined  by the total number of D5-branes.

\subsection{Calculation of D3-brane charges}

The ten-dimensional geometries on the annulus have essentially the same non-contractible cycles as the strip geometry and the following discussion on flux conservation applies to the strip as a simpler case.

\begin{figure}
\centering
\includegraphics[height=8cm,width=12cm]{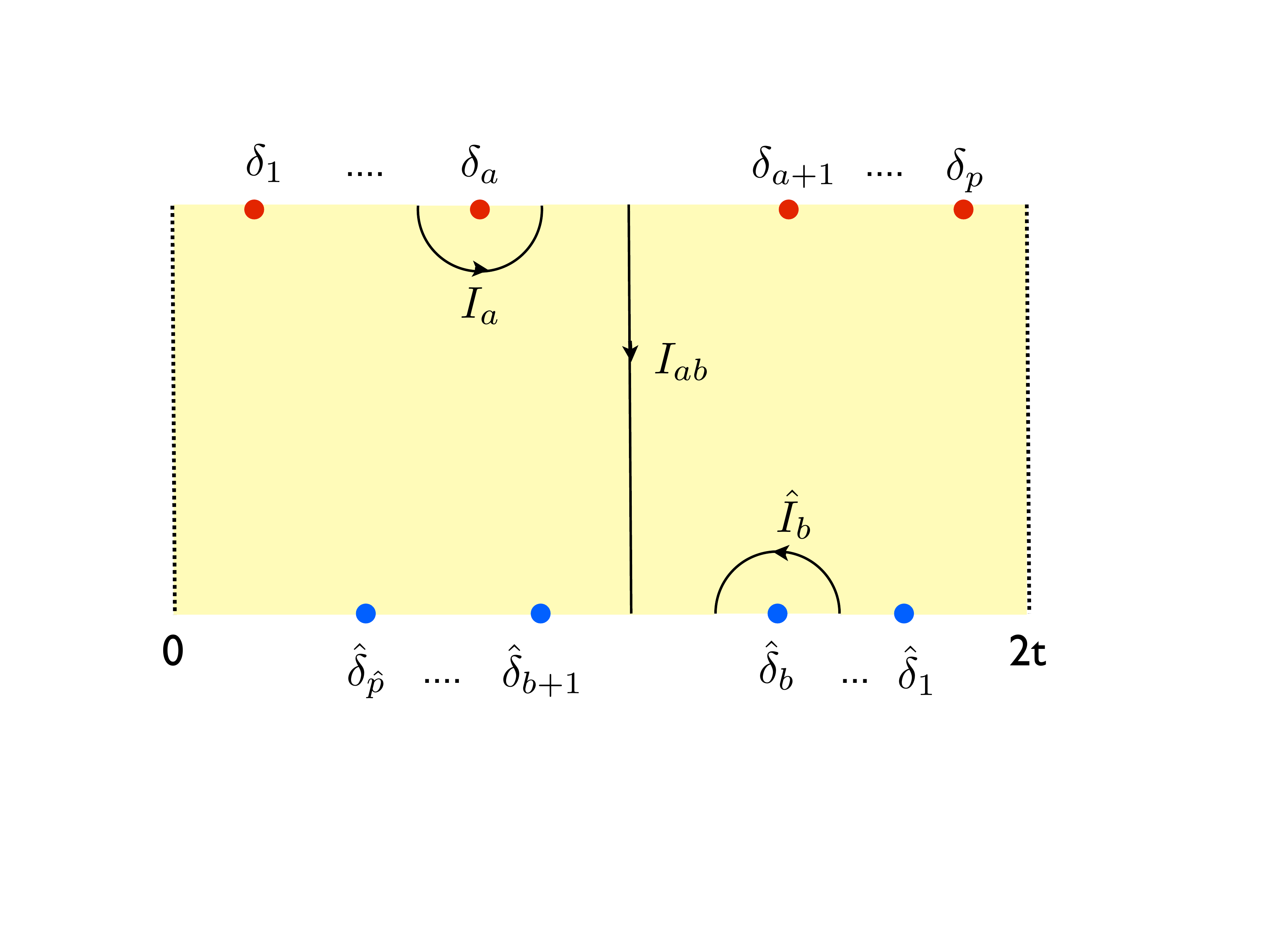}
\vskip -1.6cm
\caption{\footnotesize  The non-contractible 5-cycles in the circular-quiver geometries are fibrations of the two 2-spheres
over the  curves  shown in this figure. $\Sigma$ is an annulus, so the dotted boundaries are identified.
  }
\label{5cycles}
\end{figure}

There are the same three-cycles $I_a\times S^{\, 2}_2$
and $\hat I_b\times S^{\, 2}_1$, where $I_a$ is a semicircular curve around the $a$th singularity of $h_1$ on the upper annulus boundary,
and $\hat I_b$ is  a semicircle around the $b$th singularity of $h_2$ on the lower annulus boundary, see figure \ref{5cycles}.
These three-cycles are threaded respectively  by R-R and NS-NS three-form fluxes,  emanating from $\gamma_a$ D5-branes and from
$\hat\gamma_b$ NS5-branes (in units where $\alpha^\prime = 4$).

\smallskip
 In addition, as for the strip case, these geometries have a number of non-contractible five-cycles which can support D3-brane charge, that we expose in greater details here to be able to describe the subtelties of the charge conservation of the 5-form flux.  
These are fibrations of $S_1^2$ and $S_2^2$ over the three types of open curves
$I_a$, $\hat I_b$ and $I_{ab}$ shown in  figure \ref{5cycles}.
Recalling that $S_1^2$ shrinks to a point in the lower boundary,
and $S_2^2$ shrinks to a point in the upper boundary of the annulus, one deduces that   the  topology of these 5-cycles  is as follows:

  \begin{itemize}

\item ${\cal C}_a^5 \equiv  (S_1^2\times S_1^2) \ltimes I_a$ and $\hat {\cal C}_b^5\equiv  (S_1^2\times S_1^2) \ltimes \hat I_b$
are  topologically
 $S^3 \times S^2$;

 \item ${\cal C}_{ab}^5 \equiv  (S_1^2\times S_1^2) \ltimes I_{ab}$ are   topologically    $S^5$ .

\end{itemize}

\noindent Here $I_{ab}$ is a line segment
 which begins on the upper boundary of the annulus between the points $\delta_a$ and $\delta_{a+1}$
 and ends on
 the lower boundary   between the points $\hat \delta_b$ and $\hat \delta_{b+1}$.
  As shown in the figure, the orientation of the segments $I_{a}, \hat I_{b}$ is chosen  to be counter-clockwise, and for $I_{ab}$
from the upper annulus boundary to the lower boundary.

\smallskip

The D3-brane charges emanating from the five-brane singularities can be computed with the help of the
 general formulae  of  \S\ref{s:admissible}. Consider for example the $b$th   NS5-brane stack which corresponds
to  the $z= \hat\delta_b$  singularity on  the lower boundary of the annulus. Using $h_1^D = {\cal A}_1 + \bar {\cal A}_1$ and
the expressions   \eqref{insideNS5}, \eqref{N3hat} and
\eqref{Amany} we find
 \begin{align}
\label{D3charge1}
\hat N^{(b)}_{3} &= - \frac{2}{\pi \alpha'} \hat N^{(b)}_{5} h_1^D|_{z = {\hat \delta_b}}
\cr
&=    \hat N^{(b)}_{5} \, \sum_{a=1}^{p} N^{(a)}_{5} \, \left(  \frac{i}{2\pi}
  \ln \bigg[ \frac{\vartheta_{1}\left(\nu_{ab} \vert \qth \right)}{\vartheta_{1}\left( \bar \nu_{ab}\vert \qth \right)} \frac{\vartheta_{2}\left( \bar \nu_{ab}\vert
 \qth \right)}{\vartheta_{2}\left(\nu_{ab}\vert \qth \right)} \bigg] -   \frac{4}{\pi\alpha^\prime} \varphi_1 \right)\
\end{align}
  where
\bea\label{nuab}
   i \nu_{ab} = \frac{\delta_a - \hat \delta_b}{2 \pi} + \frac{i}{4}  \ , \qquad \qth = e^{-t}\  ,
\eea
and $\bar \nu$ is the complex conjugate of $\nu$.
Likewise, one finds for the $a$th D5-brane:
 \begin{align}
\label{D3charge2}
N^{(a)}_{3} &=  \frac{2}{\pi \alpha'}N^{(a)}_{5}  h_2^D|_{z = {\frac{i \pi}{2} + \delta_a}} \cr
&= N^{(a)}_{5} \, \sum_{b=1}^{\hat p} \hat N^{(b)}_{5} \, \left(
- \frac{i}{2\pi}  \ln \bigg[ \frac{\vartheta_{1}\left(\nu_{ab} \vert \qth\right)}{\vartheta_{1}
\left(\bar \nu_{ab}\vert  \qth  \right)} \frac{\vartheta_{2}\left( \bar \nu_{ab} \vert \qth \right)}{\vartheta_{2}\left(\nu_{ab}\vert \qth \right)} \bigg]
-   \frac{4}{\pi\alpha^\prime} \varphi_2 \right) \ ,
\end{align}
where  the arguments $\nu_{ab}$ are defined again by  \eqref{nuab}.

\smallskip

As has been discussed in  the previous section, the D3-brane (Page) charge suffers from a gauge ambiguity
which corresponds,   in the above expressions,   to the freedom in choosing the  constants $\varphi_1$ and $\varphi_2$.
 In what follows, and until  otherwise specified,
we fix  the  gauge so that the potentials are continuous inside the fundamental domain $0 \leq {\rm Re} z < 2t$, and furthermore
\begin{align}
\label{canongauge}\nonumber
C_{(2)} &= 0 \qquad \rm{in}\ \  \,[0 ,\delta_{1}] \, \ \ \ \textrm{on the upper boundary}, \\
B_{(2)} &= 0 \qquad \rm{in} \,\ \  [\hat \delta_{1},2t] \,\ \ \  \textrm{on the lower boundary}.
\end{align}
The above  choice can be motivated by considering the pinching limit
 $t \longrightarrow +\infty$ with $\delta_a$ and $\hat\delta_b$ kept fixed. In this  limit the
 geometry degenerates to that of a linear quiver, and our gauge fixing agrees with the one adopted
 for linear quiver geometries.

 \smallskip
  Using the infinite-product expressions for the $\vartheta$-functions in \eqref{D3charge1} and \eqref{D3charge2},
  and fixing as  just described $\varphi_1$ and  $\varphi_2$,  leads to the expressions
 \bea\label{emanateD5}
  { N^{(a)}_{3}  }  =   N^{(a)}_{5} \sum_{b=1}^{\hat p} \hat N^{(b)}_{5}  \Big[  \sum_{n=0}^{+\infty} f(\hat \delta_b - \delta_a -2nt ) - \sum_{n=1}^{+\infty} f(-\hat \delta_b + \delta_a -2 nt)
 \Big] \ ,
 \eea
and
\bea\label{emanateNS5}
   {\hat N^{(b)}_{3} } =  \hat N^{(b)}_{5} \sum_{a=1}^{p} N^{(a)}_{5} \Big[   \sum_{n=1}^{+\infty} f(-\hat \delta_b + \delta_a - 2 n t )
-\sum_{n=0}^{+\infty} f(\hat \delta_b - \delta_a - 2 n t )
 \Big]  \ ,
\eea
   where $ { N^{(a)}_{3}  } $ is the  D3-brane charge   in the $a$th stack of D5-branes, $  {\hat N^{(b)}_{3} }$
   is  the  D3-brane charge  in  the $b$th stack of NS5-branes, and
\bea
   f(x)= \frac{2}{\pi} \arctan(e^x) \in [0, 1] \ .
\eea
It   can be easily verified that the above charges obey   the sum rule
\begin{align}\label{sumrule2}
  \sum_{a=1}^p N^{(a)}_{3}  =  - \sum_{b=1}^{\hat p} \hat N^{(b)}_{3}  \ \equiv\  N \ .
\end{align}
   In  the pinching limit, where  only the $n=0$
 terms survive in the sums,  all  the $N^{(a)}_{3}$ are positive
 and all the $ \hat N^{(b)}_{3}$ are negative numbers. For finite $t$, on the other hand,  the numbers in each set  can  have
 either sign.

  \smallskip

 Next we consider the 5-cycles ${\cal C}_{ab}^5$.  To associate to these 5-cycles  a Page charge we must decide which (gauge-variant) 5-form
 to integrate. Take for instance the 5-form $\tilde F_{(5)} \equiv  F_{(5)} +  C_{(2)}\wedge H_{(3)}$, which obeys the
  non-anomalous Bianchi identity $d\tilde F_{(5)}=0$.  This is globally defined only on the cycles ${\cal C}_{0b}^5$, since for all other
  choices of $a$,  the gauge potential $C_{(2)}$ has a Dirac string singularity at the upper endpoint of $I_{ab}$. Put differently,
  $\int \tilde F_{(5)}$
   would depend on the precise location of this upper endpoint   unless
  $C_{(2)} =0$ in the corresponding boundary segment.  By a similar reasoning one concludes that $\tilde F_{(5)}^\prime  \equiv  F_{(5)} - B_{(2)}\wedge F_{(3)}$
 should be only integrated on the 5-cycles ${\cal C}_{a0}^5$. Both of these modified 5-forms can be integrated on the
  5-cycle ${\cal C}_{00}^5$,  which is picked out by our gauge fixing \eqref{canongauge}. Furthermore,
  the  Page charge for this cycle does not depend on the choice of the modified 5-form  since
 \bea
 \int_{{\cal C}_{00}^5} (\tilde F_{(5)} - \tilde F_{(5)}^\prime) =   \int_{{\cal C}_{00}^5}  d\, ( C_{(2)}\wedge B_{(2)}) = 0\ .
 \eea

\begin{figure}
\centering
\includegraphics[height=8cm,width=12cm]{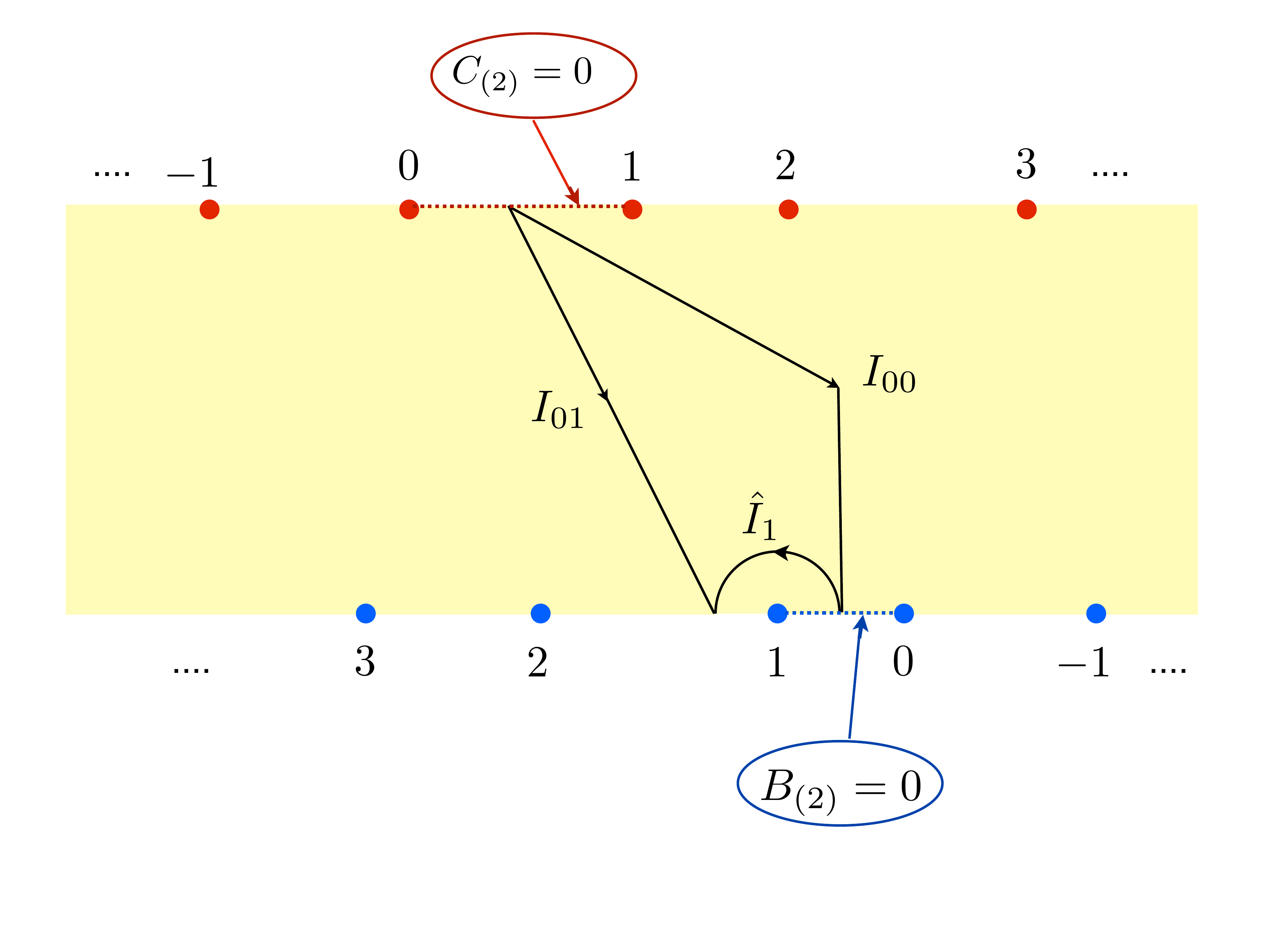}
\vskip -1.cm
\caption{\footnotesize  The infinite array of 5-brane singularities on the universal cover of the annulus.
 The D5-branes on the upper boundary are labelled from left to right, and the NS5-branes on the lower boundary
  from right to left. The choice of gauge
  determines a fundamental domain,  and a  special 5-cycle ${\cal C}_{00}^5 = I_{00}\times S_1^2\times S_2^2$.  The D3-brane charge
  supported by this cycle  is obtained  by  summing over all pairs of singularities with positive labels, and all pairs with non-positive labels, see
 equation \eqref{Mcharge2}.
  }
\label{mnem}
\end{figure}

Let us  denote  the D3-brane charge for this  special 5-cycle  by $M$. If
normalized appropriately, as in equation \eqref{insideNS5},
$M$ must be an  integer charge. We will now argue that this D3-brane charge is given by
the following expression:
 \begin{align}
\label{Mcharge2}
M  = \sum_{a,b >0}  N^{(a)}_{5} \hat N^{(b)}_{5}
  f(\hat \delta_b - \delta_a) + \sum_{a,b \leq 0}  N^{(a)}_{5} \hat N^{(b)}_{5}
  f( \delta_a- \hat \delta_b)\ ,
 \end{align}
where we  here considered  the universal cover of the annulus (i.e. the infinite strip),  and extended the range of  five-brane labels
so that   $-\infty < a< \infty$ is a label for  the infinite
array of D5-brane singularities from left to right, while  $-\infty <b < \infty$ labels  the corresponding  array of NS5-brane singularities from right to left.
Furthermore in this  notation, $\delta_{a+np} \equiv \delta_a + 2nt$ is the position of the $n$th image of the $a$th singularity on the upper
strip boundary; likewise $\hat\delta_{b+m\hat p} \equiv \hat\delta_b - 2mt$ corresponds to the $m$th  image of the $b$th singularity on the
lower strip boundary.
The expression \eqref{Mcharge2} can thus be written more explicitly as follows:
 \begin{align}
\label{Mcharge1}
M &= \sum_{a=1}^p \sum_{b=1}^{\hat p} N^{(a)}_{5} \hat N^{(b)}_{5}
 \Bigl[  \sum_{m,n=0}^{\infty}   f(\hat \delta_b - \delta_a -2nt - 2m t)
  + \sum_{m,n=1}^{\infty}   f(-\hat \delta_b + \delta_a -2nt  - 2 mt) \Bigr]
   \no \\
 &= \sum_{a=1}^{p} \sum_{b=1}^{\hat p} N^{(a)}_{5} \hat N^{(b)}_{5}  \sum_{s=1}^{ \infty} s  \Big[ f (\hat \delta_b - \delta_a -2(s-1) t )
 + f (\delta_a - \hat \delta_b -2(s +1)t ) \Big] \, .
\end{align}
 A  schematic explanation of  the above  expression is given in  Figure \ref{mnem}.
  \smallskip

To see that  \eqref{Mcharge2} is indeed right, let us consider a change of gauge which makes $B_{(2)}$ vanish
on the boundary segment  between the $b=1$ and the $b=2$ singularities.
The privileged 5-cycle is now ${\cal C}_{01}^5$, and the
corresponding D3-brane charge $M^\prime$ reads
\bea
\label{Mcharge3}
M^\prime  = \sum_{a>0 ,b > 1}  N^{(a)}_{5} \hat N^{(b)}_{5}
  f(\hat \delta_b - \delta_a) + \sum_{a  \leq 0, b\leq 1}  N^{(a)}_{5} \hat N^{(b)}_{5}
  f( \delta_a- \hat \delta_b)\ .
\eea
The difference $M^\prime - M$ is equal to $\hat N^{(3)}_1$,   the number of D3-branes in
the first NS5-brane stack, as one can check  with the help of  equation  \eqref{emanateNS5}.
This should be so  since
$I_{01} = I_{00} \oplus \hat I_1$, as illustrated in  Figure \ref{mnem}, and furthermore  the corresponding Page charges,  $M^\prime$
and $M+\hat N^{(3)}_1$,   are given by
 integrals  of  the modified form $F_{(5)} +  C_{(2)}\wedge H_{(3)}$ which does not depend on the choice of $B_{(2)}$ gauge.
\smallskip

This simple consistency check fixes almost uniquely the  expression  \eqref{Mcharge2} for the charge $M$.
To remove all doubts, we have  also verified this formula numerically.

\begin{figure}
\centering
\includegraphics[height=9cm,width=12cm]{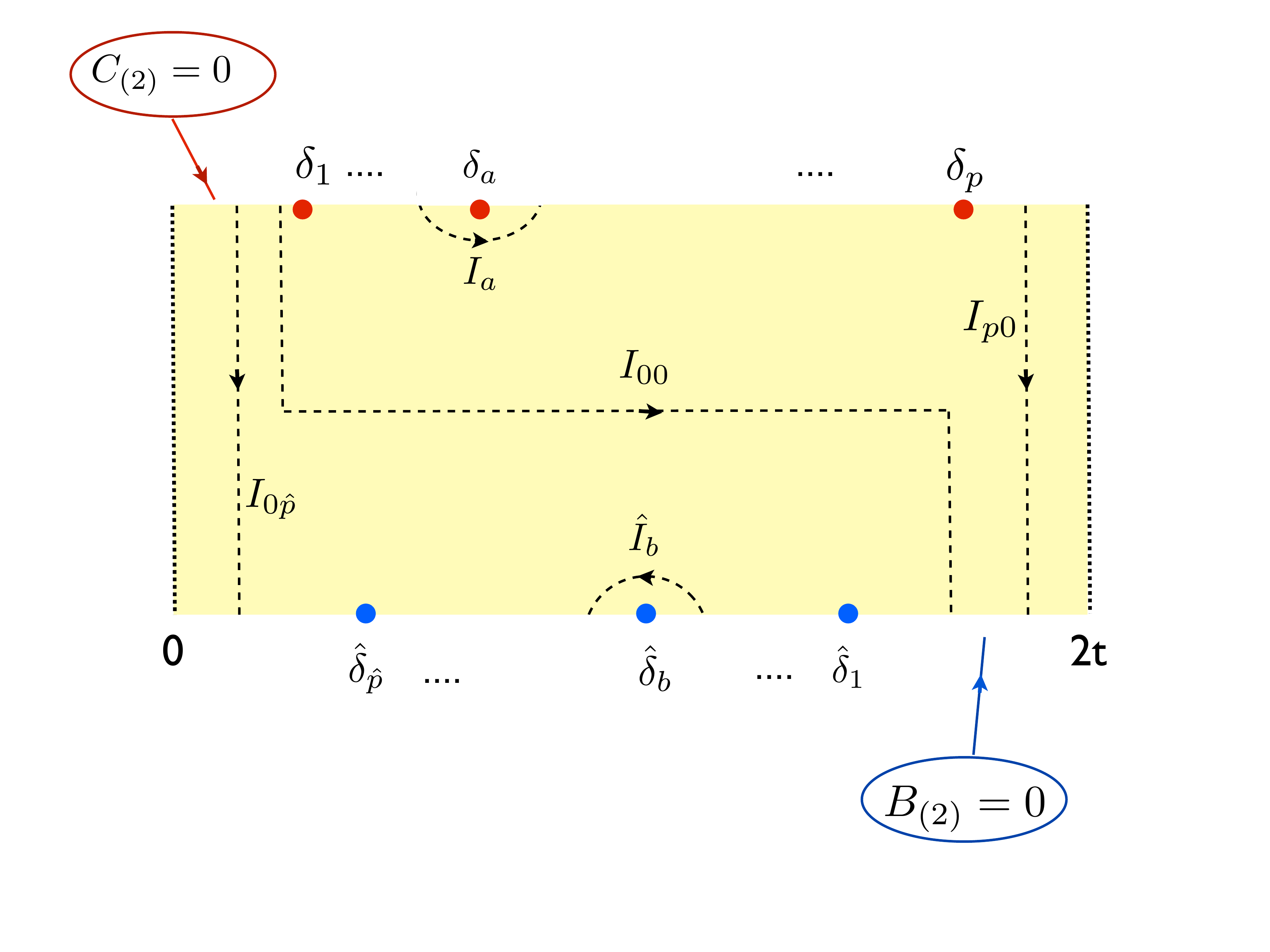}
\vskip -1cm
\caption{\footnotesize
A fundamental domain and the segments  $I_{0\hat p}$ and $I_{p0}$  which correspond to the Page charge $L$.  This
is the number of winding D3-branes, which  vanishes in the (pinching)  limit of a linear quiver.  }
\label{L}
\end{figure}

\smallskip

It will be  convenient for our purposes here to trade $M$ for   $L \equiv M - N$,
where $N$ is the total charge carried by the D5-branes, see  \eqref{sumrule2}.  The charge $L$ corresponds to the 5-form flux through the
 cycle ${\cal C}_{p0}^5$,  or equivalently the  cycle ${\cal C}_{0\hat p}^5$,  depicted  in Figure \ref{L}. Simple manipulations give
   \begin{align}
\label{Lcharge}
L  \ &=  \sum_{a=1}^p \sum_{b=1}^{\hat p} N^{(a)}_{5} \hat N^{(b)}_{5}  \sum_{n =0}^{\infty} \sum_{m =1}^{\infty}
 \Bigl[   f(\hat \delta_b - \delta_a -2nt - 2m t)
  +  f(-\hat \delta_b + \delta_a -2nt  - 2 mt) \Bigr]
   \no \\
 &= \sum_{a=1}^{p} \sum_{b=1}^{\hat p} N^{(a)}_{5} \hat N^{(b)}_{5}  \sum_{s=1}^{ \infty} s  \Big[ f (\hat \delta_b - \delta_a -2s t )
 + f (\delta_a - \hat \delta_b -2st ) \Big] \, .
\end{align}
 Below, we will identify $L$ with the number of winding D3-branes in a circular quiver.  Consistently  with this interpretation, $L$ can be seen to  vanish  in the
 pinching limit,  $t\to\infty$  with $\delta_a - \hat\delta_b$,  for  all $a=1, \cdots   p$ and $b= 1, \cdots  \hat p$,  held finite and fixed.

\vspace{8mm}

To summarize the discussion, the 5-form flux as defined above on the various 5-cycles is conserved when deforming the segment of integration along the annulus. The segment $I_{00}$ plays a special role as one can trade the 5-form $F_{(5)} +  C_{(2)}\wedge H_{(3)}$ for $F_{(5)} - B_{(2)}\wedge F_{(3)}$ or vice-versa, without changing the flux, to deform further the integration segment through the whole annulus. The 5-form fluxes or D3-charges defined this way are the $N^{(a)}_{3}$ and $\hat N^{(b)}_{3}$ charges for the D5 and NS5 singularities satisfying \ref{sumrule2}(as in the case of the strip), plus a charge $L$ wrapping the annulus. These D3-charges, together with the D5-charges $N^{(a)}_{5}$ and NS5-charges $\hat N^{(b)}_{5}$, repackage all the $2(p+ \hat p) -1$ parameters of the supergravity solution \ref{ellipticharm}, as explained below.


\subsection{Correspondence and 5-brane moves}
\label{s:match}

In analogy with the linear quiver case, we define the linking numbers of the fivebranes as the Page charge
per five-brane in each given stack:
\begin{align}\label{matchconstraint}
l^{(a)} \equiv \frac{N^{(a)}_{ 3}}{N^{(a)}_{ 5}} \  , \quad \hat l^{(b)} \equiv - \frac{\hat N^{(b)}_{ 3}}{\hat N^{(b)}_{ 5}} \  , \qquad {\rm with}\ \ \
\sum_{a=1}^{p} N^{(a)}_{ 5}l^{(a)} = \sum_{b=1}^{\hat p} \hat N^{(b)}_{ 5}\hat l^{(b)} = N \ .
\end{align}
We here assume that these linking numbers are integer. Strictly-speaking, Dirac's  quantization condition only requires
integrality of the total charge for each five-brane stack, so  solutions with fractional linking numbers
cannot be ruled out a priori as inconsistent.
We will  nevertheless discard this possibility,
 because we  have  no candidate SCFTs on the holographically dual side with fractional linking numbers.
 But the question  deserves   further scrutiny.

  \smallskip

Next let us identify the above liking numbers with those in the brane construction of the circular quivers
 described in    \S\ref{sec:branes},  by defining the following two partitions of $N$:
 \begin{align}\no
\rho &= \Big( \overbrace{l^{(1)},l^{(1)},..,l^{(1)}}^{N_{ 5}^{(1)}},\overbrace{l^{(2)},l^{(2)},..,l^{(2)}}^{N_{ 5}^{(2)}},...,
\overbrace{l^{(p)},l^{(p)},..,l^{(p)}}^{N_{ 5}^{(p)}} \Big) \ , \\
\hat \rho &= \Big( \overbrace{\hat l^{(1)},\hat l^{(1)},..,\hat l^{(1)}}^{\hat N_{ 5}^{(1)}},\overbrace{\hat l^{(2)},\hat l^{(2)},..,\hat l^{(2)}}^{\hat N_{ 5}^{(2)}},...,\overbrace{\hat l^{(\hat p)},\hat l^{(\hat p)},..,\hat l^{(\hat p)}}^{\hat N_{ 5}^{(\hat p)}} \Big) \ .
\end{align}
Together with the additional parameter $L$, we thus have the exact same data that was
used to  define   the circular-quiver gauge theories   $C_{\rho}^{\hat \rho}(SU(N),L)$ .
Put differently,  the supergravity parameters $\{ \gamma_a, \delta_a \}$ can be used to
vary the charges $\{ N_{ 5}^{(a)}, N_{ 3}^{(a)} \}$, the parameters  $\{ \hat\gamma_b, \hat\delta_b\} $ can be used to vary
  $\{ \hat N_{ 5}^{(b)}, \hat N_{ 3}^{(b)} \}$, and the annulus modulus $t$ controls the number $L$ of winding D3-branes.
  One of the charges is not independent because of the sum rule \eqref{matchconstraint}, but this agrees precisely with the fact that
    the supergravity solution is invariant under a common translation of all five-brane singularities on the boundary of the annulus.

  \smallskip

  The parameter counts on the supergravity and gauge-theory sides  therefore  match.
  The quiver data, on the other hand, had  to obey a set of  inequalities in order for the theory to flow to a non-trivial IR fixed point, see section \ref{chap:quivers}.
    We will  show that the same inequalities are also  obeyed on the supergravity side.

\smallskip

Note first that from the expressions \eqref{emanateD5} and \eqref{emanateNS5},  and  the fact that   $f(x)$ is  a monotonic function,
   it follows that the linking numbers of the supergravity solutions are automatically arranged  in decreasing order:
  \begin{align}
\label{ineqauto}
l^{(1)} > l^{(2)} > ... > l^{(p)} \qquad {\rm and}\qquad
\hat l^{(1)} > \hat l^{(2)} > ... > \hat l^{(\hat p)} \ .
\end{align}
From the brane-engineering point of view, it is possible to order the linking numbers by
moving five-branes of the {\it same type}   around each other  in   transverse space (this is
obvious in the configuration of Figure  \ref{separate}).
We have argued in section \S\ref{chap:quivers} that these  moves do not change the infrared limit of the theory,
up to  decoupled free sectors.  Such moves
 should thus be
indistinguishable on the supergravity side.\footnote{Unlike \eqref{orderedD5} and  \eqref{NS5ordered},
the inequalities  \eqref{ineqauto} are strict because they
refer to {\it stacks} of five-branes. Members of a given stack have identical linking numbers, so the linking numbers of individual five-branes are not decreasing
but only  non-increasing.}
 \smallskip

 \begin{figure}
\centering
\includegraphics[height=8.4cm,width=12cm]{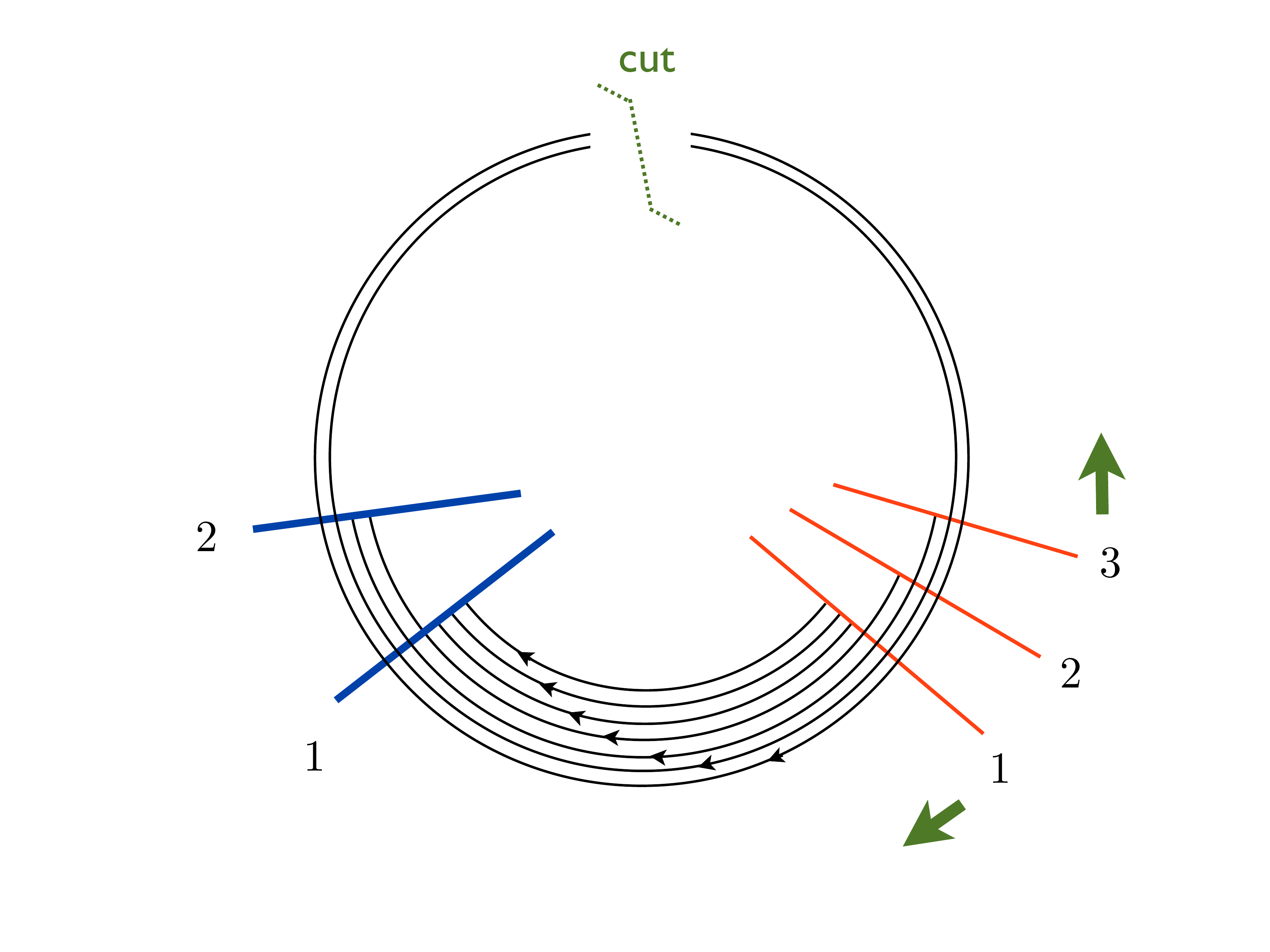}
\vskip -0.7 cm
\caption{\footnotesize  Brane engineering of a circular quiver.
 Cutting open the circle  on its high side leads to the linking-number
assignements $\rho= (3,1,1)$ and $\hat\rho = (3,2)$ with $L=2$; the  corresponding  theory is  $C_\rho^{\hat\rho}(SU(5), 2)$.
The green arrows indicate the elementary D5-brane moves described in the text. For instance,  a rotation of the $3$rd D5-brane
changes these assignments to $\rho^\prime = (3,3,1)$ and $\hat\rho^\prime = (4,3)$ with $L^\prime=3$.
  }
\label{cycle}
\end{figure}

 Besides being arranged  in decreasing order, the linking numbers of the field-theory side could
be furthermore chosen to lie in the intervals $(0, \hat k]$ and $(0, k]$, with
$k$ and $\hat k$   respectively  the total numbers of D5-branes and NS5-branes, see
  \eqref{orderedD5} and  \eqref{NS5ordered}. As was explained in  \S\ref{sec:branes},  these inequalities
  were automatic if one chose to cut open the circular chain at a link of locally-minimal rank.
  We will now explain why the same argument goes through on the supergravity side.

\smallskip

  To this end, consider the circular quiver
  of  Figure \ref{cycle} defined by the triplet data $(\rho, \hat\rho, L)$. Following the discussion in  \S\ref{sec:branes}, to
    assign   linking numbers to the five-branes we  cut   open the circular chain of D3-branes
    and then use the definitions  \eqref{defnlinking}.  Clearly, the  assignment is not unique since
    we are  free to  move one or several five-branes  around the circle before cutting  the chain.
    Let us focus,  in particular,  on the following two ``elementary"  moves:

    \begin{itemize}
    \item  Move the (right-most) $k$th D5-brane anticlockwise, which produces  the  changes
    \bea\label{move1}
    \Delta l_k = \hat k \ , \ \ \Delta \hat l_j = 1\ \ \ \forall \ j=1, \cdots \hat k \ , \ \  \Delta L = l_k\ ;
    \eea

    \item  Move the (left-most) $1$rst  D5-brane  clockwise, which leads to  the  changes
    \bea\label{move2}
    \Delta l_1 = - \hat k \ , \ \ \Delta \hat l_j = -1\ \ \ \forall \ j=1, \cdots \hat k\ , \ \  \Delta L = \hat k -  l_1\ .
    \eea

    \end{itemize}
    These formulae translate the well-known  fact that when a D5-brane crosses a NS5-brane it creates or destroys a D3-brane
 \cite{Hanany:1996ie}.\footnote{The
   linking numbers are actually  invariant under such Hanany-Witten moves, but they change in the way  indicated above
    when the D5-brane crosses the  cutting point.}
    Similar formulae clearly hold for the mirror-symmetric moves of NS5-branes.
     The main point for us here is that
    the inequalities $l_k >  0$ and $\hat k \geq   l_1$  imply that $L$ is a ``local" minimum with respect to  elementary D5-brane moves.
    Likewise,  $\hat l_{\hat k} >  0$ and $ k \geq  \hat l_1$ imply that $L$ is a  minimum with respect to  elementary NS5-brane moves.
    One can thus impose the bounds \eqref{orderedD5} and  \eqref{NS5ordered} by choosing to cut the chain  at a minimum of $L$.

 \smallskip

\begin{figure}
\centering
\includegraphics[height=8.4cm,width=12cm]{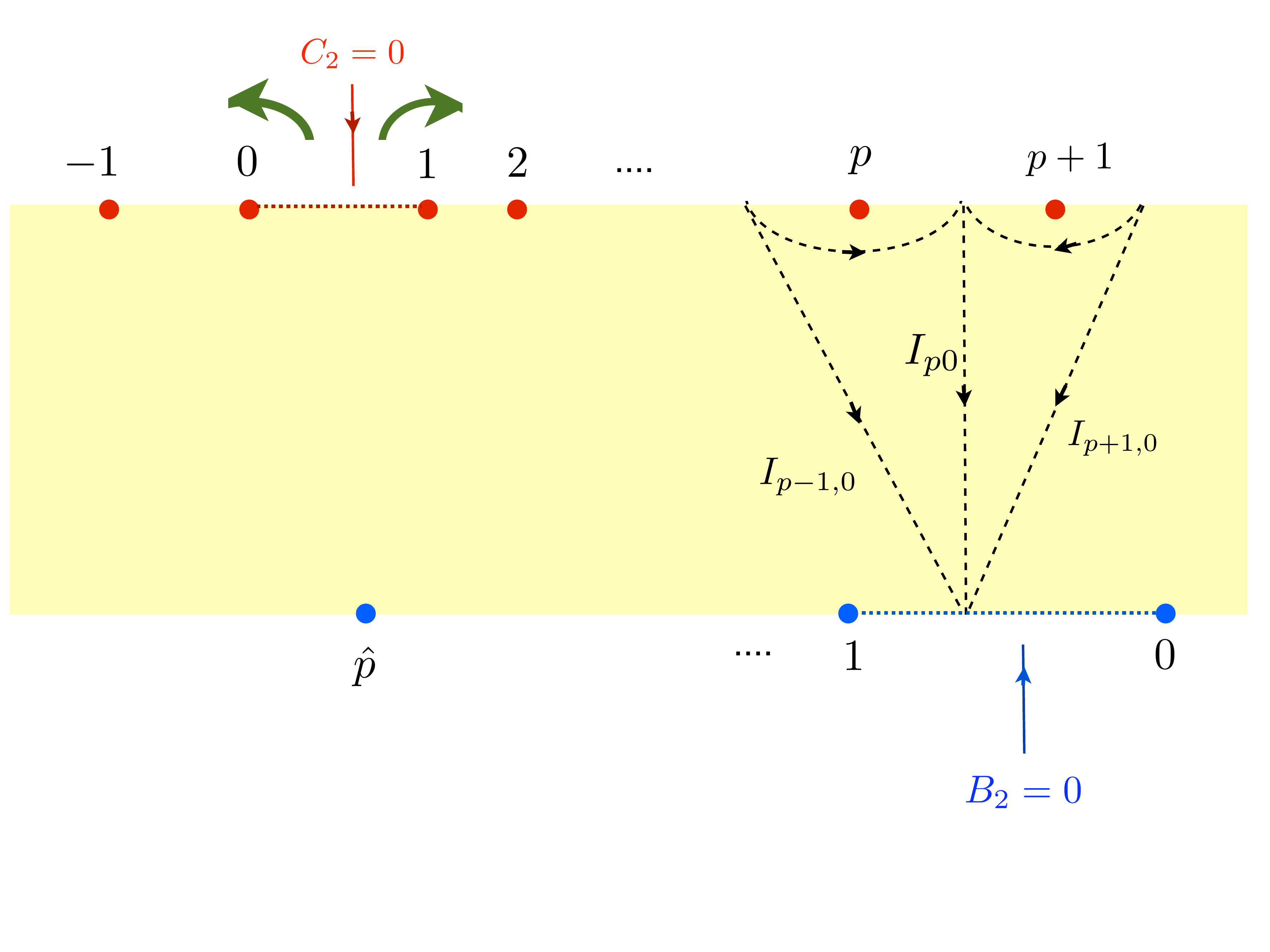}
\vskip -1.cm
\caption{\footnotesize  Global gauge transformations corresponding to the elementary D5-brane moves described in the text.
 Pushing the boundary segment on which $C_{(2)}=0$ one step to the right corresponds to moving the first stack of D5-branes
 around the circular quiver clockwise  once. Pushing this boundary segment to the left corresponds to moving the last D5-brane
 stack once in the anti-clockwise direction.
  }
\label{moves}
\end{figure}

     This same line of argument applies to the supergravity side, where five-brane moves across the cut correspond to large
     gauge transformations.  The  elementary D5-brane moves are illustrated in Figure \ref{moves}.  They correspond to
     shifting the boundary segment on which $C_{(2)}=0$ to a neighboring segment,  on the right or   left.
     Pushing for example this segment  to the left  leads to the following transformations of charges:
        \bea\label{mov1}
    \Delta l^{(p)} = \hat k \ , \ \ \Delta \hat l^{(b)} =  N_5^{(p)}\ \ \ \forall \  b   \ , \ \  \Delta L = N_5^{(p)} l^{(p)} \ .
    \eea
The last two equations follow from the expression for the linking numbers (see \S\ref{s:admissible}) and from the argument
illustrated in Figure \ref{mnem}. As for the first equation, it comes  from the fact the $p$th D5-brane stack is replaced in the fundamental
domain by the $0$th stack. On the universal cover of the annulus linking numbers
(defined as the integrals over the 5-cycles ${\cal C}^5_a$
    and $\hat C_b^5$) obey the  periodicity conditions:
      \bea
      l _{a+np} = l_a -n \hat k\ , \quad  \hat l_{b+n\hat p} = \hat l_b -nk\ .
      \eea
      Thus replacing the $p$th stack by the $0$th stack  changes the associated linking number by
      $\hat k$.\footnote{The notation in \eqref{mov1}
      is slightly abusive, because the change of the fundamental domain should be followed by a relabeling of the D5-branes. Strictly speaking
      $ \Delta l^{(p)} \equiv    l^{(1)}_{\rm new} -  l^{(p)}_{\rm old}$.
      }
    Likewise,  pushing the segment on which $C_{(2)}=0$ one step to the right
leads to the following changes:
     \bea\label{mov2}
    \Delta l^{(1)} = - \hat k \ , \ \ \Delta \hat l^{(b)} =  - N_5^{(1)}\ \ \ \forall \  b   \ , \ \  \Delta L = N_5^{(1)} (\hat k - l^{(1)}) \ .
    \eea
In this case the first D5-brane stack is replaced in the fundamental domain by the $(p+1)$th stack, as   in  Figure \ref{moves}.

\smallskip

Equations \eqref{mov1} and \eqref{mov2} are the same as  \eqref{move1} and \eqref{move2} when $N_5^{(p)} = N_5^{(1)} = 1$.
The  large gauge transformations are in this case the counterpart of the elementary D5-brane moves.
More generally,  they describe the effect of  moving the first and last {\it stacks} of D5-branes around the circular quiver.
Requiring that   $L$ be minimum under these   moves implies that $\hat k - l^{(1)} \geq 0$ and $ l^{(p)} >  0$, as advertized.\footnote{If $l^{(p)}=0$
we push  the selected line segment   to the left until the second inequality becomes strict.} Likewise one shows that
 $  k - \hat l^{(1)} \geq 0$ and $\hat  l^{(p)} >  0$, by requiring minimality under changes of the $B_{(2)}$ gauge.
 That such a minimum exists is guaranteed by the fact that  $L$  is bounded below, and goes to infinity along with the
 separation  $\delta_1 - \hat\delta_1$.
 Note that in general  there are several minima, so different
 triplets of data $(\rho, \hat\rho, L)$ may  correspond to one and the same supergravity solution, reflecting the redundancy we obtained in the circular quiver description when $L$ is only a local minimum (and not global).

\vskip 1mm

Having established the inequalities  \eqref{orderedD5} and  \eqref{NS5ordered}, we now need to prove the
 inequalities  \eqref{fixedpointcircA} for the associated Young tableaux. In the brane constructions of \S\ref{sec:branes}
these  inequalities guaranteed that all gauge groups have positive rank, i.e. that they are realized on D3-branes
rather than  anti-D3-branes. This is a condition for supersymmetry, so we expect it to be automatically satisfied on the supergravity side.
The proof
is straightforward but tedious, and we relegate  it to   appendix \ref{app:ineq2}.

\vspace{10mm}

\section{Limiting geometries}
\label{sec:lim}

In this section we discuss the solutions  described above on the strip and the annulus,  in regions of the  parameters
 where the surface $\Sigma$ with the marked points on the boundary degenerates. \\
As a first case one may take the limit $(\delta_a - \delta_{a+1})\to 0$ with the other parameters held fixed.
This simply merges the $a$th and $(a+1)$th stacks of D5-branes. Modulo the subtle issue of  linking number  quantization,
 this limit is thus rather  dull. The more interesting limits are those of an infinite separation between stacks for the strip geometries and of an infinitely-thin ($t\to\infty$) or infinitely-fat annulus ($t\to 0$).

 In the limit when the separation between 5-brane stacks is large, the size of the geometry in the ''deserted`` (or middle) region tends to zero and may be called a wormbrane geometry. The infinite separation corresponds to a ''pinching limit`` in which the wormbrane closes. In this case the strip solutions is split in two strip solutions, whereas the annulus degenerates to a strip ($t\to\infty$). We will show that these limits arise when one of the inequalities \ref{fixedpoint}, \ref{fixedpointcircA} for the partitions $\rho, \hat\rho$ is saturated.

The other limit that we describe in the large $L$ limit ($t\to 0$) for the annulus solutions which corresponds to a large number of D3-branes wrapping the circle in the brane picture. We call it the fat annulus because the volume of the annulus diverges in this limit. As we will see, this limit operates the smearing of the 5-branes.
\smallskip

After discussing these limits we analyse the regime of parameters in which the supergravity description can be used.
\bigskip

\subsection{Wormbrane limits}

\vspace{5mm}

\underline{{\bf The splitting strip}} :

\vspace{5mm}

 The limits for the linear quiver geometries in which one or more of  the inequalities  contained in the statement $\rho^T > \hat\rho$
  become equalities, are of special significance.

As explained at the end of appendix \ref{app:ineq2} one inequality can be saturated in two different limits: \\
 \indent (i) when $\delta_a \rightarrow +\infty$ for $a=I+1, I+2, ..., p$ and   $\hat \delta_b \rightarrow +\infty$ for $b=1, 2, ..., J$,
 or\\
  \indent (ii) when $\delta_a \rightarrow -\infty$ for $a=1, 2, ..., I$ and $\hat \delta_b \rightarrow -\infty$ for $j=J+1, J+2, ..., \hat p$.\\
  In the supergravity solution, these two  limits are related by a
   singular coordinate transformation corresponding to a large (infinite) translation of the strip.

 This limit corresponds to detaching  a subset of  fivebrane singularities  and moving them off to infinity on the strip.
  On the field theory side, on the other hand,  the quiver gauge theory breaks up into two (or more) pieces,
  which are  connected
  by a  ``weak node", {\rm i.e.} a node of the quiver diagram  for which the gauge group has  rank  much  smaller
  than the ranks of all  other gauge groups.  We will now make this statement  more explicit.

Consider the limit (i) in which 
$\delta_a \rightarrow +\infty$ for $a=I+1, I+2, ..., p$ and   $\hat \delta_b \rightarrow +\infty$ for $b=1, 2, ..., J$
(the limit (ii) is as we have just argued equivalent). In this limit the charges (\ref{ginvN3})  for the fivebrane stacks
at finite $z$ reduce to: 
\begin{align}
N^{(a)}_{D3} &= k^{(a)} \sum_{b=J+1}^{\hat p} \hat N^{(b)}_{NS5} + N^{(a)}_{D5} 
\sum_{b=1}^J \hat N^{(b)}_{NS5} \frac{2}{\pi} \arctan (e^{\hat \delta_b - \delta_a})  \ , \qquad a=I+1, ..., p \no \\
\hat N^{(b)}_{D3} &= \hat \hat k^{(b)} \sum_{a=1}^{I} N^{(a)}_{D5} \frac{2}{\pi} \arctan (e^{\hat \delta_b - \delta_a}) 
 \ , \qquad \qquad   \qquad b=1,..., J \ . 
\end{align}
The extra contribution in $N^{(a)}_{D3}$ coming from the branes located at $\infty$ is actually irrelevant, 
as it can be removed by an appropriate gauge transformation of $B_2$. 
 This corresponds to choosing the gauge so that $B_2 =0$  on the segment $(\hat\delta_{J+1},\hat\delta_J)$. 
In this way,  a  solution   with $I$   D5-branes stacks and $(\hat p- J)$  NS5-brane stacks is detached from the rest
of the geometry. 
\smallskip
 
\begin{figure}
\center
\vskip -1.2cm
\includegraphics[width=13cm]{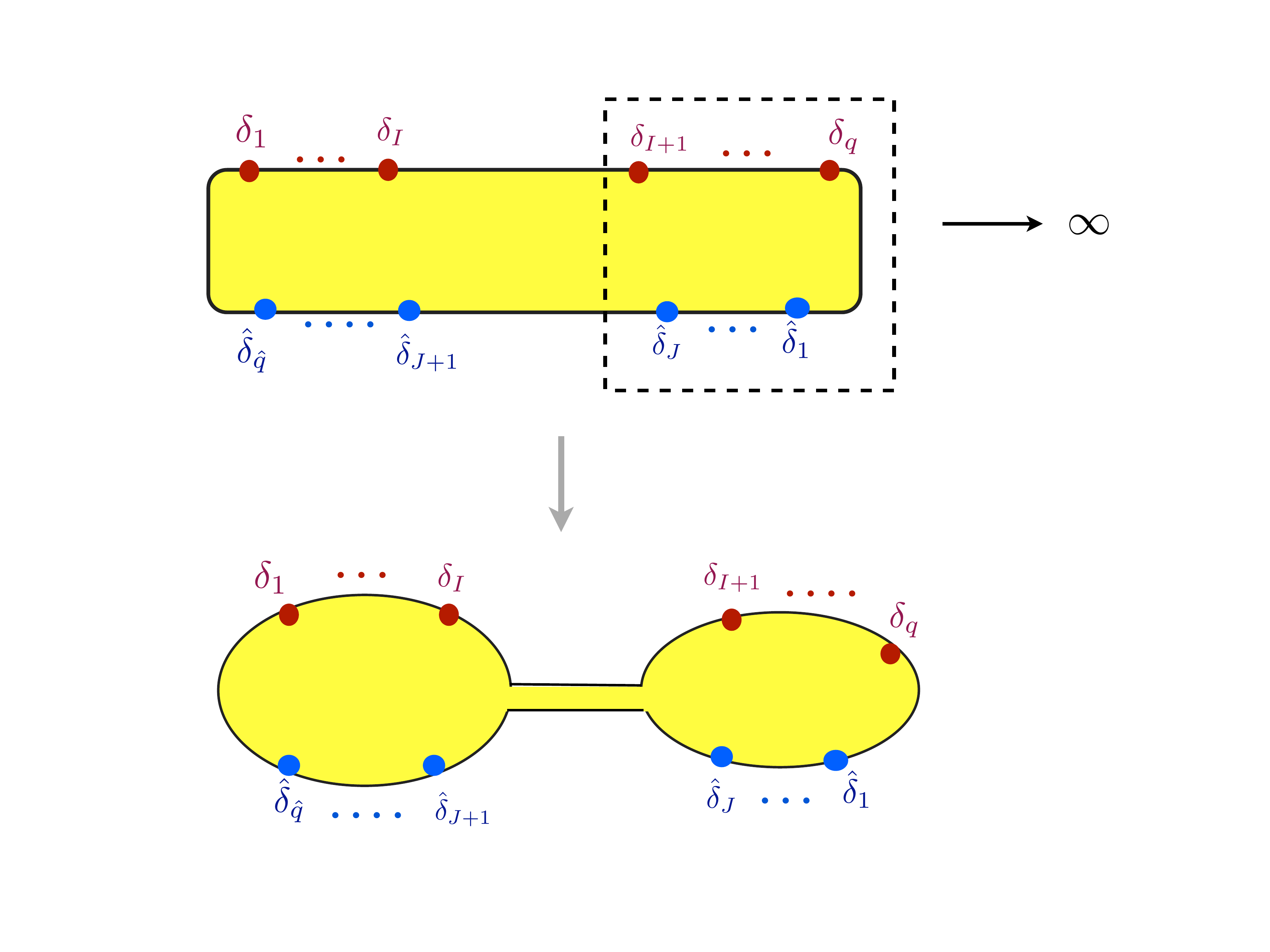}
\vskip -1.cm
\caption{\footnotesize
Schematic drawing of the factorization limit of fivebrane singularities discussed in the text. The lower picture 
is meant to show the actual size of the strip geometry.  The background consists of two,    $AdS_4\ltimes K$
and   $AdS_4\ltimes K^\prime$,   solutions 
coupled through a narrow $AdS_5\times S^5$ bridge.  The curvature of the narrow bridge is larger 
than the curvature in the rest of the  geometry, but can be small enough so as  to ignore quantum gravity corrections.
The configuration resembles therefore a wormhole (or worm-brane).
  }
\label{factorize}
\end{figure}

 More generally, if we keep  also  track of the fivebranes moving off to infinity, we find a supergravity solution
 which consists of two geometries of type $AdS_4\ltimes K$ and 
 $AdS_4\ltimes K^\prime$, connected by a narrow bridge, as illustrated in  figure \ref{factorize}.  
 The space   $AdS_4\ltimes K$ corresponds to  
  keeping  only the stacks $a = 1 , 2 , ... , I$ , $ b = J+1, J+2, ..., \hat p$,
  while  the space   $AdS_4\ltimes K^\prime$  is the solution obtained if we only keep 
  the fivebrane stacks $a = I+1 , I+2 , ... , p$ ,  and $ b = 1, 2, ..., J$.  Saturating the relation $\rho^T \geq \hat \rho$
  corresponds to eliminating all  D3-branes in the intermediate region.  It can be checked indeed that,  in the
  limit,  the D3-brane charge is separately conserved in the two regions.

We can check that the partitions corresponding to these two solutions are exactly the ones obtained by the splitting of ($\rho$ , $\hat \rho$) into two subpartitions by the saturation of the condition (\ref{niceform}). These partitions are explicitly :
\bea
\rho_{L} &=& \Big( \underbrace{l_1 -\sum_{b=1}^{J} \hat N_b,...,l_1 - \sum_{b=1}^{J} \hat N_b}_{N_1} , ... , \underbrace{l_I - \sum_{b=1}^{J} \hat N_b,...,l_I - \sum_{b=1}^{J} \hat N_b}_{N_{I}} \Big)
\nonumber\\
\hat{\rho}_{L} &=& \Big(\underbrace{\hat{l}_{J+1},...,\hat{l}_{J+1}}_{\hat{N}_{J+1}}, ... ,\underbrace{\hat{l}_{\hat{p}},...,\hat{l}_{\hat{p}}}_{\hat{N}_{\hat{p}}}  \Big)
\eea
and
\bea
\rho_{R} &=& \Big( \underbrace{l_{I+1},...,l_{I+1}}_{N_{I+1}} , ... , \underbrace{l_{p},...,l_{p}}_{N_{p}} \Big)
\nonumber\\
\hat{\rho}_{R} &=& \Big(\underbrace{\hat{l}_{1} - \sum_{a=1}^{I} N_{a},...,\hat{l}_{1} - \sum_{a=1}^{I} N_{a}}_{\hat{N}_{1}}, ... ,\underbrace{\hat{l}_{J} - \sum_{a=1}^{I} N_{a},...,\hat{l}_{J} - \sum_{a=1}^{I} N_{a}}_{\hat{N}_{J}}  \Big)
\eea
where the indices L, R refer to the left and right parts of the split quiver.
 The   linking numbers have been here gauge transformed  so as   to make them agree, for each sub-quiver separately,  with our
 earlier conventions.
  So the splitting of the  quiver  corresponds precisely to the factorization of the bulk geometry, confirming
  once again the holographic duality map.

\vspace{5mm}

\underline{{\bf The pinched annulus}} :

\vspace{5mm}

In the case of the annulus we study the limit of infinite length $t\to\infty$.\\
When  taking the limit $t\to\infty$  one must decide what to do with the positions,  $\{ \delta_a\} $ and $\{ \hat\delta_b\}$,  of the  five-brane
singularities.  If the number of singularities is kept fixed   then, since $\delta_a - \delta_{a+p} = \hat\delta_b - \hat\delta_{b+\hat p} = 2t$,
at least one of the  intervals  $\delta_a - \delta_{a+1}$ with
$a\in [1, p]$, and at least one interval
$\hat\delta_b - \hat\delta_{b+1}$ for some $b\in [0, \hat p -1]$ should become infinite in the limit.
  Without loss of generality, we take these divergent separations  to correspond to $a=p$ and $b=0$.
 From the expression \eqref{Lcharge} we conclude that $L \to 0$ in this limit, so that  the circular quiver  degenerates to
 a linear quiver.  If more than one interval diverges, the linear quiver breaks up further into disjoint linear quivers as just described.

    \smallskip

\begin{figure}
\centering
\includegraphics[height=9.5cm,width=12cm]{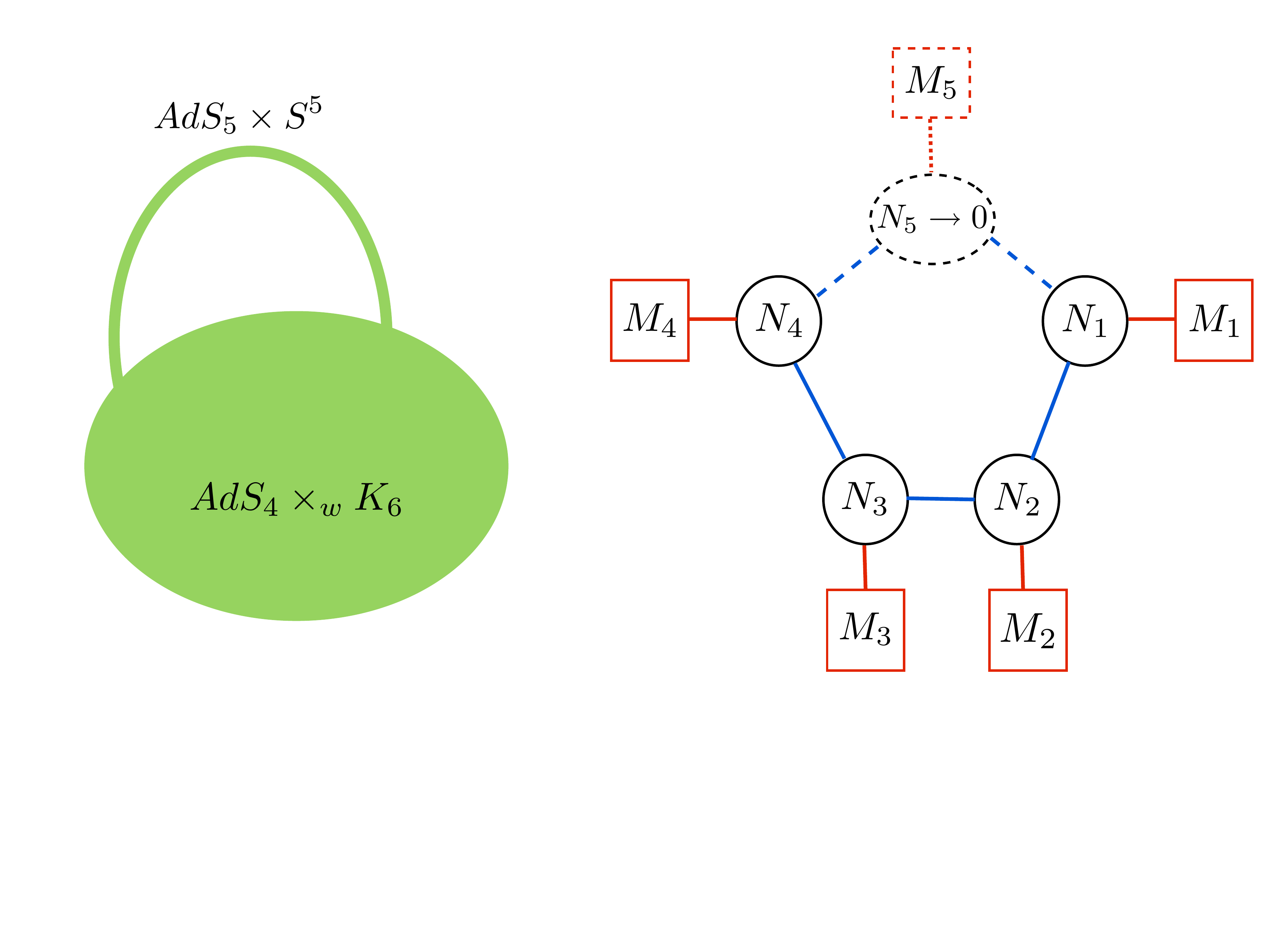}
\vskip -1.9cm
\caption{\footnotesize
Breaking up of a circular quiver as described in the text. On the left, the ten-dimensional geometry describes  a wormhole
 whose entrances are extremal D3-branes. An example of a  dual gauge theory is illustrated on the right:
 a gauge-group factor with vanishing rank opens up
 the chain into a linear quiver.
 }
\label{degenerate}
\end{figure}

In the two limits we have described,  the  geometry  describes what one  may call
 a ``worm-brane".  A schematic representation of this space-time  is given in  Figure \ref{degenerate} for the for the $L \rightarrow 0$ case.
A highly-curved $AdS_5\times S_5$  throat
\footnote{By scaling up homogeneously all charges, we can keep the curvature small enough so that
 the supergravity approximation stays valid in the $AdS_5\times S_5$ throat (though  of course not near the five-brane singularities).}
 from a bridge between either two compact spaces $K_6, K'_6$ (strip), or between two  distinct  points  of  the same compact  space $K_6$(annulus), forming a handle.
 The wormhole entrances are three-dimensional extended objects, whence the name ``worm-brane". 


  \smallskip

          From the perspective of the  gauge theory, the pinching-limit   geometries describe quivers with a gauge-group
         factor whose rank is much smaller than all other ranks.   Taking this rank formally to zero opens up the circular chain or cut the linear chain in two pieces, and decouples
         the corresponding fundamental hypermultiplets, see Figure \ref{degenerate}.  In general, the limiting geometries are smooth except when one sends a  set of stacks
         of the same type  infinitely far from all  other  stacks. This corresponds in gauge theory to the decoupling of free hypermultiplets from the
         end-points of a linear quiver.  The geometry with   five-branes of only one type is singular (it corresponds to having either $h_1=0$ or $h_2=0$), consistently with the fact that
   free hypermultiplets should have no smooth supergravity dual.

\vspace{8mm}

We have not discussed the case of domain wall solutions corresponding to defect SCFT. The case is identical to the linear quiver geometries. The same wormbrane limit is possible, corresponding to having a weak node. The two separated solutions obtained in this limit have only one $AdS_5 \times S^5$ asymptotic region (the other is capped off) and corresponds to solutions described in \cite{Aharony:2011yc}, with the gauge theory duals being SYM on a half-space with a 3d boundary SCFT. 

Let's just mention that the intuition that one could obtain two separated defect SCFT in a wormbrane limit is wrong. The wormbrane limit corresponds to a weak node in the quiver, not to a separation in the direction transverse to the defect. The separation between ''different`` 3d defects in SYM is actually irrelevant as we consider only the infrared limit of such theories.


\subsection{Large-$L$ limit and M2 branes}
\label{secLargeL}

 The second interesting limit of the circular-quiver solutions of section  \S\ref{s:striptoannulus} is the limit  $t\to 0$, $\delta_a, \hat\delta_b \to 0$ with $\delta_a/t, \hat\delta_b/t$ held fixed.
 As we will see, this is the limit of a very large number, $L$,  of winding D3-branes, in which
  the five-branes are effectively smeared, and the solution reduces
 to the near-horizon geometry of M2-branes   at a $Z_k \times Z_{\hat k}$ orbifold singularity.
   \smallskip

 To compute  the geometry in this limit
we use  the asymptotic behavior of the theta functions when $e^{i \pi   \tau} = e^{-t} \to 1$, or equivalently
$e^{i \pi   \tilde \tau} = e^{ -i \pi /  \tau}  =  e^{-\pi^2/t} \to 0$. One finds in this limit
\bea\label{thetaasymptotic}
{\vartheta_1(\nu \vert  \qth)  \over \vartheta_2(\nu \vert  \qth) } = -i {\vartheta_1(\nu \tilde \tau \vert  \tilde \qth)  \over \vartheta_4(\nu \tilde \tau \vert  \tilde \qth) } \,
= \   -2 \,    e^{-\pi^2/4t}\,  {\rm sinh} ({  \pi^2 \nu /  t })  + {\cal O}(e^{-9\pi^2 /4t})  \ ,
\eea
where the second equality  follows from the  expressions of the theta functions as infinite sums.
The formula   simplifies further if ${\rm Re} (\nu)\not= 0$, in which case the hyperbolic sine  can be replaced by
an exponential.  Inserting  \eqref{thetaasymptotic}
  in \eqref{Amany}, and recalling that
 $2\pi {\rm Re} (\nu_a) = \pi/2 - {\rm Im} (z)$  and $2\pi {\rm Re}( \hat \nu_b)= {\rm Im} (z)$, finally   gives
     \begin{align}
\label{Alimit}
{\cal A}_1  =   \sum_{a =1}^p \gamma_a  \,  {\pi \over 2t}  z
 + \varphi_1   \  , \  \quad
{\cal A}_2  =  i \sum_{b=1}^{\hat p} \hat \gamma_b \,    {\pi \over 2t} (z- {i \pi\over 2} )
+ i \varphi_2  \
 ,
\end{align}
where we have absorbed some irrelevant constants
in the arbitrary phases $\varphi_1$ and $\varphi_2$.
 This approximation breaks down at a distance   $\sim t$  from the annulus boundaries, where the linear dependence
is  replaced by the rapidly-oscillating $\log  \, {\rm sinh}$ function.
 \smallskip

    The first thing to note is that, away from the boundaries, the harmonic functions depend only on three parameters: $t$ and
    the total numbers of five-branes, $k = \sum \gamma_a $ and $\hat k = \sum \hat\gamma_b$. The precise locations of
    the five-brane singularities do not matter, as if these   were smeared.  It is convenient to scale out the $t$-dependence
    by redefining the annulus coordinate as follows: $2\pi z= 2t  x +i \pi^2  y$, so that $x\in [0, 2\pi)$ and
    $y\in [0,1]$. In terms of these coordinates, the holomorphic functions read
      \begin{align}
\label{Alimitxy}
{\cal A}_1  =  k  \, (   {x \over 2 } + i {\pi^2 y \over 4t} )
   \  , \  \quad
{\cal A}_2  =  i \hat k   \,    \left(   {x \over 2 } - \pi  + i {\pi^2 (y-1)  \over 4t} \right)
   \
 ,
\end{align}
where we have here chosen $\varphi_1$ and $\varphi_2$  so as to impose the canonical gauge condition \eqref{canongauge}, which is really $C_2|_{(x,y)=(0,1)}=0$ and $B_2|_{(x,y)=(2\pi,0)}=0$ in this smearing limit.
Inserting these functions in the general form of the solution, see  \S\ref{s:localsolutions},  gives  the
   Einstein-frame metric  (we recall that $\alpha^\prime =4$):
   \begin{align}
ds^2 =  R^2 g(y)^{1\over 4}  \left[   ds^2_{AdS_4} +  y\,  ds^2_{S^2_1} +   (1-y) ds^2_{S^2_2} \right] +
 R^2 g(y)^{-{3\over  4} } \left[{4t^2\over \pi^4}  dx^2 +  dy^2 \right]  \ , \no
\end{align}
\vskip -8mm
 \bea
 {\rm with} \quad R^4 =  \pi^4  \, { {k \hat k}\over t^2}  \ , \qquad {\rm and}\quad g(y) = y(1-y)\ .
 \label{IIBlargeL1}
 \eea
 Furthermore, the dilaton  and the non-vanishing gauge fields read:
 \begin{align}
e^{2 \phi} =  \frac{\hat k}{k} \sqrt{ \frac{1 - y}{y}}  \ ,   \qquad
C_{(4)} = R^4 \left( {6t x \over \pi^2}  \,  \omega^{0123} + y^2(y- {3\over 2})\,  \omega^{4567} \right) \ , \no
\end{align}
 \begin{align}
B_{(2)}  = 2\hat k(2\pi -x)\,  \omega^{45} \ ,
  \qquad
C_{(2)} =  -2kx\,  \omega^{67}  \  .
\label{IIBlargeL2}
\end{align}
 As already noted, this solution only depends on three integer parameters: the numbers $k$ and $\hat k$ of five-branes,
 and the modulus $t$ of the annulus which can be traded for the number of winding D3-branes via the formula \eqref{Lcharge},
 \begin{align}
L\, =\,  \frac{k \hat k}{2t^2} \, \int_{0}^{+\infty} du \, u\,  \frac{2}{\pi} \arctan(e^{-u}) \, = \, \frac{\pi^2 }{32  }\, \frac{k \hat k}{t^2} \ .
\end{align}
One may also compare \eqref{Lcharge} to the formula  \eqref{Mcharge1} for the charge $M= L+N$, where $N$ gives the
number of D3-branes emanating from five-branes. Since the summands in these two expressions differ by terms of order $t^2$,
we conclude that $N \sim k\hat k$ as $t\to 0$. Thus   the
 number of winding D3-branes  far exceeds, in this limit,  the number of D3-branes that  end on the  five-branes.
 \smallskip

 Not surprisingly, after having effectively smeared the five-branes, the  solution has a Killing isometry under translations of  the coordinate $x$. To be sure,  $x$
  enters in the expressions for $B_{(2)}$ and $C_{(2)}$  but this is a gauge artifact since the 3-form field strengths
 are $x$-independent.  One may thus T-dualize the circle parametrized by  $x$,  using Buscher's rules
 \cite{Buscher:1987qj}, to  find a solution of type-IIA supergravity. This can be then  lifted to eleven dimensions --   the  details of these
  calculations  are given  in appendix \ref{Tduality}.
 The final result for the eleven-dimensional metric is
 \begin{align}
\label{AdS4S7}
ds_{{\rm M-theory}}^2 &=  {\bar R^2}  ds^2_{AdS_4} + \bar R^2 \Big[ 4 d\alpha^2 + \sin^2 \alpha \, ds^2_{S^3/\mathbb Z_{\hat k}} + \cos^2 \alpha \, ds^2_{S^3/\mathbb Z_k} \Big] \ , \no\\
ds^2_{S^3/\mathbb Z_{\hat k}} &=   d\theta_1^2 + d\phi_1^2 + 4 dx^2 - 4 \cos \theta_1 dx d\phi_1    \ , \no\\
ds^2_{S^3/\mathbb Z_{k}} &=  d\theta_2^2 + d\phi_2^2 + 4 dv^2 - 4 \cos \theta_2 dv  d\phi_2   \ ,
\end{align}
where $x$ and $v$ are angle coordinates with periodicities $x= x+ 2\pi/ \hat k$ and $v= v+ 2\pi/ k$,  while the radius of $AdS_4$ is
 $\bar R^2 = (2^{5} \pi^2 k\hat k L)^{1/3}$.

  This is the metric of $AdS^4 \times S^7/ (\mathbb{Z}_k\times  \mathbb{Z}_{\hat k})$   with the two orbifolds acting on the two 3-spheres  in $S^7$.
  The solution furthermore carries $L$ units of four-form flux. It  can be recognized as the near-horizon geometry of $L$ M2-branes sitting
  at the fixed point of the orbifold
  $(\mathbb C^2/\mathbb Z_{\hat k} )\times (\mathbb C^2/\mathbb Z_k )$, where the orbifold identifications are
  $$
  (z_1, \bar z_2) = e^{2i\pi/\hat k} (z_1, \bar z_2)\quad {\rm and}\quad
   (z_3, \bar z_4) = e^{2i\pi/k} (z_3, \bar z_4)\ .
  $$
  Note that the two-forms $B_{(2)}$ and $C_{(2)}$ become,   after the T-duality and the lift, part of the metric. This is in
  line with the fact that D5-branes transform (to D6-branes and then) to Kaluza-Klein monopoles, while T-duality in a transverse dimension maps the NS5-branes
  to ALE spaces with  singularities of $A_n$ type \cite{Ooguri:1995wj,Tong:2002rq}.

  \smallskip

   The superconformal field theories that are dual to M theory on  $AdS^4 \times S^7/ (\mathbb{Z}_k\times  \mathbb{Z}_{\hat k})$ are
 close relatives of the  ABJM theory \cite{Aharony:2008ug,Aharony:2008gk} that  have been analyzed by many authors,
  see for example  \cite{Hosomichi:2008jd,Benna:2008zy, Imamura:2008nn,Imamura:2008ji,Herzog:2010hf,Dey:2011pt}. We  will discuss them in more detail in the following
  section. Let us here only quote their free energy  $F = - \log |Z|$ on the 3-sphere.  Using the general formula of \cite{Herzog:2010hf} one finds
  \begin{align}\label{herzogfree}
F \, =\,  L^{3/2}\sqrt{\frac{2 \pi^6}{27 \, {\rm Vol}_7}}  \, =  \,   \frac{\pi }{3} \sqrt{2 k  \hat k} \, L^{3/2}  \ ,
\end{align}
where ${\rm Vol}_7$ is the volume of the compact (Sasaki-Einstein)  manifold  whose metric is normalized so that $R_{ij} = 6 g_{ij}$.
In the case at hand, this is the unit-radius seven sphere with orbifold identifications, so that
  ${\rm Vol}_7 =  {\pi^4}/{3k \hat k}$.
  \smallskip

  As a check of our formulae, we may  compute this free energy   on the type-IIB side.
  
   The on-shell IIB action can be computed via a consistent truncation to pure four-dimensional
    gravity with unit $AdS_4$ metric  multiplied by a 6d volume factor.  We defer the explicit computation of the regularized action to the next chapter, where it plays a central role, and just quote the explicit formula
\begin{align}
\label{acteff}
S_{\rm IIB} = -\frac{1}{(2\pi)^7 (\alpha')^4} \left( \frac{4}{3} \pi^2 \right) (-6)  {\rm vol}_6 \ ,
\end{align}
where for the solutions of interest
\begin{align}
{\rm vol}_6 = -16 (4 \pi)^2 t  \int_0^{2\pi}\hskip -1mm dx \int_0^1\hskip -1mm  dy \, h_1 h_2 \, \p_z \p_{\bar z} (h_1 h_2) \ .
\label{vol6}
\end{align}
 Plugging in the harmonic functions $h_1 = -i( {\cal A}_1 - \bar  {\cal A}_1 )$ and $h_2 = {\cal A}_2 + \bar  {\cal A}_2$, and performing
 the integrals  gives
\beq
S_{\rm IIB} = \frac{\pi^4}{48} \frac{k^2 \hat k^2}{(2t)^3} = \frac{\pi}{3}\sqrt{2 k \hat k}\,  L^{3/2}
\label{SIIB}
\eeq
   in perfect agreement with the   result of M theory.
\smallskip

To summarize, we have shown here  that when $L$ is large our solutions approach smeared backgrounds
dual to M theory on $AdS^4 \times S^7/ (\mathbb{Z}_k\times  \mathbb{Z}_{\hat k})$.
 In this limit  the information about the positions of the
  five-branes is lost, and following \cite{Gregory:1997te,Harvey:2005ab} its reinstatement would require non-trivial backgrounds for the
   wrapped-membrane field.  The essential topological features of the background can be,  however,  
   in principle encoded more simply,  as  3-form torsion of the M-theory orbifold
\cite{Imamura:2008ji, Witten:2009xu,Aharony:2009fc,Dey:2011pt}.  Note that,  contrary to the $\N \geq 6$  ABJ(M) cases studied in \cite{Aharony:2008ug,Aharony:2008gk},
 the orbifolds considered here are not freely-acting on S$\hskip -0.5mm \,^7$, and one would need  to resolve their singularities. 
It would be interesting to work out the precise match 
 of the torsion with the quiver data, and see how the constraints  on the triplet $(\rho, \hat\rho, L)$ arise from  the M-theory side.

\subsection{Supergravity regimes of parameters}
\label{subsec:parameters}

A interesting question to ask is : In which regime of parameters can we trust the supergravity solutions described in this chapter ?  The answer should naïvely be that the supergravity description always breaks down, due to the presence of D5-brane singularities where the curvature diverges and NS5-brane singularities where the dilaton diverges (to $+\infty$). However we know that these singularities must be cured by string corrections. It is sensible to assume that as long as the regions of large curvature or large dilaton are confined to the close vicinity of the 5-branes in $\Sigma$ the supergravity description can be used to answer some questions and do computations in which the 5-branes contribution is subdominant.

Having evacuated the question of the 5-branes we want to find the regime of parameters where the radius of curvature is large compared to the Planck length (and string length) and the dilaton diverges to $-\infty$ (small string coupling). These are the conditions for the quantum loop corrections and string corrections to be supressed. This goes down to demanding
\begin{align}
 R_{r.c.} >> 1  \quad , \quad e^{2\phi} << 1 \quad ,
\end{align}
where $R_{r.c.}$ is the radius of curvature in string units (and $\phi$ is the dilaton). 

\vspace{6mm}

Verifying these conditions is not straightforward, because of the complexity of the parameter dependence of the supergravity solutions. By analogy with the Maldacena setup (see \S\ref{sec:corresp}) we may demand a large 5-form flux or large number of D3-branes in the geometry. Fo the linear quivers $T^{\rho}_{\hat\rho}(SU(N))$ it corresponds to the large $N$ limit. For circular quivers $C_{\hat\rho}^\rho(SU(N), L)$ it can also be the large $N$ limit, however in this limit the circular quiver geometry degenerates into a linear quiver geometry (when $N>>L$ the node of the quiver with rank $L$ is a weak node and the circular quiver breaks into a linear one, as discussed above). The interesting limit of large D3-brane flux for circular quivers is then the large $L$ limit studied in the last subsection.

\smallskip

Taking the large $N$ limit for linear quiver geometries, the formulas \ref{conserv}\ref{ginvN3} implies that there is a couple $(a,b)$ with  $N_5^a \sim N^{\alpha}$, $\hat N_5^b \sim N^{\beta}$ and $e^{\hat\delta_b - \delta_a} \sim N^{1-\alpha-\beta}$ with $0 \le \alpha , \beta \le 1 \le \alpha + \beta$. This in turn implies that in the region between the two 5-brane stacks $a$ and $b$
the harmonic functions scale like 
\begin{align}
 h_1 \sim N^{\frac{1+\alpha-\beta}{2}} \quad , \quad h_2 \sim N^{\frac{1+\beta-\alpha}{2}} \ . 
\end{align}
Then using the formulas of \S\ref{s:localsolutions} describing the solutions, we obtain that the Einstein metric scales like $g_{\mu\nu} \sim N^{1/2}$ independently of $\alpha, \beta$ and the dilaton scales $e^{2\phi} \sim N^{\beta-\alpha}$. The string frame metric $g_{\mu\nu}^{(srt)} = e^{\phi} g_{\mu\nu}$ scales $g_{\mu\nu}^{(srt)} \sim N^{\frac{1+\beta-\alpha}{2}}$. We obtain
\begin{align}
 R_{r.c.}^2 \sim N^{\frac{1+\beta-\alpha}{2}} \quad , \quad  e^{2\phi} \sim N^{\beta-\alpha} \ . \no
\end{align}
The condition $R_{r.c.} >> 1 $ is always satisfied, except in the special case $(\alpha,\beta)=(1,0)$, whereas the condition $e^{2\phi} << 1$ needs $\alpha > \beta$, which can be traced back to the condition that the number of D5-branes in the stack $a$ has to be (hierarchicaly) larger than the number of NS5-branes in the stack $b$.\\
This is not surprizing : the large $N$ conditions ensures that the radius of curvature is large, as in the Maldacena setup, and the larger number of D5-branes ensures that the string coupling is small in the geometry (remember that the dilaton goes to $+\infty$ near the NS5-branes and to $-\infty$ near the D5-branes).

Using S-duality one obtains a dual supergravity solution with NS5-brane and D5-brane stacks exchanged. The parameters $\alpha$ and $\beta$ are exchanged in the process. This means that if $\alpha < \beta$, we can use the S-dual solution where the conditions for the supergravity regime are verified. This leaves us with only two problematic regimes : $(\alpha,\beta)=(1,0)$ and $\alpha=\beta$. As we will see in chapter \ref{chap:GKPW}, in the latter ''bad`` regime one may use the supergravity solutions and obtain some correct computations.

\bigskip

We conclude that the supergravity regime for linear quiver theories is generically the large $N$ limit and small ratio $\frac{\hat k}{k}$ of the total number of NS5-branes $\hat k$ over the total number of D5-branes $k$.
\begin{align}
 N >> 1 \quad , \quad \frac{\hat k}{k} << 1 \ ,
\end{align}
with the possibility to use S-dual solutions if $ \frac{\hat k}{k} >> 1$. \\
Note that the parameters $k, \hat k$ and $N$ are not really independent. They obey for instance $k, \hat k \le N \le k \hat k$, as can be seen from the explicit expressions \ref{ginvN3} \ref{conserv}. This shows that the condition 
$ \frac{\hat k}{k} << 1$ is actually enough because it implies $N >> 1$.

\smallskip

The existence of a regime in which the supergravity calculations can be trusted (when the 5-brane contributions can be ignored) is to be contrasted with the situation on the gauge theory side where there is a priori no accessible regime (infinitely strongly coupled gauge theories).

\vspace{8mm}

The situation for the $C_{\hat\rho}^\rho(SU(N), L)$ circular quiver solutions in the large $L$ limit (with $N << L$) is different. The number $L$ of D3-branes wraping the annulus is independent from the numbers of 5-branes. From the analysis of the last subsection \S\ref{secLargeL}, we see that the radius of curvature (in string frame) is given by $R_{r.c.}^4 \sim L\hat k/k$ in all directions, except in the $x$ direction wrapping the annulus (and on a small vicinity of the upper boundary where the smeared D5-branes sit) with $R_{r.c.}^4 \sim L\hat k/k$, $R_{r.c.}^{x \ 4} \sim \hat k^4/L$. The dilaton goes like $e^{2\phi} \sim \hat k/k$ everywhere. This implies that the regime of parameters when supergravity can be used is 
\begin{align}
\label{regime:annulus}
 \frac{\hat k}{k} << 1 \quad , \quad 1 << \frac{L \hat k}{k} << \frac{\hat k^5}{k}  \  .
\end{align}
Because of S-duality one can also use the supergravity solution in the regime \ref{regime:annulus} with $k \leftrightarrow \hat k$.
Again there is a priori no weak-coupling regime on the gauge theory side.

\vspace{8mm}

The experience of the AdS/CFT correspondence lets us think that the presence of supergravity regimes for the solutions dual to the linear and circular quivers may be an indication that there is a corresponding weak-coupling regime on the gauge theory side, for dual SCFTs that remains to find. The orbifold duality between Yang-Mills theories and Chern-Simons-Matter theories that we present in chapter \ref{chap:SL2R} may be the answer to this question, however the situation is not really clear as the orbifold duality is not a duality at the quantum level.

%

\chapter{Testing the correspondence : free energy calculations}
\addcontentsline{lot}{chapter}{ Testing the correspondence : free energy calculations}
\label{chap:GKPW}

In this chapter we provide further quantitative consistency checks of this $AdS_4/CFT_3$
correspondence by verifying the GKPW relation \ref{GKPW3} (reviewed in section \ref{sec:corresp}) for the partition function  \cite{Witten:1998qj,Gubser:1998bc} in the leading large $N$ limit :

\begin{align}
\left | Z_{\rm CFT} \right |& =e^{-S_{\rm gravity}} \ , \quad
\textrm{i.e.}
\quad
F_{\rm CFT}=S_{\rm gravity} \ ,
\end{align}
where $Z_{\rm CFT}$ is a CFT partition function on $S^3$,
$F_{\rm CFT}:=-\ln |Z_{\rm CFT}|$ is the free energy, and
$S_{\rm gravity}$ is the action for the type IIB supergravity
holographic dual to the CFT.

We concentrate on
a class of 3d $\scN=4$ linear quiver SCFTs
$T^{\rho}_{\hat{\rho}}[SU(N)]$ in which the numbers of 5-branes grow as fractional powers of $N$ (see below).

On the CFT side, we take the large $N$ limit of the $S^3$ partition
functions of \cite{Benvenuti:2011ga,Nishioka:2011dq}, evaluated at the conformal point. 
On the gravity side, we evaluate the gravity action on the linear quiver solutions presented in section \ref{sec:gravitysolns}. The action is evaluated on the (regularized) euclidean $AdS_4$ spacetimes, whose conformal boundary is $S^3$. The radius $r$ of $S^3$ provides a (IR) cutoff and makes it possible to compute the partition function using the localization techniques (\cite{Kapustin:2009kz}), however the $r$ dependence disappears from the final result.  
We find that in both cases the leading contribution of the free energy in the large $N$
 limit scales as
\begin{equation*}
F\sim N^2 \ln N+{\cal O}(N^2) \ .
\end{equation*}
As we will see,
on the CFT side $N^2 \ln N$ comes from the asymptotic behavior of the
Barnes $G$-function.
On the gravity side, a factor of $N^2$ comes from the
local scaling of the supergravity Lagrangian, and an extra $\ln N$
comes from the size of the geometry.

\vspace{8mm}

\noindent\underline{{\bf Summary of results }} :
\vspace{5mm}

Our findings are summarized as follows.

\begin{itemize}
\item The simplest prototypical example is the $T[SU(N)]$ theory, which
      is a $T^{\rho}_{\hat{\rho}}[SU(N))]$ theory with
\beq
\rho = \hat\rho = \big[\overbrace{1,1,...,1}^{N}\big] \ .
\label{rhospecial}
\eeq

In this case we find
\beq
F_{\rm CFT}=S_{\rm gravity}=\frac{1}{2} N^2 \ln N +\scO(N^2) \ .
\label{Fspecial}
\eeq

\item More generally we consider the case $\hat{p}=1$, i.e.,
\begin{align}
\begin{split}
\rho &= \Big[ \overbrace{l^{(1)},l^{(1)},..,l^{(1)}}^{N_{5}^{(1)}},\ \overbrace{l^{(2)},l^{(2)},..,l^{(2)}}^{N_{5}^{(2)}},\ ...\ ,
\ \overbrace{l^{(p)},l^{(p)},..,l^{(p)}}^{N_{5}^{(p)}} \Big] \ ,   \\
\hat \rho &=
\Big[ \overbrace{\hat l,\hat l,..,\hat l}^{\hat N_5} \Big] \ .
\end{split}
\label{rhogeneral}
\end{align}
We take the scaling limit
\beq
N_5^{(a)} =N^{1-\kappa_a} \gamma_a,\quad l^{(a)} = N^{\kappa_a}
\lambda^{(a)},\quad
\hat N_5 = N \hat \gamma \ ,
\label{scaling}
\eeq
where we take $N$ large, while keeping $\kappa_a, \lambda^{(a)}, \gamma_a, \hat{\gamma}$
finite.
\footnote{Notice that in this chapter $\gamma_a$ and $\hat\gamma_b$ are not directly the numbers of 5-branes, as it was the case in chapter \ref{chap:sugra}, but are only proportional to these numbers. We hope this will note add confusion.}

We require
\beq
\kappa_{a-1}\ge \kappa_a,\quad  0\le \kappa_a<1,   \quad \textrm{for all  } a
\ .
\label{kappa}
\eeq
The first condition is necessary for $\rho$ to be a partition as defined in chapter \ref{chap:sugra}, that is with non-increasing linking numbers $l^{(a)}$, and the
      second ensures that the $N_5^{(a)}$ becomes large, simplifying de computations.
We also have, from the sum rule \ref{sumrule} \ref{conserv}, the constraint
\beq
\sum_{a=1}^p \gamma_a \lambda^{(a)}=\hat{\gamma}\, \hat{l}=1 \ .
\label{sumN2}
\eeq
In this more general case we find (the CFT analysis will be provided only for $\hat{l}=1$
      and the gravity analysis for general $\hat{l}$):
\begin{align}
F_{\rm CFT}=S_{\rm gravity}= \frac{1}{2} N^2\ln N \left[  (1-\kappa_1)
+ \sum_{i=2}^p \left( \sum_{a=i}^p \gamma_a \lambda^{(a)} \right)^2\left( \kappa_{i-1}-\kappa_i\right)
\right] + {\cal O}(N^2).
\label{Fgeneral}
\end{align}
In particular when all $\kappa_a=0$, i.e. when all $l^{(a)}$ are finite,
      the leading large $N$ behavior
      coincides with that in \eqref{Fspecial}.
\end{itemize}

Note the number inside the bracket in \eqref{Fgeneral} is a non-negative number smaller than $1$ due to \eqref{kappa}.
Motivated by this result we conjecture
\beq
F_{T^{\rho}_{\hat{\rho}}[SU(N)]}\le F_{T[SU(N)]} \ ,
\label{conj}
\eeq
for all $\rho, \hat{\rho}$ satisfying the supersymmetry inequalities \eqref{fixedpoint}.
We will explain at the end of the chapter how  \ref{conj} can be explained in terms of the F-theorem \cite{Jafferis:2010un,Jafferis:2011zi}
and the RG flows between the fixed points.
Before that, we will derive the announced results \eqref{Fspecial} and \eqref{Fgeneral}.

\section{CFT Analysis}\label{sec.gauge}

\subsection{The $S^3$ Partition Function}

%

\begin{figure}[htbp]
\vspace{-1.7cm}
\centering
\includegraphics[width=0.5\textwidth]{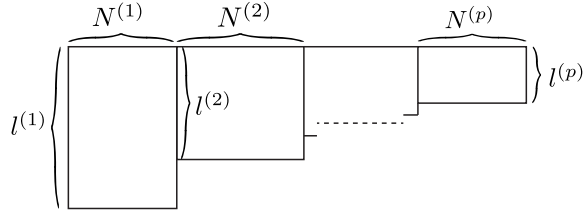}
\vspace{-1.7cm}
\caption{We decompose the young diagram corresponding to $\rho$ into $p$
 blocks, see \ref{linpartitions}.}
\label{fig.rule}
\end{figure}

The partition function of the deformed $T^{\hat{\rho}}_{\rho}[SU(N)]$ theories given in \ref{Zformula} is
\beq
Z_{S^3}[T^{\rho}_{\hat{\rho}}[SU(N)]](m, \xi)=\frac{\sum_{w\in \mathfrak{S}_N}
(-1)^w e^{2\pi i m_{\rho}\cdot w(\xi_{\hat{\rho}})}}{\Delta_{\rho}(m) \Delta_{\hat{\rho}}(\xi)}.
\label{Zformula2}
\eeq
Here $m_{\rho}, \xi_{\hat{\rho}}$ are the $N$-deformation vectors defined in section \ref{sec:Zexpr}, and each of their
 components
is associated with a box of the Young diagram
corresponding to the deformed partitions $\rho, \hat{\rho}$.
For later purposes let us describe them by
dividing the boxes of $\rho$ into $p$
blocks, where the $a$-th block is a rectangle with rows of length $N_5^{(a)}$
and columns of length $l^{(a)}$ (recall \ref{linpartitions}, and
see fig. \ref{fig.rule}). A box of
$\rho$ could then be labeled by
a triple $(a, i, \alpha)$ with
$1\le a\le p, 1\le i \le N_5^{(a)}, 1\le \alpha
\le l^{(a)}$, where $a$ is the label for the block
and $i$ ($\alpha$) is the label for the column (row) inside the $a$-th block.
The same applies to $\hat{\rho}$.
In this notation, we have
\beq
\label{notation3}
(m_{\rho})_{(a,i, \alpha)}=i(w_{l^{(a)}})_{\alpha}+ m_{a,i} , \quad
(\xi_{\hat{\rho}})_{(a,i, \alpha)}=i(w_{\hat{l}^{(a)}}
)_{\alpha}+ \xi_{a,i} \ ,
\eeq
where $w_N$ is a Weyl vector of the $su(N)$ Lie algebra defined by
\bea
w_N=\left(\frac{N-1}{2}, \frac{N-3}{2},\ldots,
-\frac{N-1}{2}\right)\ .
\label{rhoW}
\eea
Also, $\Delta_{\rho}(m)$ and $\Delta_{\hat{\rho}}(\xi)$ are defined by (\ref{Delta})
\begin{equation}
\begin{split}
\Delta_{\rho}(m)& =\prod_p \prod_{q<r} 2 \sinh \pi ((m_{\rho})_{[p,q
]}-(m_{\rho})_{[p,r]}) ,
\\
\Delta_{\hat{\rho}}(\xi)& =\prod_p \prod_{q<r} 2 \sinh \pi
((\xi_{\hat{\rho}})_{[p,q]}-(\xi_{\hat \rho})_{[p,r]}) \ ,
\end{split}
 \end{equation}
where $[p,q]$ represents a box inside the Young tableau $\rho$ (or $\hat{\rho}$) at row $p$ and column $q$.
Note that the $(m_{\rho})_{[p,q]}$ are simply a relabeling of the $(m_{\rho})_{(a,i,\alpha)}$ introduced previously.

\subsection{$T[SU(N)]$}
\label{subsec.largeN}

Let us study the large $N$ behavior of our partition functions.

For clarity, let us begin with the $T[SU(N)]$ theories,
whose partition function is given in \eqref{ZTSUN}. Let's rewrite it 

\begin{align}
\label{ZTSUN2}
 Z^{T(SU(N))} &= (-i)^{\frac{N(N-1)}{2}} \frac{ \sum_{w \in  \mathfrak{S}_N} (-1)^w \, e^{2i\pi \sum_j^N \xi_j m_{w(j)}}}{\prod_{j<k}^N \sh(\xi_j - \xi_k)\sh(m_j - m_k)} 
\end{align}

When the parameters $m_j$ and $\xi_j$ are
generic and kept finite in the limit,\footnote{By generic we mean that there are no cancellations in the sum
in the numerator of
\eqref{ZTSUN2}.}
we have $\sum_{w\in \mathfrak{S}_N} \sim \scO(N!)$, whose
logarithm contributes $\scO(N \ln N)$ to $F_{\rm CFT}$.
The remaining contributions come from the two sinh Vandermonde determinants,
each of which involves roughly speaking ${N\choose 2}\sim \scO(N^2)$ terms.
This gives
\beq
F_{\rm CFT}\sim \mathcal{O}(N^2) \ .
\eeq
This is not really surprising since after all our theories are standard gauge
theories with gauge group ranks of order $N$.

However, the scaling behavior could change if we consider non-generic values of
$m$ and $\xi$.
This is exactly happens to our CFT case, where we need to take the limit
$m, \xi\to 0$ of \eqref{ZTSUN2}:
\beq
Z_{\rm CFT}=\lim_{m, \xi\to 0} \big| Z_{S^3} \big| \ .
\eeq

We choose to take the limit in two steps.
First, let us take the $\xi\to 0$ limit of \eqref{ZTSUN2} with $\hat{\rho}=[1,\ldots,
1]$.
This is conveniently done by setting
$\xi=\epsilon w_N$ and by taking $\epsilon\to 0$,
where $w_N$ is defined in \eqref{rhoW}.
Using the Weyl denominator formula, we have
\begin{align}
\sum_{w\in \mathfrak{S}_N} (-1)^w e^{2\pi i \epsilon w_N \cdot m_{\rho}} 
&=\prod_{\alpha>0} 2i \sin\left(\pi \epsilon \alpha\cdot  m_{\rho}\right) \
 = \ \prod_{j<k} 2i \sin\left(\pi \epsilon (m_j-m_k)\right)   \no\\
&\simeq (2 i \pi \epsilon)^{\frac{N(N-1}{2}} \prod_{j<k} (m_j-m_k) \ .
\end{align}
In the limit $\epsilon_2\to 0$, this combines with the factor
$\Delta_{\hat\rho}(\epsilon w_N)=\prod_{j<k} 2\sinh \pi (\epsilon (j-k)) \simeq (2\pi \epsilon)^{\frac{N(N-1}{2}} \prod_{j<k}(j-k) $ in the denominator and the prefactor phase of \ref{ZTSUN2}, giving
\beq
 \left| \prod_{j<k} \frac{1}{(j-k)} \frac{\prod_{j<k} (m_j-m_k)}{\Delta_{\rho}(m)}\right|
=
 \frac{1}{G_2(N+1)}
\left|
 \frac{\prod_{j<k} (m_j-m_k)}{\Delta_{\rho}(m)}
\right| \ .
\label{halflimit}
\eeq

We next need to take the limit $m \to 0$.
This is easy for our case, $\rho=[1,\ldots, 1]$;
\[
  \frac{\prod_{j<k}
 (m_j-m_k)}{\Delta_{\rho}(m)}=
  \prod_{j<k}
 \frac{(m_j-m_k)}{2 \sinh \pi (m_j-m_k)}\to (2\pi)^{-\frac{N(N-1)}{2}},
\]
which gives
\beq
Z_{\rm CFT}=\frac{1}{(N-1)!(N-2)! \ldots 2!
1!}\left(\frac{1}{2\pi}\right)^{\frac{N(N-1)}{2}} =\frac{1}{G(N+1)} \left(\frac{1}{2\pi}\right)^{\frac{N(N-1)}{2}} \ ,
\label{ZTSUNzero}
\eeq
where $G_2(x)$ is the Barnes $G$-function defined in Appendix \ref{app:Barnes}.
From the asymptotics of $G_2(x)$ \eqref{Gasymptotics}, we have
\beq
F_{\rm CFT}=\frac{N^2}{2} \ln N +\left[-\frac{3}{4}-\frac{1}{2} \ln
\left(\frac{1}{2\pi} \right) \right] N^2 + \mathcal{O}(N \ln
N) \ ,
\label{cftparttsun}
\eeq
which gives \eqref{Fspecial}.

\subsection{$T^{\rho}_{\hat{\rho}}[SU(N)]$}

Let us consider the more general case given in \eqref{rhogeneral}.

As long as $\hat{\rho}=[1,\ldots, 1]$
the argument of the previous subsection works
up until \eqref{halflimit} with the result for the limit $\xi_j \to 0$ given by
\begin{align}
Z_{\rm CFT} &= \frac{1}{G_2(N+1)}
\left|
 \frac{\prod_{j<k} (m_{\rho})_j-(m_{\rho})_k  }{\Delta_{\rho}(m_{\rho})}
\right| \ .
\label{halflimit2}
\end{align}

In \eqref{halflimit2} we already have a factor of $G_2(N+1)$.
Just as in the $T[SU(N)]$ case, this contributes
\begin{equation}
\frac{1}{2}N^2 \ln N \ ,
\label{part1}
\end{equation}
to the free energy.
Next, let us send the mass parameters $m_j$ to zero in \eqref{halflimit2}.
The denominator $\Delta_{\rho}(m)$ goes to zero in the limit,
but it can be combined with a subset of the numerator factors given by the $(m_{\rho})_{(a,i,\alpha)}-(m_{\rho})_{(b,j,\alpha)}$ with $a \le b$, $i<j$,  in the notation of \ref{notation3} (namely we pick the couples involved in the definition of $\Delta_{\rho}(m)$), yielding a finite answer.
We obtain powers of $2\pi$ in this process from the limit of
$\Delta_{\rho}(m)$,
however this only gives a subleading contribution of order $N^2$.

There are still contributions from the numerator $\prod_{j<k}
\left[(m_{\rho})_j-(m_{\rho})_k\right]$, which we have not yet
taken into account. In the notation of the previous
section
the limit of this contribution is
\begin{equation*}
(m_{\rho})_{(a,i,\alpha)}-(m_{\rho})_{(b,j,\beta)}
=i\left[(w_{l^{(a)}})_{\alpha}-(w_{l^{(b)}})_{\beta}\right]=i(\alpha-\beta) \ ,
\label{alphabeta}
\end{equation*}
where $1\le \alpha\le l^{(a)}, 1\le \beta\le l^{(b)}$ and $\alpha \ne \beta$.

When the two boxes are in the same block,
this contributes a factor
\begin{align*}
\left(N_5^{(a)}\right)^2 \ln \left[ (l^{(a)}-1)! (l^{(a)}-2)! \ldots 1!
			     \right] \ ,
\end{align*}
where the factor $\left(N_5^{(a)}\right)^2$ accounts for the degeneracy from the
column labels $i$.
This contributes, under the scaling \eqref{scaling},
\begin{align}
-\frac{1}{2} \left[\kappa_a
 (\lambda^{(a)}\gamma_a)^2\right]N^2 \ln N  +\mathcal{O}(N^2)
\ ,
\label{part2}
\end{align}
to the free energy.
When the two boxes are in the different blocks $a, b$
with $l^{(a)}\ge l^{(b)}, \kappa_a\ge \kappa_b$,
the contribution to the free energy is
\begin{align*}
-2\left(N_5^{(a)} N_5^{(b)}\right) \ln \left[
 \left(\frac{l^{(a)}+l^{(b)}}{2}-1\right)!
 \left(\frac{l^{(a)}+l^{(b)}}{2}-2\right)!  \ldots \left(\frac{ l^{(a)}-l^{(b)}}{2}\right)!
			     \right]  \ .
\end{align*}
The expression inside the bracket gives
\begin{equation*}
\ln \left[ G_2\left(\frac{l^{(a)}+l^{(b)}}{2}+1\right)\right]
-\ln
 \left[G_2\left(\frac{l^{(a)}-l^{(b)}}{2}+1\right)\right]
\sim \frac{1}{2}l^{(a)} l^{(b)} \ln l^{(a)} \ .
\end{equation*}
Thus the contribution amounts to
\begin{align}
-2\left(N_5^{(a)} N_5^{(b)}\right)
\frac{1}{2}l^{(a)} l^{(b)} \ln l^{(a)} =-2\frac{1}{2}\left[(\lambda^{(a)}
 \gamma_a \lambda^{(b)} \gamma_b) \kappa_a\right]
 N^2 \ln N \ .
\label{part3}
\end{align}
Collecting all the contributions \eqref{part1}, \eqref{part2} and \eqref{part3}, we have
\beq
F_{\rm CFT}=\frac{1}{2} N^2 \ln N\left[1-\sum_{a=1}^p
(\lambda^{(a)} \gamma_a)^2 \kappa_a -2\sum_{a\ne b,\, l^{(a)}>l^{(b)}}
(\lambda^{(a)} \gamma_a \lambda^{(b)} \gamma_b) \kappa_a \right]
+\scO(N^2) \ .
\eeq
From \eqref{sumN2} we can show that this coincides with \eqref{Fgeneral}.


In all of the examples above, the leading contribution to the partition
function comes from the Barnes $G$-functions. 
The same $N^2 \ln N$ type behavior appears in a number of different
contexts, such as Gaussian matrix models, $c=1$, topological string on
the conifold or more recently in the weak coupling expansion of the
ABJM theory \cite{Drukker:2010nc}.

\section{Gravity Analysis}\label{sec.gravity}

In this section we analyze the type IIB supergravity action $S_{\rm gravity}$ on the gravity side. We explain how to regularize the action and provide an explicit formula for any supergravity solution corresponding to linear and circular quiver geometries. Then we evaluate it in the large $N$ limit for the solutions described by the scaling limit \ref{scaling}. We find perfect agreement with the gauge computations of the previous section.

\subsection{The Gravity Action}\label{sec:action}

The type IIB action in Einstein frame is\footnote{We use the convention $|F_{(a)}|^2 = { 1 \over a!} F_{(a) \, M_1 M_2 .. M_a} F_{(a)}^{\, M_1 M_2 .. M_a}$.}
\begin{align}
\label{IIBaction}
\begin{split}
S_{\rm IIB} &=
-{1 \over 2 \kappa_{10}^2} \int d^{10}x \sqrt{g} \bigg \{
R - {4 \over 2} \p_M \phi \p^M \phi
- \half e^{4 \phi} \p_M \chi \p ^M \chi
- { 1 \over 2} e^{-2 \phi} |H_{(3)}|^2
\\& \qquad
- {1 \over 2} e^{2 \phi} |F_{(3)} + \chi H_{(3)}|^2
- {1  \over 4} | \hat F_{(5)}|^2 \bigg \}
+ {1 \over 4 \kappa _{10}^2} \int d^{10}x \ C_{(4)} \wedge H_{(3)} \wedge F_{(3)} \ ,
\end{split}
\end{align}
where one imposes the self-duality condition $\hat F_{(5)} = * F_{(5)}$ as a supplementary equation.  The coupling $\kappa_{10}$ is related to the string scale $\alpha'$ by $2\kappa_{10}^2 = (2\pi)^7 (\alpha')^4$.

Due to the presence of the self-duality condition, the action \eqref{IIBaction} cannot be directly used to compute the on-shell value of the action.  One way to deal with this is to relax the requirement of Lorentz invariance of the action.  In this case an action principle could be obtained along the lines of \cite{Henneaux:1988gg}.  As suggested in \cite{Giddings:2001yu}, perhaps the easiest way to implement this for the full type IIB supergravity action is to make a T-duality transformation of the type IIA action. The prescription of \cite{Giddings:2001yu} (footnote on p.8) consists in reducing $F_5$ to its ``electric'' part (the part along $AdS_4$ in our case) and doubling its contribution in the supergravity action. This also corresponds to the alternative non-Lorentz invariant action of \cite{Henneaux:1988gg}, which is the true action in a sense, as it need not be supplemented by a self-duality condition.\footnote{Lorentz invariance in that case is not a symmetry of the action but it is preserved by 
the equation of motions.}

A simpler method is to first dimensionally reduce the theory to 4-dimensions.  After carrying out the dimensional reduction, one can then truncate the theory to the 4-dimensional graviton.  To see this is consistent, one may check that the solutions of \cite{D'Hoker:2007xy,D'Hoker:2007xz} can be extended by replacing the AdS$_4$ space with any space which obeys the same Einstein equations.  Thus truncating to the 4-dimensional graviton is a consistent truncation.\footnote{To see this more 
explicitly, first consider the 10-dimensional metric $ds^2 = f_4^2 ds_{(4)}^2 + f_1^2 ds^2_{S_1^2} + f_2^2 ds^2_{S_2^2} + 4 \rho dz d\bar z^2$, where $ds_{(4)}^2$ is an arbitrary 4-dimensional metric.  This is a solution to the type IIB supergravity equations of motion as long as the 4-dimensional Ricci tensor satisfies $R_{(4) \mu\nu} = -3 g_{(4)\mu\nu}$.  One can then write the 10-dimensional Ricci scalar as $R = f_4^{-2} R_{(4)} + ...$, where the omitted terms do not depend on $ds_{(4)}^2$.  The action then takes the form $S =- \frac{1}{2 \kappa_{10}^2} \int d^{10}x (f_4 f_1 f_2)^2 4 \rho^2 \sqrt{g_{(4)}} (R_{(4)} + ...)$, where again the omitted terms do not depend on $ds_{(4)}^2$.  Requiring the variation with respect to $ds_{(4)}^2$ to now reproduce the correct equation of motion yields the effective action \eqref{acteff0}.}

The effective action for this mode is given by
\begin{align}
\label{acteff0}
S_{\rm eff} =- \frac{1}{2 \kappa_{10}^2} {\rm vol}_6
 \int_{\textrm{AdS}_4} \! d^4x \sqrt{g_{(4)}} (R_{(4)} + 6) \ ,
\end{align}
where the cosmological constant has been chosen so that the unit AdS$_4$ space is a solution.  The subscript $(4)$ reminds us that $g_{(4)}$ is the 4-dimensional metric and $R_{(4)}$ is the associated Ricci scalar.  The quantity ${\rm vol}_6$ follows from the initial dimensional reduction and is the volume of the internal space dressed appropriately with the warp factor of AdS$_4$
\begin{align}
{\rm vol}_6 = (4 \pi)^2 \int_{\Sigma} d^2x (f_4 f_1 f_2)^2 4 \rho^2
= 32 (4 \pi)^2 \int_{\Sigma} d^2x (-W) h_1 h_2 \ .
\label{vol6}
\end{align}
The specific solution we are interested in is AdS$_4$ with Ricci scalar $R_{(4)} = -12$.  Thus the on-shell action becomes simply
\begin{align}
\label{acteff}
S_{\rm eff} = -\frac{1}{(2\pi)^7 (\alpha')^4} {\rm vol}_6 \left( \frac{4}{3} \pi^2 \right) (-6) \ ,
\end{align}
where we have used the regularized volume of AdS$_4$, ${\rm vol}_{AdS_4} = (4/3) \pi^2$, whose derivation was presented in section \ref{sec:HR} using the method of holographic renormalization.

\bigskip

We emphasize here that the formula \ref{acteff} with \ref{vol6} provides a remarkably simple exact expression for all the supergravity solutions that describe linear and circular quiver fixed points.

The domain wall solutions (\ref{sec:defectsugra}), corresponding to defect SCFTs, have a non-compact internal volume, so $vol_6$ is infinite and needs further regularization involing (probably) boundary counterterms at $x = \pm \infty$.

\subsection{$T[SU(N)]$}

Let us first consider the gravity dual for $T[SU(N)]$.  The harmonic
functions describing the supergravity solution for $\rho = \hat \rho = (1,1,...,1)$ (see section \ref{sec:linsugra}) are:
\begin{align}
\label{exactTSUN}
\begin{split}
h_1 &= - \frac{\alpha' N}{4} \ln \bigg[ \tanh \bigg( \frac{i \pi}{4} - \frac{z - \delta}{2} \bigg) \bigg] + c.c. \ , \\
h_2 &= - \frac{\alpha' N}{4} \ln \bigg[ \tanh \bigg( \frac{z + \delta}{2} \bigg) \bigg] + c.c. \ ,
\end{split}
\end{align}
with
\beq
\delta = - \half \ln\left[\tan\left(\frac{\pi}{2N}\right)\right] \ ,
\eeq
where we have used a translation to set $\hat \delta = - \delta$.  There is one stack of $N$ D5-branes at the position $z = i \frac{\pi}{2} - \half \ln[\tan(\frac{\pi}{2N})]$ and one stack of $N$ NS5-branes at $z = \half \ln[\tan(\frac{\pi}{2N})]$ with $N$ D3-branes stretched between them ($N$ units of 5-form flux going from one singularity to the other).

\begin{figure}
\centering
\includegraphics
[width=8cm]
{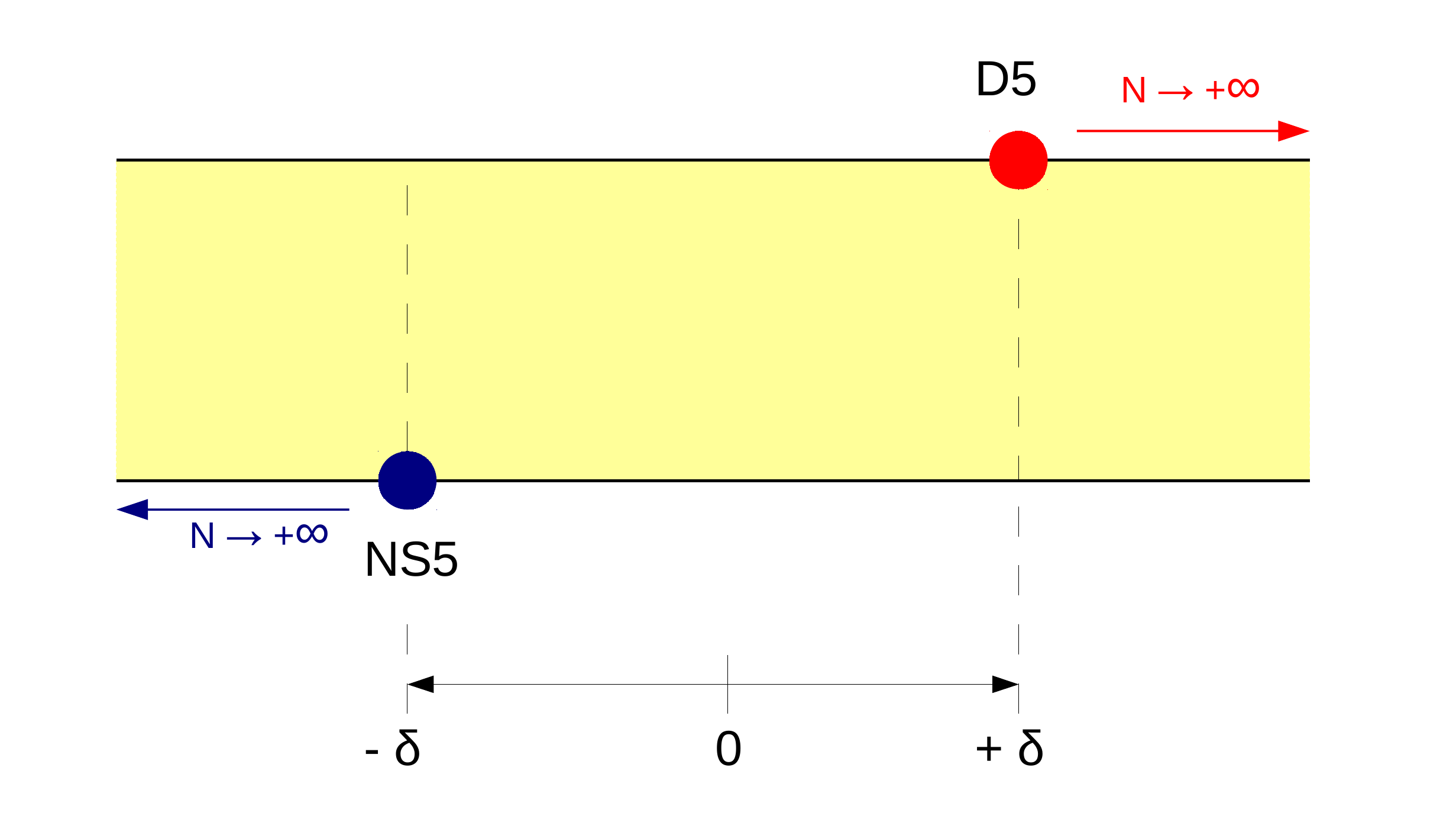}
\caption{ \footnotesize Geometry of the $T[SU(N)]$ dual background represented by the
 strip with two 5-brane singularities at positions $\pm \delta \sim
 \pm \half \ln N$. In the large $N$ limit the stacks go to $\pm$
 infinity.}
\label{strip_tsun}
\end{figure}

We now wish to take the large $N$ limit of this configuration.  It will turn out that locally the Lagrangian density will scale with a factor of $N^2$ at leading order in $N$.  Secondly, as $N$ goes to infinity, the positions $\delta$ of the 5-brane stacks are sent to infinity in opposite directions (see fig. \ref{strip_tsun}).  This leaves a large region of geometry between $-\delta$ and $\delta$ of size $\ln N$, which will reproduce the $\ln N$ behavior of the partition function.  Thus one can understand the leading behavior of the $T[SU(N)]$ partition function as coming from the geometry located between the two stacks of 5-branes.

To make this more explicit and also compute the exact numerical coefficient, we now work out the large $N$ expansion.  First we re-scale the $x$ coordinate so that $z = \delta x + i y$ and then expand the harmonic functions $h_1$ and $h_2$ around large $N$. At leading order we obtain
\begin{align}
\begin{aligned}
h_1 &= \alpha' \sin(y) N\, e^{\delta(x-1)} + ...&  &\textrm{if} \ \ x<1 \ ,\cr
    &= \alpha' \sin(y) N\, e^{\delta(1-x)} + ...&  &\textrm{if} \ \ x>1 \ ,\cr
h_2 &= \alpha' \cos(y) N\, e^{\delta(1+x)} + ...&  &\textrm{if} \ \ x<-1\ , \cr
    &= \alpha' \cos(y) N\, e^{-\delta(1+x)} + ...&  &\textrm{if} \ \ x>-1 \ .
\label{h1h2tsun}
\end{aligned}
\end{align}
From \eqref{h1h2tsun} we find that the only contribution to the action at this order comes from the central region $-1 < x < 1$.  In this region $W$ is given by $W = - \frac{1}{2} e^{-2 \delta} N^2 (\alpha')^2 \sin(2y)$.  Computing the volume of the internal space, \eqref{vol6}, and plugging into the expression for the effective action, \eqref{acteff}, we find
\begin{align}
\label{tsungravpart1}
S_{\rm eff} &=  \frac{4 N^4 \delta e^{-4 \delta}}{\pi^2} + ... \cr
&= \frac{1}{2} N^2 \ln N + {\cal O}(N^2) \ .
\end{align}
This reproduces exactly the leading order behavior of the CFT partition function \eqref{cftparttsun}.  Finally we note that including higher order terms in the expansions of the harmonic functions will give additional contributions of order $N^2$.

\vspace{8mm}
\noindent\underline{ {\bf Validity of the computation}} :

\vspace{5mm}

Since we have explicit D5-brane and NS5-brane singularities in the geometry, one may worry about the validity of our approximation.  We shall argue that the corrections due to the 5-brane singularities are at most of order $N^2$ and do not contribute to the leading $N^2 \ln N$ behavior.  To do so, we first examine the geometry in the central region in the large $N$ limit.  The metric factors are given by
\begin{align}
\begin{aligned}
f_4^2 &= \sqrt{2} \alpha' N e^{-\delta} [(2-\cos(2y))(2+\cos(2y))]^{\frac{1}{4}} \ , \cr
f_1^2 &= 2 \sqrt{2} \alpha' N e^{-\delta} \sin(y)^2  \left[ \frac{2+\cos(2y)}{(2-\cos(2y))^3} \right]^{\frac{1}{4}} \ , \cr
f_2^2 &= 2 \sqrt{2} \alpha' N e^{-\delta} \cos(y)^2  \left[  \frac{2-\cos(2y)}{(2+\cos(2y))^3} \right]^{\frac{1}{4}} \ , \cr
4 \rho^2 &= 2 \sqrt{2} \alpha' N e^{-\delta} [(2-\cos(2y))(2+\cos(2y))]^{\frac{1}{4}} \ ,
\label{tsungeom}
\end{aligned}
\end{align}
while the dilaton and fluxes are given by (see section \ref{s:localsolutions})
\begin{align}
\begin{aligned}
e^\phi  &= e^{-\delta x} \left( \frac{2+ \cos(2y)}{2-\cos(2y)} \right)^{\frac{1}{4}} \ , \cr
b_1 &= 8 \alpha' N e^{-\delta(1+x)} \frac{\sin^3(y)}{2-\cos(2y)} \ , \cr
b_2 &= -8 \alpha' N e^{\delta(x-1)} \frac{\cos^3(y)}{2+\cos(2y)} \ , \cr
j_1 &= - e^{-2 \delta} N^2 (\alpha')^2 (3 x \delta + \cos(2y)) \ .
\end{aligned}
\label{tsunfluxes}
\end{align}
It is interesting to note that this is exactly the limiting geometry of Janus found in \cite{Bachas:2011xa} and described by a domain wall solution without 5-brane stacks, so it is simply a domain wall between to $AdS_5 \times S^5$ regions with the same radius but different values for the dilaton. The limit we obtain here looks like the Janus case with an infinite jump in the coupling.\footnote{The supersymmetric Janus solution is dual to ${\cal N} = 4$ super-Yang-Mills with a jumping coupling at a 3d interface.}  The radius $L$ of the Janus space is related to $N$ by $L^2 = 2 \sqrt{2} \alpha' N e^{-\delta}$.  In the case we consider here, the $\Sigma$ space comes with a natural cutoff at $|x| = \delta$, while for Janus the space is unbounded.

We now consider curvature corrections.  Using the above formulas for the metric factor and dilaton, the string frame Ricci scalar in the central region, $-1 \leq x \leq 1$, is given by
\begin{align}
\label{riccscl}
\alpha' R &= \frac{1}{\pi 2^{1/2}} \bigg( \frac{2N}{\pi} \bigg)^{\frac{x-1}{2}} \frac{419 - 60 \cos(4y)+ \cos(8y)}{(7-\cos(4y))^2 (2+\cos(2y))^{1/2}} \ .
\end{align}
Due to the large $N$ limit, throughout most of the region we have
$\alpha' R \ll 1$.  However, due to the presence of D5-branes, as one approaches $x = 1$, $\alpha' R$ is of order one and one expects higher curvature corrections to play a role.  Since these corrections are localized only in the region near $x=1$, we expect that they do not receive the $\ln N$ enhancement and therefore contribute only at order $N^2$.  A similar argument can be made when one examines the geometry near the D5-branes using \eqref{exactTSUN} before taking the large $N$ limit, showing that the DBI action naively scales like $N^2$.

Due to the presence of N5-branes, the second issue for our calculation is to understand if the string coupling, $g_s$, is small so that string loop corrections can be ignored.  The dilaton in the central region, $-1 < x < 1$, is given by
\begin{align}
g_s = e^{2\phi} &= \bigg( \frac{2N}{\pi} \bigg)^{-x} \, \sqrt{\frac{2+\cos(2y)}{2-\cos(2y)}} \ .
\end{align}
We observe that the dilaton is small in the region $0 < x < 1$ but is big in the region $-1 < x < 0$.  We first focus our attention on the region $0 < x < 1$.  In the large $N$ limit, the string coupling is small except in the neighborhood of $x = 0$, where it is of order one.  Thus we expect string loop corrections to be important, but again we argue that since they are localized near $x=0$, they will give contributions at most of order $N^2$.

For the region $-1 < x < 0$, we find that the string coupling is generically large and one might expect string loop corrections to modify the leading $N^2 \ln N$ behavior.  From this point of view, the exact match between gravity and CFT partition functions is surprising and we do not have a good a priori argument for why string loop corrections do not modify the $N^2 \ln N$ behavior.  One possible explanation can be given in terms of a local S-duality transformation in this region.  To be more precise, we divide the manifold into three regions $-1 < x < -\epsilon$, $-\epsilon < x < \epsilon$ and $\epsilon < x < 1$ with $\epsilon \ll 1$.  In the first region, we make an S-duality transformation, while in the third region the theory is already weakly coupled.  The middle region then has to interpolate between two different S-duality frames and we do not know how to compute the action there.  However, since the $\ln N$ enhancement requires the entire internal space and patching only needs to occur locally in 
the region near $x=0$, one might hope that the middle region does not receive the $\ln N$ enhancement.  Of course this argument is only heuristic and it would be interesting to either make it more precise or determine the exact mechanism for why the loop corrections are suppressed.\\
As for the NS5-branes action, a naive counting of its scaling with $N$, the geometry near the NS5-branes being given by \eqref{exactTSUN} before taking the large $N$ limit, leads again to a $N^2$ behavior. In the end the justification for all these arguments comes a posteriori from the match with the gauge computation.

\subsection{$T^{\rho}_{\hat{\rho}}[SU(N)]$}

We now consider more general partitions which take the form \eqref{rhogeneral}.
In this case, there is a single NS5-brane stack and the charge relations, (\ref{ginvN3}), can be easily inverted to express the phases $\delta_a$ and $\hat \delta$ in terms of the partitions $\rho$ and $\hat \rho$:
\begin{align}
\delta_a - \hat \delta &= - \ln \left[\tan \left( \frac{\pi}{2}
 \frac{l^{(a)}}{\hat N_5} \right)\right]  \ .
\end{align}
To analyze the large $N$ behavior, we proceed analogously to the $T[SU(N)]$ case and consider the limit where $\hat \delta \rightarrow - \infty$ and the $\delta_a \rightarrow \infty$.  In this case, we approximate the harmonic functions by the following expressions
\begin{align}
\begin{aligned}
h_1 &= \alpha' \sin(y) \sum_{a=1}^p N_5^{(a)} e^{x-\delta_a} + ...& &\textrm{if}\ \ x < \delta_1\ ,\cr
    &= \alpha' \sin(y) \sum_{a=i}^p N_5^{(a)} e^{x-\delta_a} + ...& &\textrm{if}\ \ \delta_i < x
 < \delta_{i+1} \ , \cr
h_2 &= \alpha' \cos(y) \hat N_5 e^{-(x - \hat \delta)} +
 ...& &\textrm{if}\ \ x>\hat \delta \ ,
\end{aligned}
\end{align}
while the regions with $x > \delta_p$ and $x<\hat \delta$ will give only subleading contributions.
In this approximation we find that $W = - h_1 h_2$ so that
\begin{align}
\begin{aligned}
-W h_1 h_2  &=  \frac{1}{4} (\alpha')^4 \hat N_5^2 \left( \sum_{a=1}^p
 N_5^{(a)} e^{-(\delta_a - \hat \delta)} \right)^2 \sin^2(2y)&
&\textrm{if}\ \ \hat{\delta}<x < \delta_1 \ ,\cr
&=  \frac{1}{4} (\alpha')^4 \hat N_5^2 \left( \sum_{a=i}^p N_5^{(a)}
 e^{-(\delta_a - \hat \delta)} \right)^2 \sin^2(2y)& &\textrm{if} \ \
 \delta_i < x < \delta_{i+1} \ .
\end{aligned}
\end{align}
Using this in \eqref{vol6} we find
\begin{align}
\textrm{vol}_6&=32(4\pi)^2 \int_{\hat \delta}^{\delta_p}\! dx \int_0^{\frac{\pi}{2}}
\! dy \,(-W h_1 h_2) \nonumber \\
&=
32 \pi^3 (\alpha')^4 \hat{N}_5^2 \sum_{i=1}^p \left( \sum_{a=i}^p N_5^{(a)} e^{-(\delta_a - \hat \delta)} \right)^2 (\delta_{i} - \delta_{i-1})
\end{align}
where we define $\delta_0 \equiv \hat \delta$.  Plugging into \eqref{acteff} and combining all of the numerical factors, we obtain
\begin{align}
S_{\rm eff} &=  \frac{2}{\pi^2} \hat N_5^2 \sum_{i=1}^p \left(
 \sum_{a=i}^p N_5^{(a)} e^{-(\delta_a - \hat \delta)} \right)^2
 (\delta_{i} - \delta_{i-1}) + ... \ .
\end{align}

We now consider the scaling behavior defined by \eqref{scaling}, which introduces separations between the $\delta_a$ which are of order $\ln N$.  In this case each region between a given $\delta_a$ and $\delta_{a+1}$ will contribute to the action at order $N^2 \ln N$.  In terms of this scaling the action becomes
\begin{align}
S_{\rm eff}
&= \frac{1}{2} N^2 \left[  \left( \sum_{a=1}^p \gamma_a \lambda^{(a)} \right)^2 \ln \left( \frac{2}{\pi} \frac{\hat \gamma}{l^{(1)}} N \right)
+ \sum_{i=2}^p \left( \sum_{a=i}^p \gamma_a \lambda^{(a)} \right)^2 \ln \left( \frac{l^{(i-1)}}{l^{(i)}} \right) \right]
+ {\cal O}(N^2) \ ,\nonumber \\
&=  \frac{1}{2} N^2\ln N \left[  (1-\kappa_1)
+ \sum_{i=2}^p \left( \sum_{a=i}^p \gamma_a \lambda^{(a)} \right)^2\left( \kappa_{i-1}-\kappa_i\right)
\right] + {\cal O}(N^2) \ ,
\end{align}
which coincides with \eqref{Fgeneral}.

Computing the $N$ dependence of the metric and dilaton we find here $g_{\mu\nu} \sim N^2$ and $e^{2\phi} \sim N^{\kappa_1}$. This implies that the curvature is small everywhere (except in the close vicinity of the D5-branes) as in the $T[SU(N)]$ case, but now the dilaton is very large on the whole region between the 5-branes. This is different from the $T[SU(N)]$ case. Now our computation is justified, because we can use S-duality to dualize to obtain a solution with small $e^{2\phi}$ everywhere. This is in agreement with our analysis of \S\ref{subsec:parameters}. 

\subsection{Subleading Terms}\label{sec:sublead}

So far we have concentrated on the leading $N^2 \ln N$ contributions to the free energy and it is a natural question to ask about the subleading $N^2$ contributions.  Comparing the CFT and gravity partition functions, we find that the subleading $N^2$ contributions do not match.\footnote{We have checked this numerically for $T[SU(N)]$ using the full expressions for the harmonic functions \eqref{exactTSUN}.}  However, this is not surprising since the gravity solution contains 5-brane singularities around which supergravity approximation breaks down.  Additionally, as already mentioned, there are regions in the bulk of $\Sigma$ where the string coupling becomes large.  It would be interesting to interpret and if possible match the subleading contributions to the CFT partition function with higher curvature corrections, coming from both string and loop corrections, on the gravity side.  For the $T[SU(N)]$ theory, we note that near the D5-brane singularity, the Ricci scalar, \eqref{riccscl} does not depend on 
$N$ and so all powers of 
$R$ will contribute at order $N^2$.  Similarly, one may check that other contractions of the Riemann tensor will also contribute at order $N^2$.  Thus even at order $N^2$, the CFT partition function contains information about all orders of the higher curvature corrections.

\section{Consistency with F-theorem}\label{sec:F-theorem}

The results \ref{Fgeneral} we have found for the free energy of $T^{\rho}_{\hat{\rho}}[SU(N)]$ SCFTs in the large $N$ limit obey the inequality
\begin{align}
F_{T^{\rho}_{\hat{\rho}}[SU(N)]}\le F_{T[SU(N)]} \ .
\label{Ftheorem}
\end{align}
Our results suggest that this inequality is true for any $T^{\rho}_{\hat{\rho}}[SU(N)]$ SCFTs.

This is to be compared with the hypothetic F-theorem (\cite{Jafferis:2011zi,Gulotta:2011si,Amariti:2011xp,Klebanov:2011gs,Closset:2012vg}), whose weaker version stipulates that when two SCFTs are connected by a RG flow the free energy of the UV theory is bigger than the free energy of the IR theory $F_{UV} \geq F_{IR}$. A stronger version was formulated in \cite{Amariti:2011xp} where the free energy was defined along the RG-flow and the proposed F-theorem says that it is monotonicaly decreasing along the flow. A relation of proportionality between the free energy on the 3-sphere and a certain entanglement entropy was shown in \cite{Casini:2011kv} and used in \cite{Casini:2012ei} to argue for the F-theorem.
Non-supersymmetric examples have also been studied \cite{Klebanov:2011gs}. The F-theorem would be the analogue of the $c$-theorem in 2 dimensions (\cite{Zamolodchikov:1986gt}) and $a$-theorem in 4 dimensions (\cite{Komargodski:2011vj}).

\vspace{5mm}

It is actually possible to argue that \ref{Ftheorem} is a manifestation of the F-theorem. To show it we have to find a deformation of the $T[SU(N)]$ SCFT initiating a flow to an arbitrary $T^{\rho}_{\hat{\rho}}[SU(N)]$ SCFT.

\begin{figure}[t]
\centering
\includegraphics[scale=0.5]{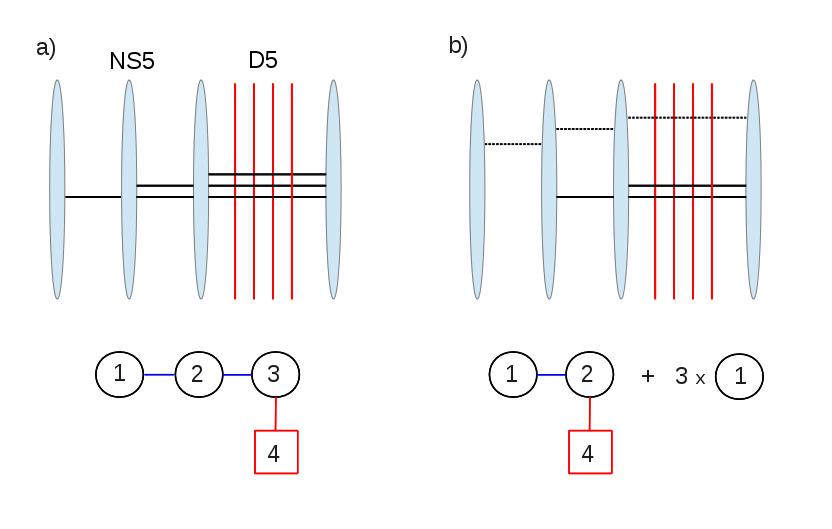}
\vskip -0.5cm
\caption{\footnotesize  a) Brane picture and quiver for $T[SU(4)]$ SCFT. b) After separating 3 D3-segments in transverse space (doted segments) the theory flow to the SCFT described by $\hat\rho' = (2,1,1)$ plus 3 decoupling abelian vector multiplets, as described by the quiver below the figure.}
\label{T(SU(4))}
\end{figure}

The deformations that we need to consider can be understood from the brane picture. Let's consider the brane picture associated to the quiver of $T[SU(N)]$, shown in figure \ref{T(SU(4))} for $N=4$. We can modify the partition $\hat\rho = (1,1,...,1)$ to any partition $\hat\rho'$ by separating D3-segments connecting NS5-branes in transverse space (along $x^4,x^5,x^6$) (see figure \ref{T(SU(4))}) and let the theory flow to the IR. The IR SCFT resulting from this flow should be composed of decoupled pieces : the SCFT $T^{(1,1,..,1)}_{\hat{\rho}'}[SU(N)]$ and a number of decoupled $U(1)$ vector multiplets corresponding to the separated D3-segments.

In gauge theory the brane manipulation that consists in moving a D3-segment in transverse space corresponds to moving on the Coulomb branch of vacua. So the RG-flow that we are looking for is initiated by giving vevs to the scalars in the vector multiplets.

Let's consider the minimal example of separating one D3-segment in a $T^{\rho}_{\hat{\rho}}[SU(N)]$ SCFT with 
\begin{align}
\rho &= ( l_1 ,  l_2, ... ,  l_{ k}) \no\\
\hat\rho &= (\hat l_1 , \hat l_2, ... , \hat l_{\hat k}) .
\end{align}
 Separating a D3-segment from the $N_J$ segments of the $J$-th node $U(N_J)$ amounts to giving a vev to the scalars in the vector multiplet of this node, so that the gauge symmetry is broken to $U(N_J -1) \times U(1)$. If the (adjoint) scalars are represented by $N_J \times N_J$ matrices, this means giving a vev to a corner element of the matrices. When flowing to the infrared, the $U(N_J)$ gauge symmetry is higgsed to $U(N_J -1) \times U(1)$. The $U(1)$ vector multiplet decouples from the matter fields (as it was coupled only through massive modes), so we end up with the SCFT $T^{(1)}_{\hat{\rho}'}[SU(N)]$ which is the same as $T^{\rho}_{\hat{\rho}}[SU(N)]$ except that the node $U(N_J)$ is replaced by $U(N_J -1)$, plus a free $U(1)$ vector multiplet. \\
The new partition $\hat\rho'$ is given by
\beq
\hat\rho' = \lp  \hat l_1 \, ,  \, ...  \, ,  \, \hat l_{J-1} \, ,  \, \hat l_J +1 \, ,  \, \hat l_{J+1} -1 \, ,  \, \hat l_{J+2} \, ,  \, ... \,  ,  \, \hat l_{\hat k} \rp \ .
\eeq 

The new partition $\hat\rho'$ may not be ordered, in which case the IR SCFT is not irreducible and is believed to flow to the IR irreducible fixed point of the theory with ordered partition, plus decoupling hypermultiplets, as explained in section \ref{sec:quiv}. The hypermultiplets in question are actually twisted hypermultiplets (transformation properties under $SU(2)_L \times SU(2)_R$ interchanged) that are exactly the $U(1)$ vector multiplets that we just saw. Actually the reordering of the NS5-branes, which is the reordering of the partition $\rho'$, can be effectively done by separating more D3-segments from the brane configuration (moving the Coulomb branch of other nodes) and flowing to the infrared. In the process we obtain again decoupled $U(1)$ vector multiplets, but in 3-dimension abelian vector multiplets can be dualized to twisted hypermultiplets (see \S\ref{sec:3dN4}), through the relation $F_{\mu\nu} = \epsilon_{\mu\nu\sigma} \p^{\sigma} \gamma $, where $\gamma$ is a real scalar named {\it 
dual photon}. The decoupling of abelian vector multiplets is thus the decoupling of the twisted hypermultiplets of \cite{Gaiotto:2008ak}.

\vspace{6mm}
The dual minimal example consists in moving a D3-segment that is stretched between two D5-branes. For instance let's consider the case of $M_J$ D5-branes intersecting the $N_J$ D3-segments of the $J$-th node and we move a D3-segment stretched between two D5-branes, as shown in figure \ref{Higgsing} (we assume $M_J \geq 2$).
From our understanding of mirror symmetry, which exchanges the Coulomb branch and the Higgs branch of the theories, we guess that the correct flow will be initiated this time by moving on the Higgs branch of the theory. The brane situation that we have described corresponds to having a $U(N_J)$ gauge node with $M_J$ fundamental hypermultiplets. Let's consider the scalars in two hypermultiplets (2 complex in each). They can be combined in a couple of complex matrices $A,\tilde A$ of size $N_J \times 2$ and $2 \times N_J$ respectively, where $A$, resp. $\tilde A$, contains the scalars transforming in the fundamental, resp. anti-fundamental, representation of $U(N_J)$. Setting to zero the vevs of the other fundamental hypermultiplets and possible bifundamental hypermultiplets, the matrices $A$ and $\tilde A$ have to satisfy the constraints $A\tilde A = 0$ (critical point of the superpotential) and $AA^{\dagger} - \tilde A^{\dagger}\tilde A = 0$ (D-term). Preserving these constraints we may move on the Higgs 
branch by turning on constant vevs  $a$ and $-a$ for the first row of $A$ and the first column of $\tilde A$ : 
\begin{align}
A = 
\begin{bmatrix}
a & a \\
0 & 0 \\
 .. & .. \\
0 & 0
\end{bmatrix}
\quad , \quad
\tilde A =
\begin{bmatrix}
a & 0 & ... & 0 \\
-a & 0 & ... & 0
\end{bmatrix}
\end{align}
The $U(N_J)$ gauge transformations act by left multiplication on $A$ and right multiplication on $\tilde A$, so
at this point on the Higgs branch the gauge group is broken to $U(N_J-1)$. From the massless degrees of freedom of the $M_J$ hypermultiplets, some are eaten in the process and some decouple from the quiver theory (in the infrared). In total there are   $M_J- 1$ decoupling hypermultiplets.  
In the infrared the gauge group is thus higgsed to $U(N_J-1)$, with $M_J -2$ hypermultiplets in the fundamental representation. The bifundamental hypermultiplet in $U(N_{J-1})\times U(N_J)$, resp. $U(N_{J})\times U(N_{J+1})$, splits into a bifundamental of $U(N_{J-1})\times U(N_J -1)$, resp. $U(N_{J}-1)\times U(N_{J+1})$, plus a fundamental hypermultiplet in $U(N_{J-1})$, resp. $U(N_{J+1})$.\\
This is in complete agreement with the brane picture : moving a D3-segment between to D5-branes to infinity and displacing the D5-branes so that the net number of D3-brane ending on them is zero (one D5 moves to the node on the left, while the other D5 moves to the node on the right), we get a brane picture corresponding to the quiver theory we just described (see figure \ref{Higgsing}). There are $M_J-1$ decoupling hypermultiplets corresponding to the D3-segment (that we moved away) splitting in $M_J-1$ D3-segments inbetween the $M_J$ D5-branes.

\begin{figure}[h]
\centering
\includegraphics[scale=0.5]{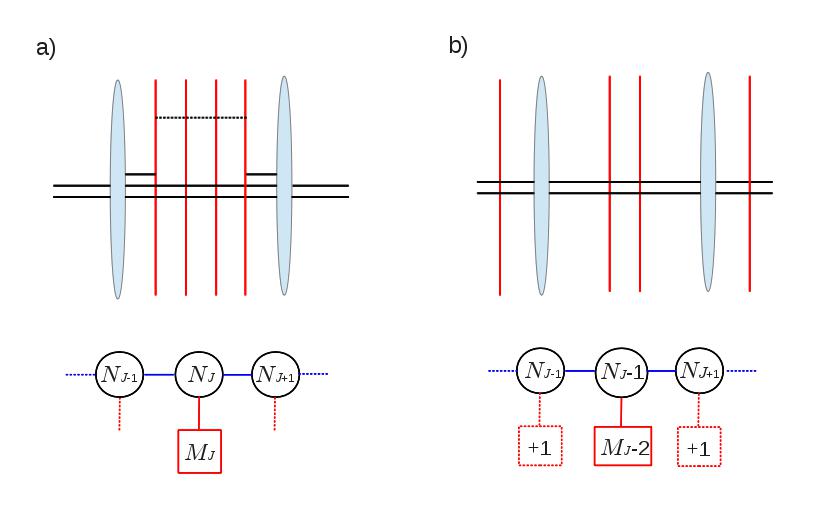}
\caption{\footnotesize  a) Brane picture corresponding to moving a D3-segment stretched bewteen two D5-branes. b) Final brane configuration after moving the D5-branes (and D3 segment at infinity) and corresponding quiver gauge theory.}
\label{Higgsing}
\end{figure}

The partition $\rho'$ obtained for the infrared linear quiver is 
\begin{align}
 \rho' = \big( ... \ , \  \overbrace{ l^{(J)}+1 \ , \ l^{(J)} \ , \ ... \ , \ l^{(J)} \ , l^{(J)} -1 }^{N^{(J)}_5} \ , \ ... \big) \ ,
\end{align}
which corresponds to the minimal case for moving on the Higgs branch.

\vspace{8mm}

This shows that, by moving on the Coulomb branch and Higgs branch, we can flow from $T[SU(N)]$ to any $T^{\rho}_{\hat{\rho}}[SU(N)]$ SCFT plus decoupling hypermultiplets. The F-theorem then predicts
\begin{align}
F_{T[SU(N)]} \ \geq \  F_{T^{\rho}_{\hat{\rho}}[SU(N)]} \, + \,  F_{hypers} \ > \ F_{T^{\rho}_{\hat{\rho}}[SU(N)]} \ , \no
\end{align}
where the second inequality follows from the fact that the free energy of a free hypermultiplet (computed via localization techniques on the 3-sphere \cite{Kapustin:2009kz}) is $\half$.

This prediction is confirmed by our results \ref{Ftheorem}.

%

\vspace{8mm}

One may wonder if it is possible to derive inequalities comparing the free energy of to arbitrary SCFTs $T^{\rho_1}_{\hat{\rho}_1}[SU(N)]$ and $T^{\rho_2}_{\hat{\rho}_2}[SU(N)]$. Reasoning along the same lines (D3-segments decoupling) one can understand that we cannot derive inequalities for any two such SCFTs but only in some cases :
\begin{align}
\label{Fineq}
\left\{
\begin{array}{c}
\rho_1 \geq \rho_2 \\
\hat{\rho}_1 \geq \hat{\rho}_2 
\end{array}
\right.
  \quad  \Longrightarrow  \quad   
  F_{T^{\rho_1}_{\hat{\rho}_1}[SU(N)]}  \leq F_{T^{\rho_2}_{\hat{\rho}_2}[SU(N)]}  \ .
\end{align}
An easy way to derive this inequality is to consider the brane configuration corresponding to a quiver when the 5-branes are separated (NS5-branes on the left, D5-branes on the right, \ref{separate}). In this brane configuration the partitions of $N$ are directly visible with the D3-branes ending on the 5-branes. The inequality $\hat{\rho}_1 \geq \hat{\rho}_2$ means that we can transform $\hat\rho_2$ into $\hat\rho_1$ by decoupling D3-segments stretched between NS5-branes from the brane configuration of 
$T^{\rho_2}_{\hat{\rho}_2}[SU(N)]$. Similarly, if $\rho_1 \geq \rho_2$ we can transform $\rho_2$ into $\rho_1$ by decoupling D3-segments stretched beween D5-branes. Decoupling the D3-segments as indicated, one is left with the brane configuration of $T^{\rho_1}_{\hat{\rho}_1}[SU(N)]$  leading to the inequality \ref{Fineq}.

The result that was found for the free energy \ref{Fgeneral} confirms the prediction \ref{Fineq} from the F-theorem in simple cases : let's consider two theories of the form \ref{rhogeneral}, \ref{scaling}, with $\rho_1 = \rho_2$, identical parameters $\gamma_a, \lambda^{(a)}$ but different scalings $\kappa_a^{(1)}$ $\kappa_a^{(2)}$. Imposing $\kappa_a^{(1)} \geq \kappa_a^{(2)}$ for all $a$ ensures that $\hat\rho_1 \geq \hat\rho_2$. In this case we have $ F_{T^{\rho_1}_{\hat{\rho}_1}[SU(N)]}  \leq F_{T^{\rho_2}_{\hat{\rho}_2}[SU(N)]}$ because the coefficient in front of each $\kappa_a$ in \ref{Fgeneral} is negative.

\vspace{8mm}

It would be interesting to see if \ref{Fineq} can be checked on the supergravity side using the general formula \ref{acteff} at finite $N$. This is not really expected to work since the supergravity regime is at large $N$, but it might be that the regularized on-shell supergravity action is a protected quantity, as our computations tend to suggest.

\chapter{ Solutions with $(p,q)$-5branes and Chern-Simons SCFTs}
\addcontentsline{lot}{chapter}{ Solutions with $(p,q)$-5branes and Chern-Simons SCFTs}
\label{chap:SL2R}

%

Classical type-IIB supergravity has a continuous global $SL(2,\mathbb{R})$ symmetry
\cite{Schwarz:1983wa} which transforms the axion-dilaton field,  $S = \chi + i e^{-2 \phi}$,  and the NS-NS and R-R  three-form  field strengths  as follows:
\begin{align}
\label{sl2rtrans}
 S^\prime =  {a S +  b\over  c S + d}\ , \qquad
  \left(   \begin{array}{c}
\, \, H_{(3)}^\prime  \\
F_{(3)}^\prime
 \end{array}
  \right) =
   \left(   \begin{array}{cc}
 d & -c \\ - b & a
  \end{array}
  \right)
  \left(   \begin{array}{c}
\, H_{(3)}  \\
F_{(3)}
 \end{array}
  \right) \ ,
 \end{align}
where $a,b,c,d$ are real numbers with $ad-bc=1$. The transformations leave invariant the
Einstein-frame metric, and the gauge-invariant five-form field strength.
\smallskip

 As is well known, only the integer subgroup $SL(2,\mathbb{Z})$ is a  symmetry of the full string theory \cite{Hull:1994ys}, whereas continuous transformations can be
 used to generate  inequivalent  solutions.
    The authors of  \cite{D'Hoker:2007xy} have indeed used such   $SL(2,\mathbb{R})$ transformations
     to bring the general  solution  of the Killing-spinor equations
  to the local  form given in  \S\ref{s:localsolutions}.  Conversely, acting with  the transformations \eqref{sl2rtrans}  generates
  new solutions from the ones of  section 3,  with singularities that  correspond to general $(p,q)$  five-branes.\footnote{The symbol $p$, which usually
   indicates the NS5-brane charge of a $(p,q)$ five-brane, was   also used for the
   number of  five-brane singularities in the upper boundary of $\Sigma$. We hope the context will make it clear in which sense this symbol is being used.
   The same comment applies to the lower-case Latin letters  which   label the five-brane stacks;   following standard notation we also use them for the
    elements of the $SL(2,\mathbb{R})$ matrix. }
 We will now discuss  briefly  these
  new solutions. In this section we focus on the solutions on the annulus, however the discussion is directly applicable to the solutions on the strip (linear quiver solutions or defect solutions).



\section{Solutions with $(p,q)$ five-branes}

   The solutions given by the harmonic functions \eqref{harm1} or \eqref{hmany}
      have  singularities on the upper boundary of the infinite strip or the annulus that correspond to D5-branes,  and
      singularities   on the lower boundary that correspond to NS5-branes. The  charges
    are,
    respectively,  $\gamma^{(e)}$ and  $ \hat\gamma^{(f)}$ for the  stacks labeled by $e$ and $f$.   Since the metric
    is invariant, the $ SL(2, \mathbb{R})$ transformations
    do not change the  positions and the  total number of five-brane stacks. It transforms, however, their charges as follows
       \beq
  \gamma^{(e)} (0, 1) \to   \gamma^{(e)} (-c, a)\qquad {\rm and}\qquad
   \hat\gamma^{(f)} (1,0) \to   \hat\gamma^{(f)}(d, -b )\ ,
   \eeq
where the NS5-brane and D5-brane charges are arranged as usual in a doublet.
    Let us write $(-c, a) = w (p, q)$ and $(d, -b) = \hat w (\hat p, \hat q)$, where $p, q$ and $\hat p, \hat q$ are pairs of relatively-prime
    integers. Charge quantization requires that
   \bea\label{D5NS5prime}
   N_5^{(e)\, \prime} =  w \gamma^{(e)}\quad {\rm  and } \quad    \hat N_5^{(f)\, \prime}  = \hat w \hat\gamma^{(f)}
  \eea
     be integer for all
    $e$ and $f$. Since the $\gamma$'s and $\hat\gamma$'s  are arbitrary parameters, this can always be arranged to get
    any desired number of five-branes in each  stack. The only conditions are that all five-branes on the upper boundary are of  the
    same kind, including the sign, that the same is true for all five-branes on the lower boundary, and that furthermore these two
    kinds are different, $p\hat q -  q\hat p \not=  0$. This last constraint follows from  the fact that the $SL(2, \mathbb{R})$ matrix has determinant one.

 \smallskip

 It should be stressed that the  $SL(2, \mathbb{R})$ transformations take us, in general, outside the ansatz of \S\ref{s:localsolutions};
 they generate in particular a non-vanishing R-R axion field.
 The only exception is  S-duality
  ($S \to -1/S$) which interchanges the harmonic functions, and acts as mirror symmetry on the holographically-dual  SCFT.

 \smallskip

   Consider next the D3-brane charges. These are not affected by  $SL(2, \mathbb{R})$   transformations, provided
   one transforms the gauge choice   covariantly. More explicitly, let us consider the D3-brane charge of the $(p,q)$ singularities in the upper
   boundary. The 2-form that has no component on $S_2^2$ [and is therefore well defined on a patch containing the whole upper boundary where this 2-sphere shrinks]  is $ B_{(2)}= a  B_{(2)}^\prime + c C_{(2)}^\prime$.
   The D3-brane charge of a $(p,q)$ five-brane stack  is given therefore by the integral of the following closed five-form
   \bea
   N_3^{(e)\, \prime} = {1\over (4\pi\alpha^\prime)^2}  \int_{{\cal C}_e^5} \left[ F_{(5)} -  (a  B_{(2)}^\prime + c C_{(2)}^\prime)\wedge (b H_{(3)}^\prime + d F_{(3)}^\prime)  \right]\ ,
   \eea
   with the gauge choice $a  B_{(2)}^\prime + c C_{(2)}^\prime=0$ in the lower-boundary segment
   $[\hat \delta_1, 2t]$.  This is identical to the integral   in the non-transformed solution,  so that , in the case of the annulus solutions for instance,
   \bea\label{D3prime}
   N_3^{(e)\, \prime} =     \gamma_e
 \sum_{f=1}^{\hat p}  \hat \gamma_f  \, \left(
- \frac{i}{2\pi}  \ln \bigg[ \frac{\vartheta_{1}\left(\nu_{ef} \vert \qth\right)}{\vartheta_{1}
\left(\bar \nu_{ef}\vert  \qth  \right)} \frac{\vartheta_{2}\left( \bar \nu_{ef} \vert \qth \right)}{\vartheta_{2}\left(\nu_{ef}\vert \qth \right)} \bigg]
-   \frac{4}{\pi\alpha^\prime} \varphi_2 \right) \ ,
   \eea
which is the same result as   \eqref{D3charge2}.   The quantization of this charge puts the same constraints on the continuous parameters
 as in the untransformed solution. This is not however  the case  for the quantization of individual linking numbers,  since the number $w \gamma^{(e)}$ of
 $(p,q)$ five-branes    depends,  via  $w$,  on the $SL(2, \mathbb{R})$   transformation.
     \smallskip

\begin{figure}[h]
\centering
\includegraphics[scale=0.3]{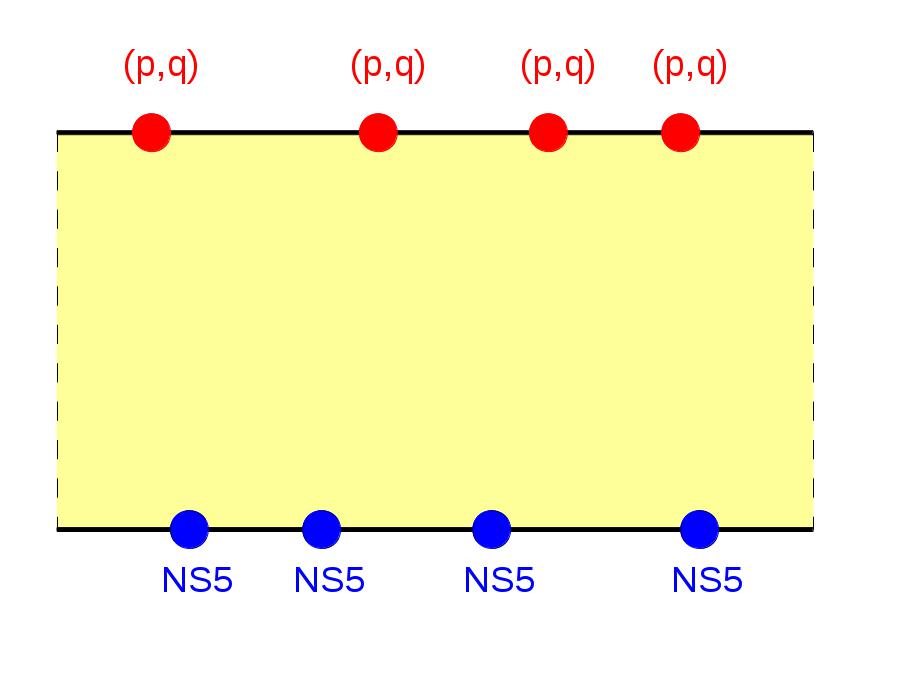}
\vskip -1cm
\caption{\footnotesize Canonical solution on the annulus with NS5-branes on the lower boundary and $(p,q)$-5branes on the upper boundary, $0 \le p < |q|$.
  }
\label{annulusSL2R}
\end{figure}

 Among all the solutions discussed here, those  related by $SL(2,\mathbb{Z})$ transformations are physically equivalent \cite{Hull:1994ys}.
 To characterize inequivalent solutions, we may  perform a $SL(2,\mathbb{Z})$ transformation that maps $(\hat p, \hat q)$ to $(1,0)$, so that
 the singularities on the lower boundary correspond to pure NS5-branes. Using then the
  shift symmetry $(p, q) \to  (p + ql, q)$, which leaves invariant the NS5 branes,  we can   bring the  second type
of five-branes to a canonical  form $(p, q)$ with  $0\leq p < \vert q\vert$ (see figure \ref{annulusSL2R}).

 The $SL(2, \mathbb{R})$  transformation from the ansatz of  \S\ref{s:localsolutions}
 to the above  canonical form of the general  solution is effected by the following matrix
\bea
  \left(   \begin{array}{cc}
 \hat w  & -wp  \\ 0  & wq
  \end{array}
  \right)\quad {\rm with} \quad w\hat w q = 1\ .
\eea
 Multiplying \eqref{D3prime} with $w\hat w q$,    using   \eqref{D5NS5prime} and the infinite-product expressions for
 the $\vartheta$-functions gives 
\footnote{The charges are given for the solutions on the annulus, but the discussion is the same in the case of strip solutions. One can always recover the case of the strip by taking the $t \rightarrow \infty$ limit in the formulas.}
 \bea\label{emanateD5p}
  { N^{(a)\, \prime}_{3}  }  =
   q N^{(a)\, \prime}_{5} \sum_{b=1}^{\hat p} \hat N^{(b)\, \prime}_{5}  \Big[  \sum_{n=0}^{+\infty} f(\hat \delta_b - \delta_a -2nt ) - \sum_{n=1}^{+\infty} f(-\hat \delta_b + \delta_a -2 nt)
 \Big] \ ,
 \eea
and likewise
\bea\label{emanateNS5p}
   {\hat N^{(b)\, \prime}_{3} } =  q \hat N^{(b)\, \prime}_{5} \sum_{a=1}^{p} N^{(a)\, \prime}_{5} \Big[   \sum_{n=1}^{+\infty} f(-\hat \delta_b + \delta_a - 2 n t )
-\sum_{n=0}^{+\infty} f(\hat \delta_b - \delta_a - 2 n t )
 \Big]  \ .
\eea
A similar expression can be written for the winding charge $L^\prime$. Integrality of the linking numbers,  $ l'_a = N^{(a)\, \prime}_{3} /
   N^{(a)\, \prime}_{5} $ and $ \hat l'_b = \hat N^{(b)\, \prime}_{3} /  \hat N^{(b)\, \prime}_{5}$,  constraints the positions of the singularities on
   the boundary of $\Sigma$ and the modulus $t$ (in the case of the annulus). 
When  $q\not= 1$ there are more inequivalent allowed solutions than in the case of pure D5-branes and NS5-branes, corresponding to different choices of $p$, with $0\le p < |q|$.

  \smallskip

   The
   charges \eqref{emanateD5p} and \eqref{emanateNS5p} obey the sum rule \eqref{sumrule2}, and they thus still define two partitions $\rho$ and $\hat\rho$
   of some integer $N$. \\
To summarize, the inequivalent solutions are classified by the two partitions $\rho, \hat\rho$, the winding charge $L$ and the type of 5-branes on the upper boundary ($(p,q)$-5branes with $0\le p < |q|$).

Furthermore, these partitions still satisfy the basic inequalities $L + \rho^T > \hat\rho$ \eqref{fixedpointcircA}. In general, we have no clear argument for why these
   conditions should be obeyed on the gauge-theory side. Indeed, for arbitrary $(p,q)$ there is no known
  Lagrangian description of the  field theory (we refer the reader to section 8 of \cite{Gaiotto:2008ak} for more details).
   Such a description only exists for the configurations involving $(1,k)$ 5-branes  \cite{Gaiotto:2008sa,Gaiotto:2008sd,Hosomichi:2008jd} :
  the  $U(N)$ gauge theory  living on a stack of $N$ D3-branes has level $k$ or $-k$ Chern-Simons terms
  depending on whether the  D3-branes end on the $(1,k)$ five-brane from  the left or the right.

\vspace{8mm}

\subsection{IIB dual of ABJM gauge theory}

\vspace{5mm}

The famoust example of such Chern-Simons SCFT is the ABJM theory presented at the end of section \ref{sec:corresp}.
The gauge theory is of circular type, with Chern-Simons gauge group $U(N)_{M}\times U(N)_{-M}$ and two bifundamental hypermultiplets. The indices $M$ and $-M$ refer to the Chern-Simons levels of the two gauge factors. The corresponding brane configuration was pictured in figure \ref{ABJM} for the more general ABJ theory (nodes of different ranks).

Let's describe the IIB dual solutions of ABJM as an example. \\
The data of the ABJM theory are the type of 5-brane on the upper boundary, which is the $(1,M)$ 5-brane, the number $N$ of winding D3-brane charge
\footnote{Here we switch from the notation $L$ to $N$ for the winding 5-form flux, to be closer to the notations in the literature.} and the two partitions (with a single entry) $\rho = \hat\rho = [0]$, reflecting the fact that in the brane confiration no D3-branes end on the 5-branes. Here we took a liberty with our description of circular quivers, which are supposed to have positive linking numbers. The correct partitions are obtained after a Hanny-Witten move (winding of one 5-brane around the circle), which creates $M$ D3-branes stretched between the NS5 and the $(1,M)$ 5-brane, and are given by $\rho = \hat\rho = [M]$ 
\footnote{The carefull reader certainly noticed that for the quivers involving only NS5 and D5-branes, that we have described in great details, the linking numbers were also bounded from above by the total number of 5-branes of the other type. This was a manifestation of the $s$-rule \cite{Hanany:1996ie}. For Chern-Simons quivers the upper bound is changed to $|M|$ times the number of opposite 5-branes, as can be deduced from the transformation of charges under $SL(2,\bR)$, providing a larger $s$-rule.}
. Our non-standard choice of partitions corresponds to the same quiver theory and will be described by the same supergravity solution, but with a different gauge fixing (see section \ref{s:match} for details). 

To find the supergravity solution we need to understand the $SL(2,\bR)$-related solution with vanishing axion field, then give the associated harmonic functions $h_1, h_2$ and implement the $SL(2,\bR)$-tranformation.

The $SL(2,\bR)$-related solution with vanishing axion field is the solution with $N$ winding D3-branes, one NS5-brane and one stack of $M$ D5-branes. The non-standard partitions are $\hat\rho = [0]$ and $\rho = [0,0,...,0]$ ($M$ entries).
\footnote{The standard partitions would be $\hat\rho = [M]$ and $\rho=[1,1,...,1]$.} \\
The parameters of the IIB solution are the period $t$ of the annulus and the distance $\Delta = \hat\delta - \delta$ on the $x$ axis between the D5 and NS5-branes. They are determined by the equations (\ref{Lcharge},\ref{emanateNS5}) for the charges
\begin{align}
 N &=   M  \sum_{n=1}^{ \infty} n  \Big[ f (\Delta -2n t )
 + f (-\Delta -2nt ) \Big]\no\\
 0 &=   M \Big[   \sum_{n=1}^{+\infty} f(-\Delta - 2 n t )
-\sum_{n=0}^{+\infty} f(\Delta - 2 n t ) \ ,
 \Big]
\end{align}
with $f(x) = (2/\pi) \arctan(e^x)$. The second equation express the D3-charge of the NS5-brane or equivalently the D3-charge of the D5-branes. It is solved for the $\Delta = -t$ and the parameter $t$ is fixed by the first equation, which (unfortunately) is not easily inverted.
\begin{align}
 \Delta = -t  \quad , \quad N = M \, \sum_{n=0}^{\infty} (2n+1) \, f \big[-(2n+1)t \big] \ .
\end{align}

\noindent Setting $\hat\delta=0$, the harmonic functions describing the solution are (\ref{hmany})
\begin{align}
\label{solABJM}
h_1 &= - M \ \ln \bigg[ \frac{\vartheta_{1}\left(\nu \vert \qth  \right)}{\vartheta_{2}\left(\nu  \vert \qth  \right)} \bigg]
 + c.c.    \  , \  \qquad {\rm with}\ \ \
  i\, \nu  = - \frac{z - t}{2 \pi } + \frac{i}{4} \ , \ \tau = it/\pi \ , \\
h_2 &= \ - \ \ln \bigg[ \frac{\vartheta_{1}\left(  \hat \nu \vert \qth \right)}{\vartheta_{2}\left(  \hat \nu \vert \qth \right)} \bigg]
+ c.c.  \ ,
\ \  \qquad {\rm with}\ \ \    i\,  \hat \nu =   \frac{z}{2 \pi } \ .
\end{align}
If we parametrize the $x$-axis of the annulus by $-t \le x \le t$, the free constants in the holomorphic functions $\scA_1, \scA_2$ \ref{Amany} are fixed so that $B_2=0$ on the lower boundary $(0, t]$ and $C_2=0$ on the upper boundary $[-t,t)$.

The supergravity solution given by \ref{solABJM} describe the metric and 5-form of the IIB dual of ABJM.
The dillaton  and 3-forms are obtained by acting with the $SL(2,\bR)$ transformation
\begin{align}
\label{SL2RABJM}
   \left(   \begin{array}{cc}
  1  &    M^{-1}  \\  0   &  1
  \end{array}
  \right) \ .
\end{align}
In terms of the axion-dilaton $S = i e^{-2 \phi}$ and the 3-forms $H_3, F_3$ of the (vanishing axion) solution \ref{solABJM}, we get, using \ref{sl2rtrans}, 
\begin{align}
\label{solABJM2}
 S' =  \frac{M}{1+ M^2 \, e^{4\phi}} \big( -1 + i M \, e^{2\phi} \big)  \ , \qquad  H'_3 = H_3 + \frac{1}{M} \, F_3   \ , \  F'_3 =  F_3 \ .
\end{align}

It is nessecary to pause here to comment about the solution we found. The metric is given in terms of the harmonic functions \ref{solABJM} and it is way more complicated than the ABJM type IIA or M-theory metrics \ref{ABJMIIA}, \ref{ABJMMth}. When going from type IIA to type IIB using T-duality (see appendix \ref{Tduality}) one end up with the smeared solution that we got in the large $N$ ($L$) limit in \S\ref{secLargeL}, which is indeed much more simple.\\
This reveals that the ususal rules for T-duality are not enough to get the full type IIB solution.
Even the simpler problem of T-dualizing pure NS5-branes  is notoriously subtle
\cite{Gregory:1997te, Tong:2002rq, Harvey:2005ab,Okuyama:2005gx}. As explained in these references, 
the contributions of  world-sheet instantons are 
 responsible for the localization of the NS5-branes on the type-IIB side \cite{Tong:2002rq},  and
for creating the dual throats in winding space on the type-IIA side \cite{Harvey:2005ab}.
The correct T-dual backgrounds have localized singularities, breaking the invariance under translation along the annulus, and encodes the full data $(\rho,\hat\rho,L)$ describing the quiver theory. It would be very interesting to understand this T-duality precisely and to be able to relate the corrections to the metric (compared to the smeared case) to dual quantities in type IIA string theory and M-theory.

\vspace{8mm}

\noindent\underline{ {\bf Comment about supersymmetry and brane configurations }}

\vspace{5mm}

In the above description we considered brane configurations with $(1,k)$-5branes orthogonal to NS5-branes. Such brane configurations preserve $\N=4$ supersymmetry in 3-dimensions. However it is known that Yang-Mills Chern-Simons gauge theories in 3-dimensions have only up to $\N=3$ supersymmetry, and the brane configuration preserving $\N =3$ has the $(1,k)$-5brane and the NS5-brane at an angle ($\neq \pi/2$) \cite{Kitao:1998mf}. The same issue is raised in \cite{Aharony:2008ug}, where it is argued that the to brane configurations (at angle or orthogonal) have the same low energy SCFT living on the D3-branes, the ABJM theory in that case. Moreover in the infrared limit the Yang-Mills coupling diverges and the Yang-Mills kinetic term can be dropped. The effective low energy Lagrangian contains only the Chern-Simons kinetic term for the gauge field and has at least $\N=4$ supersymmetry (although naively only $\N=3$).

\vspace{1cm}


\section{Orbifold equivalences and free energies}

    An interesting  corollary of the holographic dualities that we have presented  in this work  is the orbifold equivalence
of different    $\N=4$   superconformal      gauge theories in three dimensions.  Orbifold  equivalences translate the fact
that quantities which  are sensitive only to the untwisted sector,  are not affected by an orbifold operation \cite{Kachru:1998ys,
 Lawrence:1998ja, Bershadsky:1998mb}. Such quantities usually exist in the classical limit of string theory,  and in the large-$N_c$ (planar)
limit of gauge theories.\footnote{For a discussion  of when the equivalence is  exact  see \cite{Armoni:2004ub, Kovtun:2004bz}.}
An example of orbifold equivalence  for the ABJM  theory was analyzed
 recently in \cite{Hanada:2011zx, Hanada:2011yz}. Here we will   present some more examples relating  $\N=4$ circular-quiver theories. The same kind of orbifold equivalence apply to the linear quiver theories and the defect theories.

\smallskip

  The   theories that we will discuss are related by  $SL(2,\mathbb{R})$ transformations with rational entries, i.e. by elements of
  $SL(2,\mathbb{Q})$.  Two  theories related in this way are clearly  equivalent in the limit where the supergravity approximation is valid, since
$SL(2,\mathbb{R})$ is a symmetry of  type-IIB supergravity. A similar rational extension of the perturbative T-duality group
$O(d,d,\mathbb{Z})$ has been discussed recently in  \cite{Bachas:2012bj}. As explained in this reference,
$O(d,d,\mathbb{Q})$ transformations can be seen as
 orbifold operations\footnote{If $x=x+2\pi$ parametrizes the orbits of a Killing isometry, then the orbifold identification $x \equiv x + 2\pi \kappa$ for rational $\kappa$
 changes the radius of the Killing orbits,  and can thus be viewed as a $O(1,1,\mathbb{Q})$ transformation. Rationality ensures that the orbifold
 group is of finite order. These observations  generalize  to
 $O(d,d,\mathbb{Q})$.}
  which lead to equivalences that are valid at any order  in
the $\alpha^\prime$ expansion. One may likewise  view the $SL(2,\mathbb{Q})$ transformations as orbifold operations on the F-theory torus.
This formal  interpretation does not, however, imply in any obvious way that the equivalences presented here extend beyond the
supergravity approximation.

\smallskip

The simplest example of  ``equivalent"  theories are theories related by the $SL(2,\mathbb{Q})$ transformation
\bea
   \left(   \begin{array}{cc}
  r/s   & 0   \\ 0  &  s/r
  \end{array}
  \right)\qquad {\rm with}\quad (r,s)  \quad {\rm relatively\, prime \ integers}\, . \no
\eea
Such diagonal transformations do not modify the five-brane types, but they change
  the  number of five-branes in each stack. They also transform  their linking numbers,  so as to leave unchanged the  D3-brane charges:
\bea
\hat N^{(b)\, \prime}_{5} = {r\over s}\,  \hat N^{(b) }_{5}\, , \quad
\hat l_j^{\, \prime} = {s\over r}\, \hat l_j\,  , \quad\quad
N^{(a)\, \prime}_{5} = {s\over r}\,   N^{(a) }_{5}\, , \quad
l_i^{\, \prime} = {r\over s}\,  l_i\, .
\eea
Consistency with charge quantization requires of course that    $ \hat N^{(b) }_{5}$ and  $l_i$ be multiples of $s$, and
  that $N^{(a) }_{5}$ and $\hat l_j$  be  multiples of $r$. \\
We first note that, since the number $L$  of winding D3-branes does not transform,  whereas the total number of 5-branes change as
  \bea
k \to {s\over r}\, k\   \qquad {\rm and}\quad  \hat k \to {r\over s}\, \hat k\ ,
\eea
  the supergravity  free energy 
\eqref{herzogfree}, that we conputed in the large $L$ limit,  is invariant, as expected.

Remark that even these simple $SL(2,\mathbb{Q})$ transformations act highly non-trivially on the field theory side.
For instance, the number of gauge-group factors is multiplied by $r/s$, while the total number of fundamental hypermultiplets
is  multiplied by $s/r$.

  \smallskip

\begin{figure}[t!]
\centering
\includegraphics[height=9.6cm,width=15cm]{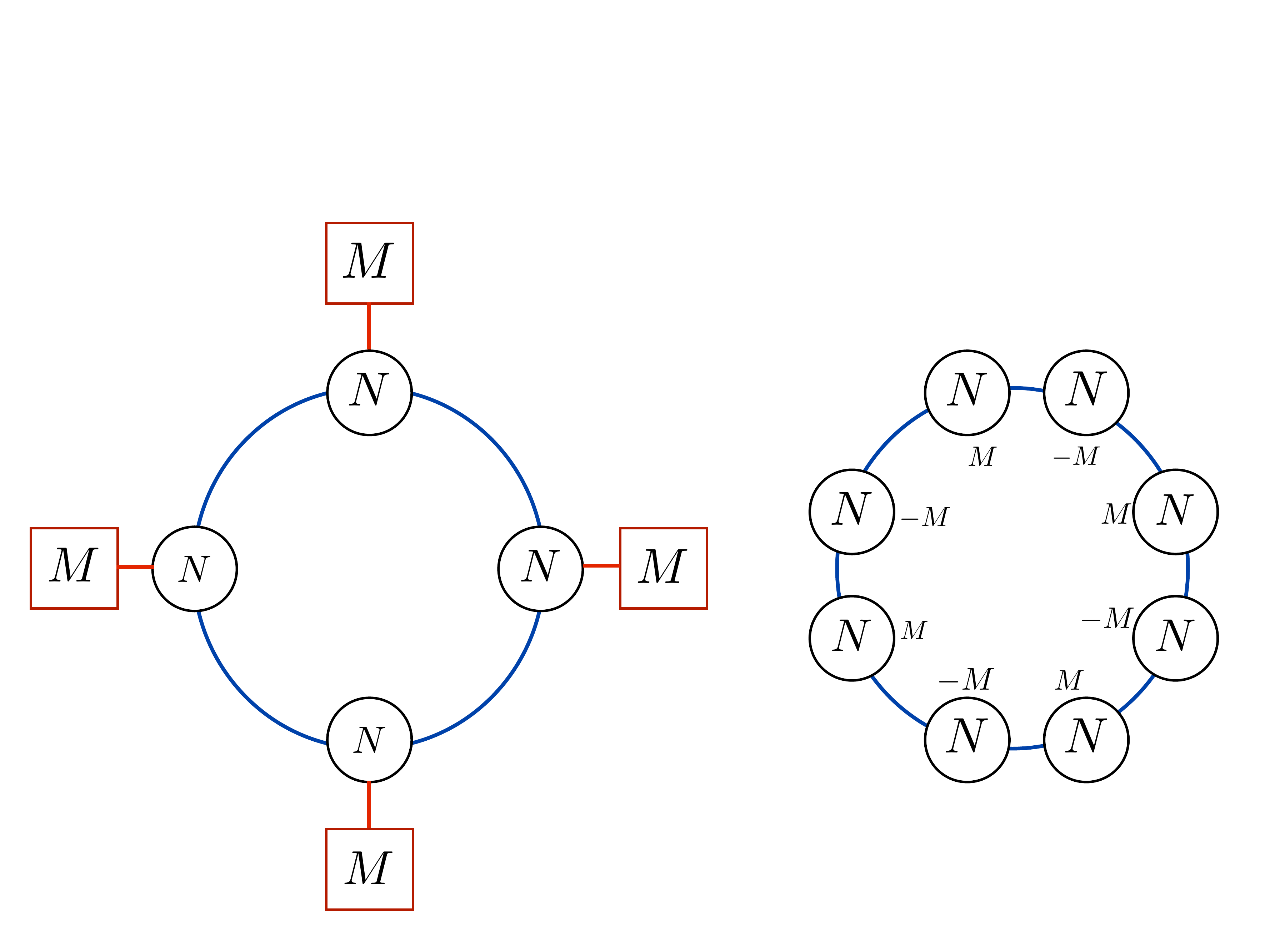}
\caption{ \footnotesize
The two circular-quiver gauge theories related by the $SL(2,\mathbb{Q})$ transformation \eqref{SL2Q}.
  The theory on the right is obtained from the one on the left by doubling the number of gauge-group factors, removing the fundamental hypermultiplets
  and adding   Chern-Simons  terms with alternating sign.}
\label{slQ}
\end{figure}

\bigskip

As another example of $SL(2,\mathbb{Q})$ equivalence, we consider the transformation
\bea\label{SL2Q}
   \left(   \begin{array}{cc}
  1  &    M^{-1}  \\  0   &  1
  \end{array}
  \right)\qquad {\rm with}\quad M\in \mathbb{N} \, .
\eea
This  transformation leaves the NS5-branes invariant, while it  converts a stack of $M$ D5-branes into a
single $(1,M)$ five-brane. Recall that the worldvolume theory of a stack of
$N$ D3-branes intersecting a stack of
 $M$ D5-branes is a $U(N)$ gauge theory with $M$
  fundamental hypermultiplets and one adjoint hypermultiplet. Replacing  the D5-branes by a $(1,M)$ five-brane
     leads to a $U(N)_M \times U(N)_{-M}$ gauge theory with one bifundamental hypermultiplet
     and level $M$ (respectively $-M$) Chern-Simons terms (see e.g. \cite{Aharony:2008ug}).
The  transformation  \eqref{SL2Q} can be used therefore to relate  the following two   theories:\\ \vskip -1.2 mm
\indent (i)  a $U(N)^{\hat k}$ gauge theory, with $M$ fundamental hypermultiplets for every gauge-group factor,
and a bifundamental for each neighboring pair;\\ \vskip -1.2 mm
\indent
(ii) a $U(N)^{2\hat k}$ gauge theory with bifundamentals  for each neighboring pair,  and Chern-Simons terms of alternating
level $\pm M$. \\ \vskip -1.2 mm
\noindent
The corresponding
circular quivers  are illustrated in  Figure \ref{slQ}.
As a test of their  $SL(2,\mathbb{Q})$ equivalence we will
 conclude this section by comparing the free energies of these two gauge field theories in the limit  $N\gg 1$.

\smallskip

Let us first recall the result  \eqref{herzogfree} for the
  free energy on the supergravity side.
  Replacing the number  of winding D3-branes  by $N$,   and the total number
  of D5-branes by $M \hat k$,   leads to  the expression
 \begin{align}
\label{Fsugra}
F_{\rm sugra}  \,  =  \,   \frac{\pi \sqrt{2} }{3}\hat k  M^{1/2} \, N^{3/2}  \ .
\end{align}
This should be compared to the result on the field-theory side.  For
  the necklace quiver of theory  (ii)  the calculation  has been performed
   in  \cite{Herzog:2010hf}. These authors  used the localization techniques of \cite{Kapustin:2009kz}
  to reduce the calculation to a matrix-model  integral, which they then evaluated for large-$N$  by the saddle-point method.
  Their result agrees precisely with \eqref{Fsugra},  confirming the AdS/CFT correspondence. What we need to do  is to
 also  recover  this   result   from the original gauge theory (i).
 \smallskip

  Since     for theories with $\N \geq 4$ supersymmetries the free energy does not run
 \cite{Kapustin:2009kz},  we
 may perform the calculation near  the  (ultraviolet) Gaussian  fixed point.  Using the standard localization techniques, one  reduces the partition function
 of theory (i) to the following matrix-model  integral:
\beq
Z_{(i)} = {1\over (N!)^{\hat k}} \int \prod_{a=1}^{\hat k} \frac{d^N \sigma_a}{(2\pi)^N} \frac{\prod_{i<j} 4\sinh^2 \big(\frac{\sigma_{a}^i - \sigma_{a}^j}{2} \big) }{\prod_{i,j} 2\cosh \big(\frac{\sigma_{a+1}^i - \sigma_{a}^j}{2} \big) } \frac{1}{\big[\prod_{j} 2\cosh \big(\frac{\sigma_{a}^j}{2} \big) \big]^M}\ ,
\eeq
where $i,j$ run from $1$ to $N$. This can be written as  $Z_{(i)} = \int e^{-F (\sigma_a)}$ with
\begin{align}
F (\sigma_a)\,  =\,  - 2 \sum_{a\,;\,i<j} \log\Big[ 2\sinh \big(\frac{\sigma_{a}^i - \sigma_{a}^j}{2} \big) \Big] + \sum_{a\,;\,i,j} \log\Big[ 2\cosh \big(\frac{\sigma_{a+1}^i - \sigma_{a}^j}{2} \big) \Big] \no\\
+ \sum_{a\,;\,j} M \log\Big[ 2\cosh \big(\frac{\sigma_{a}^j}{2} \big) \Big] + \hat k \log(N!) + \hat k N \log(2\pi)\ .
\label{F_A}
\end{align}
\smallskip
Following reference \cite{Herzog:2010hf},  we let  $\sigma_a^j = N^{\beta} x_a^j$, and fix $\beta$ so that
at the saddle point
 the $x_a^j$ are of order one.  Contrary to  this reference,  we do not introduce an imaginary part for the $x_a^j$.
 Indeed,   the saddle point equations are invariant under complex  conjugation, so we are entitled to look   for real solutions.

\smallskip

In the limit $N\gg 1$, we may replace the variables $x_a^i$ by a continuous density $\rho_a(x) $ normalized so that   $\int dx \rho_a(x) = 1$.
The expression \ref{F_A} can be written as
 \bea
F(\rho_a) =\sum_{a=1}^{\hat k} &{1\over 2} \Big[ {\pi^2}  N^{2-\beta} \int dx_a   \rho_a(x_a)^2 +  {M}  N^{1+\beta} \, \int dx_a |x_a| \rho_a(x_a) \Big] \no  \\
&  + O(N^{2-2\beta},N \log N)\ .
\label{F_continuum}
\eea
The details of the computation are subtle, at least to obtain the first term in \ref{F_continuum}, and can be found in appendix A of \cite{Herzog:2010hf}.
\smallskip

The saddle-point equations are non-trivial when the
two terms in this expression are of the same order, so that  $\beta = \half$. Furthermore, thanks to the  symmetries of the problem, we may look
for saddle points with $\rho_a(x) = \rho(x)$ for all $a$,\footnote{The authors of \cite{Herzog:2010hf} arrive to this same ansatz after some approximation
of the saddle point equations.}
 and $\rho(x) = \rho(-x)$.  With these assumptions the above free energy  reduces to
\begin{align}
F(\rho) = \frac{\hat k}{2} N^{\frac{3}{2}} \Big[ \pi^2  \int dx \,  \rho(x)^2 + M  \int dx \, |x| \rho(x) - \gamma \int dx \, \rho(x) +  \gamma  \Big]\ ,
\label{F_A_last}
\end{align}
where the Lagrange multiplier $\gamma$  imposes  the constraint $\int dx \rho(x) = 1$. The ensuing saddle point equation,
\beq
2\pi^2 \rho(x) + M  |x| = \gamma \ ,
\eeq
 is solved by the eigenvalue density
\begin{align}
\rho(x) &= \frac{1}{2\pi^2}\big( \gamma - M  |x| \big) \quad \textrm{for} \quad |x| < x_0 , \no\\
&= 0 \quad \textrm{for} \quad |x| > x_0 .
\end{align}
 The constraint $\int dx \rho(x) = 1$ fixes the Lagrange multiplier
 \bea
 \gamma = \frac{M  x_0}{2} + \frac{\pi^2}{x_0}\ ,
 \eea
whereas the positivity of $\rho$ implies $x_0 \leq \pi\sqrt{\frac{2}{M}}$. Combining all these formulae gives
\begin{align}
F(x_0) = \hat k N^{\frac{3}{2}} \Big[ \frac{\pi^2}{4 x_0} + \frac{M  x_0}{4} - \frac{M^2 x_0^3}{48 \pi^2} \Big]\ .
\end{align}
We now need to minimize this expression with respect to $x_0$ which takes values in $(0, \pi\sqrt{ {2/M}} ]$.
The minimum is achieved at the rightmost endpoint, leading to the final result for the gauge theory (i):
 \begin{align}
F_{(i)}  = \frac{\pi \sqrt{2}}{3}\hat k \sqrt{M} N^{\frac{3}{2}}\ ,
\end{align}
in perfect agreement with both the necklace-quiver and the supergravity calculations.
Note that although the final results agree, the three calculations  differ greatly in their specific details.

 \chapter*{Perspectives}
\addcontentsline{toc}{chapter}{ Perspectives }
\label{chap:ccl}

The AdS/CFT proposals that have been presented cover all the fixed points of 3d $\N=4$ linear quivers, circular quivers and $\half$-BPS defect SCFTs preserving the supergroup $OSp(4|4)$. All the quarter-BPS brane configurations involving D3-branes, D5-branes and NS5-branes in type IIB string theory have been associated to a supergravity solution and quiver gauge theory. In this sense our classification seems complete. However new three dimensional $\N=4$ SCFTs associated to star-shaped quivers have been proposed in \cite{Benini:2010uu}, raising the question of possible other supergravity solutions with the same symmetries. Although the construction of \cite{Benini:2010uu} is not completely clear to us, it is interesting to notice that the data needed for these star-quivers might match the parameters of supergravity solutions on a surface $\Sigma$ with a richer structure of boundary singularities, namely more boundary segments with D5 or NS5 singularities. The essential problem of these solutions is the 
presence of line singularities on $\Sigma$ or conical point singularities on $\Sigma$ with $2\pi$ deficit angle. The question of the existence of such solutions without singularity in the interior of $\Sigma$ is not settled. The question of the possible interpretation of the conical singularities is also open. 

\vspace{6mm}

The question of dualities between the IIB solutions and IIA or M-theory solutions is also an interesting direction of investigations. For circular quiver geometries we have seen that the naive T-duality and lift to M-theory is related to the smeared IIB solution, where the surviving data is are just the total numbers of branes of each type. Recovering the 5-branes localization may follow from corrections to the metric due to worldsheet instantons as in \cite{Tong:2002rq}, breaking the isometry of the T-duality. On the IIA side (resp. M-theory side) the information characterizing the quiver seems to be encoded in $B_2$ holonomies (resp. $C_3$ torsion fluxes) around the two-cycles (resp. three-cycles) of the geometry (\cite{Imamura:2008ji}). This picture is not clear. The precise rules for the dualities should be clarified. The ABJM gauge theory might be a good place to start because we know its IIA, M-theory and now IIB supergravity duals.

\vspace{6mm}

We have also noticed that the domain wall supergravity solutions provide the string theory arena to explore the Karch-Randall scenario of localization of gravity in a non-compact internal space \cite{Karch:2000ct,Karch:2001cw}. Their model is based on 5d Einstein gravity with a negative cosmological constant in the presence of a 4-dimensional ``thin brane''. The solution to the equations of motions for small enough thin-brane tension is an $AdS_4 \ltimes \bR$ fibration with a peak of the warp factor at the position of the thin brane. In this setup the first 4d graviton mode has a small mass and has its wavefunction localized near the thin-brane. Furthermore the rest of the graviton mass spectrum is separated from the lowest mass by a ``large'' mass gap. This provides an effective realization of 4d ``almost massless'', ``almost flat'' gravity with a non-compact internal space, which is phenomenologically promising, except that the small 4d cosmological constant is negative. 
There was hope that this scenario can be realized in string theory with a brane configuration of D3-branes intersecting D5-branes. The near horizon geometries of such configurations are the one we have studied.\\
In \cite{Bachas:2011xa} the fluctuations of the metric corresponding to the 4d gravitons were studied in detail for the case of the BPS Janus domain wall solution, which connects to $AdS_5 \times S^5$ regions with different values of the dilaton. This corresponds to the absence of 5-branes in the geometry. It was shown that the Janus geometry does not reproduce the good features of the Karch-Randall model. The analysis for a situation with D5-branes and NS5-branes was essentially left for future work. \\
Our computations in the case of one stack of D5 and one stack of NS5-branes tend to show that the Karch-Randall scenario is again not reproduced. The main problem is that the favorable growth of the $AdS_4$ warp factor in the region near the 5-branes result in the formation of a quasi-flat central region of increasing size. Our analysis is presented in appendix \ref{app:KR}. Further work is still needed to evaluate the graviton mass spectrum and possibly explore richer geometries.

\chapter{  Appendices}
\addcontentsline{lot}{chapter}{ Appendices}
\label{appendices}

\setcounter{section}{\value{appendices}}
\renewcommand{\thesection}{\Alph{section}}


\section{Mirror symmetry of inequalities}
\label{app:ineq}

We will here show that the inequalities
 \eqref{fixedpointcircA} are invariant under the mirror map, i.e. that
 \beq
\label{ineq2a}
L+ \rho^{T}    > \hat \rho\  \ \Longleftrightarrow \ \   L+  \hat \rho^{T}    >   \rho\ .
\eeq
The proof of mirror symmetry for the linear quiver inequalities \ref{fixedpoint} is then obtained simply by setting $L=0$.

\begin{figure}[t]
\centering
\includegraphics[scale=0.3]{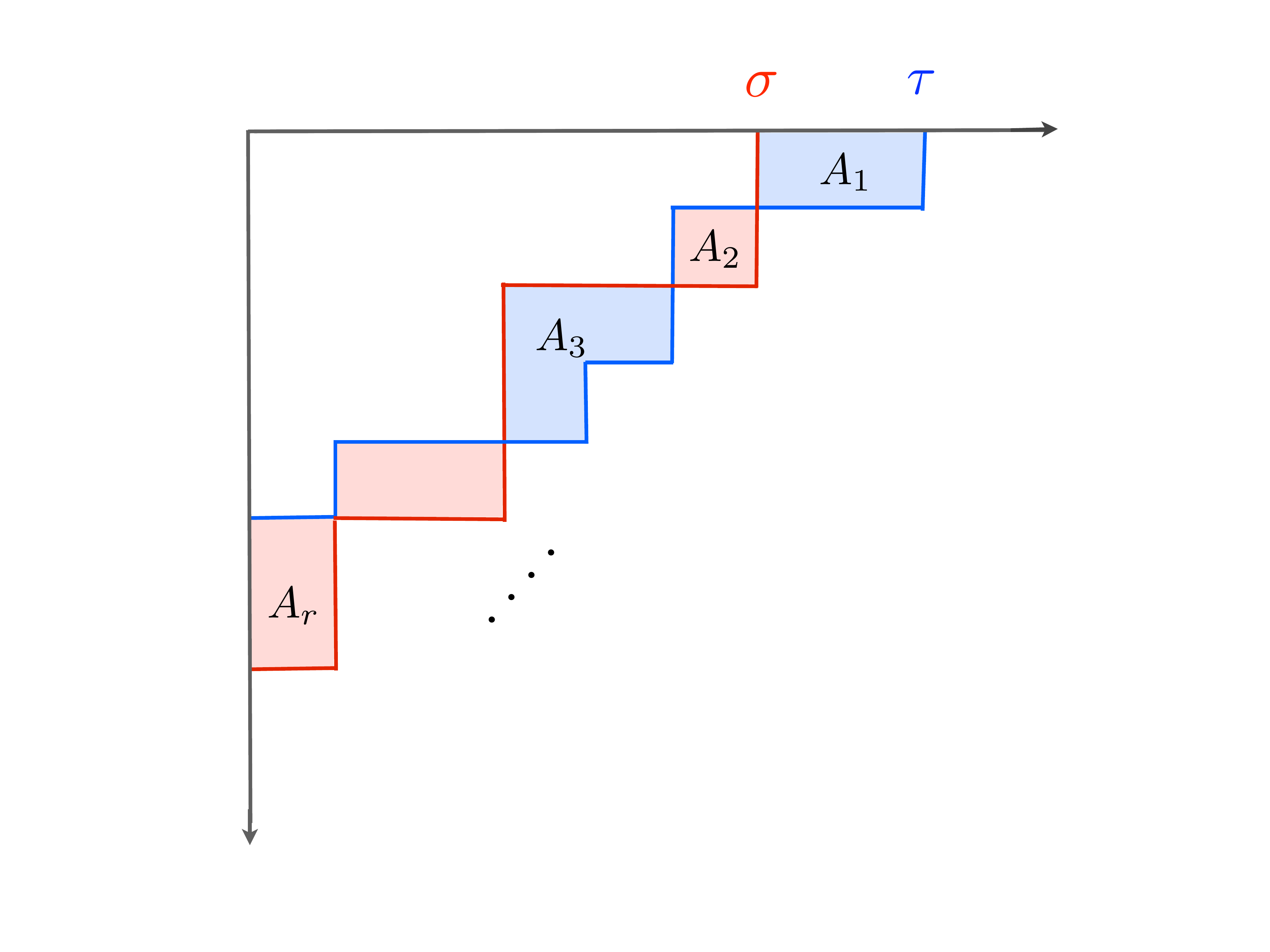}
\caption{\footnotesize   The difference of two  Young tableaux defines an alternating  sequence  $\{A_1, A_2, \cdots , A_r\}$
where $\vert A_i\vert$  counts  the number of boxes  in the $i$th  region enclosed by  the two histograms of the Young tableaux.
In this  example   $A_1 = 2 , A_2 = -1, A_3 = 3 , \cdots$.
  The difference of the transposed
 tableaux, obtained by transposing the figure ($180^{\rm o}$ rotation around the descending diagonal), defines  the inverse opposite sequence  $\{-A_r, \cdots , -A_2, -A_1\}$. }
\label{mirror}
\end{figure}

Let us first recall that if $\tau = (a_1,a_2,...,a_t)$ and $\sigma = (b_1,b_2,...,b_s)$ are two partitions of the same number $N$,
expressed as vectors with non-increasing positive components,
then $L+ \tau    > \sigma$  is a shorthand notation for the set of inequalities
\begin{align}
 L+  \sum_{i=1}^{n} a_i > \sum_{i=1}^{n} b_i \quad\quad \textrm{for\ all } \quad n=1,...,\textrm{max}(t,s).
\end{align}
These can be visualized more easily in the
diagrammatic representation of figure \ref{mirror}, which defines a sequence $\{A_1, A_2, \cdots , A_r\}$ of
areas with alternating signs. In terms of this sequence, the inequalities read
\bea\label{reverse}
L+ A_1 >0\ , \quad L+(A_1+A_2) >0\ , \quad \cdots \quad , \quad   L+\sum_{s=1}^{r-1} A_s >0\ , \quad  L >0\ ,
\eea
where the last inequality follows from  the fact that $A_1 + A_2 \cdots  + A_r =0$.
Reversing the order, one may put these inequalities  in the following form:
\bea\label{reverse2}
  L  > 0 \ , \quad L-A_r  >0\ , \quad  L- A_r - A_{r-1} > 0\ , \cdots \quad , \quad  L - \sum_{s=2}^{r} A_s  >0\ .
\eea
This is exactly the set of inequalities corresponding to $L + \sigma^T > \tau^T$, as is evident if one transpose the figure \ref{mirror}.\\
Setting   $\tau\equiv \rho^T$ and $\sigma\equiv \hat \rho$
proves the mirror equivalence  \eqref{ineq2a}, as claimed.

\vspace{10mm}

\section{Proof of the inequalities in supergravity}
\label{app:ineq2}

In this appendix we prove the inequalities on the partitions $\rho$ and $\hat\rho$ required for the positivity of the ranks $N_j$ in the quiver dual description, which can be thought of as supersymmetry preserving conditions. We prove the inequalities for the supergravity solutions on the annulus, corresponding to circular quivers, keeping in mind that the inequalities for the solution on the strip, corresponding to linear quivers, are obtained in the limit $t \rightarrow \infty$ ($L \rightarrow 0$) as a (simpler) subcase. The computation can be easily adapted to the verification of inequalities for the domain wall supergravity solutions.

\vspace{5mm}

We have already shown  in \S\ref{s:match} that,  with an appropriate choice of gauge, the linking numbers of the
supergravity solution can be confined to the intervals  $\ l^{(a)}\in (0, \hat k]$ and $\hat l^{(b)}\in (0,  k]$.
In particular, the linking numbers are positive, and we  demand  that they be quantized. Thus the Young tableaux
$\rho$ and $\hat\rho$ are well defined, and the inequalities $L+ \rho^{T}    > \hat \rho$ make  sense. We will now
prove that these inequalities are automatically obeyed on the supergravity side.\footnote{In the graphic form of
Figure \ref{mirror} the inequalities actually make sense for any pair of monotonic functions with equal  definite integral,
and with transposition  of the Young tableau being replaced by   function inversion. This should make it possible to prove the
inequalities without using quantization and the partial gauge fixing that was required to define the Young tableaux.
We will not pursue   this approach further here.
}
\smallskip

Let us recall the explicit expressions of the five-brane  linking numbers and of  $L$:
\begin{align}
\label{formulae}
l^{(a)} &= \sum_{b=1}^{\hat p} \hat N_b  \Big[  \sum_{n=0}^{+\infty} f(\hat \delta_b - \delta_a -2 n t) - \sum_{n=1}^{+\infty} f(-\hat \delta_b + \delta_a - 2 n t)  \Big] \ , \nonumber \\
\hat l^{(b)} &= \sum_{a=1}^{p} N_a \Big[  \sum_{n=0}^{+\infty} f(\hat \delta_b - \delta_a -2 n t ) - \sum_{n=1}^{+\infty} f(-\hat \delta_b + \delta_a -2 n t )  \Big] \ , \\
L  &= \sum_{a=1}^{p} \sum_{b=1}^{\hat p}\sum_{k=1}^{+\infty} k \ N_a \hat N_b \Big[ f(\hat \delta_b - \delta_a -2 k t ) + f(\delta_a - \hat \delta_b - 2k t ) \Big] \ , \nonumber
\end{align}
where  $f(x)= \frac{2}{\pi} \arctan(e^x)$, and we use in this appendix a lighter notation for the five-brane charges,  $N_a \equiv  N^{(a)}_{5}$
and $\hat N_b \equiv  \hat N^{(b)}_{5}$.
In terms of these linking numbers and the five-brane charges the partitions   $\hat \rho$ and $\rho^T$ read:
\begin{align}
 \hat \rho &= (\underbrace{\hat l^{(1)},...,\hat l^{(1)}}_{\hat N_1},...,\underbrace{\hat l^{(b)},...,
\hat l^{(b)}}_{\hat N_b},...,\underbrace{\hat l^{(\hat p)},...,\hat l^{(\hat p)}}_{\hat N_{\hat p}}) \ ,
\end{align}
and
\begin{align}
\rho^T = (\underbrace{\sum_{a=1}^p  N_a,...,\sum_{a=1}^p  N_a}_{l^{(p)}},\underbrace{\sum_{a=1}^{p-1} N_a,...,\sum_{a=1}^{p-1} N_a}_{l^{(p-1)} - l^{(p)}},...,\underbrace{\sum_{a=1}^A N_a,...,\sum_{a=1}^A N_a}_{l^{(A)} - l^{(A+1)}},...,\underbrace{N_1,...,N_1}_{l^{(1)} - l^{(2)}}) \ .
\end{align}
We need now to establish the set of  inequalities
\bea
\label{ineq3}
\sum_{s=1}^r m_s \ + L  \ &>&\  \sum_{s=1}^r \hat l_s\qquad \forall r = 1,\ldots , \max(k,\hat k) \, .
\eea
where $\hat \rho = (\hat l_1,\hat l_2, ..., \hat l_{\hat k})$ and $\rho^T = (m_1,m_2,...,m_k)$ are the above two partitions.

\smallskip

The last  inequality, the one for $r=\max(k,\hat k)$,  implies that
 $L  > 0$. This is obeyed automatically, as seen  from the explicit expression \eqref{formulae}  and the fact that $f$ is strictly positive.

Let us  show now that it is sufficient to 
prove  the inequalities in (\ref{ineq3}) for the corners of the histogram $\hat\rho$, i.e. for the values  
\begin{align}
\label{defr}
r = \sum_{b=1}^{J} \hat N_b \qquad {\rm where} \qquad J = 1, 2 ,..., \hat p \ .
\end{align}
 To see why, assume that  $r$ is in the   range $\sum_{b=1}^{J-1}\hat N_b < r \leq \sum_{b=1}^{J}\hat N_b$, 
 for some $J = 1, 2 ,..., \hat p$.  Then 
 if   (\ref{ineq3}) is satisfied for all $r'<r$ but not  for $r$, it  will not be satisfied for
  $r'' = \sum_{b=1}^{J} \hat N_b$ either.  This is because 
   $\hat l_s$  is constant for $s$ in the range $\sum_{b=1}^{J-1}\hat N_b < s \leq \sum_{b=1}^{J}\hat N_b$, while the
 integer $m_s$, which belongs to a
 non-decreasing sequence of integers, 
 does not increase as $s$ ranges over the values $\sum_{b=1}^{J-1}\hat N_b < s \leq \sum_{b=1}^{J}\hat N_b$.  
 Conversely, if the constraint is satisfied for $r''$ then
  it will be satisfied also  for $r$. 
   We remark here that the limit of decoupled quivers, corresponding to disjoint brane configurations, is reached when the inequality is saturated for some value of $r$, with the saturation preserved for $r' > r$.  Following the logic 
   of the previous argument, such an $r$ must be of the form $r = \sum_{b=1}^{J}\hat N_b$.
\smallskip

Let us now take a fixed $J$ with $1 \leq J \leq \hat p$.
 By summing over the number of rows in $\rho^T$, we can always find an integer $I$ such that
\begin{align}
l^{(I)} > r \geq l^{(I+1)}\ .
\end{align} 
 We may then write the sum over $m_s$ as
\begin{align}
\sum_{s=1}^r m_s &= \sum_{A=I+1}^p \sum_{a=1}^{A} N_a \left(l^{(A)}-l^{(A+1)}\right) + \left(r - l^{(I+1)}\right) \sum_{a=1}^{I} N_a \cr
&= \sum_{a=I+1}^p l^{(a)} N_a + \left(\sum_{b=1}^{J} \hat N_b\right) \left(\sum_{a=1}^{I} N_a\right) \ ,
\end{align}
where we have used (\ref{defr}) to replace $r$.
The inequality (\ref{ineq3}) then becomes
\begin{align}
\label{niceform}
\sum_{b=1}^{J}\hat l^{(b)} \hat N_b  \ < \ L + \sum_{a=I+1}^{p} l^{(a)} N_a 
+ \left(\sum_{a=1}^{I} N_a \right)\left(\sum_{b=1}^{J} \hat N_b\right) \ .
\end{align}
This is the form of the inequality that we will now prove using the supergravity calculation of the charges. 

 \smallskip

Let us give a name to the infinite sum that enters in the supergravity expressions for the linking numbers:
\bea
F(x, 2t ) \equiv   \sum_{n=0}^{\infty} f(x - 2n t) - \sum_{n=1}^{\infty} f(-x -2 n t )\ .
\eea
 In terms of the function $F$ the inequalities \ref{niceform}  can be written as
 \begin{align}
\sum_{a = 1}^{p} \sum_{b=1}^{J} N_a \hat N_b \,  F( \hat{\delta}_{b}-\delta_{a}, 2t ) \
&< \ L  + \sum_{a=I+1}^{p} \sum_{b = 1}^{\hat{p}} N_a \hat N_b\,  F ( \hat{\delta}_{b}-\delta_{a}, 2t )  + \sum_{a=1}^{I}\sum_{b=1}^{J} N_a \hat N_b \ . \nonumber
\end{align}
 Splitting the sums,
 simplifying and rearranging terms gives:
\begin{align}
\sum_{a = 1}^{I} \sum_{b=1}^{J} N_a \hat N_b \, F( \hat{\delta}_{b}-\delta_{a}, 2t)
- \sum_{a=I+1}^{p} \sum_{b = J+1}^{\hat{p}} N_a \hat N_b \, F(\hat{\delta}_{b}-\delta_{a}, 2t) \
&< \ L   +  \sum_{a=1}^{I}\sum_{b=1}^{J} N_a \hat N_b \ . \nonumber
\end{align}
We show that this is automatically satisfied by putting the following successive bounds on
  the left hand side:
\begin{align}
\label{calculus}
& \ \hskip 1.5 cm  \sum_{a = 1}^{I} \sum_{b=1}^{J} N_a \hat N_b \, F( \hat{\delta}_{b}-\delta_{a}, 2t)
- \sum_{a=I+1}^{p} \sum_{b = J+1}^{\hat{p}} N_a \hat N_b \, F(\hat{\delta}_{b}-\delta_{a}, 2t)   \nonumber \\
  & \ \hskip 0.3cm  < \ \sum_{a = 1}^{I} \sum_{b=1}^{J} N_a \hat N_b \, \sum_{n=0}^{\infty} f(\hat \delta_b - \delta_a - 2nt)
+ \sum_{a=I+1}^{p} \sum_{b = J+1}^{\hat{p}} N_a \hat N_b \, \sum_{n=1}^{\infty} f(-\hat \delta_b + \delta_a - 2nt) \nonumber\\
< &  \  \sum_{a = 1}^{I} \sum_{b=1}^{J} N_a \hat N_b \ f(\hat \delta_b - \delta_a) \ + \  \sum_{a = 1}^{p} \sum_{b=1}^{\hat p}
N_a \hat N_b \ \sum_{n=1}^{\infty}  \Big[ f(\hat \delta_b - \delta_a - 2nt) + f(-\hat \delta_b + \delta_a - 2nt) \Big] \nonumber \\
 &   \  \hskip 3.8cm   < \  L + \sum_{a = 1}^{I} \sum_{b=1}^{J} N_a \hat N_b \ .
\end{align}
In the first inequality we have dropped terms that are explicitly negative.
The second inequality is obtained by extension of  the sums.  Finally, in
 the third inequatlity we  used, in addition to  the bound  $0< f(x) <1$,
  the  expression  (\ref{formulae}) for the winding charge $L$.  This completes the proof.

\vspace{8mm}

One can saturate the inequality $L>0$ by sending $t\rightarrow +\infty$ in which case $L \rightarrow 0$, obtaining a linear quiver geometry.
In this limit ($t=+\infty$) we can saturate the inequality \ref{niceform} in two different manners: \\
 \indent (i) when $\delta_a \rightarrow +\infty$ for $a=I+1, I+2, ..., p$ and   $\hat \delta_b \rightarrow +\infty$ for $b=1, 2, ..., J$,
 or\\
  \indent (ii) when $\delta_a \rightarrow -\infty$ for $a=1, 2, ..., I$ and $\hat \delta_b \rightarrow -\infty$ for $j=J+1, J+2, ..., \hat p$.\\

 This limit corresponds to detaching  a subset of  fivebrane singularities  and moving them off to infinity on the strip.

\section{From IIB to M theory for  large $L$}
\label{Tduality}

  We give here the detailed  T-duality transformation of the type-IIB solution for large winding number $L$ to a solution of type-IIA
  supergravity, and the subsequent uplift to eleven dimensions.  We will follow the metric, dilaton and two-form gauge fields,
  which all become part of the metric in eleven dimensions. The  four-form potential of the IIB theory   transforms to the three-form potential of M theory,
  which at leading order  has a field strength  proportional to the ${\rm AdS}_4$ volume form. 
  The way in which the 3-form field may encode the information on the five-brane throats  is a very subtle issue, as already noted in
  the main text. We will not discuss it further in this appendix.    

 \smallskip

   The type-IIB backgrounds in the large-$L$  limit are given by the expressions
  \eqref{IIBlargeL1} and \eqref{IIBlargeL2}.  In order to use the standard Buscher rules, we make a gauge transformation
  that removes the $x$-dependence from the gauge potentials. The new two-form potentials read
     \begin{align}
B_{(2)} &= - 2 \hat k \,  \cos(\theta_1) dx\wedge d\phi_1 \ , \no\qquad
C_{(2)} = - 2  k \,  \cos(\theta_2) dx\wedge d\phi_2  \ ,
\end{align}
where we recall that $x$ is periodic with period $2\pi$.  We also transform the Einstein-frame to the string-frame metric, $G_{MN}  = e^\phi g_{MN}$,
in terms of which  Buscher's rules read  \cite{Buscher:1987qj}:
 \begin{align}
G^\prime_{\mu \nu} &=  G_{\mu \nu} - \frac{G_{x \mu}G_{x \nu}-B_{x \mu}B_{x \nu}}{G_{x x}}  \ , \qquad
G^\prime_{0 \nu} = \frac{B_{x \mu}}{G_{x x}}  \ , \qquad  G^\prime_{x x} =\frac{1}{G_{x x}} \ , \no\\
  B^\prime_{\mu \nu} &= B_{\mu \nu} - \frac{G_{x \mu} B_{x \nu}-B_{x \mu} G_{x \nu}}{G_{x x}}  \ , \qquad
  B^\prime_{0 \nu} = \frac{G_{x \mu}}{G_{x x}} \ , \qquad e^{4   \phi^\prime}  = \frac{e^{4 \phi}}{G_{x x}} \ ,
 \end{align}
 where the prime indicates the type-IIA fields in string frame,  and the lower-case Greek indices $\mu,\nu$ run over all dimensions other than $x$.
  In addition, the 2-form R-R potential transforms to a one-form potential,
\bea
  C^\prime_{(1)\mu} &= C_{(2) x \mu} \ .
  \eea
\noindent Since the IIB metric had no $(x\mu)$ components  $B^\prime$ is zero, while the original 2-form NS-NS gauge field becomes an off-diagonal component of
the IIA  metric. In string-frame this latter reads:
   \begin{align}
dS_{{\rm IIA}}^{\, 2} =  & { { \pi^2 \hat k}\over t }   \sqrt{1-y}  \left[   ds^2_{AdS_4} +  y\,  ds^2_{S^2_1} +   (1-y) ds^2_{S^2_2} \right]    \no \\  &
 +    {4 \pi^2  \over   t }    \sqrt{1-y}  \left[ {y\over \hat k}  (dx - {\hat k\over 2} {\rm cos}\theta_1 d\phi_1)^2 +    \hat k {dy^2\over y(1-y) } \right]     \ ,
\end{align}
  whereas  the R-R gauge field and the transformed dilaton field  are given by
\bea
C_{(1)}^\prime = - 2k\,  \cos \theta_2 d \phi_2 \ , \qquad e^{4 \phi^\prime}  =     {4\pi^2  \over t  } {\hat k\over k^2}  (1-y) ^{3/2} \ . \\ \ \no
\eea
\vskip -3mm
 \indent  Finally we uplift the solution  to M theory,  whose metric  (denoted here by a bar)
is  given in terms of the type-IIA backgrounds by the following  relations \cite{Witten:1995ex}
\bea
 \bar g_{MN}  = e^{- 4    \phi^\prime/3 } ( G^\prime_{MN}  + {1\over 4} e^{4\phi^\prime} C^\prime_M C^\prime_N ) \ , \quad
  \bar g_{Mv} = e^{ 8  \phi^\prime/3 }C^\prime_M\ , \quad
   \bar g_{vv} =  4 e^{ 8  \phi^\prime/3 }\ ,
\eea
where $v = v+2\pi$ parametrizes  the eleventh dimension.
Redefining the coordinates $x \rightarrow \hat k x$ , $v \rightarrow k v$ and $y= \sin^2 \alpha $
gives, after some straightforward algebra,  the $AdS^4 \times S^7/ (\mathbb{Z}_k\times  \mathbb{Z}_{\hat k})$ metric, equation \eqref{AdS4S7}.

\section{Barnes $G$-function}
\label{app:Barnes}

Let us briefly summarize the properties of the Barnes $G$-function.
Barnes $G$-function $G_2(z)$ satisfies
\beq
G_2(z+1)=\Gamma(z) G_2(z), \quad G_2(1)=1 \ .
\eeq
From the definition it follows that
\beq
G_2(N)=(N-2)!(N-3)!\cdots 1!, \quad N=2,3,\cdots \ .
\eeq
Its asymptotic expansion is given by
\begin{equation}
\label{Gasymptotics}
\ln G_2(N+1)=\frac{N^2}{2} \ln N - \frac{3}{4} N^2+\mathcal{O}(N) \ .
\end{equation}

\section{Realization of the Karch-Randall model in domain wall supergravity solutions}
\label{app:KR}

In this appendix we explore the possibility of reproducing the Karch-Randall scenario of localization of a nearly massless graviton mode in our supergravity domain wall solutions (non-compact internal space). This short analysis follows the work of Bachas and Estes in \cite{Bachas:2011xa}. First we present the essential features of the Karch-Randall model and then we study the first graviton mode in a simple domain wall solution. We explain qualitatively that the localization of gravity is not reproduced in this case, despites the presence of a nearly massless mode, because the region of localization decompactifies in the relevant limit.

\subsection{The Karch-Randall scenario}

In \cite{Karch:2000ct} Karch and Randall studied the graviton fluctuations in $AdS_5$ spacetime in the presence of a  4-dimensional "thin brane" embedded in a $AdS_4$ slice. 
Their model is based on the effective 5-dimensional action
\begin{align}
S_{KR} &= - \frac{1}{2 \kappa_5} \ \int d^4xdy \ \sqrt{g} \, \left( R + \frac{12}{L^2} \right) + \lambda \ \int d^4x \ \sqrt{[g]_4} \ ,
\end{align}
where $[g]_4$ is the induced metric at $y=0$. The coordinates $x^{\mu}$, $\mu=0,1,2,3$ parametrize the unit $AdS_4$ which is fibered over the direction $y$. $y=0$  is the position of the thin-brane. The parameters of the model are the 5D gravitational coupling $\kappa_5^2$, the bulk cosmological constant $\Lambda = -\frac{6}{L^2}$ and the tension $\lambda$ of the thin-brane.

Einstein equations are solved by the metric
\begin{align}
ds^2 &= L^2 \, \cosh^2 \left( \frac{y_0 - |y|}{L} \right) \ \bar g_{\mu\nu} dx^{\mu} dx^{\nu} + dy^2
\end{align}
with $ \bar g_{\mu\nu}$ the metric of the unit radius $AdS_4$ and $y_0 = L \ \textrm{arctanh}\left(\frac{\kappa_5^2 \lambda L}{6}\right)$.

\begin{figure}[h]
\centering
\includegraphics[scale=0.7]{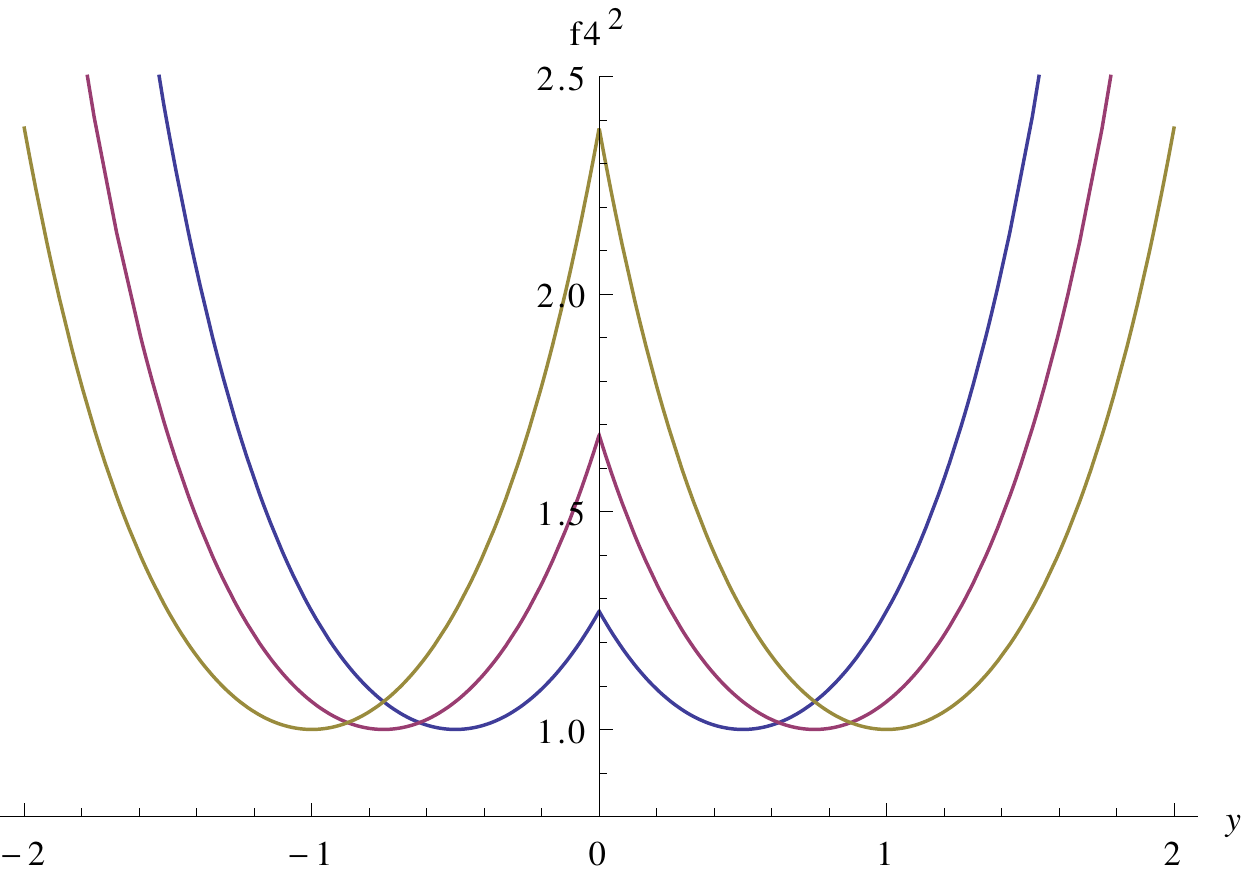}
\caption{\footnotesize $AdS_4$ warp factor $f_4^2(y) \equiv L^2 \cosh^2[(y_0-|y|)/L]$ of the Karch-Randall geometry for $L=1$ and $y_0 = 0.5 , 0.75 , 1$ from the smaller to the higher central pick respectively.}
\label{KRwarp}
\end{figure}

The geometry is characterized by the profile of its warp factor, which has two wells glued together at a distance $y_0$ from their centers, as shown in figure \ref{KRwarp}. The limit $y_0 \rightarrow 0$ corresponds to having no thin-brane ($\lambda=0$) and the two wells fusionning to reconstruct the $AdS_5$ spacetime. The limit $y_0 \rightarrow \infty$ corresponds to the two wells pushed apart far from each other and the spactimes splitting into two $AdS_5$. This limit corresponds to  the thin-brane tension approaching (from below) a finite value $\lambda \rightarrow 6/\kappa_5^2 L$. When $\lambda \geq 6/\kappa_5^2 L$ the solutions to Einstein equations are 4-dimensional Minkovski or dS fibrations over the $y$ direction (see \cite{Karch:2000ct}).

\vspace{5mm}

The limit of phenomenological interest of the KR-model is $l^2 >> L^2$ where
\begin{align}
 l^2 = L^2 \ \cosh^2 \lp \frac{y_o}{L} \rp
\end{align}
is the $AdS_4$ warp factor at the location of the thin-brane $y=0$. This limit corresponds to $y_0$ larger than $L$ but not hierarchically larger.
The graviton of 4-dimensional mass $m$ (in units of $1/L$) is defined by excitations of the unit $AdS_4$ metric 
$g_{\mu\nu} = \bar g_{\mu\nu} +  h_{\mu\nu}$ and the ansatz
\bea
h_{\mu\nu} = h_{\mu\nu}^{[tt]}(x^{\sigma})\psi(y) \\
\bar{\Box}_{AdS_4}h_{\mu\nu}^{[tt]} = \left( m^2 -2 \right) \,  h_{\mu\nu}^{[tt]}
\eea
$[tt]$ stands for ``transverse traceless''. The case $m=0$ corresponds to a reduced number of polarization (massless or partially massless graviton) \cite{Deser:2001wx}.

The mass spectrum contains excitations of mass $m = O(1)$ localized in the $AdS_5$ wells plus an additional zero mode localized on the thin-brane at $y=0$ of mass 
$m_0^2 \simeq \frac{3 L^2}{2 l^2}$ (\cite{Miemiec:2000eq}). In the limit $L^2 << l^2$ this mode becomes nearly massless, with a mass gap with the other modes of order $\delta m = O(1)$. \\
At low energies the effective gravitational force is 4-dimensional Newton gravity localized near $y=0$. The corrections due to the other gaviton modes are suppressed by the presence of the mass gap, but also by the fact that their wavefunctions decrease exponentially fast apart from the $AdS_5$ wells, so that they are exponentially suppressed at $y=0$. 

The interest of the KR model is that it realizes an effective theory containing 4-dimensional Newton's gravity despites the presence of a non-compact internal space. This opens a new window for phenomenological models.

\subsection{First graviton mass in a simple background}

\begin{figure}[h]
\centering
\includegraphics[scale=0.4]{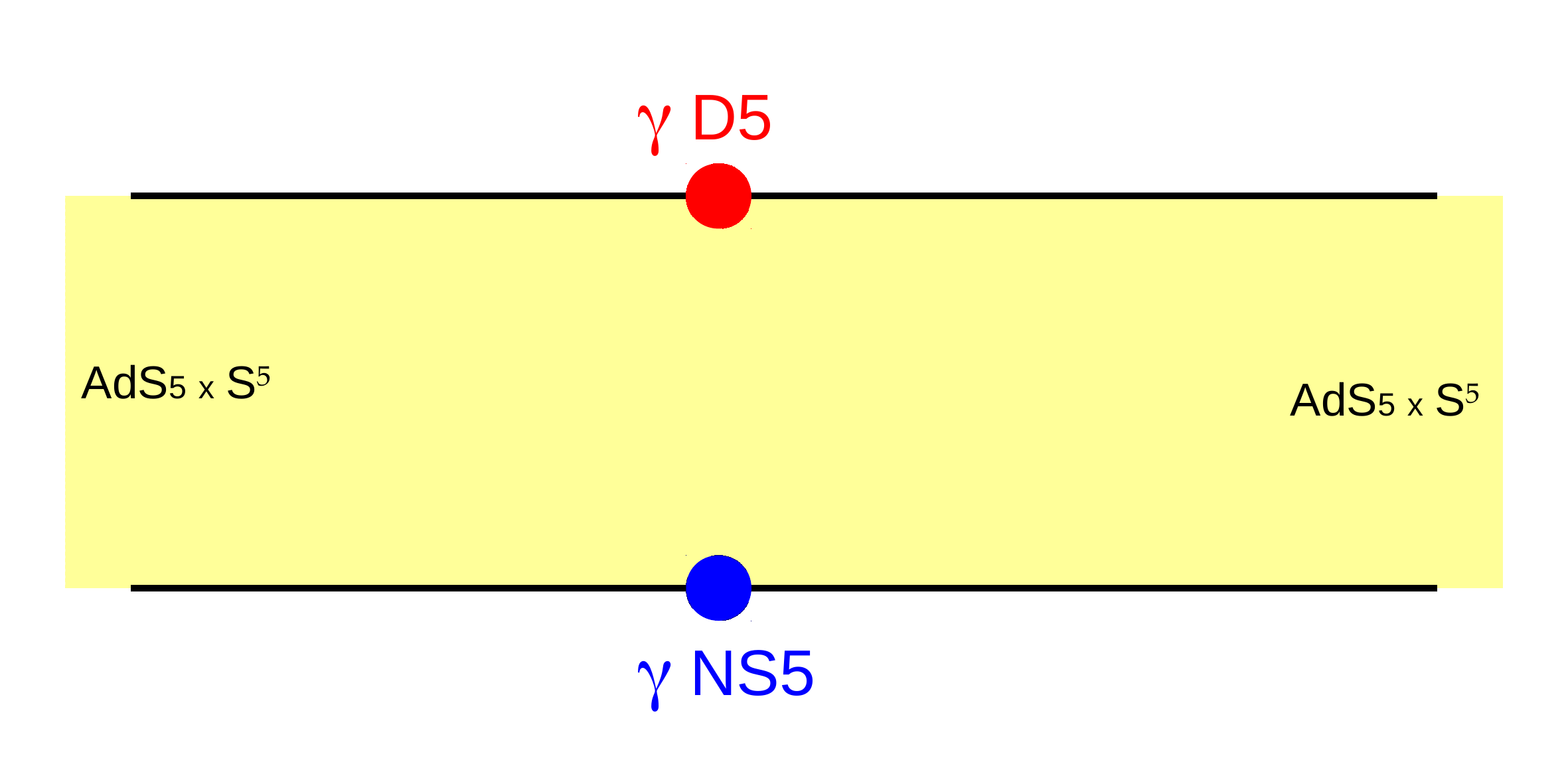}
\caption{\footnotesize Strip geometry with asymptotic $AdS_5 \times S^5$ regions ($x=\pm\infty$) of same radii, one stack of $\gamma$ D5-branes and one stack of $\gamma$ NS5-branes sitting in front of each other at $x=0$.}
\label{stripKR}
\end{figure}

A simple background to explore the KR scenario is the near-horizon geometry
  of a set of D3-branes intersecting one stack of NS5-branes and one stack of D5-branes.
More precisely we choose the supergavity solution on the strip with identical asymptotic $AdS_5 \times S^5$ regions at $x = \pm \infty$, one stack of $\gamma$ D5-branes at $z= i\pi/2$ and one stack of $\gamma$ NS5-branes at $z=0$. We take the same number of 5-branes of each type, sitting at the same position in $x$ to stabilize the dilaton field in the region near the 5-branes, where the first graviton mode might be localized. \\
This is a 2-parameter solution, corresponding to the following choice of harmonic functions ($\alpha >0, \gamma>0$):
\begin{align}
\label{hNS5D5}
h_1 &=    \, \left[ -i \alpha \ {\rm sinh} (z) -    \gamma \  {\rm ln}\left(  {\rm tanh} \left({i\pi\over 4} - {z \over 2}\right)\right)
\right]
 +  {\rm c.c.} \ ,  \no \\
h_2 &=  \, \left[ \alpha  \  {\rm cosh}(z)   -   \gamma \
 {\rm ln}\left({\rm tanh} \left({z\over 2}\right)\right)\right]  +  {\rm c.c.}   \  .
\end{align}
The corresponding geometry on the strip is depictedin figure \ref{stripKR}.

The asymptotic $AdS_5 \times S^5$ radii are $L_{+}^4 = L_{-}^4 = 16( \alpha^2 + 4 \alpha\gamma) \equiv L^4$ (with $\alpha'=4$) and the numbers of 5-branes are $N_{D5}=\N_{NS55}=\gamma$. Note that there is also a D3-flux $\pm N_{D3} = \gamma/2$ escaping from each 5-brane singularity. This means that in the flat brane picture we also have $\gamma/2$ D3-branes stretched between the NS5-branes and D5-branes. It would be interesting to study the case when these D3-branes are not there.
\vspace{4mm}

The limit of interest consists in having the asymptotic $AdS_5$ regions with fixed radius $L$, and to increase the number of 5-branes $\gamma$. This means $\alpha\gamma$ constant and $\gamma >> 1$. The qualitative features of this limit are captured by the less restrictive limit $\gamma >> \alpha$, which is the limit we study.

\begin{figure}
\centering
\includegraphics[height=7cm,width=10cm]{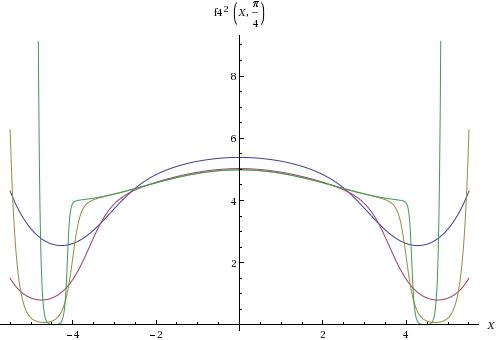}
\caption{\footnotesize  Warp factor $f_4^2$ of the $AdS_4$ metric as a function of the invariant distance $X[x]= \int_{0}^{x} 2\rho(u,\pi /4) du$ evaluated on middle line $y=\pi /4$ of the strip, for $\gamma = 1$ and $\alpha = 10^{-1},10^{-2}, 10^{-4}, 10^{-6}$. When $\alpha$ decreases the two wells become narrower and the central region gets flatter.}
\label{f4}
\end{figure}

One essential feature of the geometry \ref{hNS5D5} is the $AdS_4$ warp factor $f_4^2$ illustrated in figure \ref{f4}. In the limit $\gamma>>\alpha$ there is a central region
of size $ \sim \gamma ^{1/2} \ln(\gamma/\alpha)$, which seems to become flat, connected on both sides to two fixtures, which are
$AdS_5\times S^5$ wells with radius $L \simeq (\alpha\gamma)^{1/4}$, much smaller than the size 
of the central region.\\
The warp factor at the origin is given by $l^2 \equiv f_4^2(x=0) \sim \gamma$. The ratio $L^2/l^2$ giving the scaling of the lowest mass of the KR model is then 
\begin{align}\label{mKR}
m_{KR}^2 \sim \frac{L^2}{l^2} \sim \lp \frac{\alpha}{\gamma} \rp^{\half} \ . 
\end{align}

As we will see now, the analysis of the first graviton mode does not seem to reproduce the same scaling.

\vspace{6mm}

Solving directly the spectral problem on the strip for the graviton in the presence of the fivebrane sources is a difficult issue.
 In order to understand the mass spectrum of gravitons, we chose to use variational tools, giving bounds on the masses. The discreteness of the spectrum, despites the non-compactness of the space, is a well-known property of $AdS$ spaces that acts like a box. What is of particular interest to us is the estimation of the lowest graviton mass and the mass gap between this {\it first graviton } and the rest of the spectrum. The phenomenologically interesting situation should combine the property of localization in space on the strip and the Karch-Randall mass hierarchy, which consists in a first mass much smaller than the mass gap between modes. 

Here we only provide a numerical bound on the scaling of the first mass and comment qualitatively the situation, leaving a more serious analysis for a future work.

\bigskip
The graviton modes correspond to excitations of the 4-dimensional AdS part of the metric. The excitations on the two 2-spheres can be decomposed into Kaluza-Klein modes. Selecting the lowest mass mode means that we take the graviton to be constant on these 2-spheres.
Following reference (\cite{Bachas:2011xa}) we consider only perturbations $h_{\mu\nu}$ of the $AdS_4$ part of the metric:
\beq
ds^2 = f_4^2 \Big( \bar{g}_{\mu\nu} + h_{\mu\nu} \Big) dx^{\mu}dx^{\nu} + f_1^2 ds^2_{S_1^2} + f_2^2 ds^2_{S_2^2} + 4 \rho^2 dzd\bar{z}
\eeq
where $\bar{g}_{\mu\nu}$ is the $AdS_4$ metric with unit radius and $f_4,f_1,f_2,\rho$ are the functions of $z,\bar{z}$ introduced in \S\ref{s:localsolutions}.
We look for factorizable fluctuations with $AdS_4$ mass $m$ :
\bea
h_{\mu\nu} = h_{\mu\nu}^{[tt]}(x^{\sigma})\psi(z,\bar{z}) \\
\bar{\Box}_{AdS_4}h_{\mu\nu}^{[tt]} = \left( m^2 + \frac{2}{3} \Lambda \right) \,  h_{\mu\nu}^{[tt]}
\eea
$[tt]$ stands for ``transverse traceless'' and $\Lambda$ is the cosmological constant of $AdS_4$ ($\Lambda = -3$ for unit radius). The mass in the 4d equation of motion for $ h_{\mu\nu}^{[tt]}$ is defined so that the case $m=0$ corresponds to a reduced number of polarization  (massless graviton) \cite{Deser:2001wx}.

\smallskip

The linearized Einstein equations with this ansatz for the graviton fluctuations have been worked out in \cite{Bachas:2011xa}. It turned out to be (universally) independent of the matter fields that the theory may contain and translates into a differential equation for $\psi$ on the strip. For the precise backgrounds given by the supergravity solutions that are studied in this presentation, the differential equation is given in terms of the two harmonic functions $h_1,h_2$ by (see \cite{Bachas:2011xa} for details)
\beq
 2\frac{h_1 h_2}{W} \ \partial\bar{\partial} \, \tilde{\psi}(z,\bar{z}) = (2 + m^2)\tilde{\psi}(z,\bar{z})
\eeq
where $W = \p \bar\p  (h_1 h_2)$ and $\tilde{\psi}$ is related to the wavefunction on the strip $\psi$ by the relation $\tilde{\psi} = h_1 h_2 \, \psi$. The constant mode corresponds to $\tilde{\psi} = h_1 h_2 $ and is a local solution of this equation for $m = 0$. Because of the unbounded asymptotic regions, this constant graviton mode is not normalizable.

\smallskip

One can show that the mass of the lightest mode $m_0$ depends only on the parameter $\frac{\alpha}{\gamma}$ because it is the only parameter appearing in the wave equation.

\smallskip

For any normalizable test function $\chi$, we have the following inequality 
\footnote{ The inequality follows from the fact that $O$ is a hermitian operator of lowest eigenvalue $2+m_0^2$. The proof is done by expanding the normalizable function $\chi$ in the basis of the eigenfunctions of $O$.}
 with appropriate metric factor on the strip $\Sigma$ (with $z=x+iy$) :
\beq
2 + m_0^2 \quad \leq \quad \frac{\int\limits_{\Sigma} \chi^{\ast} O\chi |\frac{W}{h_1 h_2}| dxdy}{\int\limits_{\Sigma} \chi^{\ast}\chi |\frac{W}{h_1 h_2}| dxdy} \equiv 2 + m_{00}^2
\eeq
where $O \equiv  2\frac{h_1 h_2}{W} \ \partial\bar{\partial}$ and the volume factor on the strip is given by $|\frac{W}{h_1 h_2}|$ (\cite{Bachas:2011xa}).

\noindent We can try to use this inequality with test functions that are close to the constant mode ($\tilde{\psi} = h_1 h_2 $) in the region between the two $AdS_5$ wells and close to zero outside, as the following test functions :
\beq
\chi(x,y) = \Big( \frac{1 + \tanh(p(x+x_0))}{2} \Big) \Big( \frac{1 + \tanh(p(-x +x_0))}{2} \Big) h_1 h_2 (x,y) \nonumber
\eeq
where $p$ can be adjusted for minimization and $x_0 = \ln{\sqrt{\frac{\gamma}{\alpha}}} $ is the position of the well on the right of the strip (the left one being at position $-x_0$).

\smallskip

\begin{figure}[h]
\centering
\includegraphics[scale=0.7]{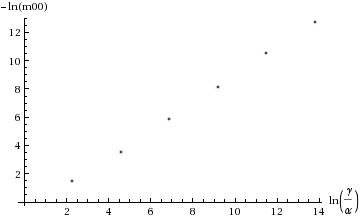}
\caption{Numerical evaluation of an upper bound $m_{00}$ for the lightest mode for different values of $\gamma / \alpha$}
\label{m00}
\end{figure}

With the help of these test functions we obtained the graphic presented in figure \ref{m00}.

\bigskip

The graphic is very close to a linear function of slope $1$, so we can infer the relation $m_{00} \propto \frac{\alpha}{\gamma}$, which means
\beq
m_0 =  O\lp\frac{\alpha}{\gamma}\rp
\eeq

The conclusion from this simple analysis is that the first graviton mode has a mass which is even smaller than the na\"ive scaling of the Karch-Randall model \ref{mKR}.

We recover the property that there is very light graviton localized in the central region. A complete analysis requires much more work. There are several types of other graviton modes : the modes that are localized in the $AdS_5$ wells will have (dimensionless) masses of order one as in the KR model, the Kaluza-Klein modes from the 2-spheres needs to be analysed, finally the presence of a nearly flat central region between the two $AdS_5$ wells indicates that there will be other graviton modes with their wavefunction localized in this region. To obtain an effective 4-dimensional Newton gravity, The masses of the modes localized in the central region should be large compared to the mass of the lowest mode.

The essential difference with the KR-model is the phenomenon of {\bf decompactification} of the central region in the limit of large $\gamma$ : the first graviton mode does not seem to be localized near the position of the 5-branes, but rather it has a wavefunction quasi-constant on the whole region between the two $AdS_5$ wells. The size of this region is proportional to $\gamma ^{1/2} \ln(\gamma/\alpha)$, so it decompactifies in the limit $\gamma >> 1$, $\alpha\gamma$ fixed. Qualitatively we expect that in this limit the localization region gets larger and the coupling of the first mode to the other graviton modes tends to reproduce 5-dimensional gravity. This indicates that the KR scenario is not reproduced.

\pagebreak 
\printindex

\pagebreak 

\bibliographystyle{JHEP}
\bibliography{mybib}

\end{document}